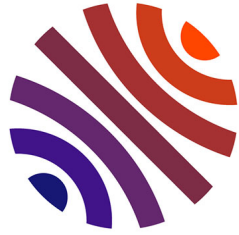

# Ultra-compact plasmonic modulator for optical inteconnects

Nicolás Mario Abadia Calvo Abadía Calvo

▶ **To cite this version:**

Nicolás Mario Abadia Calvo Abadía Calvo. Ultra-compact plasmonic modulator for optical inteconnects. Other [cond-mat.other]. Université Paris Sud - Paris XI, 2014. English. NNT : 2014PA112353 . tel-01154052

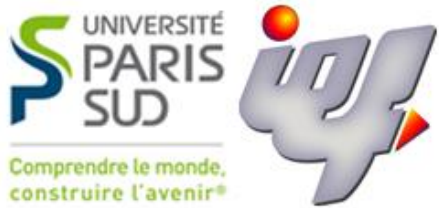
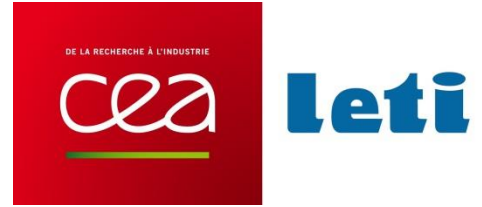

UNIVERSITE PARIS-SUD

ÉCOLE DOCTORALE : STITS

L'Institut d'Électronique Fondamentale et CEA-Leti

*DISCIPLINE* PHYSIQUE

THÈSE DE DOCTORAT

soutenue le 02/12/2014

par

### Nicolás Mario ABADÍA CALVO

# Modulateur plasmonique ultra-compact pour les interconnexions optiques sur silicium

**Directeur de thèse :**     Laurent VIVIEN     Université Paris-Sud
**Encadrante de thèse :**     Ségolène OLIVIER     Commissariat à l'Energie Atomique

<u>Composition du jury :</u>
*Président du jury :*     Renaud BACHELOT     Université de Technologie de Troyes
*Rapporteurs :*     Alain MORAND     Université de Grenoble
     Gérard COLAS DES FRANCS     Institut Carnot de Bourgogne
*Examinateurs :*     Stéphane COLLIN     Centre National de la Recherche Scientifique

## Acknowledgement

I would like to thank my supervisor Dr. Laurent Vivien and Dr. Ségolène Olivier for giving me the opportunity of doing this doctoral thesis at the Paris-Sud 11 University and CEA-Leti.

I want to mention the people with whom I worked in the MASSTOR project: Prof. Dr. Jean-Claude Weeber, Dr. Alexandre Bouhelier and Dr. Roch Espiau de Lamaëstre.

I also would like to acknowledge the collaboration of the researchers and engineers of Paris-Sud 11 University: Dr. Delphine Marris-Morini and CEA-Leti: Dr. Jean-Marc Fedeli, Dr. Sylvie Menezo, Dr. Daivid Fowler, Eng. Philippe Grosse, Dr. Jean-Michel Hartmann and Dr. Karim Hassan.

I appreciate the work done by the examiners of this work Prof. Dr. Gérard Colas des Francs and Dr. Alain Morand and the reviewers Dr. Renaud Bachelot and Dr. Stéphane Collin.

I am also thankful for the personal funding received from CEA through a CEA doctoral fellowship and SPIE under a SPIE Optics and Photonics Education Scholarship. I also would like to acknowledge ANR for funding the project MASSTOR which was part of my thesis.

I also appreciate the moments I passed with the people I met in Paris, Dijon and Grenoble. I also want to mention the efforts and support of my mother María Antonia Calvo.

## Abstract

This work aims to design a CMOS compatible, low-electrical power consumption modulator assisted by plasmons. For compactness and reduction of the electrical power consumption, electro-absorption based on the Franz-Keldysh effect in Germanium was chosen for modulation. It consists in the change of the absorption coefficient of the material near the band edge under the application of a static electric field, hence producing a direct modulation of the light intensity. The use of plasmons allows enhancing the electro-optical effect due to the high field confinement. An integrated electro-optical simulation tool was developed to design and optimize the modulator. The designed plasmonic modulator has an extinction ratio of 3.3 dB with insertion losses of 13.2 dB and electrical power consumption as low as 20 fJ/bit, i.e. the lowest electrical power consumption reported for silicon photonic modulators. In- and out-coupling to a standard silicon waveguide was also engineered by the means of an optimized Si-Ge taper, reducing the coupling losses to only 1 dB per coupler. Besides, an experimental work was carried out to try to shift the Franz-Keldysh effect, which is maximum at 1650 nm, to lower wavelength close to 1.55 µm for telecommunication applications.

## Résumé

Ce travail vise à concevoir un modulateur optique assisté par plamsons, compatible CMOS et à faible consommation électrique. L'électro-absorption, basée sur l'effet Franz-Keldysh dans le germanium, a été choisie comme principe de modulation pour réduire la taille du dispositif et la consommation d'énergie électrique associée. L'effet Franz-Keldysh se traduit par un changement du coefficient d'absorption du matériau près du bord de bande sous l'application d'un champ électrique statique, d'où la production d'une modulation directe de l'intensité lumineuse. L'utilisation de plasmons permet en principe d'augmenter l'effet électro-optique en raison du fort confinement du mode optique. Un outil de simulation électro-optique intégré a été développé pour concevoir et optimiser le modulateur. Le modulateur plasmonique proposé a un taux d'extinction de 3.3 dB avec des pertes d'insertion de 11.2 dB et une consommation électrique de seulement 20 fJ/bit, soit la plus faible consommation électrique décrite pour les modulateurs photoniques sur silicium. Le couplage du modulateur à un guide silicium standard en entrée et en sortie a également été optimisé par l'introduction d'un adaptateur de mode Si-Ge optimisé, réduisant les pertes de couplage à seulement 1 dB par coupleur. Par ailleurs, un travail expérimental a été effectué pour tenter de déplacer l'effet Franz-Keldysh, maximum à 1650 nm, à de plus faibles longueurs d'onde proches de 1.55 µm pour des applications aux télécommunications optiques.

# Content

# List of Figures

# List of Tables

# 1. Introduction

## 1.1 Context

In order to increase the operational frequency of the transistors, the semiconductor industry downscaled the size of the devices. There is an empirical law, called the Moore's law [1], saying that for a constant fabrication cost, the number of transistors in the same surface of an electronic integrated circuit doubles every two years. Since the invention of the transistor [2], this empirical law seems to be true [3] but nowadays it is reaching its limit [4]. The number of transistors in a chip versus the year of invention is represented in Figure 1. The Moore's law is satisfied since the first Intel 4004 [5] until the current Intel Core i7 [6]. This allows a reduction in the fabrication cost due to the mass production of the chips in the foundries.

There are different physical effects that happen in the downscaled transistors and that limit the performances of modern circuits [4]. The first one is that when the transistor is downscaled, the electrical wires within the circuit are also downscaled. This produces an increment in the resistance of such a wire. The increased resistance combined with the capacitance seen by the wire produces an increase of the RC delay [7] as well as the increase of heat dissipation due to the increment of the resistivity. A big delay causes a problem in the signal that propagates through a chip, for example, the clock signal on synchronous integrated circuits. Furthermore, the wires are closer and closer to each other when the electrical interconnect wires are downscaled. This produces an increment of the parasitic capacitance between wires which causes signal interference. This downgrades the performances. It means, this capacitance plays a role in the crosstalk of the wires. As a consequence of this, there is interference in one wire due to the proximity of the adjacent ones.

Furthermore, the cited delay reduces the operational frequency of the chip because the computational delay is equal to the propagation delay through the chip. This problem is known as the interconnect bottleneck and it is a major problem in nowadays high performance computer integrated circuits.

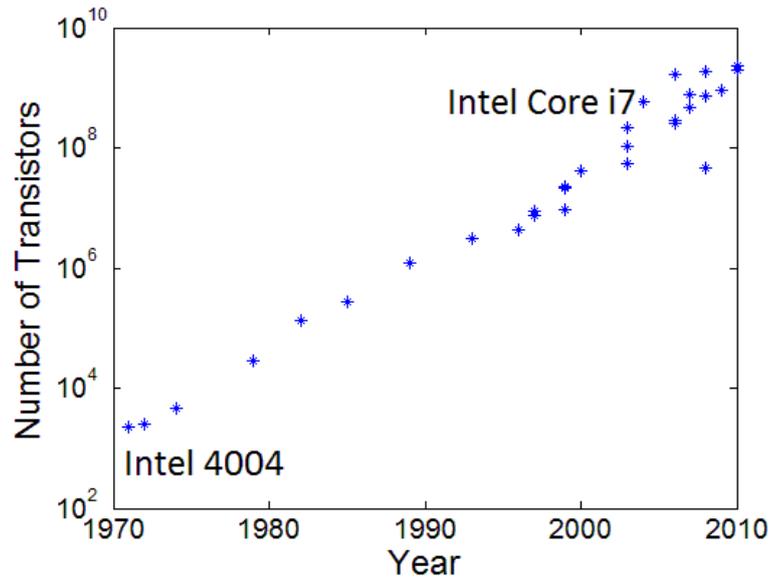

**Figure 1: Number of transistors versus the year of appearance. It follows the Moore's law** [8]

Regarding the problem of increased resistivity due to the shrink of the interconnect wires it can be reduced using copper (Cu) instead of the well-established aluminum (Al). Cu has half of the resistivity of Al. Furthermore, the employment of low-κ dielectrics (E.g.: porous silica materials [9], SiOF, SiOC, polymers, etc.) is used to reduce the parasitic capacitances between wires.

The main problem of using Cu as a material for the intra-chip interconnects is that it is not a CMOS compatible material. Cu is a quick diffusion material into Si [10]. This fact can create a short circuit in the transistor between the source and the drain. Nevertheless, to avoid this, insulators barriers like $SiO_2$, $Si_3N_4$ or $HfO_2$ are used between the metal and the semiconductor. This invention allows the Moore's law to still remain a little bit more time [8]. Nevertheless, a long term solution needs to be addressed to circumvent the interconnect bottleneck.

One of the long term solutions to the problem of the interconnect bottleneck could be achieved by the use of silicon photonics [11], [12], [13], [14] whose main applications are to create optical links [15] to connect the different parts of the electronic integrated circuit. In this case the information is carried by photons instead of electrons; this leads to a reduction in the energy consumption as well as an increase in the operational frequency of the link [16]. For this, a set of silicon devices are needed. Namely: emitters, modulators, waveguides and detectors.

Another advantage of Si is that it is a widely available material and it has a low cost. It possesses a wide window with low optical losses (the main reduced losses are around 1.3 and 1.55 μm), a high optical damage threshold (the supported intensity is around 3 GW/cm² and depending on the doping and the fabrication processes it can reach 5 GW/cm²) and good thermal conductivity. The good thermal conductivity is important to well dissipate the heat from the integrated circuit.

Nowadays there are many optical communication systems [17], [18] that are commercially available and that are substituting electrical links in long distances (from 1 meter to many kilometers). This example of long-haul optical links overcomes the drawback of electrical ones. They can offer better bandwidth and fewer losses. Furthermore such long-haul optical interconnects are lighter with respect to the metal ones. Another advantage is that they are immune to electromagnetic noise.

One recent trend of silicon photonics is to carry the optical communications from the long-haul distances to intra-chip communications. It is believed that optical interconnects acting as a bus in a chip can overcome the limitations of metal interconnects and increase the performance.

Furthermore, another of the most important arguments of the technology is that the main used material is Si which is a well-known material in the semiconductor industry. Being CMOS compatible, it allows using the well-established foundries [19] and fabrication methods. This enables a massive fabrication of devices with the consecutive reduction in cost. One of the milestones of silicon photonics is to integrate in the same chip electronic and photonic devices.

# 1.2    Silicon Photonics

Probably Si is one of the most studied materials in the history. In this case silicon photonics [20] can take advantage of the well understood semiconductor Si.  Furthermore, manufacturer machines in the foundries are well developing to fabricate small patterns of few nanometers. Taking into account that silicon photonics' sizes are bigger than electronic features this allow a well precision in the fabrication process. Furthermore, the fact that the foundries are well established allows the mass production of photonic devices reducing the global cost of the product.

The use of the silicon-on-insulator (SOI) technology [21], [22] facilitates the integration with the electronic fabrication techniques and processes. It consists in a crystalline Si on silicon dioxide ($SiO_2$). A SOI waveguide is represented in Figure 2. This technology has the advantage that crystalline Si can be put over $SiO_2$ using the SmartCut® technique [23]. With this method, the Si is crystalline. The main use of this technology is to fabricate high contrast photonic waveguides. Such waveguides confine the light in small areas [24], [25], [26] due to the large refractive index difference between Si and $SiO_2$. This allows the fabrication of sharp bends and devices. Due to this fact the modal area can be as small as 400 nm for the width and 200-300 nm for the height.

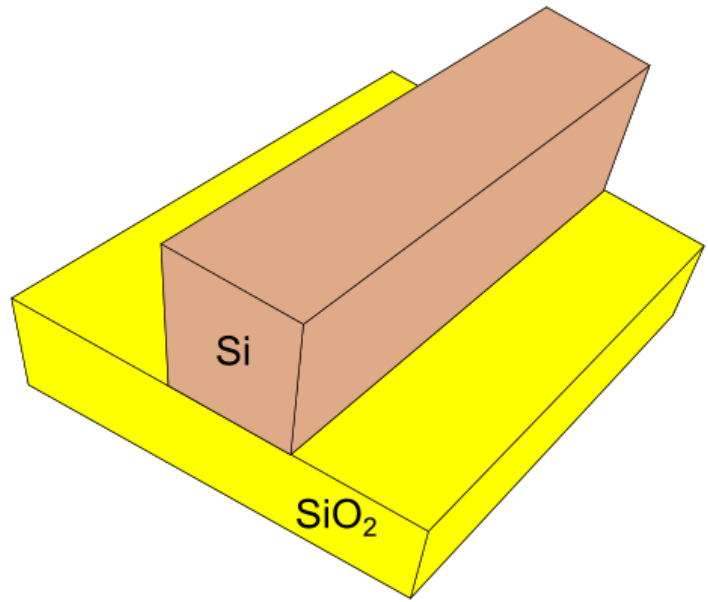

**Figure 2: Example of a strip waveguide with SOI technology**

One of the main disadvantages of SOI waveguides, and in optics in general, is that they are limited in size reduction by the diffraction limit of light. The field of plasmonics tries to overcome the diffraction limit of light by using metals in the active region of the devices. This allows the design of smaller devices. It is expected that photonics devices reduces their dimensions enough to be integrated into an electronic integrated circuit. For doing this, the size of photonic devices and electronic ones are needed to be very close.

For setting an intra-chip optical interconnects, several elements are needed. Mainly, a source to produce the light within the integrated circuit, a modulator to encode the continuous wave (CW) light, a multiplexer to combine several wavelengths in one beam, waveguides to propagate the light from one point of the integrated circuit to another one, a demultiplexer to separate the wavelength and photodetectors to transform the information carried by the optical signal to the electrical one.

A schematic view of an optical link with its elements is presented in Figure 3. There are well-known several passive devices like waveguides but recently the main effort is taken to design sources [27], modulators [28] and photodetectors [29]. The first commercial silicon photonic-based optical data interconnect with integrated lasers is presented in [18] and [30].

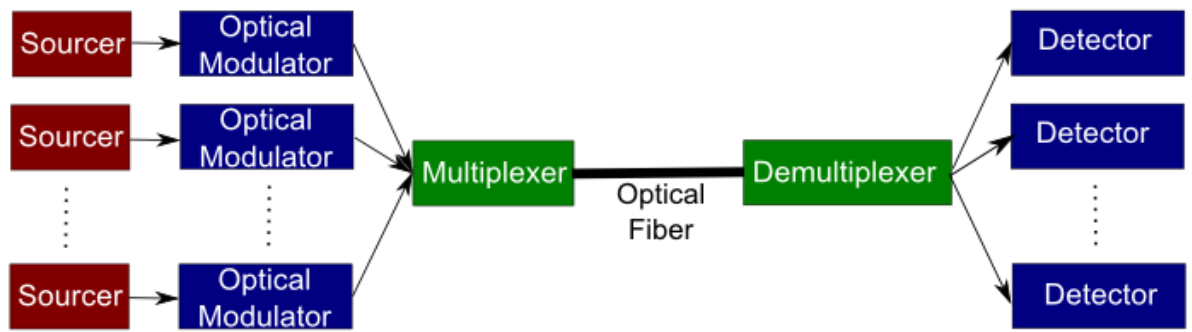

**Figure 3: Schematic of an optical interconnect**

One important component to achieve an optical interconnect is the optical modulator [28] which converts the electrical signal that carries the information into an optical one. This device is the subject of this thesis.

It is worth to mention that Si is a centro-symmetric material making it difficult to design a modulator due to the absence of the Pockel's effect. Furthermore, Si possesses an indirect band-gap. This is a main difficulty in the design of optical sources since the electron excitation from the valence band to the conduction one requires the presence of a phonon to conserve energy and momentum. The fact that Si is transparent at the wavelength of 1.55 µm does not allow the design of photodetectors at this wavelength. This is why other materials are being introduced in the field of silicon photonics. One of those examples is the use of Ge.

But even optical interconnects have its own challenges [31]. Some comparison between optical and electrical interconnects were found in [32].

## 1.3    Objective and Dissertation Outline

In the framework of this work, I want to develop a low power consumption modulator based on an electro-absorption effect. It will use the Franz-Keldysh effect (FKE) in germanium as the active principle of the deivce. To further reduce its power consumption we use a plasmonic structure as the optical waveguide to concentrate the light into the germanium core and reduce the size mismatch between photonic and electronic devices. The CMOS compatiblility is another main goal. Regarding the power consumption of the device we want that it is below 50 fJ/bit as stated in several roadmaps [4], [33] for future optical links.

Taking this brief summary of the device we propose, we introduce the outline of the chapters of this thesis in which we explain how we will design it, the rational of why electro-absorption modulator, why the FKE, why plasmonic structures, etc.

In the second chapter we introduce the basics and the state-of-the-art of silicon photonic modulators. For this, we do a small introduction to the device and its main characteristics. We explain the different effects that are used to perform the modulation and we do a state-of-the-art of the different photonic modulators that use the FKE. We also present all the silicon photonic plasmonic modulators. Since we are interested in low power consumption we explain at the end of the chapter the formulas used to calculate the energy consumption of modulators.

In the third chapter we introduce the field of plasmonics. We explain its advantages and disadvantages. We also discuss on CMOS compatibility and the metals that can be used. The metal-semiconductor (MS) and metal-insulator-semiconductor (MIS) waveguides are also explained.

We also explain the rationale of using a plasmonics waveguide which we advance here. Since the appearance of integrated optics [34], optical devices tried to reduced their dimension and footprint in order to be integrated into an electronic chip. Nevertheless there is a miniaturization limit of photonic devices due to the diffraction limit of light. Nowadays there is a dimensional mismatch between electrical and photonic integrated components. This mismatch difference hinders the integration of both electronic and photonics devices in the same wafer. This mismatch in size is due to the fact that photonic devices are limited in size by the diffraction limit of light. It is demonstrated that using metals the diffraction limit of light can be overcome and the dimensions of the devices can be reduced. This is why in the optical structure of our modulator we will use a plasmonic waveguide.

Regarding the power consumption, designing the modulator using metals to propagate light at the interface will allow a reduction of the dimensions of the device with respect to the photonic ones. The reduction of the device will also decrease the energy consumption of this. Furthermore, the reduction in the dimensions of the device will increase the interaction of the effects (FKE, carrier dispersion plasma effect, etc.) in Ge due to a bigger intensity of the electromagnetic fields within the material. Additionally, there is a trade-off between the losses of the propagating modes in a plasmonic waveguide and the spatial concentration of the electromagnetic field. It means, the smaller the spatial concentration of the electromagnetic field into a plasmonic waveguide, the bigger the propagation losses. Consequently, in our design, we want to have a small footprint but we also need to take care about the propagation losses of the modulator.

In the fourth chapter we explain the physics behind the FKE. We also derive mathematically the formulas to model the FKE. We present a simply generalized model and a more complete one. This model will be useful to calculate the main benchmarks of the device using the integrated electro-optic simulator that we introduce in the fifth chapter.

Regarding the low power consumption that we want to achieve we advance that, we want to design an electro-absorption modulator rather than an electro-refraction one. The main advantages of both electro-absorption over electro-refraction effects are that it allows the design of smaller devices (small footprint) and consequently it will have lower power consumption. This is why as a first glance we will discard electro-refraction over electro-absorption. The active principle of the modulator we want to design is the FKE. This effect is present in both silicon and germanium. Nevertheless the effect is stronger in germanium rather than in silicon. It is worth to mention that germanium can be integrated well in a silicon photonic fabrication process. Furthermore, the use of the FKE does not induce any limitation in the speed of the device since the FKE is an instantaneous effect (it is in the sub-picosecond regime [35]). The FKE will be analyzed and presented in the fourth chapter.

In chapter five we present the design of the plasmonic modulator which is the main objective of this thesis. We establish the benchmarks of the modulator that we want to achieve. We also explain the numerical tools that we are going to use to model the device and we present many structures to perform the modulation. We analyze the performance of each structure and we derive the main characteristics of the modulator mainly: the extinction ratio, the propagation losses, the energy consumption, etc.

For this we developed an opto-electronic integrated simulator. We use a commercial electrical simulator called ISE-dessis to calculate the static electric field distribution in the proposed structure of the modulator. With this information we apply a Franz-Keldysh effect model to calculate the change in the absorption of the material (in our case germanium). Knowing the absorption of the material we use a finite-difference method (FDM) mode solver to know how the effective losses of the mode are changed due to the mentioned Franz-Keldysh effect. With this tool we can obtain the main figure of merit of the modulator, it means, we can know the extinction ratio, the propagation losses, the bandwidth and the power consumption.

Using this tool we simulated many structures to act as a modulator and we selected the best performance. Knowing the good structure we optimized the device in order to increase the extinction ratio, decrease the propagation losses and reduce the power consumption.

The initial idea of the device is also discussed in the fifth chapter. We advance that we will use a plasmonic waveguide that can guide the light in the interface between the metal and a semiconductor. It is called a metal-semiconductor (MS) plasmonic waveguide. Placing germanium in the semiconductor part of the MS waveguide we will be able to exploit the FKE to produce the modulation of the plasmon. We want that a significant part of the plasmonic mode is in germanium. The germanium will be deposited in the core of the waveguide. We also want that the waveguide is connected to two electrodes in order to induce a static electric field in the germanium in order to change the absorption of the material and modulate the plasmon supported by the MS plasmonic waveguide. One electrode can be placed in the metal of the MS waveguide and the other below the semiconductor of the MS structure using a doped region. Inducing an electric field in the germanium will allow changing the optical absorption of the material. Since the plasmonic mode is also present in the germanium this will change the effective losses of such a mode, hence, producing the modulation of the plasmon. We will show that using a MS waveguide in the core of the modulator leads to high propagation losses of the plasmon. Instead of this we can use a traditional metal-insulator-semiconductor (MIS) waveguide. This waveguide, due to the use of an insulator layer, reduces the propagation losses of the plasmon.

Finally, in chapter six we design the coupling to the optimized structure designed in chapter five from a standard Si rib waveguide. The objective is to excite the proper mode of the modulator from the mentioned Si rib waveguide. For this we use 3D Finite-difference time-domain (FDTD) simulations. We used the commercial tool called Lumerical®. In this chapter we analyzed many different coupling structures (butt-coupling, tapers, etc.), we optimized them to increase the coupling efficiency into the modulator and finally we select the best one to place it in the device designed in chapter five.

In the last chapter we present the conclusions, the outline and the future perspective of this work.

# 2 Silicon Optical Modulation

In this chapter we introduce the main characteristics of Si optical modulators. We explain the main effect for modulation in silicon photonics. We also do a state-of-the-art of photonic FKE modulators. Plasmonic modulators are also explained. Finally, we present the calculation of the electrical power consumption of these devices.

## 2.1    Introduction to Optical Modulators

Nowadays, the main application of silicon photonics is to try to substitute electrical interconnects by optical ones. The main objective is to increase the bandwidth and reduce the energy consumption. Furthermore, the main components of an optical interconnects are the light source, the modulator, the waveguide and the photodetector. One of the most important elements is the modulator. It can determine the bandwidth and it is the device which consumes a big portion of the power dissipated in the link. In this thesis we aim to design a compact and low-energy consumption modulator.

The modulator is a device which transforms an electrical signal that carries information into an optical one that also carries the same encoded information. In this case photons are used to transmit the information through a waveguide instead of electrons. For this, the modulator changes the phase, the intensity, the frequency or the polarization of a continues-wave (CW) light beam. The structure of an optical modulator is represented in Figure 4. In this example the information is encoded in the intensity of the light beam.

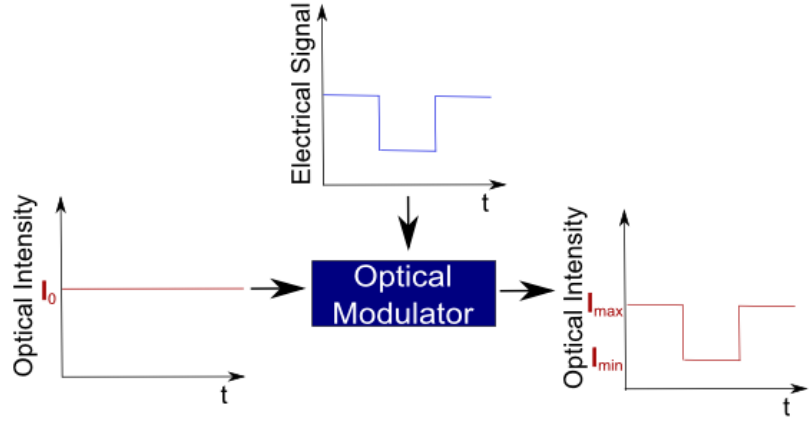

**Figure 4: Schematic view of an optical modulator principle. In this case the information is encoded in the intensity of the light beam using an electrical signal**

The main characteristics of an optical modulator are the insertion losses (IL), the extinction ratio (ER), the power consumption, the operational frequency or bandwidth (BW), the modulation voltage ($V_{DD}$), the device footprint and the working spectrum. A good modulator has low insertion losses, a high extinction ratio, consumes low energy and has a small footprint. We will specify the typical values and the objective for our modulator once we presented the state-of-the-art at the end of this chapter.

The insertion losses are the optical losses from the input until the output of the device. They consist in both coupling and propagation losses through the device. Equation 1 quantifies this magnitude,

$$IL = 10\log\left(\frac{I_{max}}{I_0}\right)$$

Equation 1

Where $I_{max}$ is the maximum intensity of the encoded light beam at the output of the modulator and $I_0$ is the input intensity of the encoded light beam. It is represented in Figure 4.

The extinction ratio is related to the difference between the maximum intensity $I_{max}$ and the minimum intensity $I_{min}$ of the encoded beam. It is represented in the following formula:

$$ER = 10\log\left(\frac{I_{max}}{I_{min}}\right)$$

Equation 2

Where $I_{max}$ and $I_{min}$ are the maximum and minimum intensities of the encoded light beam. It is represented in Figure 4.

The power consumption is normally the electrical power consumption that the modulator employs to encode the light beam. It is given by the effort done by charging and discharging the electrical capacitance of the device. The electrical power consumption in energy per bit is given by,

$$\Delta E_{bit} = \frac{1}{4}CV_{DD}^2$$

Equation 3

Where $\Delta E_{bit}$ is the energy per bit, C is the intrinsic capacitance of the device and $V_{DD}$ is the driving voltage of the modulator. A deduction of Equation 3 is given at the end of the chapter.

The optical power consumption is not normally given since it depends on the intensity $I_0$ of the input light beam. Nevertheless, the insertion loss is a good measure of the optical power consumption since it is proportional to it.

The operational frequency of the modulators measures the bandwidth of the modulator. It is a measure of the transmission velocity of the information.

The operational voltage is the voltage that the modulator requires in order to perform the modulation. The device footprint is given by the area that the modulator needs in the wafer. Finally, the working spectrum is the wavelength range within which the modulator can work.

In the following section we will describe the different effects that can be used in Si to modulate a CW beam of light from an electrical signal.

## 2.2     Optical Modulation Effects

Depending on the physical effect that the modulator uses to encode the information it can be classified as electro-absorption or electro-refraction device. In the case of electro-absorption devices the effect used changes the optical absorption coefficient of the material and consequently modulates the intensity of the optical light beam. On the other hand, electro-refraction devices change the refractive index of the material. In this case the phase of the optical CW beam is modified. The use of structures like ring resonators (RR) or Mach-Zenhder interferometers (MZI) can transform the change in phase into the change in the optical intensity.

In the following sections, we explain the different effects used in silicon photonics to perform the modulation. They are divided on electro-refraction effects and electro-absorption effects.

### 2.2.1 Electro-Refraction Effects

In this section we explain the electro-refraction effect used in silicon photonics. Those effects change the refractive index of the material in order to change the phase of the optical propagating signal. The main effects are the thermo-optic effect, the Pockels and Kerr effects and the plasma dispersion effect.

## 2.2.1.1  Thermo-Optic Effect

In the thermo-optic effect the application of heat to the material is used to change the refractive index of it. Neither an electric field nor the injections of carriers are used in this effect. The thermo-optic coefficient of Si [36] is:

$$\frac{dn}{dT} = \frac{1.86 \times 10^{-4}}{K}$$                                          Equation 4

A refractive index change of the order of $10^{-3}$ happens when a temperature variation of 10°C is applied to Si. There are some problems like controlling the changes of temperature uniformly [37] in all the volume of the material.

Another issue is that the operational frequency of the device is not big since heating and cooling down the material is a slow process. The operational frequency of thermo-optic devices is around 100 kHz. Some modulators [36] and switches have been proposed [38].

## 2.2.1.2  Pockels and Kerr Effect

When a static electric field is applied in a medium it may change its refractive index of the material. This effect is called the Pockels [39] effect when the change is proportional to the static electric field. If the change is quadratic then it is called the Kerr effect or the Quadratic Electro-Optic effect (QEO effect) [40]. The change in the refractive index as a function of the applied static electric field is given by,

$$\Delta \left( \frac{1}{n_r^2} \right) = rE + sE^2 + \cdots$$                                          Equation 5

In the previous equation $n_r$ represents the refractive index of the material, E is the electric field, r is the electro-optic Pockels coefficient and s is the electro-optic Kerr coefficient. The Pockels and the Kerr effects change the polarizability of the material; consequently, the change in the refractive index depends on the polarization of the electric field with respect to the optical axes of the material.

Pockels effect is present in crystals without inversion symmetry. Some materials which exhibit good Pockels effect are: indium phosphide (InP), gallium arsenide (GaAs) or lithium niobate (LiNbO$_3$). On the other hand, the Kerr effect is present in any material.

As explained before, the Pockels effect is not present in Si because it is a centro-symmetric material. Recent research showed that straining the Si, the centro-symmetry is broken and consequently a Pockels effect can be measured [41]. A modulator is proposed in [42] using this technique. Normally, to break the symmetry, a silicon nitride (Si$_3$N$_4$) layer is deposited over the Si. This causes a strain in the material and consequently, the appearance of the effect on it.

Regarding the Kerr effect, it can be measured in Si but it is a weak effect [43]. The breakdown static electric field of Si must be exceeded in order to obtain a significant change in the refractive index of the material. The breakdown static electric field of Si is around 3x10^5 V/cm. The range of the change in the refractive index at the wavelength of 1.3 μm is around Δn=10^{-3}-10^{-4} [43]. Where, Δn is the change in the refractive index of the material. In conclusion, it is not a good effect to design an optical modulator.

### 2.2.1.3  Plasma Dispersion Effect

When the concentration of carriers is changed into a semiconductor it modifies both the refractive index and the optical losses of the material due to the Kramers–Kronig relations. This is called plasma dispersion effect or free carrier dispersion effect. When the number of carriers is increased in a semiconductor like Si, the refractive index decreases while the optical losses of the material are increased. Inversely, if the concentration of carriers decreases we have the opposite effect.

The theoretical change in the refractive index and in the optical losses against the concentration of electrons and holes is giving by the following theoretical equations:

$$\Delta n = \frac{-e^2 \lambda_0^2}{8\pi^2 c^2 \varepsilon_0 n}\left(\frac{N_e}{m_{ce}^*} + \frac{N_h}{m_{ch}^*}\right)$$

Equation 6

$$\Delta \alpha = \frac{e^3 \lambda_0^2}{4\pi^2 c^3 \varepsilon_0 n}\left(\frac{N_e}{\mu_e (m_{ce}^*)^2} + \frac{N_h}{\mu_h (m_{ch}^*)^2}\right)$$

Equation 7

Where Δn is the change in the refractive index of the medium, Δα is the change in the optical absorption of the medium, e is the electric charge of the electron, $\lambda_0$ is the wavelength of the incoming light, c is the velocity of the light in vacuum, $\varepsilon_0$ is the electrical permittivity of vacuum, n is the refractive index of the medium, $N_e$ and $N_h$ are the concentrations of electrons and holes in the medium, $\mu_e$ and $\mu_h$ are the mobility of electrons and holes and $m_{ce}^*$ and $m_{ch}^*$ are the reduced effective mass of electrons and holes.

There is also an experimental study of the change of the refractive index and the optical absorption of the material in [43]. The authors studied those changes for many different carrier concentrations at the wavelengths 1.3 μm and 1.55 μm. The results are in good agreement with the theoretical formulas given by Equation 6 and Equation 7. The derived empirical formulas at $\lambda_0$=1.55 μm are:

$$\Delta n = \Delta n_e + \Delta n_h$$
$$= -[6.2x10^{-22}\Delta N_e$$
$$+ 6.0x10^{-18}(\Delta N_h)^{0.8}]$$

Equation 8

$$\Delta\alpha = \Delta\alpha_e + \Delta\alpha_h$$
$$= 6.0x10^{-18}\Delta N_e$$
$$+ 4.0x10^{-18}\Delta N_h$$

<div align="right">Equation 9</div>

Where $\Delta n_e$ and $\Delta n_h$ are the changes in the refractive index caused by free electrons and free holes that are present in the material. $\Delta\alpha_e$ and $\Delta\alpha_h$ are the changes in the optical absorption caused by free electrons and free holes that are present in the medium. It is expressed in cm$^{-1}$. From those equations we can see that for a carrier concentration of 5x10$^{17}$ cm$^{-3}$ we can obtain a change of $\Delta n$ in the order of 10$^{-3}$ for 1.3 μm.

From Equation 8 and Equation 9 we can see that holes are more efficient than electrons in changing the refractive index of the material $\Delta n$ for a free carrier concentration $\Delta N<10^{19}$ cm$^{-3}$. This effect can be achieved by carrier injection, depletion or accumulation [28]. Normally this effect is used to change the phase of light into a waveguide. To change the amplitude of the incoming light beam structures like MZI and RR are used.

## 2.2.2  Electro-absorption Effects

### 2.2.2.1  The Franz-Keldysh Effect

The Franz-Keldysh effect (FKE) [44], [45] is the change of the optical material absorption for wavelengths close to the direct energy bandgap (E.g.: Si and Ge) when an external static electric field is applied to the bulk material. The energy diagrams used to explain the FKE are represented in Figure 5. Under no applied static electric field into the material, the conduction and valence band are like in Figure 5 (a). Then a photon of energy bigger than $E_g$ (energy bandgap of the material) can excite an electron from the valence band to the conduction band. It is represented in the Figure 5 (a). Nevertheless, when an electric field is applied into the material both the valence and the conduction bands are tilted like in Figure 5 (b). In this case the wave-function of the electrons in the valence band can penetrate into the forbidden band by tunneling. Due to this tunneling, a photon with less energy than the bandgap of the material $E_g$ can excite an electron from the valence band to the conduction one producing the absorption of light. It is represented in Figure 5 (b). This increases the optical absorption of the material just below the energy bandgap $E_g$. This is known as the FKE. Due to the Kramers–Kronig relation when the absorption of the material is changed then the refractive index of it is also modified. So, the FKE produce electro-absorption and electro-refraction.

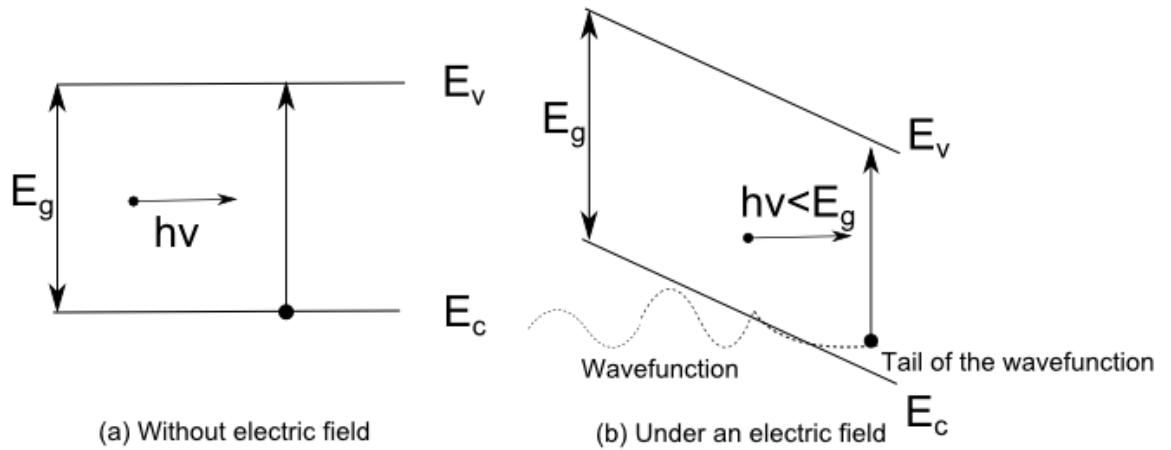

**Figure 5: Energy diagram of the FKE. In (a) there is not a static electric field while in (b) a static electric field is applied**

The FKE is both present in Si and Ge. It is bigger in Ge than in Si due to the sharper band-edge in Ge. Near the band-edge of Si there is absorption due to phonons that softs the band-edge of Si. The FKE was measured experimentally in Si in [43]. They measured the optical absorption of the material due to the FKE and using the Kramers–Kronig relation they calculate the electro-refraction. Note that this electro-refraction is cause by the FKE and not due to plasma dispersion effects.

As a conclusion, the FKE in Ge is stronger than in Si for the reasons explained in the previous paragraph. Several photonic modulators were proposed using the FKE in Ge. In this thesis we will use the FKE to produce the modulation of a plasmon. This effect will be detailed in chapter four.

The FKE also depends on the strain that the Ge has. The first observation of FKE in Ge over Si is reported in [46]. The structure studied is a PIN junction made of Ge over a Si wafer. When the Ge is deposited over the Si there is a biaxial stress caused by the difference in the thermal expansion coefficients between Ge and Si during deposition of Ge over the Si at a high temperature (750 °C) the Ge experiences a compressive strain. Nevertheless, when it is cooled down it presents a biaxial tensile strain. There is a thermal process that allows the creation of a tensile strain around 0.20 % [47]. This produces the narrowing of the direct bandgap of Ge. Furthermore, due to the strain there is a degeneration of the heavy and light holes in the valence band. Without strain the valence band of heavy and light holes is degenerated. The tensile strain also reduces the energy difference between the valence and conduction bands. This shifts the band-edge (and consequently the FKE) to longer wavelengths. The energy band of Ge is represented in the following Figure 6,

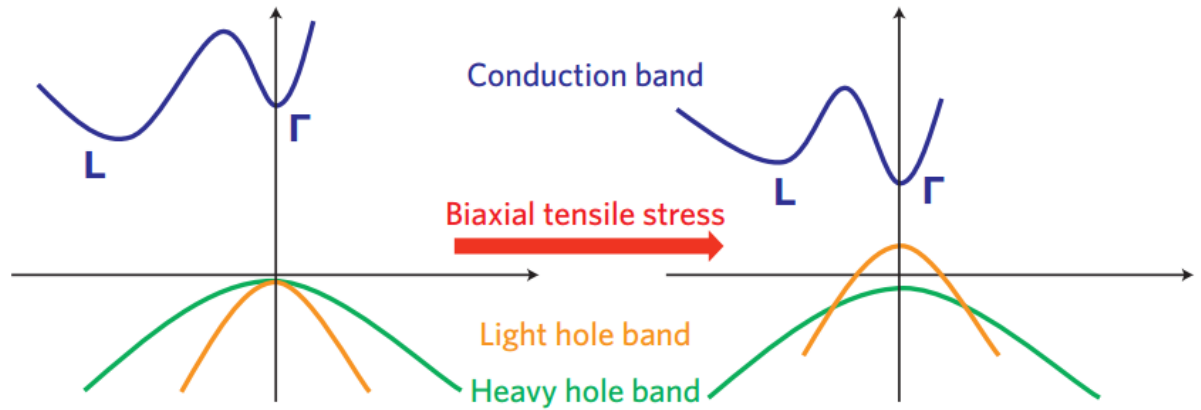

**Figure 6: Energy band structure of Ge and the influence of biaxial tensile strain** [29]

In the left energy diagram of Figure 6 we see the direct band-gap in the Γ point. The indirect band-gap occurs at the point labeled L. Furthermore, the indirect band-gap L is less than the direct band-gap Γ. It is worth noting that in the valence band the light hole band and the heavy hole band are non-degenerated in the strained case.

Furthermore, in this strained case the direct band-gap Γ is now smaller than the indirect band-gap L for enough biaxial tensile strain. It will be interesting to know how the gap energy of the direct band-gap $E_g(\Gamma)$ and the indirect one $E_g(L)$ changes with the biaxial tensile strain of the material. It is represented in Figure 7,

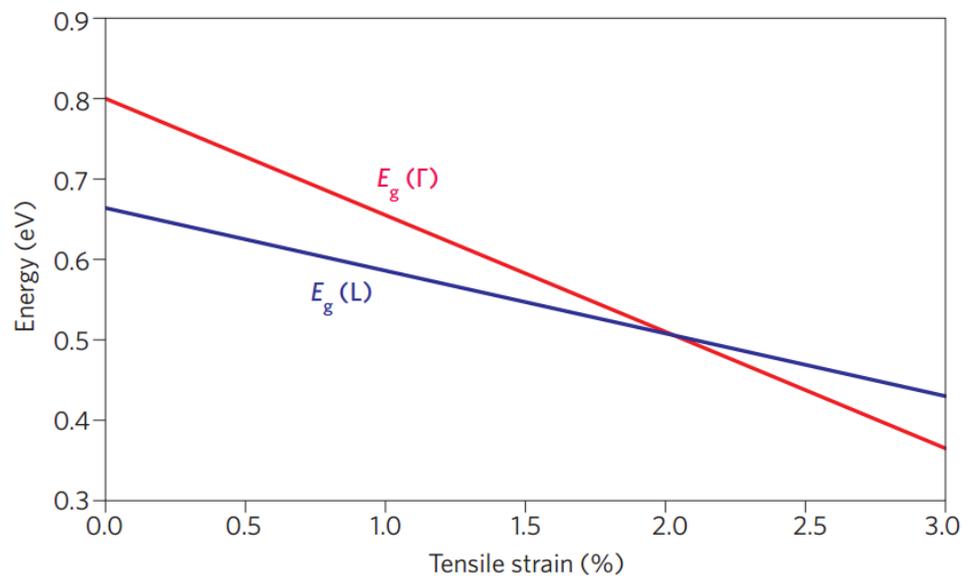

**Figure 7: Influence of the biaxial tensile strain in the direct and indirect band-gaps in Ge** [29]

From Figure 7 we see the influence of the biaxial tensile strain on the direct energy gaps $E_g(\Gamma)$ and the indirect energy gap $E_g(L)$. We will model the FKE in a Ge which is strain 0.2%. Consequently, the energy gap of the direct band-gap $E_g(\Gamma)$ will be larger than the energy gap of the indirect one $E_g(L)$. As we explained using Figure 6 we see that the direct energy gap $E_g(\Gamma)$ is decreasing as the biaxial tensile strain increases. Using the values of Figure 6 we can predict the FKE for different strains using the FKE model that we will describe in chapter five.

The strain electro-optical effects of the Ge-on-Si PIN junctions are proposed to fabricate a modulator in [46]. Reverse biasing the Ge PIN structure produces a static electric field in the intrinsic region. Due to the FKE this changes the absorption coefficient and the real part of the refractive index of the intrinsic region. These changes can be used to fabricate an electro-absorption or an electro-refraction modulator. The performances derived from the former has a figure of merit of $\Delta\alpha/\alpha$ of 3.03, while the later can have a $\Delta n/F$ of 280 pm/V [46]. Where $\Delta\alpha$ is the change of the absorption coefficients due to the static electric field, $\alpha$ are the losses, $\Delta n$ is the change of the refractive index due to the static electric field and F is the mentioned static electric field. A configuration of a 100 µm long photonic device has an extinction ratio of 7.5 dB with 2.4 dB insertion losses according to the predictions in [46]. Regarding the $\Delta n/F$ of the electro-refractive modulator, the parameter is better than the one of the bulk Ge ($\Delta n/F$ =160 pm/V), InP ($\Delta n/F$=240 pm/V) and LiNbO$_3$ ($\Delta n/F$=164 pm/V). The strained Ge devices work at the wavelength of 1647 nm where the FKE was found to be maximum [46].

In this thesis we are going to design a FKE plasmonic modulator. As stated in [46] we will work at the wavelength of 1647 nm in order to exploit the maximum effect of FKE. Furthermore, we will use the values of [46] to model the FKE in the simulation tool that we will describe in chapter five. We will propose an electro-absorption plasmonic modulator.

## 2.2.2.2   The Quantum Confined Stark Effect

The Quantum Confined Stark effect (QCSE) [48] is an electro-absorption effect like the FKE described in the previous section. It changes the material absorption near the bandgap of the semiconductor upon application of a static electric field. The main difference is that the QCSE happens in quantum wells (QW) while the FKE happens in bulk material. A QW is a structure in which a thin film (5-10 nm) of a material of lower bandgap is put in the middle of another material of larger bandgap.

An example, QW can be formed by a stack of SiGe/Ge/SiGe or InGaN/GaN/InGaN. The total thickness of one period is around 20 nm. The band-energy structure of a QW is represented in Figure 8 (a).

When the QW is at equilibrium (in the absence of a static electric field) the wave-function is symmetrical for both electrons and holes. Furthermore, due to the quantum confinement electrons and holes are linked by Coulomb forces and form excitons. The absorption of the exitons occurs at photon energy smaller than the bandgap energy of the QW material (as example Ge or GaN). This energy is given by:

$$\hbar\omega = E_g + E_{el} - E_{hl} - E_{ex}$$                    Equation 10

$E_g$ is the bandgap energy of the material in the middle of the QW. $E_{el}$ and $E_{hl}$ are the electron and holes sub-band energy and $E_{ex}$ is the binding energy of the exciton.

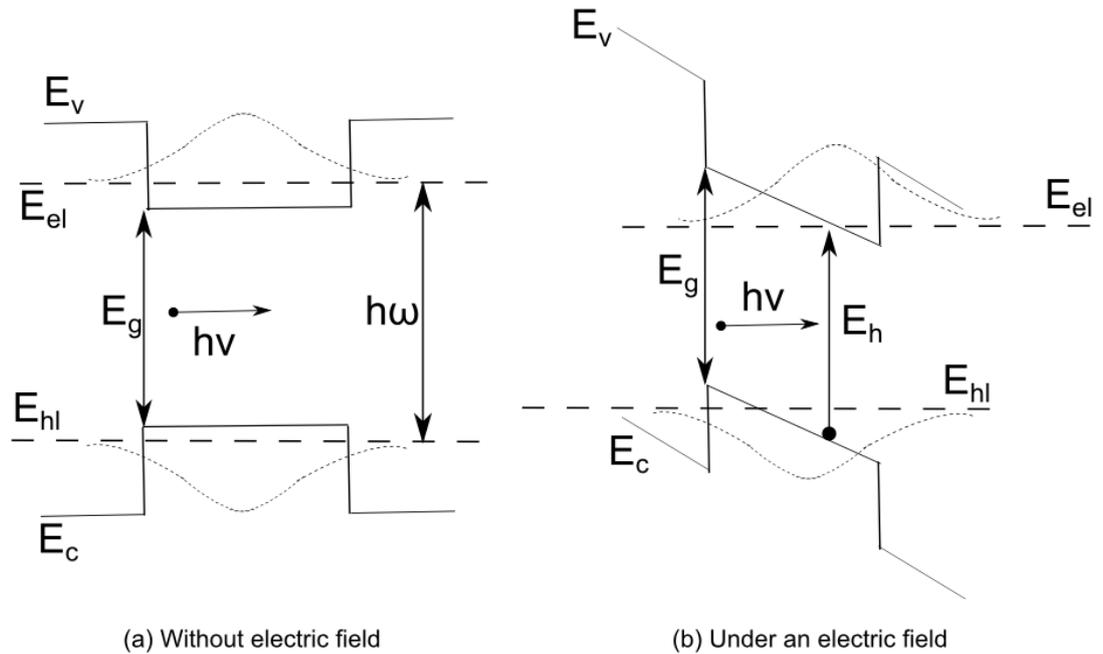

(a) Without electric field  (b) Under an electric field

**Figure 8: Energy diagram of the QCSE. In (a) there is not an electric field while in (b) an electric field is applied**

When an electric field is applied perpendicular to the QW it tilts the energy bands as shown in Figure 8 (b). As a consequence the energy between $E_{el}$ and $E_{hl}$ is reduced since $E_{el}$ decreases while $E_{hl}$ increases. The binding energy of the excitons $E_{ex}$ is also decreased with an increased static electric field. The electric field pushes the wave-functions of the electrons and holes to the sides of the QW. This reduces the overlap integral between the wave functions in the valence and conduction band. Consequently, the recombination efficiency is also reduced. In this case the bandgap energy is reduced. This increases the absorption of the material in the band-edge of the absorption spectrum. It means, the absorption band-edge is shifted to longer wavelengths. The effect is represented in Figure 9,

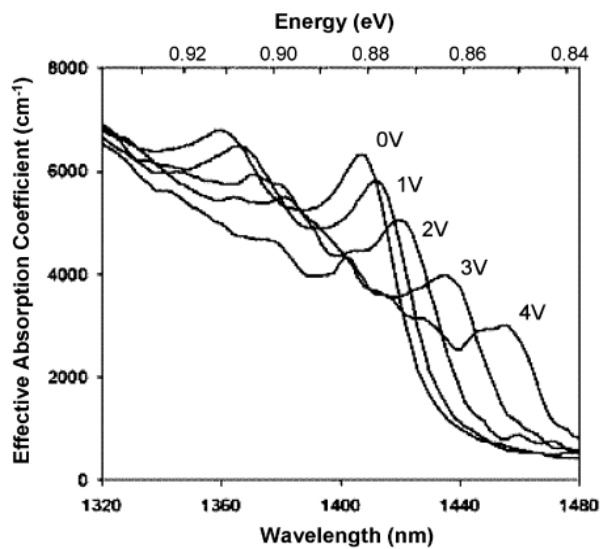

**Figure 9: Quantum confined Stark effect absorption coefficient [49]**

In Figure 9 the effective optical absorption of the QCSE is present versus the optical wavelength for different applied voltages that go from 0 to 4 V every 1 V. We see that as the applied voltage is increased the excitonic peak is reduced.

The range of absorption of the QCSE is around 4000-6000 cm$^{-1}$ [49]. This value is bigger than the one we have in the FKE which is around 200-300 cm$^{-1}$ [46]. Although the QCSE is stronger than FKE it is present in QW.

### 2.2.3  Conclusion

As a conclusion of both electro-refraction and electro-absorption effects we explained that Si is a centro-symmetric material and consequently does not support second order nonlinear effects like the Pockels effect [39]. This difficults the design of electro-absorption modulator. Despite this fact, third order nonlinear effects [50] can be used to perform the modulation like the Kerr effect [40] although as explained before it is also weak.

The most used effect in Si modulators is the plasma dispersion effect [43]. It is used to change the phase of the signal since the refractive index of the material is changed upon the application of a static electric field which modifies the carrier concentration on the material. There are other effects like the Franz-Keldysh effect (FKE) [44], [45] or the Quantum Confined Stark effect (QCSE) [48] that leads to electro-absorption modulators. These effects are quite strong in Ge (FKE) and SiGe/Ge/SiGe quantum wells. Bulk Ge can be implemented in a silicon photonic application using CMOS standard processes. It means: Ge is compatible with Si. Si has also the FKE but it is weak and it is not enough to perform modulation.

In the following section we present the state-of-the-art of silicon photonics electro-absorption modulators. We concentrate on the FKE photonic modulators since the active principle of the device that we design uses this effect as the active principle.

## 2.3    State-of-the-art of Electro-Absorption Silicon Optical Modulators

The purpose of this thesis is to design an electro-absorption plasmonic modulator based on the FKE. In this section we present the current state-of-the-art of photonic modulators based on the FKE and plasmonic modulators with the objective of comparing our device with the previous state-of-the-art.

## 2.3.1 State-of-the-art of Franz-Keldysh Effect Photonic Modulators

Since we will design a plasmonic modualtor that uses the FKE it will be interesting to compare our device with the state-of-the-art of photonic modulators that uses also the FKE. With this, we will be able to compare plasmonic modulator with a photonic one.

After the work described in [46] the next FKE electro-absorption modulator is described in [35]. A monolithically integrated GeSi electro-absorption modulator and a photodetector are presented in [35]. The modulator uses the FKE in GeSi. This is the continuation of the previous work but instead of using bulk Ge, the Ge is mixed with Si in order to have a maximum performance ($\Delta\alpha/\alpha$) around 1550 nm instead of 1647 nm obtained in bulk Ge [46]. For such a configuration the proper material used is $Ge_{0.9925}Si_{0.0075}$. As a conclusion, we can shift the wavelength of our modulator to 1550 nm using GeSi rather than bulk Ge.

The structure of the monolithically integrated modulator and photodetector is represented in Figure 10. It consists of a SOI waveguide butt-coupled to the Ge modulator. At the end of the device a photodetector is added. Since this is not in the scope of this thesis, the photodetector part is not explained. The modulator is composed of a vertical PIN structure. When it is reverse biased, a static electric field is formed in the intrinsic GeSi region. Consequently, the loss of the material is changed. In this way the intensity of the optical mode is modulated.

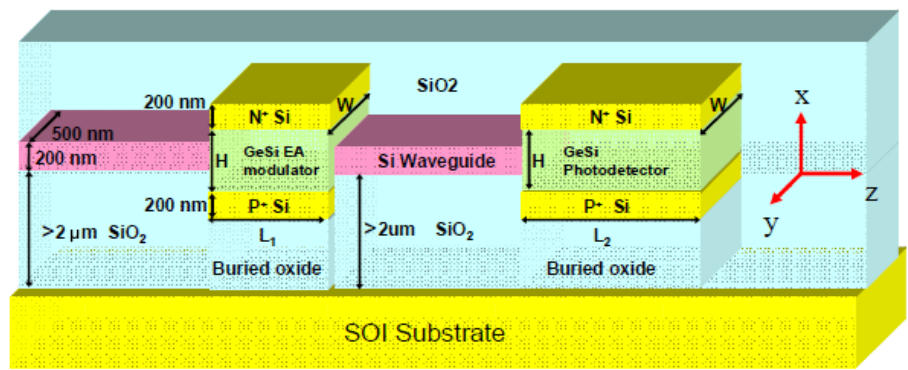

**Figure 10: Structure of the monolithically integrated modulator and photodetector [35]**

Like other electro-absorption modulators, it is intrinsically high speed and has low power consumption. Due to the sub-picosecond response time [35] of the FKE it is RC limited. It can be designed to have a 3 dB cut-off frequency larger than 50 GHz for an extinction ratio of 3 dB. For a device length of 70 μm the insertion losses were around 5 dB.

An ultra-compact GeSi electro-absorption modulator is presented in [51]. It is CMOS compatible and uses the FKE. It is claimed that it has a footprint as small as 30 μm². The structure of the device is represented in Figure 11. It consists of a SOI waveguide that evanescently couples to an amorphous Si (a-Si) one. The active region of the modulator is composed of a vertical p-Si/GeSi/n-Si PIN structure with crystalline Si p-doped region and poly-Si n-doped regions. The intrinsic region is made of $Ge_{0.9925}Si_{0.0075}$ as explained in the previous monolithically integrated modulator.

The footprint of the device is ultra-compact. It is 600 nm wide and 50 μm long, giving an area of 30 μm². As in the previous case, the modulator can operate in the wavelength range from 1539 to 1553 nm. The insertion loss for a 50 μm long device is 7.5 dB and the extinction ratio is around 1.3 times the insertion losses. The power consumption of the device is around 50 fJ/bit. It is CMOS compatible and has a 3 dB bandwidth of 1.2 GHz.

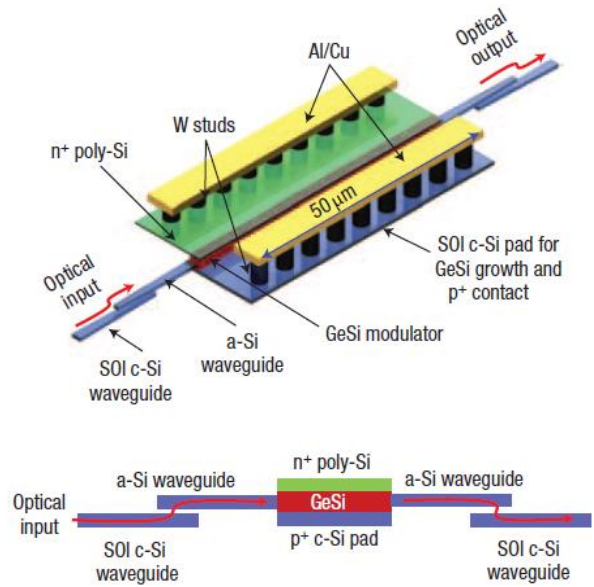

**Figure 11: Structure of the SiGe waveguide-integrated modulator [51]**

An evanescent coupled electro-absorption modulator is presented in [52]. The structure of such a modulator is illustrated in Figure 12. It consists of a Ge rib waveguide on a Si slab one. Around the Ge core of the modulator a PIN structures is done in order to induce a static electric field in the Ge rib waveguide to produce the FKE. The optical mode is evanescently coupled to the Ge rib waveguide. Consequently, changing the absorption of the Ge material, the optical mode is modulated. This evanescently coupled structure facilitates the integration in a CMOS compatible process.

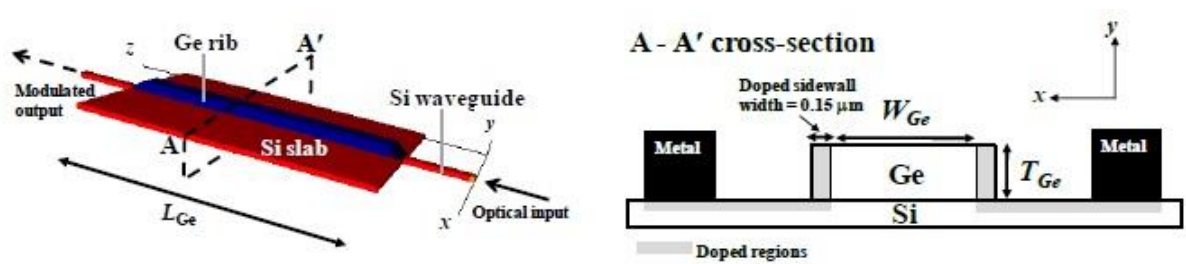

**Figure 12: Structure of evanescent coupled Ge electro-absorption modulator** [52]

Regarding the metrics of the device it has an extinction ratio of 10 dB under a -5 V reverse bias voltage in the wavelength range of 1580-1610 nm. It has a modulation efficiency of 2 dB/V. It also has a small active region of 16 μm². Finally, the modulator is designed such that ER/IL>1. A power consumption of about 1 pJ/bit can be achieved.

A Ge electro-absorption modulator integrated with a SOI waveguide is presented in [53]. The schematic of the structure is represented in Figure 13. It consists of a SOI waveguide that is butt-coupled with a taper to a Ge waveguide that is formed by a horizontal PIN structure. As in the other cases, reverse biasing the PIN a static electric field is induced in the Ge core in order to change the absorption coefficient due to the FKE. The induced strained Ge core is deposited over the Si waveguide.

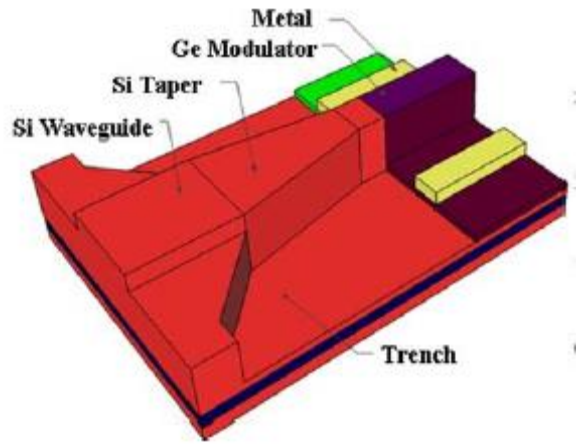

**Figure 13: Ge electro-absorption modulator integrated with a SOI waveguide [53]**

The 3 dB bandwidth of the device allows an operation frequency of 30 GHz for 4 $V_{pp}$. This modulator can have an extinction ratio between 4-7.5 dB with insertion losses between 2.5-5 dB. The operational wavelength of the device is between 1610-1640 nm. The compact active region is in the range of 1.0 ~ 45 $\mu m^2$. The electrical power consumption is around 100 fJ/bit.

The first demonstration of a FKE electro-absorption modulator working at 1550 nm is presented in [54]. The structure of the device is similar to the previous one (Figure 13). But, instead of having a Ge waveguide in the active region it has a GeSi core. Doping the Ge with Si around 0.7% shifts the FKE to a wavelength around 1550 nm. The working wavelength of the device is between 1545-1580 nm.

For the metrics of the device, the extinction ratio is around 6.0 dB for a reverse bias voltage of 2.8 $V_{pp}$ with insertion losses around 5.0 dB. The frequency of operation is 40 GHz at 2.8 $V_{pp}$. The active region has a size of 1.0x55 $\mu m^2$. The modulator is designed in a 3 $\mu m$ waveguide platform.

Finally, in [55] different GeSn alloys are used to extend the operational wavelength of the device from the near-infrared to the mid-infrared. In this case the materials SiGeSn/GeSn can be used for the fabrication of the optoelectronic devices.

A set of two modulators: one is an electro-absorption modulator while the other is a resonant modulator are proposed. The former is made of a PIN structure with GeSn in the core. Reverse biasing the PIN structure induces an electric field in the GeSn material having a FKE that produces the modulation. The structure is butt coupled with a material whose bandgap is larger than the GeSn in order to guarantee transparency. In Table 1 we present a summary of the modulators described in this section.

| Metrics/Device | MIT (2006) [46] | MIT (2007) [35] | MIT (2008) [51] | A*STAR (2011) [52] | Kotura Inc (2011) [53] | Kotura Inc (2012) [54] |
|---|---|---|---|---|---|---|
| Structure | Absorption | Absorption | Absorption | Absorption | Absorption | Absorption |
| Bandwidth [GHz] | - | 50 | 1.2 | - | 30 | 40 |
| Wavelength [nm] | 1647 | 1550 | 1540 | - | - | 1550 |
| Spectrum [nm] | - | - | 1539-1553 | 1580-1610 | 1610-1640 | 1545-1580 |
| Insertion losses [dB] | - | 5 | 7.5 | 10 | 2.5-5 | 5 |
| Extinction ratio [dB] | - | 10 | 9.75 | 10 | 4-7.5 | 6 |
| $\Delta\alpha/\alpha\vert_{max}$ | 3 | 2 | 1.3 | 1 | 2-3.3 | - |
| Voltage [V] | - | 3.3 | 3 | 5 | 4 | 2.8 |
| Footprint [µm] | -x1.3x- | 0.6x0.4x70 | 0.6x-x50 | 16 µm$^2$ | 1x-x45 | 1x-x55 |
| Transient [fs] | - | - | - | - | - | - |
| Power Consumption [fJ/bit] | - | 50 | - | - | 100 | - |
| CMOS compatibility | - | Yes | Yes | Yes | - | Yes |

**Table 1: Comparative table of FKE photonic modulator**

## 2.3.2 State-of-the-art Plasmonic Modulators

In this section we described the state-of-the-art of plasmonic modulators since the invention of the plasMOStor, the first proposed silicon photonic and plasmonic modulator which came out in 2008 [56]. It will be interesting to compare the photonic devices that use the FKE and the plasmonic modulators with the device proposed in this work.

For presenting the state-of-the-art we describe the different modulators in chronological order. We also divide them in electro-refraction or electro-absorption depending on the method used for performing the modulation. We start with electro-refraction and later we continue with electro-absorption.

### 2.3.2.1   Electro-Refraction Plasmonic Modulators

The first plasmonic modulator that uses Si in its core region is presented in [56]. It achieves the first modulation of light in a plasmonic waveguide. It consists of a stack of Ag-SiO$_2$-Si-Ag. The structure is represented in Figure 14. It has a structure similar to a transistor with a gate, a source and a drain (Figure 14). A photonic mode from the source excites both a plasmonic mode at the Ag-SiO$_2$ interface and a photonic mode in the Si core. It is excited from the input source. The length of the gate of the plasMOStor is set in a way that when both the photonic and the plasmonic modes are propagating, then destructive interference occurs in the drain. By accumulating carrier applying a proper voltage in the gate the photonic mode can be set to cut-off or not. So, changing the photonic mode a modulation is performed in the drain of the plasMOStor where both the plasmonic and the photonic mode interfere.

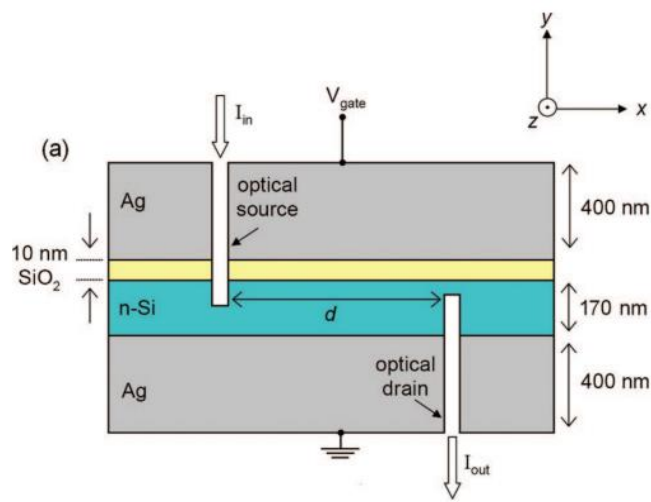

**Figure 14: Structure of the plasMOStor [56]**

The plasMOStor can achieve a frequency of operation between 3 and 15 GHz. The frequency of operation of the plasMOStor is limited by the formation of the accumulation layer in the Si-SiO$_2$ interface, so, it is transport limited. It has a total insertion loss of 20 dB and an extinction ratio of 4 dB. The operational voltage is between 0.4 and 4 V. The total footprint of the device is around 4 μm².

A Mach-Zehnder plasmonic interferometer (MZI) is presented in [57]. The device structure is represented in Figure 15. It consists of two V-shape splitter/combiner and two parallel plasmonic slot waveguides. The plasmonic slot waveguides are composed of a stack of Metal-SiO₂-Si-Metal. The modulation principle is based on the free carrier dispersion effect of the induced charge in the doped Si when applying a voltage difference between the metals.

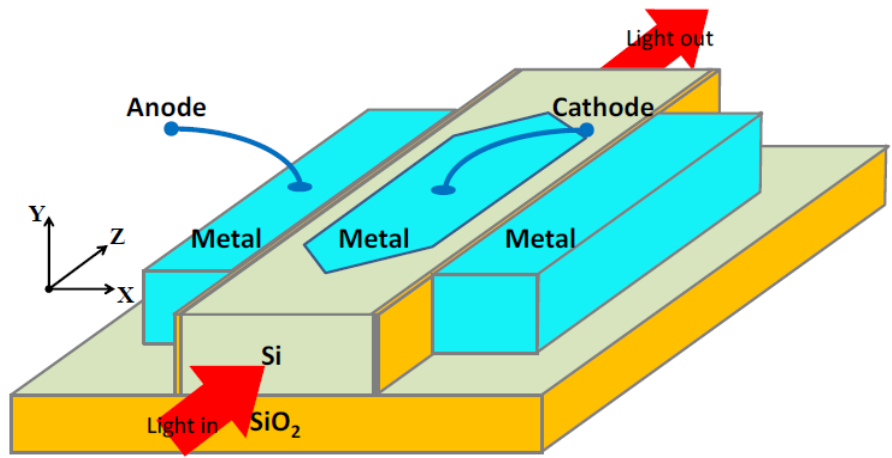

**Figure 15: Structure of the MZI plasmonic modulator [57]**

This device is CMOS compatible and has a similar process to the recent gate all around MOSFETs and/or FinFETS transistors.

A proper simulation design of the modulators yields the following metrics: a modulation speed of 500 GHz, for an insertion loss of 8 dB, and an extinction ratio of 7.6 dB. The working voltage is around 5.6 V. It is said that the operational voltage can be reduced using a high-κ dielectric instead of the SiO₂.

In [58] a plasmonic phase modulator is presented. The active principle of the device recalls in the use of the Pockels effect into a nonlinear polymer. The structure of the modulator is presented in Figure 16. The plasmonic section consists of a metal slot waveguide formed by two pieces of Au which form a nano-gap 140x150 nm. This gap is filled with a non-linear polymer which exhibits the Pockels effect. It means, the refractive index of the nonlinear polymer can be changed due to the presence of the Pockels effect when a static electric field is applied. By changing the refractive index of the nonlinear polymer the information is encoded in the phase of the plasmon that propagates through the slot waveguide.

At the input and the output of the modulator two tapers are designed to allow the conversion from a photonic mode to a plasmonic one and conversely. The gold parts at the sides of the modulator act also as a contact to induce a static electric field in the gap of the slot waveguides.

Furthermore, there is an overlap of the electromagnetic field of the plasmon and the static electric field induced in the slot. This enhanced the presence of the electromagnetic field of the plasmon and the static electric field induced by the Au side contacts into the slot.

Using the Pockels effect has the advantage that the process is ultrafast and does not have the time speed limitation like the plasma dispersion effect in Si. We remind that the carrier dispersion effect is limited by the finite mobility of the carriers through the material. Furthermore, the low capacitance of the structure (mainly in the gap) and the high conductivity of the metal used do not limit the RC speed.

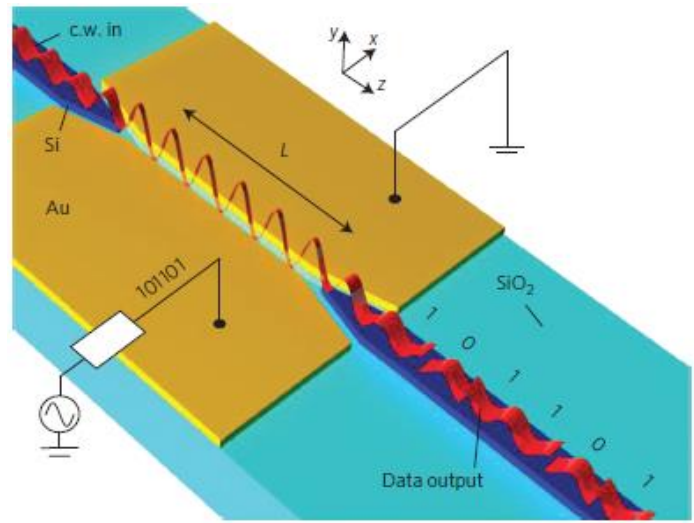

**Figure 16: Structure of the phase modulator using a nonlinear polymer [58]**

It is said that the operational frequency of the device can be scaled until 65 GHz. The bias voltage used for the device is in the order of V=0.1 V. Regarding the electrical power consumption of the device it consumes 60 fJ/bit.

In this work [58] the performance of the device was measured in similar conditions as those experienced inside a telecommunication rack, E.g.: at a temperature that can reach 85 °C. At this temperature, the performance of the device was not degraded. Consequently, it can be integrated in a telecommunication rack.

The device has an operational frequency around 40 GHz for a length of 29 μm. The wavelength range of the modulator is around 120 nm and it is centered on 1.55 μm. the performance of the device was tested at different temperatures of operation until 85 °C showing a low bit error rate (BER). It is worth noting that in this modulator the information signal is carried out in the phase of the plasmonic mode supported by the structure. For a 29 μm length the device has an insertion loss of 12 dB.

## 2.3.2.2   Electro-Absorption Plasmonic Modulators

In [59] a novel approach inspired from the microwave domain is proposed to achieve a plasmonic modulator using a plasmonic stub in a waveguide is presented. The modulation is performed in a SOI rib waveguide to which a metal stub was attached. The metal-coated stub is able to control the amount of light that passes through the waveguide. A schematic of the structure is represented in Figure 17,

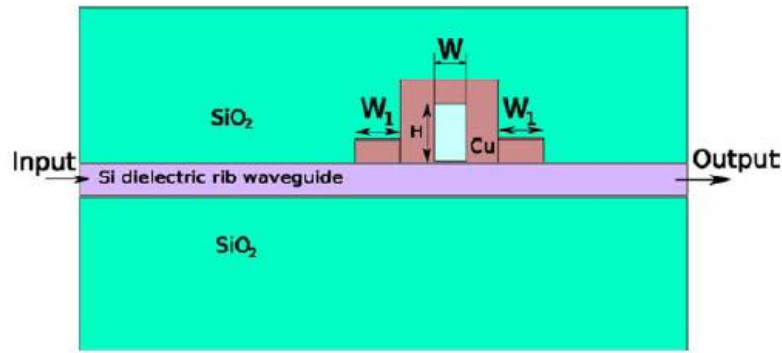

**Figure 17: Structure of a plasmonic modulator using a metal-coated stub [59]**

Using the right wing of the stub (seen in Figure 17) this mode excites a SPP mode in the interface between the Cu and the Si. Such a mode is modulated by changing the imaginary part of the material which fills the stub.

The stub is filled with a material or structure that is able to change its optical properties (E.g.: the optical losses) upon an application of a control signal (E.g.: the static electric field, etc.). In this way the quality factor Q of the stub is controlled in such a way that it produces a change in the transmission of the rib waveguide. When the stub is in resonance, the losses through the waveguide are increased. In this case, there is no transmission through the Si rib waveguide. While when it is out of resonance they are reduced and the transmission increases. Hence, producing the modulation.

The material used to fill the stub must have an externally controllable absorption in order to change the quality factor Q. The use of Ge/SiGe multiple quantum wells (MQW) in order to exploit the QCSE or GeSi Quantum Dots (QDs) were proposed. The material used for the metal is Cu.

The dimensions of the stub are critical in order to optimize the metrics of the modulator (mainly: the extinction ratio and the insertion losses). The design of the stub will also determine the amount of change of the absorption of the material filling the stub. To summarize, the design of the modulator is determined by the design of the stub. Such a design has a trade-off between the extinction ratio and the insertion losses.

From a theoretical point of view this system is analogous to a microwave waveguide with a short circuit stub attached to it. This is widely employed to adapt impedance in microwave engineering.

The authors report that which such a design it is possible to have an extinction ratio of 6.6 dB with low insertion losses around 8.5 dB. They specify that it is using SiGe/Ge/SiGe quantum wells. The length of such a design is around 0.5 μm. The power operation and the modulation speed are not specified in [59].

In [60] an all-plasmonic method of modulation was presented. It uses two surface plasmon polariton (SPP) co-propagating signals in the same direction. The signals are propagating in a medium with gain. Two signals are used, one called the SPP pump ($\lambda_{pump}$=980 nm) and the other the SPP probe ($\lambda_{probe}$=1550 nm). It is represented in Figure 18. When the SPP pump is present in the medium the SPP probe is amplified via stimulated emission due to the presence of the material $Er^{3+}$. Consequently, when the SPP pump is not present the SPP probe feels a higher absorption. So, by using the presence of the SPP pump the SPP probe signal is modulated. The amplification of SPP signals has been studied in order to try to mitigate the high propagation losses of plasmonic waveguides. E.g.: [61], [62]. Nevertheless the fact of compensating the losses with gain is far from being achieved.

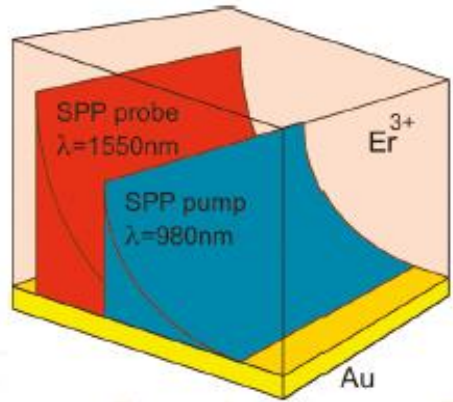

**Figure 18: Structure of the arm of the Mach-Zehnder Modulator [60]**

The material used as a metal is gold (Au) and the gain medium is $Er^{3+}$. Due to the use of Au the device is not CMOS compatible. There are many materials like semiconductor quantum dots, polymers or dye molecules that can be used as the gain medium.

This kind of device has a trade-off between extinction ratio and propagation losses. It is interesting to note that devices that use nonlinear effects need high optical intensities in order to achieve measurable effect or interaction length.

In [63], a CMOS compatible SPP absorption modulator is presented. The device structure is illustrated in Figure 19. It is formed by a stack of Ag-ITO-SiO₂-Ag layers. The innovative aspect is the use of the absorption properties of Indium-Tin-Oxide (ITO) compared with the other plasmonic modulators presented in this chapter. A Si waveguide is used to excite a plasmonic mode in the Ag-ITO-SiO₂-Ag stack. It is interesting to note that an important part of the plasmonic mode is concentrated into the ITO layer. When a voltage is applied between the two Ag layers, the absorption properties of ITO are changed and consequently the plasmonic mode is modulated. Study of ITO is presented in [64].

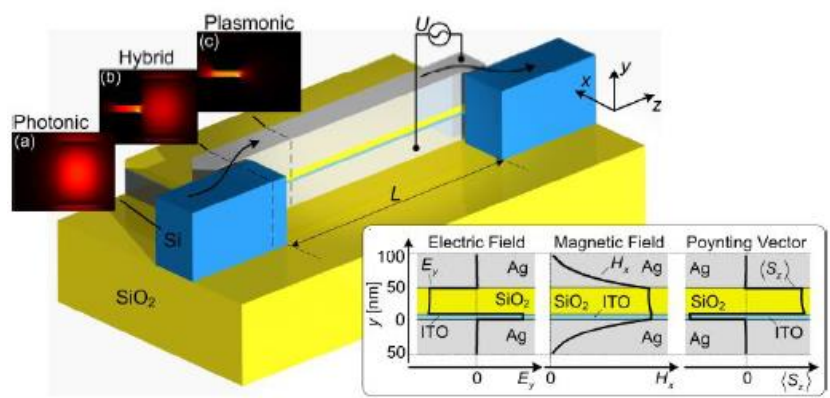

**Figure 19: Structure the elector-absorption modulator using ITO [63]**

It is not limited by carrier speed limitations. It is worth to mention that the carrier mobility is very high in ITO. Consequently the maximum frequency of operation is limited by the driving circuit (RC).

The device can work in the THz regime. Another advantage of the device is that it can work at both 1.3 and 1.5 µm. There is a trade-off between the device length, the extinction ratio and the insertion losses. In any case the device length is around a few µm.

A good configuration of the device has a length of 2 µm, an extinction ratio of 1 dB and insertion losses of 18 dB. The insertion loss also includes the coupling losses from the photonic mode to the plasmonic one. Changing the SiO$_2$ layer by Si$_3$N$_4$, the plamonic mode may be further enhanced within ITO, consequently the device length could be reduced.

In [65] a CMOS compatible electro-absorption modulator is presented. It is based on a Metal-Insulator-Semiconductor slot waveguide. The structure is represented in Figure 20. It is connected at both ends to a semiconductor pad formed of doped semiconductor material. The modulation principle relies in the accumulation of carriers at the insulator-semiconductor interface. To produce the modulation, the accumulation layer is created by translating the electrons from the doped pad to the interface and vice versa when a voltage is applied between an electrode and the lateral pad.

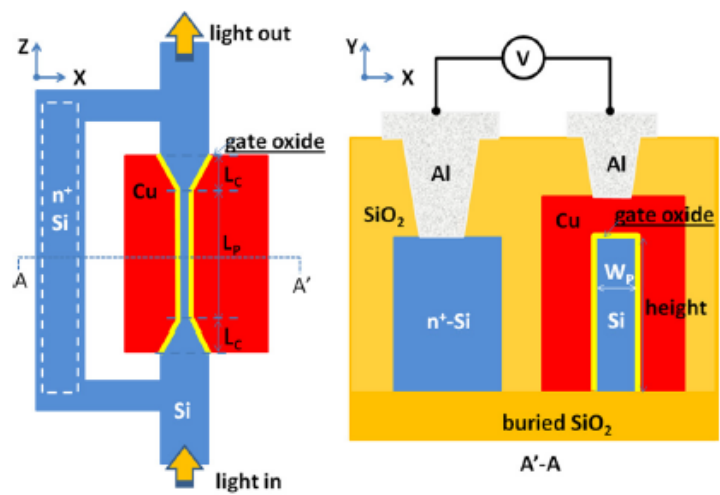

**Figure 20: Structure of the arm of the CMOS compatible elector-absorption modulator and top view [65]**

The metal used is Cu, the insulator is SiO$_2$ and the semiconductor is Si. It is mentioned that the extinction ratio of the device can be improved by using high-κ insulator.

A 4 µm long device provides an extinction ratio of 3 dB with a bias voltage of 6.5 V. The frequency operation of the device is in the sub-GHz regime and it is limited by the carrier velocity of the electrons which travel from the lateral pad to the Si-SiO$_2$ interface in the core plasmonic waveguide. It is claimed in [65] that the performance can be improved to tens of GHz by changing the position of the lateral pad.

Another device which uses indium-tin-oxide (ITO) is presented in [66]. The structure of the modulator is represented in the following Figure 21. The modulator consists of a Si strip waveguide with a MIS part made of Au-SiO$_2$-ITO-Si. The ITO layer is between the metal and the Si core and has a thickness of 10 nm. With this the propagation losses are reduced with respect to the waveguide formed by the Si and Au without the ITO slot and without the oxide. When a voltage is applied to the MIS structure then it changes the mode absorption of the plasmon. It modifies the laser beam intensity.

The electromagnetic field of the mode is concentrated in the insulator slot in the MIS. The slot is formed by insulator SiO₂ and the conductive oxide ITO. This is a material whose optical absorption can be modified by means of changing the carrier concentration into the material. This is changed using a static electric field.

The active principle of the modulator is the ITO layer in the MIS waveguide. Changing the carrier density of the material it changes its optical absorption. The efficiency of the extinction ratio of the modulator is 1 dB/μm in the wavelength range between 1.2-2.2 μm. For a device length of 5 μm the extinction ratio is 5 dB and the propagation losses are 2.8 dB. It is a compact modulator based on Si with a high 1 dB/μm extinction ratio. It is the best reported extinction ratio in the literature. This device can be integrated in a SOI technology bringing a good potential to efficient broadband communication within circuits.

This record is due to the fact that placing the ITO layer between Au and Si enhances the static electric field in the slot where the effect takes place. Furthermore, the ITO slot extracts the static electric field from the Au which produces a reduction in the propagation losses of the device with respect to the structure of Figure 21 without the slot.

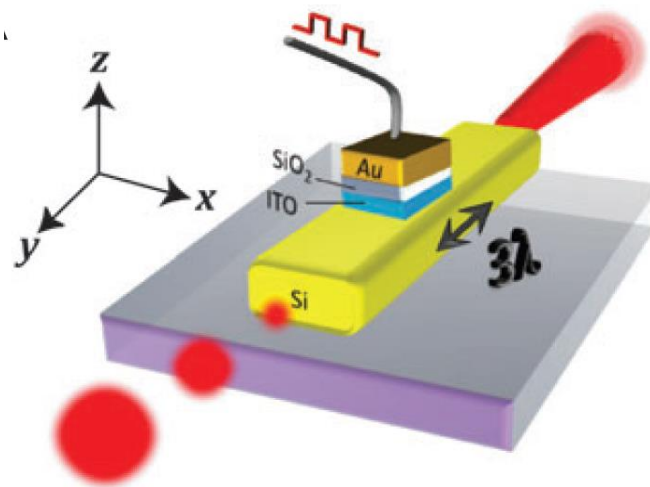

**Figure 21: Structure of the modulator presented in [60]**

Regarding the insertion losses of the device of Figure 21 it is interesting to note that there are two main contributions, one is the mismatch in the impedance while passing from a Si strip to a MIS structure. The second contribution is the propagation losses of the plasmonic mode in the MIS waveguide. This contribution is dominant due to the use of metals in the active region. This propagation loss cannot be avoided but it is reduced by the use of the slot.

The total coupling losses due to the coupling from the Si strip to the MIS waveguide is around 1 dB. The reported total insertion losses are around 4.8 dB and 19.1 dB for the device of 5 μm and 20 μm long devices.

It is worth noting that in the structure of Figure 21 has a reflection coefficient around 1 % between the mode in the input Si strip waveguide and the plasmonic mode in the MIS one. Then, a low reflection is expected due to the impedance mismatch between the waveguides.

Additionally, the broadband response of the modulator was measured and it can be seen that the modulator can operate from 1.2 μm to 2.2 μm. It has a good performance over the 1 μm long wavelength range.

Regarding the operational frequency of the device there are two limiting factors. The first one is due to the RC delay time (RC limited) and the speed at which the accumulation layer in the MIS waveguide is formed (the accumulation layer is formed in the ITO material). For a 5 μm long device the RC limitation is found to be around 300 GHz. The estimated resistance is around R=500 Ω. The speed of the formation of the accumulation layer in the layer of ITO the calculations showed that it is in the THz regime for the 10 nm layers. Consequently, the modulator is RC limited and the operational frequency is around 300 GHz.

The electrical power consumption depends on the capacitance C of the MIS waveguide. The measured C is C=7 fF for 5 μm length device. The operational voltage is around V=4 V and the energy consumption is around 56 fJ/bit. This value is close to the roadmaps presented in [4], [33] in which the estimated electrical consumption for the modulator in the optical link must be around 50 fJ/bit.

Finally, it is interesting to mention that there is a trade-off in the modulator: increasing the thickness of the ITO layer the electrical power consumption is reduced. But also the extinction ratio is reduced. It is reported that a power consumption of 23 fJ/bit is achieved for a thickness of the slot around 40 nm.

To summarize with the modulator of Figure 21, 1 dB/μm extinction ratio efficiency is reported. The bandwidth of the modulator is around 1 μm. The operational frequency is 300 GHz and the electrical power consumption is less than 56 fJ/bit for an operational bias voltage V=4 V.

An interesting plasmonic switch is presented in [67]. This switch is characterized by using a new material called vanadium dioxide ($VO_2$) to perform the switching. The structure of the device is presented in Figure 22.

The extinction ratio are shown around 23-32 dB for a length of 5 μm. The driving bias voltage is around V=60 mV and the electrical power consumption is around 9 mW.

As other plasmonic modulators and switches, this device suffers from many trade-offs between extinction ratio, insertion losses and the size (mainly the longitudinal dimension) of the device. To improve this performance it is interesting to use new materials exhibiting a large change in refractive index or absorption when a bias voltage is applied. As exposed before, good materials are ITO and $VO_2$. In this modulator [67] $VO_2$ is used.

$VO_2$ is a metal oxide that exhibits a transition from an insulating state to a conductive metallic state at the temperature of T=341 K. Using a static electric field or electromagnetic radiation allows this phase transition in the material. The transition time from metal to dielectric and from dielectric to metal depends on the direction of the phase change. The metal to dielectric transition it is in the time range of nanoseconds to milliseconds. Whereas the transition from dielectric to metal is faster and it happens at the femtoseconds scale.

The structure of the modulator of [67] (Figure 22) is composed of a $VO_2$ layer sandwiched between an Ag layer and a $SiO_2$ one. The structure is over a $SiO_2$ buffer layer. When the device is in transmission mode (called on-state), the input waveguide excites the plasmon mode and the $VO_2$ is acting as a dielectric. This allows the transmission of the mode. The plasmonic mode goes in the direction indicated in Figure 22 from the optical input until the optical output.

When a current is applied into the metal part of the structure (Figure 22) the metal is heated. Note that the direction of the current and the propagation of light are perpendicular. When the metal is heated this induces the phase change of $VO_2$ which passes from being dielectric to metallic. This increases the losses of the propagated mode. Consequently, the intensity of the outgoing plasmonic mode is decreased and it represents the off-state, of the switch. When the current is turned off the device is cooled down and the $VO_2$ passes again from being metallic to dielectric. This reduces the losses and then the device is in the on-state with a high outgoing intensity. We remind that due to this effect, the device is limited by the velocity at which it is heated up and cooling down. So, the operational frequency of the device is thermally limited in the MHz regime.

The losses of $VO_2$ are too high to use this material as the core of the waveguide. This is why a small slot is used in the design (Figure 22). Nevertheless, using thin metal films with $VO_2$ it is possible to design devices with high extinction ratio and low insertion losses. It uses the large index change of $VO_2$ and the strong static electromagnetic confinement into the structure. An example of this assets is the modulator presented in [67]. Using a current can induce the thermally phase transition of $VO_2$ which is used to perform the modulation.

In [67] the switch of Figure 22 is optimized with respect to the FoM=ER/IL. With this FoM the extinction ratio is maximized while the insertion losses are reduced. A high FoM is desirable.

Depending on the dimensions of the geometry of the switch it can work in a strongly-hybridized regime or in a weakly-hybridized one. These regimes are different in the overlap of the $VO_2$ with the light. Here, we only summarize the performance of these two regimes.

Working in a strongly-hybridized design the efficiency of the 0.9 dB/μm and the efficiency of the 6.1 dB/μm. The optimal strongly-hybridized device has a length of 5 μm. This allows having an extinction ratio of 32.1 dB and insertion losses of 13.4 dB. In the insertion losses the losses due to the coupling from a Si waveguide and the propagation losses of the plasmon are included.

On the other hand, in the weakly-hybridized mode the optimal device has an efficiency of the 0.3 dB/μm and an efficiency in the extinction ratio 3.6 dB/μm. For a length of the device around 5 μm the insertion losses are around 17 dB while the extinction ratio 22.7 dB.

In the switch of Figure 22 it is worth noting that the metal (Ag) that supports the plasmon is also the one used to pass a current and heat up and cool down the structure to induce a absorption change into the modulator. The current is passed perpendicular to the direction of propagation of the beam in order to reduce the size of the device. The length of the heater has been optimized in order to reduce the heating volume to the minimum and increase the heating distribution with the $VO_2$ layer. This also reduced the electrical power consumption since the volume close to the $VO_2$ layer is heated up and cooling down.

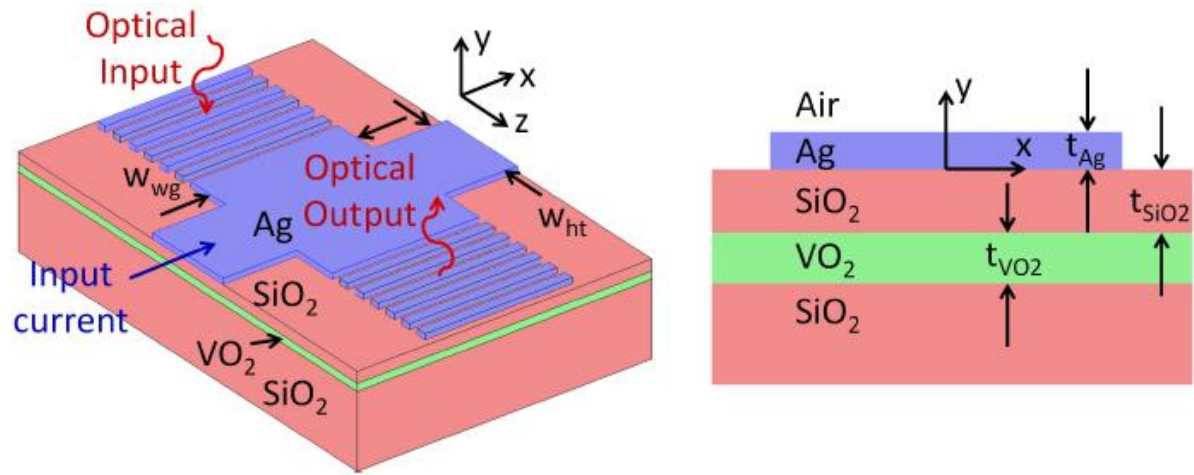

**Figure 22: Vanadium dioxide plasmonic switch structure [67]**

To summarize, this work [67] presented a novel plasmonic switch using $VO_2$. It is a thermo-optic device having an integrated heater. The modulator has its active principle in the change of the phase of $VO_2$ that can pass from metallic to dielectric upon application of an electric field. The optimized device has an extinction ratio of 32.1 dB and an insertion losses of 13.4 dB. The length of the active region is 5 μm. The operational voltage is around 0.6 V and the electrical power consumption is of 9 mW. Nevertheless the operational frequency of the device is limited by the thermo-optic effect. Consequently, the speed is in the MHz regime. To improve the operational frequency the electrical properties of $VO_2$ (induce the phase change of the material with a static electric field) can be used instead of using the thermic properties.

In [68] several plasmonic modulators configurations using alternative plasmonic materials are presented. The new materials used are transparent conducting oxides (TCO) and titanium nitride (TiN). In TCO the absorption of the modulator is changed by applying a static electric field. In [68] it is stated that they can design a modulator with an extinction ratio efficiency of 46 dB/μm. With such efficiency a 3 dB extinction ratio can be achieved with a modulator of only L=65 nm long.

Although several plasmonic modulator were proposed using metals like Au or Ag that are not CMOS compatible, the new plasmonic materials [69], [70] like TCO or TiN can be introduced into a CMOS environment.

Among these materials, TiN is a metal that can support the propagation of a plasmonic mode. Due to the intrinsic properties like: thermal stability (2900 ºC for the melting point), mechanically hard, it is compatible with biological applications and it can be growth on top of Si. Furthermore, it does not oxidize like other metals like Cu or Ag.

On the other hand, TCO materials can be used to perform a modulation of a plasmonic mode since this kind of materials can change the optical absorption upon an application of a static electric field.

The proposed plasmonic modulator in [68] has the following structure in Figure 23,

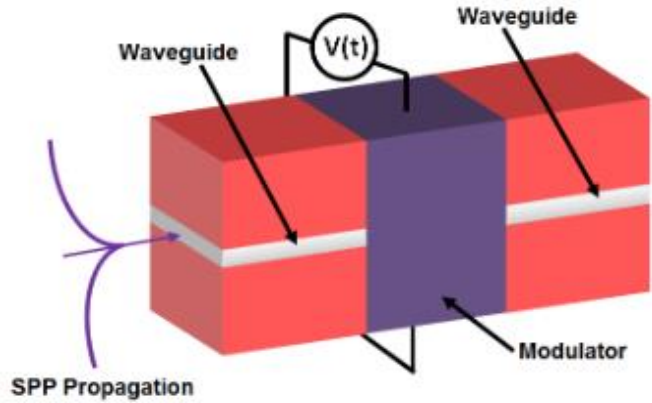

**Figure 23: Black box representation of the plasmonic modulators proposed in** [68]

In Figure 23 the input strip waveguide supports a SPP. The light blue represents a metal and the red layers represent the insulator. Once the SPP is propagated it enters the modulator, represented in violet and finally goes out in the output stripe waveguide.

A FoM that accounts for the extinction ratio, the mode confinement and the insertion losses were proposed. The materials used were indium tin oxide (ITO), aluminium-doped zinc oxide (AZO) and gallium-doped zinc oxides (GZO) were used to perform the modulation while TiN was used to support the plasmonic mode. Using these materials the operational wavelength of the modulator is at 1.55 μm.

Among all the configurations used the best one is the one presented in Figure 24,

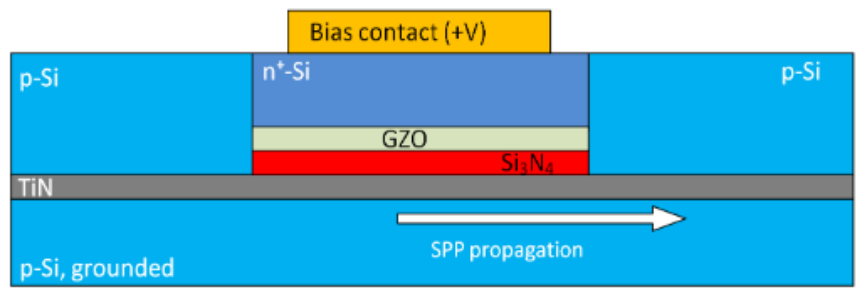

**Figure 24: Structure of the best plasmonic modulator presented in** [68]

In Figure 24, the TiN supports the plasmonic mode, when entering to the modulator the change in the absorption upon application of an electric field is done by GZO. Using this structure the extinction ratio is 3 dB for a length of L=65 nm. The propagation losses are around 0.02 dB in the modulator (for L=65 nm). The coupling efficiency is around 0.7 dB per facet. With this, the total insertion losses are 1.42 dB.

In the following Table 2 we present the summary of the plasmonic modulators explained in this section.

| Metrics/ Device | UGent (2009) | Caltech (2009) [56] | A*STAR (2011) [65] | A*STAR (2011) [57] | KIT (2011) [63] | Leeds (2011) [71] | Leeds (2011) [71] | Leeds (2011) [59] | KCL (2011) [60] | Toronto (2012) [67] | Berkeley (2013) [66] | DTU (2013) [68] | KIT (2014) [58] |
|---|---|---|---|---|---|---|---|---|---|---|---|---|---|
| Structure | MZI | MZI | Absorption | MZI | Absorption | MZI | MZI | - | Absorption | Absorption | Absorption | Absorption | Phase |
| Bandwidth [GHz] | kHz | 3-15 | sub-GHz | 500 | THz | 10 | - | 7-13 | - | - | 300 | - | 40 |
| Insertion losses [dB] | 8 | 20 | - | 8 | 12-18 | 8 | 10 | 8.5 | - | 13.4 | 5-20 | 1.42 | 12 |
| Extinction ratio [dB] | 23 | 10 | 3 | 7.3 | 1-2 | - | - | 6.6 | - | 23-32 | 5-20 | 3 | - |
| Voltage [V] | - | 0.7-4 | 6.5 | 5.6 | 10 | 7 | - | - | N/A | 0.6 | 4 | - | 0.1 |
| Footprint [μm] | -x0.21x4 | -x-x4 | -x0.5x3 | 0.05x0.05x3 | -x0.3x2 | -x-x1.2 | -x-x3.87 | -x0.5 | -x-x(25-250) | -x-x5 | - | -x-x0.065 | - |
| Transient [fs] | - | 10 | - | - | 35 | - | - | - | - | - | - | - | - |
| Energy per bit [fJ/bit] | - | 6.8 | - | - | - | 360 | - | - | - | - | 56 | - | 60 |
| Limit | Thermal | Carrier Transport | Carrier Transport | - | RC | RC | RC | - | SE Rate | Thermal | RC | - | RC |
| CMOS compatibility | - | - | Yes | Yes | No | No | No | No | No | No | No | No | No |

Table 2: Comparative table of silicon photonics plasmonic modulators

### 2.3.3 State-of-the-art of Photonic Modulators

Since the first GHz silicon photonic modulator [72] a lot of modulators appeared in the following years. The active region was based on the carrier accumulation in a MOS capacitor. This was part of a MZI. A good state-of-the-art is done in [28] and [73]. They presented state-of-the-art of key silicon optical modulators.

In this section we summarized the main characteristics of key silicon photonics modulators. We only show a table with the main characteristics of the modulator. The structure that the modulator uses like Mach-Zehnder interferometer (MZI), ring resonator (RR) or disk resonator (DR); the operational frequency; the device footprint; the power consumption; the extinction ratio; the insertion losses; the working spectrum and the operational voltage. We are interested in particular in comparing the electrical power consumption of those devices with our FKE plasmonic modulator described in chapter five.

| Metrics/Device | Intel (2005) [74] | IBM (2007) [75] | Cornell (2007) [65], [77] | Sandia (2008) [78] | Cornell (2009) [79] | Kotura Inc. (2009) [73] | u-PSud 11, CEA-Leti (2012) [80] | A*STAR (2013) [81] |
|---|---|---|---|---|---|---|---|---|
| Structure | MZI | MZI | RR | DR | RR | RR | MZI | MZI |
| Bandwidth [GHz] | 30 | 10 | 12.5 | 10 | 3 | 10 | 40 | 20 |
| Insertion losses [dB] | 7 | 12 | 10 | 1.5 | 1 | 2 | 6 | 10.5 |
| Extinction ratio [dB] | 1 | - | 3-8 | - | - | 8 | 6.6 | 13.9 |
| Voltage [V] | 6.5 | 7.6 | 3.5 | 3.5 | 0.5 | 2 | 6-7 | 4 |
| Footprint [μm] | $10^4$ | $10^3$ | $10^2$ | 20 | $10^2$ | $10^3$ | - | $6 \times 10^3$ |
| Transient [fs] | - | - | - | - | - | - | - | - |
| Energy per bit [fJ/bit] | $3 \times 10^4$ | $5 \times 10^3$ | 300 | 85 | 86 | 50 | - | - |
| CMOS compatibility | Yes | Yes | - | Yes | Yes | Yes | Yes | Yes |

**Table 3: Comparative table of key silicon photonics modulators** [28], [73]

In Table 3 we can see that the modulators that use a MZI interferometer have a large footprint and relative high power consumption. Such is the case of the works presented in [74], [75] where the power consumption is around pJ/bit. The footprint is in the order of $10^3$-$10^4$ μm². Using a RR structure the final device is more compact and consequently the electrical power consumption is reduced around 50-100 fJ/bit. Such is the case of the works presented in [73], [77], [79]. Nevertheless, they have a smaller bandwidth due to the ring resonator structure.

# 2.4    Energy Consumption in Optical Modulators and Interconnects

It is believed that photonics can reduce the energy consumption in the transmission of information [82]. It will be designed for short distance inside computers [33] and other electronic machines. When the transistors are shrinked to increase the operational frequency of them, the electrical wires that connect them are also shirked. This increases their resistance and consequently the electrical power consumption. Additionally, it is also necessary to charge the intrinsic capacitance C seen by the wires. Using a photonic interconnect avoids this loss as electrical power is needed only to drive the source and the modulator. Optical amplifiers also consumes electrical power [83]. On the other hand, light is absorbed and scattered by the waveguides. Nevertheless, it is believed that an optical interconnect leads to a low electrical power consumption.

As explained in this chapter there are many ways of performing Si integrated modulators [28]. Modulators used the plasma dispersion effect including carrier accumulation, injection and depletion with different structures like MZI, RR and DR. Furthermore other electro-absorption modulators using the FKE or the QCSE were presented. We are interested in calculating the electrical power consumption of electro-absorption modulators that uses the FKE.

As stated before the electro-absorption modulators base on the FKE and the QCSE do not need to use resonator structures (MZI, RR or DR) to modulate the intensity light and this allows both low electrical power consumption and a high operational speed due to the fact that they are more compact. On the other hand, the modulators that exploit the carrier dispersion effect need a MZI or resonant structures like a RR or DR and this increases the power consumption which is around pJ/bit. This is due to an increase in size of the structures. It means, electro-absorption modulators are more compact than electro-refraction ones. It is worth noting that RR or DR are more compact and consumes less energy than MZI's modulators.

Most of the equivalent circuits of several modulators are represented by a capacitance C and a resistance R which is in serie with the mentioned capacitance C. To model the photocurrent and leakage current of the device a parallel current source is placed. The modulator that we are going to present in this work has such an equivalent circuit. It is represented in Figure 25. For example, in the case of electro-absorption modulators that use a PIN structure in reverse bias to induce a static electric field in the intrinsic region the capacitance of the device is mainly determined by the width of the intrinsic region. This depends in a small part in the voltage due to the change in the depletion of the n and p parts.

There are several sources of dissipating energy in the modulators. The main sources of power consumption are given by:

(1) The dynamic dissipation when the capacitance of the modulator is charged and discharged. The electrical equivalent circuits of most modulator (E.g. based on a PIN structure) are represented by a capacitance C and a resistance R. The electrical power consumption is related with the energy dissipated in the resistance R when the capacitance is charged and discharged. This is the electrical power consumption.
(2) Insertion loss of the optical energy while performing the modulation. It refers to the optical energy consumed when the light passes through the modulator.

And mainly present in electro-absorption modulators,

(3) The energy consumed by the photocurrent. The optical energy absorbed in the modulator by the material generates free electrons in the conduction band and free holes in the valence band. Since in the materials there is a static electric field these carriers are swept out of the modulator. This leads to additionally electrical power consumption into the modulator.

In this section we analyze the electrical power consumption due to the charge and discharge on the capacitance C of the modulator. It means the dynamic electrical power consumption.

In the modulator that we designed in this work we have a stack of p-Si/Ge/$Si_3N_4$/Cu. The FKE and the absorption of light happen in the Ge where the optical field is present. The Ge is intrinsic. A bias voltage is applied between the p-Si (p doped Si) layer and the Cu. This induces a static electric field into the Ge and the $Si_3N_4$. This structure is similar to a capacitor in which the plates are the p-Si layer in the bottom and the Cu contact on the top. The dielectric is formed by Ge and a thin layer of $Si_3N_4$. When the light passes through the modulator some of it is absorbed due to material absorption in Ge. Since the optical field is present in Ge. This may generate free carriers. Nevertheless, despite the fact that there is a strong static electric field into the Ge, photocurrent cannot be generated due the presence of the $Si_3N_4$ insulator in the path to the electrical source. The same passes with the leakage current of the capacitor, it is reduced due to the presence of the $Si_3N_4$ layer. Consequently, we can neglect the energy consumed by the photocurrent and the leakage current in this section.

The dynamic dissipation of the modulators is due to the current that flows in charging and discharging its capacitance. We remind that in our case the capacitance will be formed by the following stack of p-Si/Ge/$Si_3N_4$/Cu. The dissipation is produced by the resistor that is in serie with the capacitance. This resistor R is formed by all the series resistances present in the device. We also include the resistance of the device contacts and the driver circuit among others that may appear.

In order to model the dynamical power consumption of the modulators we need to show a circuit which is the responsible of charging and discharging the capacitance C of the modulator together with the resistance R. For this circuit we will consider a CMOS inverter circuit. This inverter is responsible of charging and discharging the capacitance C of the modulator between 0 V and $V_{DD}$. It is represented in Figure 25 (a). We selected this circuit because is the simplest one to drive the modulator. The electrical power consumption of the modulator will be independent of the circuit used to drive it. The only exception is the addition of a contribution of the output resistance $R_{out}$ of the CMOS inverter which is added to the series resistance of the modulator R. An equivalent circuit of the CMOS inverter and the modulator is presented in Figure 25 (b). The modulator is represented by the resistance R, the capacitance C and the current source $I_R$. The output resistance of the CMOS inverter $R_{out}$ is added to R. We advance that the electrical power consumption of the modulator will be independent of both R and the circuit used to drive it.

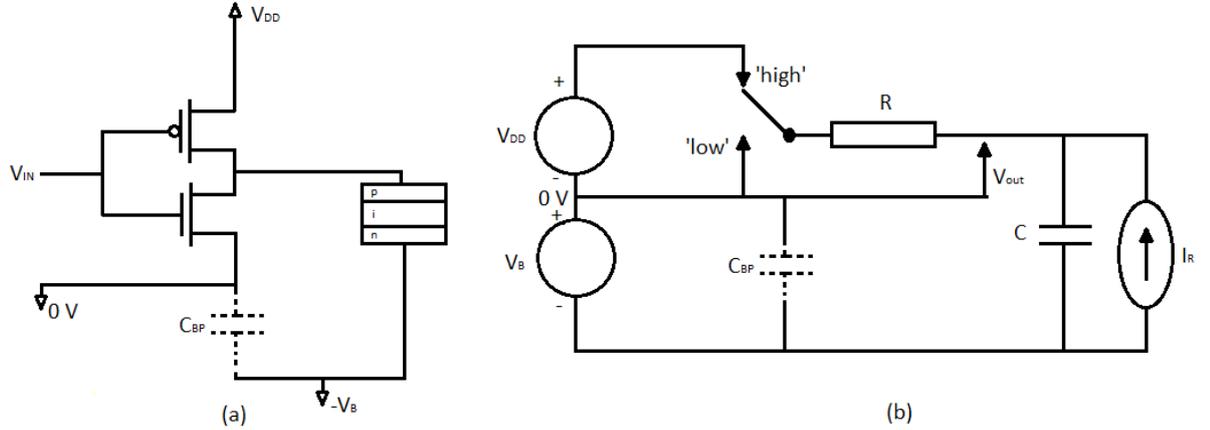

**Figure 25: Circuit to drive the modulator and its equivalent circuit**

The total charge that passes through the modulator is represented as $I_R$. It is a current source next to the capacitance C of the modulator. $I_R$ includes both the photocurrent and the leakage current. It is represented in Figure 25 (b). Although, in this derivation we neglect the leakage current and the photocurrent for the reasons explained at the beginning of this section. Briefly, it is due to the presence of the insulator layer of $Si_3N_4$ in the stack.

In Figure 25 $C_{BP}$ represents the bypass capacitance of the circuit. The purpose of the bypass capacitor $C_{BP}$ is to reduce the effective source impedance of power supplies. Furthermore, this capacitor $C_{BP}$ avoids that the energy supplied by $V_B$ is dissipated in the battery.

Another simplification done in this derivation is given by neglecting of the voltage drop $V_{drop}=I_R R$ in the resistor R due to the current $I_R$. For this we assume that the current $I_R$ is small. This is mainly due to the presence of the insulator layer of $Si_3N_4$ in the stack of the modulator.

When the driver circuit starts to modulate the circuit the voltage across the capacitance C changes from $V_B$ to $V_B+V_{DD}$. This is represented by the label 'high' and 'low' in the Figure 25 (b). In the 'high' state the voltage is $V_B+V_{DD}$ while in the 'low' state the voltage across the capacitance is $V_B$.

It is well known that the energy in a capacitance C at a voltage V is given by,

$$E = \frac{1}{2}CV^2 \qquad\qquad \text{Equation 11}$$

Then, the energy in the capacitance C of the modulator is given by $\Delta E_{CES} = \frac{1}{2}C[(V_B+V_{DD})^2 - V_B^2]$. Where $\Delta E_{CES}$ is the energy stored in the capacitor C. Operating over this equation, we obtain,

$$\Delta E_{CES} = \frac{1}{2} C V_{DD}^2 + C V_{DD} V_B$$

Equation 12

Now we would like to explain the physical reason of the first and second term of Equation 12. Note that the charge Q needed to drive the capacitance to a voltage $V_{DD}$ is given by $Q = CV_{DD}$. On the other hand, the energy given by a power supply at the voltage $V_{DD}$ is given by $\Delta E_{VDD} = QV_{DD} = CV_{DD}^2$. This is the first term that appears in Equation 12. Now we want to identify the second term of Equation 12.

Previously, the charge $Q = CV_{DD}$ was charged into the upper plate of the capacitor C by the source at $V_{DD}$. The same charge needs to appear at the bottom plate of the capacitor C. This plate is charged at a voltage of $V_B$. The energy employed by the source to charge this is given $\Delta E_B = QV_B = CV_{DD}V_B$ was $\Delta E_B$ is the energy employed by the source $V_B$ to charge the bottom plate of the capacitor C.

When a charge $\Delta Q$ passes through a resistor R at a voltage V the energy dissipated by the resistor is,

$$\Delta E_R = V \Delta Q$$

Equation 13

On the other hand, the change in voltage of a charge flowing throw a capacitor is given by,

$$\Delta Q = C \Delta V$$

Equation 14

Inserting Equation 14 into Equation 13 we obtain a new formula to calculate the energy dissipated due to charging a capacitance C from 0 to $V_{DD}$, it is: $\Delta E_R = C \int_0^{V_{DD}} V dV$. We will use this formula to calculate the energy dissipated into the resistor R. In the circuit of Figure 25 (b) the voltage V that drops in the resistor R is given by $V = V_{DD} - V_{OUT}$. So, the total energy dissipated in the resistor R due to the charge and discharge of the capacitance C from 0 to $V_{DD}$ is given by,

$$\Delta E_R = C \int_0^{V_{DD}} (V_{DD} - V_{OUT}) dV_{OUT}$$
$$= \frac{1}{2} C V_{DD}^2$$

Equation 15

It is worth noting to mention that the energy consumed in the resistor in Equation 15 is independent of the resistor R. Furthermore, the resistor can even be nonlinear.

Using Equation 15 we can see that the total energy dissipated by the resistance R both when the capacitor is charged and discharged is given by,

$$\Delta E_{DISS} = 2\Delta E_R = C V_{DD}^2$$

Equation 16

When the modulator is used to transmit information the number of charge and discharge cycles are given by the electrical signal that transmits the information. It is expected to have the same number of ones and zeros in the signal. Consequently, there is a complete charge, discharge of the modulator every 4 bits. Consequently, the electrical consumed energy per bit is given by,

$$\Delta E_{bit} = \frac{1}{4} C V_{DD}^2 \qquad\qquad \text{Equation 17}$$

Consequently, knowing the capacitance C of the modulator and the voltage V$_{DD}$ employed to perform the simulation we can calculate the electrical energy consumption per bit.

The objective of this thesis is to develop a low power consumption modulator. To achieve this reducing the electrical power consumption is a major objective. To achieve this objective first we selected an electro-absorption modulator because it consumes less than electro-refraction modulators. The main reason of this is that it is more compact.

Having an electro-absorption modulator we selected the FKE in Ge to be the active principle of the modulator. The FKE changes the optical absorption of the material close to the band-edge of it when a static electric field is applied in the bulk material. Consequently, with the design of the modulator we would like a structure that allows us to obtain a high static electric field with a small voltage. Reducing the operational voltage will reduce the electrical power consumption according to Equation 17.

To reduce the operational voltage and to have a high static electric field in the Ge we want to use a plasmonic waveguide in the core of the modulator. This will allow reducing the cross section of the device beyond the diffraction limit of light. We will have a more compact structure which is smaller than if we use a photonic waveguide. The electrical power consumption will be lower than in photonic devices due to the compactness of the plasmonic structure.

To summarize, we want to do an electro-absorption modulator that uses the FKE. Furthermore we want to design it in a plasmonic structure to reduce the electrical power consumption of the device. For this, we will introduce the field of plasmonic in the next chapter.

# 3 Plasmonics

In this section we introduce the field of plasmonics. We start describing the background of the topic with the traditional advantages and disadvantages of the field. We introduce the main plasmonic waveguides and finally we center into the metal-semiconductor waveguide and the metal-insulator-semiconductor one. The later is explained in a more detailed way. We also explain the selection of the metal and the insulator of this waveguide.

## 3.1    Background of Plasmonics

There are a lot of interests and ongoing researches in the field of plasmonics. Since the appearance of the bottleneck interconnects in the domain of electronic integrated circuits there are a lot of efforts in substituting electrical interconnects by optical ones.

The building blocks of silicon photonics are bigger than classical electronic components. The main reason is that silicon photonics devices are limited in size by the diffraction limit of light. This means that the light cannot be confined in a size smaller than $\lambda_0/(2n_{eff})$, where $\lambda_0$ is the wavelength of the optical beam in free space and $n_{eff}$ is the effective refractive index of the mode.

The relation between photonics and electronics is represented in Figure 26. The critical dimension and the operational frequency improvement are represented in the mentioned Figure 26.

Plasmonics tries to break the diffraction limit and offer subwavelength confinement. One of the first waveguides to provide a good confinement is based on the SOI technology. Due to the high difference in the refractive index between the cladding and the core they can confine the light in a small cross-section [84] but unfortunately it cannot break the diffraction limit of light. One example of the SOI technology is the strip or slab waveguides. The sizes are around 400 nm for the width and 200-300 nm for the height of them.

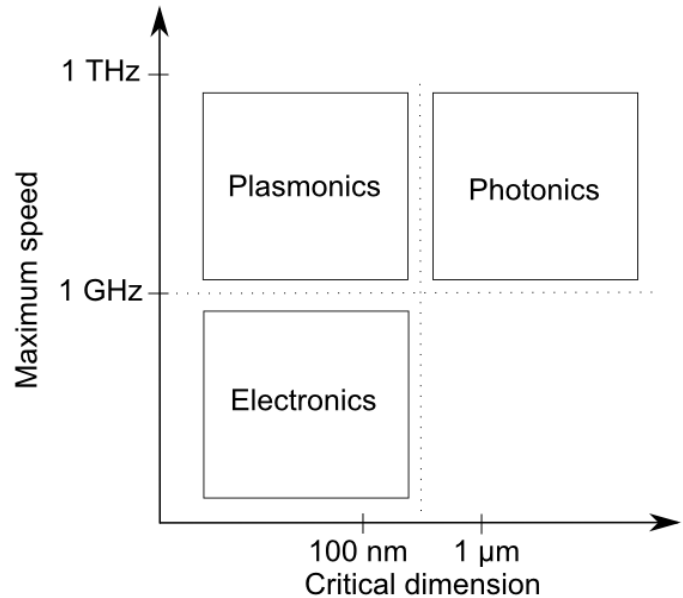

**Figure 26: Relation between electronics, photonics and plasmonics regarding the critical dimension of the feaures and the maximum speed that can be achieved**

There are also several designs of dielectric slot waveguides [85], [86], [87] in order to confine the light. They have a small dielectric slot in the middle of the slab waveguide in order to confine the optical electromagnetic field into a few nanometer dimensions. For example, the cross section of [85] is in Figure 27,

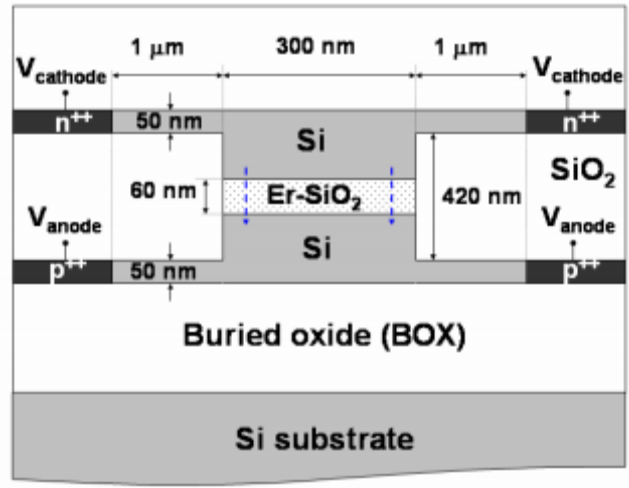

**Figure 27: Example of slot photonic waveguides to further confine the light in a smaller cross section [85]**

Combining metals with semiconductors and dielectrics may overcome the limitation of the diffractive limit of light [88] as contrast to the photonic waveguides shown in the previous paragraph. In this case, instead of having a photonic mode, one can use a plasmonic mode. A surface plasmon polariton (SPP) is a mode that propagates at the interface between a metal and a semiconductor. In a SPP the electromagnetic field couples to the longitudinal oscillations of electrons in the metal. A representation of a plasmonic mode is shown in Figure 28.

The optical electromagnetic field is well confined at the surface of the metal and decays exponentially in the dielectric and in the metal. The penetration depth in the metal is very small and it almost determines the level of losses of plasmonic modes. In the top of Figure 29 an optical fiber with a Si core and a $SiO_2$ cladding is represented. When the dimension d of the waveguide core is reduced to a size smaller than $\lambda_0$, the mode starts to be deconfined from the Si core of the mentioned waveguide. A high evanescent field appears in the $SiO_2$ cladding. In the case of the bottom of Figure 29, the waveguide has a metal core that is surrounded by $SiO_2$. When the core size d is reduced to a size smaller than $\lambda_0/(2n_{eff})$, the mode is also compressed below the diffraction limit of light. This is one of the main property of plasmonic modes.

The use of SPPs allows reducing the critical dimensions of plasmonic devices while conserving the high operational frequency of photonic ones [89]. This allows in principle the integration in the same integrated circuit with electrical components [90]. It was summarized in Figure 26.

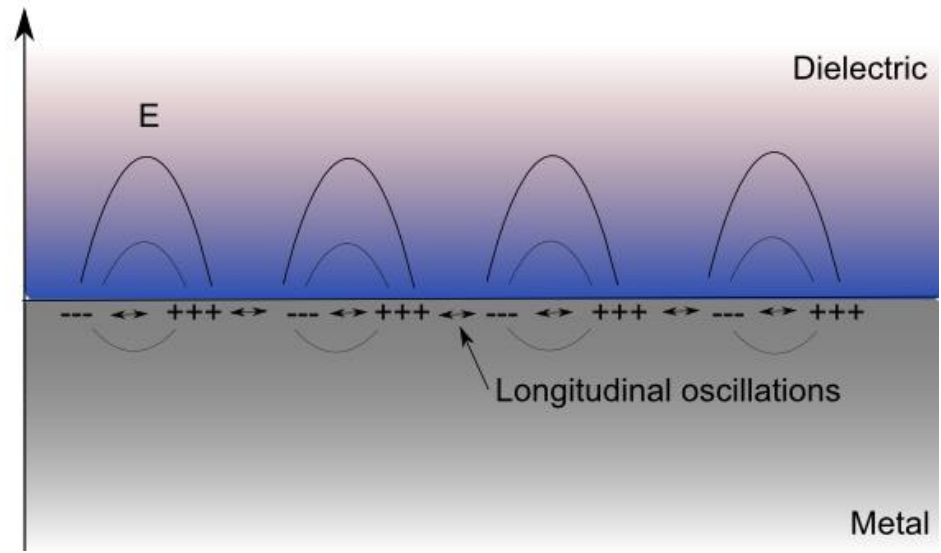

**Figure 28: Representation of the propagation of a SPP between the interface of a metal and a dielectric material. The longitudinal oscillations of the electron in the metal are represented. The electromagnetic field can penetrate more in the dielectric than in the metal**

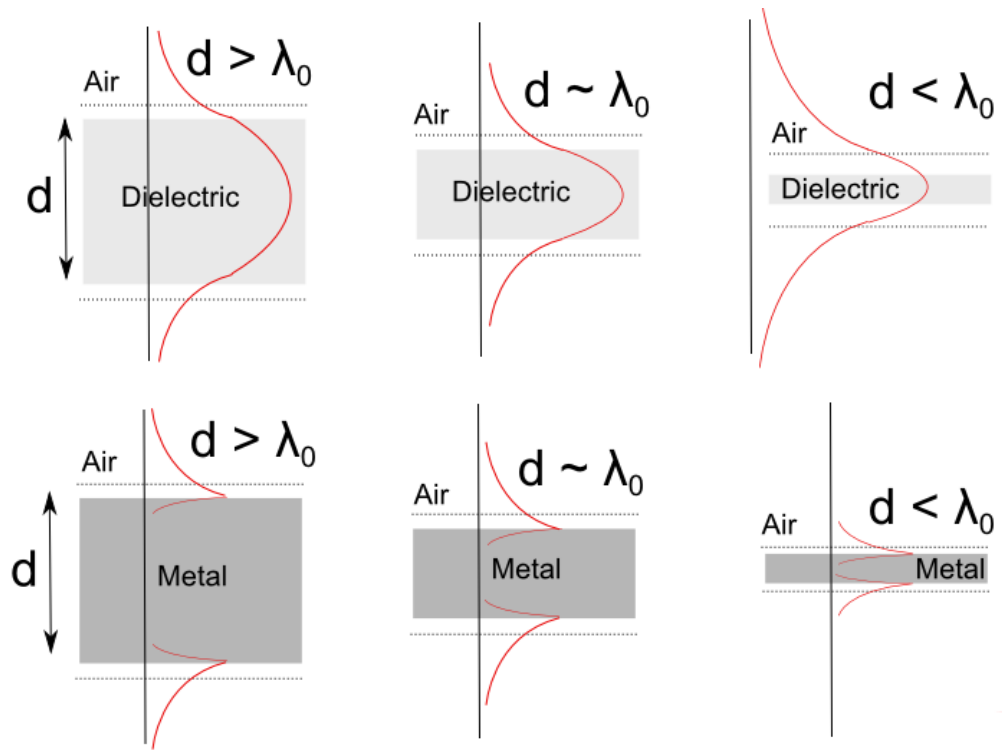

**Figure 29: Photonic and plasmonic mode shrinking and light confinement**

Since the reappearance of plasmonic in the year 2000 [91] several devices have been proposed including sources [92], [93], detectors, biosensors [94], waveguides [95], modulators (See Table 2), etc. Some of the devices use the strong optical field present between the metal and the dielectric to exploit weak effects like in nonlinear plasmonics [96] and in biosensing [94]. There are novel research directions like magneto-plasmonics [97], which combine plasmonics with magneto-optical materials; or graphene plasmonics [98].

# 3.2    Review of Plasmonic Waveguides

There are several plasmonic waveguides that were proposed [99]. It is also important that the structure is CMOS-compatible. The conductor must exhibit a negative permittivity in order to allow the subwavelength confinement of light. A detailed derivation of the permittivity of a metal and its properties is summarized in Appendix A.

There is another kind of photonic waveguides that use structures taken from microwave engineering. Some of them are like a coaxial waveguide [100]. In this kind of waveguides the dimension of the inner and outer metal of the coaxial are important. These kinds of waveguides can break the diffraction limit of light. The waveguide is illustrated in Figure 30. The dimension of this waveguide are around 224 nm for the inner radius a' and 286 nm for the outern radius h. With these dimensions the gap supports a plasmonic mode that is at subwavelength confinement.

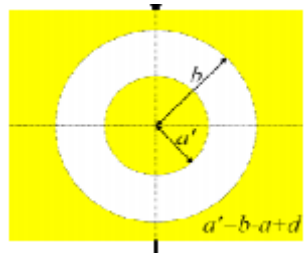

**Figure 30: Example of coaxial waveguide. The metal is represented in yellow and the dielectric in white. The inner radius is around 224 nm while the outern radius is around 286 nm for a typical configuration [100]**

The simplest waveguide consist of a metal-semiconductor (MS) structure (illustrated in Figure 141 in Appendix B). Furthermore a thin metallic film between two semiconductors can support the well known as short-range and long-range plasmonic modes [101]. Other proposed plasmonic waveguides are strip [102] (Figure 31 (a)), V-grooves [103] (Figure 31 (b)), straight groove [104] (Figure 31 (c)), and gap waveguides [104] (Figure 31 (d)). There are several gap waveguides like a slot plasmonic waveguide [105].

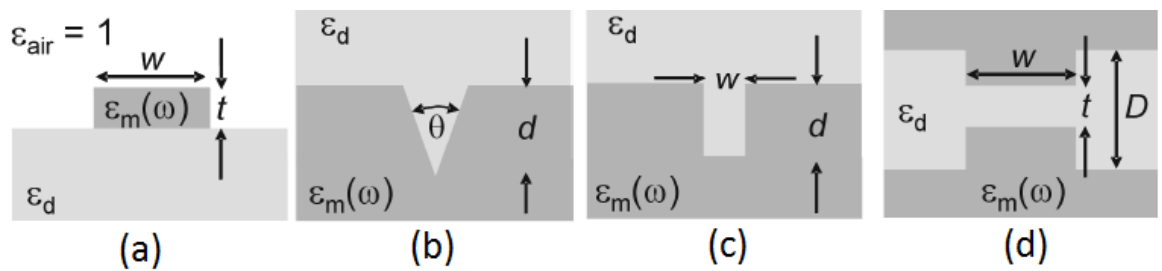

**Figure 31: Example of plasmonic waveguides [104]**

There are also slot waveguides in which a dielectric is embedded between two metals. They are known as gap plasmon polariton [106]. The main problem of this kind of waveguide is the high propagation losses although they have a pretty small confinement (the cross section may be around 50x50 nm). It is represented in Figure 32.

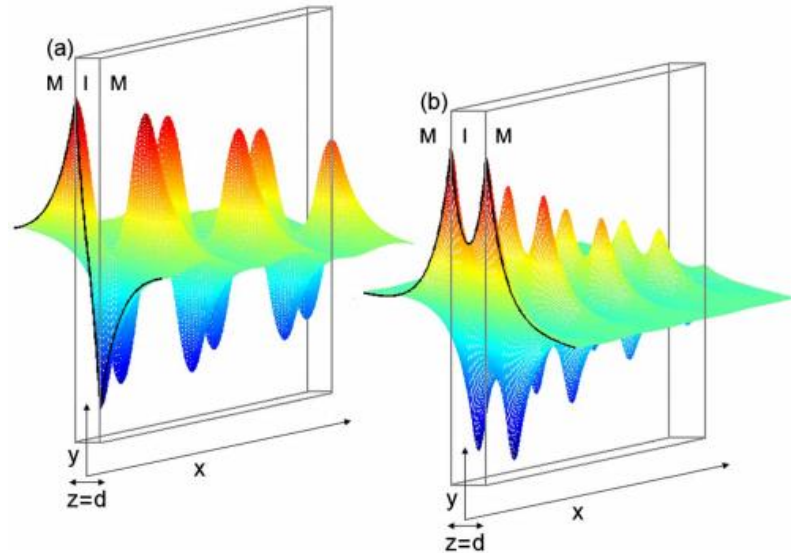

**Figure 32: Example of metal-insulato-metal (MIM) waveguide. The antisymmetric and symetric plasmons are represented [105]**

A new plasmonic waveguide that consist of a dielectric cylinder separated by a small gap from a metal surface [107] provides a new way of confining the light and reducing the propagation losses. It is illustrated in Figure 33. The presence of the small gap takes the optical electromagnetic field out of the metal. Due to this the propagation losses are reduced. The dimension of d is around 100-200 nm while h is between 5-20 nm for the gap.

Furthermore, in the small insulator gap the optical electromagnetic field is well confined. The gap has a dimension of few nanometers (5-20 nm). This waveguide [107] leads to a new kind of plasmonics waveguides that are basically composed by a few nanometer slot insulator between a metal and a dielectric [108], [109], [110], [111], [112], [113]. This structure is known as the MIS waveguide. It is represented in Figure 35.

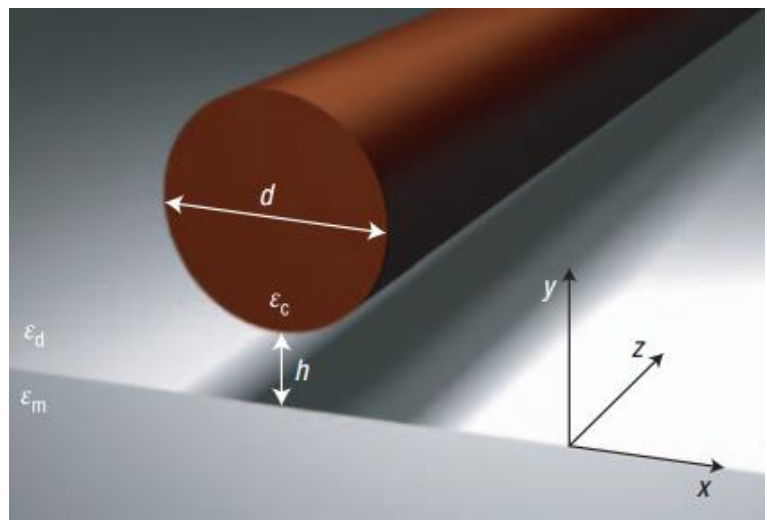

**Figure 33: First plasmonic hybrid waveguide [107]**

In the following sections we present the main trade-off of plasmonic waveguides which is the trade-off between the confinement and the propagation losses (smaller confinements provide higher propagation losses). We also introduce two interesting waveguides to perform a FKE plasmonic modulator. They are the metal-semiconductor (MS) waveguide and the metal-insulator-semiconductor waveguide (MIS).

We will select the MIS waveguide in our device due to the fact that it is CMOS compatible and that it provides relatibly low propagation losses which will lead to low insertion losses of the modulator.

# 3.3    Trade-off in Plasmonics Waveguides

Trying to merge silicon photonics with plasmonics will allow reducing the size mismatch of photonic devices and electronic ones. This will facilitate the integration of both photonic and electronic devices in the same integrated circuit [114], [115]. The mentioned integrated opto-electronic circuit using plasmonic will be formed by passive devices like plasmonic waveguides, multiplexers, demultiplexers, etc. It will also include active devices like sources, photodetectors and modulators.

The extreme localization of SPP allows reducing the size of photonic devices breaking the diffraction limit of light and presenting a new concept in opto-electronic integration. The use of metal for propagating SPP can allow its use for propagating light as well as an electrical contact.

The use of a metal to propagate light does not increase the RC delay time like in electrical interconnects. Furthermore, the current technology provides new techniques to fabricate small metallic features. This will further allow integrating both plasmonic and electronic devices in the same integrated circuit.

The relation between the optical propagation losses in a photonic and in a plasmonic waveguide is represented in Figure 34. There is a trade-off between the mode confinement and the optical propagation losses of the plasmonic mode. It means, the smaller the confinement the higher the mode losses. It means that the higher the optical confinement the higher the optical propagation losses are.

Another source of losses in plasmonic waveguides is due to the scattering produced by the roughness of the metal layer at the interface with the semiconductor or insulator. If the interface of the metal-dielectric presents a high level of roughness then this also increases the propagation losses of the plasmonic mode.

One of the leading lines of plasmonic is to try to reduce the losses of the modes. Several techniques were proposed like using a material with gain in the dielectric part of the waveguide (Figure 28) [116], [117] or using a small slot of $SiO_2$ or $Si_3N_4$ between the metal and the semiconductor [107]. Other materials like $HfO_2$ or $Al_2O_3$ are also used.

In the case of using a gain material to compensate the losses the main idea is that this material compensate the propagation losses of the mode due to the metal present in plasmonic waveguides and devices. Nevertheless, since the optical losses of the plasmons are very high it is difficult to find a material with such a big gain. Furthermore, if we want to fabricate the device with CMOS processes then this restriction reduces the possible materials that can be used for compensating the propagation losses.

In the case of using a small slot of $SiO_2$ or $Si_3N_4$ between the metal and the semiconductor in a plasmonics waveguide we form a so called metal-insulator-semiconductor (MIS) waveguide. The use of the insulator slot reduces the propagation losses since it pushes out the electromagnetic field from the metal. This kind of plasmonic waveguide will be presented at the end of this chapter.

To summarize, the main propagation losses in a plasmonic waveguide is due to the penetration of the electromagnetic field of the mode into the metal. Another source of optical loss is the scattering of the interface between the metal and the insulator or semiconductor used. Additionally, we have the absorption of the semiconductor material (E.g.: Ge) in some cases although this absorption can be neglected in the majority of the cases. Furthermore, the grain of the metal plays an important role in the optical losses of the modes. Among all these losses the main ones are the ones due to the damping factor of the metal used (penetration of the optical mode into the metal). Consequently, to design plasmonics devices a careful choice of the metal must be taken.

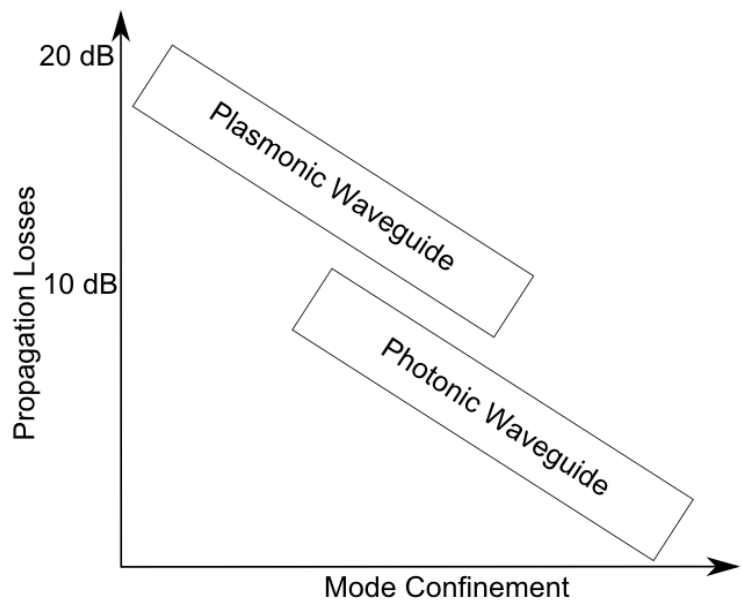

**Figure 34: Relation between propagation losses in a plasmonic and in a photonic waveguides**

# 3.4   Surface Plasmon Polaritons at a Metal-Dielectric Interface

We remind that an SPP is electromagnetic wave which is coupled to longitudinal oscillations of electrons in a metal. The simplest waveguide to propagate a SPP is the metal-semiconductor (MS) interface. An example of a MS structure is represented in Figure 141 (Apendix B).

To obtain the dispersion relationship of this plasmonic mode, one needs to solve the Maxwell's equations and apply the proper boundary conditions in the MS waveguide. It is interesting to calculate the propagation losses of the supported plasmonic mode $\alpha_{MS}$. Another interesting parameter is the penetration depth of the electromagnetic field of the plasmonic mode into the metal $l_1$ and into the semiconductor $l_2$. It is also interesting to know some reference for the propagation losses to try to reduce the value. It is explained in the next section that adding a small (3-15 nm) insulator layer between the metal and the semiconductor of the MS waveguide will reduce the propagation losses of the plasmon. This waveguide is known as a MIS plasmonic waveguide. We will present it in the next section.

The losses $\alpha_{MS}$ of an SPP in an MS waveguide (derivation in Appendix B) is given by,

$$\alpha_{MS} \approx \frac{2\pi}{\lambda} \frac{n_1^3 \varepsilon_2''}{\varepsilon_2'^2}$$

Equation 18

Where $n_1$ is the refractive index of the semiconductor, $\varepsilon_2 = \varepsilon_2' + \varepsilon_2''$ is the permittivity of the metal and $\lambda$ is the wavelength of the light in free space. Note that $\alpha_{MS}$ depends on the third-power of the dielectric's refractive index $n_1$.

The penetration depth into the dielectric (derivation in Appendix B) is given by,

$$l_1 = \frac{\lambda}{2\pi} \frac{1}{Re\left(\sqrt{n_{MS}^2 - n_1^2}\right)} \approx \frac{\lambda}{2\pi} \frac{\sqrt{|\varepsilon_2'|}}{n_1^2}$$

Equation 19

To obtain $l_1$ the distance at which the intensity of the optical electric field decays by a value of $1/e$ from the interface of the MS waveguide is calculated. The distance is measured perpendicular to the interface between the semiconductor and the metal. In a similar way the penetration depth of the SPP in the metal (derivation in Appendix B) is calculated. It is given by,

$$l_2 = \frac{\lambda}{2\pi} \frac{1}{Re\left(\sqrt{n_{MS}^2 - \varepsilon_2}\right)} \approx \frac{\lambda}{2\pi} \frac{1}{\sqrt{|\varepsilon_2'|}}$$

Equation 20

From Equation 19 and Equation 20 we see that the SPP in the MS waveguide is confined close to the interface between the metal and the dielectric since $l_1 \sim 1/n_1^2$. The penetration depth $l_2$ is even smaller. The maximum of the electromagnetic field that propagated through the waveguide is present at the interface.

We have presented the main characteristic of the MS waveguide. In the following section we introduce a small slot between the metal and the semiconductor is order to reduce the propagation losses and have a stronger confinement of the light into the waveguide. Comparing with Equation 18 we will see that the propagation losses are reduced.

# 3.5    Metal-Insulator-Semiconductor Waveguide

The MIS waveguide is a plasmonic waveguide formed by a stack of Metal-Insulator-Semiconductor (MIS). The insulator has a subwavelength thickness. Typically it is between 3 to 15 nm. The structure is represented in Figure 35. This waveguide has the property that the optical electric field is well confined in the subwavelength slot between the metal and the semiconductor.

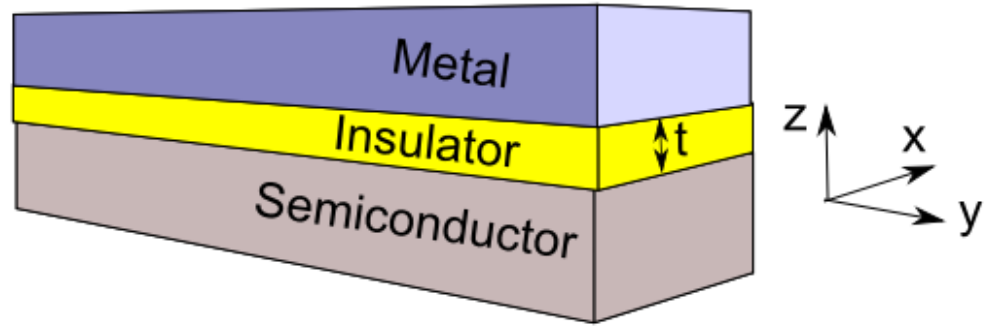

**Figure 35: Metal-Insulator-Semiconductor (MIS) waveguide. Both the semiconductor and the metal are semi-infinite**

The structure of Figure 35 is formed by an insulator of electrical permittivity $\varepsilon_{gap}$ and thickness t. It is between a semi-infinite semiconductor of electrical permittivity $\varepsilon_1$ and a semi-infinite metal of electrical permittivity $\varepsilon_2 = \varepsilon_2' + i\,\varepsilon_2''$. The only finite parameter in the structure is t. When the parameter t is set to t=0 then we have the structure of a Metal-Semiconductor (MS) waveguide that was studied in the previous section. [118] gives a complete analysis of the MIS structure. We summarize the main results in this section.

In Figure 36 we represent the effective refractive index $n_{eff}$ of the plasmonic mode supported by the MIS waveguide and the propagation effective losses $\alpha_{eff}$. Both parameters are ploted versus the parameter t. We also present $\alpha_{eff}$ versus t in Figure 36 (b). To obtain Figure 36, 2D simulations were performed in [118] to obtain these results.

From Figure 36 (b) we observe that the losses are lower when t is below a certain value $t_{min}$ and bigger when it is over another value of t denoted $t_{max}$. For some values of t there is not a supported plasmonic mode. It means for $t_{min} < t < t_{max}$ there is not supported plamonic mode by the waveguide.

In the interval $0 < t < t_{min}$ we can see that we obtain a low loss plasmonic mode confined in the gap (bottom figure in Figure 36, it is noted as a MIS mode). The effective refractive index of the mode $n_{MIS}$ is bigger than the dielectric in $0 < t < t_{min}$. This mode is called the MIS mode.

If t tends to infinite an SPP between the metal and the insulator is obtained. When t is decreased from $t_{min}$ to lower values we obtain a so-called MIS mode which effective refractive index $n_{MIS}$. When t reaches a value $t_{min}$ then this mode disappears (top figure in Figure 36). When a value called $t_{max}$ is overtaken then a LG-MIS mode (Large Gap-MIS) is supported.

We remind that in the regime $t_{min} < t < t_{max}$ there is not solution for any mode in the MIS waveguide. The effective refractive index of the mode in the MIS waveguide has a gap in the region $t_{min} < t < t_{max}$.

Note that the MIS mode has fewer losses than the LG-MIS. In this case we are interested in using t<$t_{min}$ to reduce the propagation losses of the modulator which use a MIS waveguide to propagate the light.

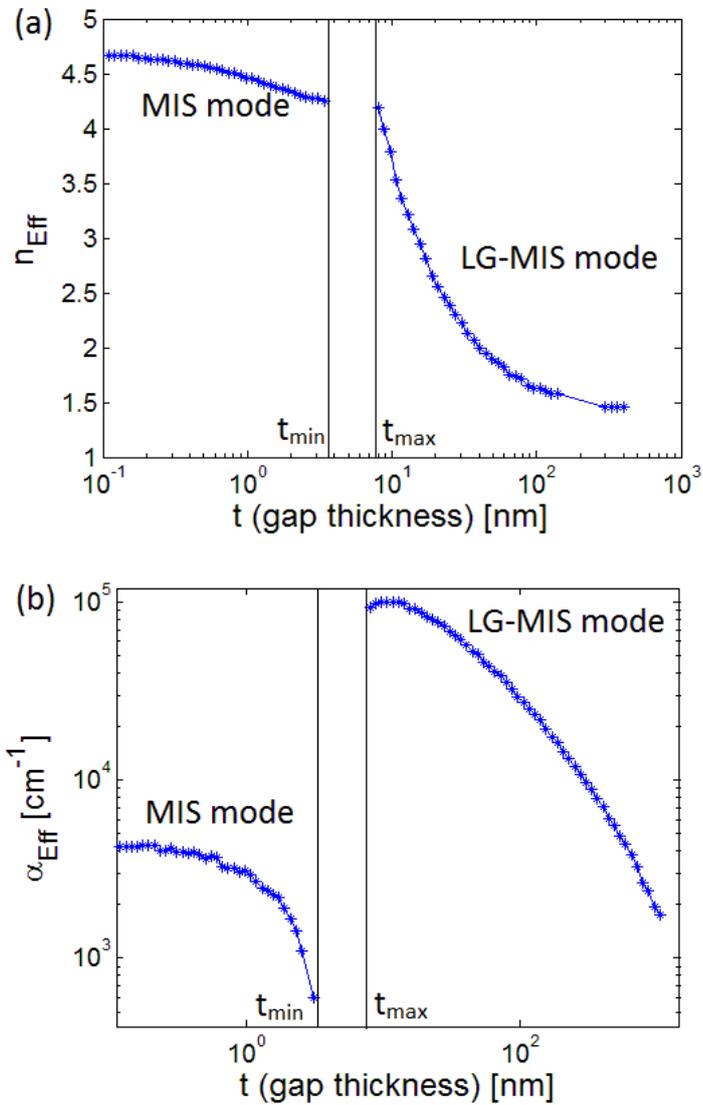

**Figure 36: Analysis of the effective refractive index and the losses of the MIS waveguide [118]**

The MIS waveguide supports only a plasmonic TM mode. Due to this mode condition and the boundary condition of the normal component of the electric displacement vector between the insulator and the semiconductor, it is expected to obtain a high optical electric field inside the slot.

The value of $t_{min}$ is given in [118],

$t_{min}$

$$\approx Re\left\{\frac{\lambda}{4\pi\sqrt{\varepsilon_1-\varepsilon_{gap}}}ln\left(\frac{\varepsilon_2\sqrt{\varepsilon_1-\varepsilon_{gap}}-\varepsilon_{gap}\sqrt{\varepsilon_1-\varepsilon_2}}{\varepsilon_2\sqrt{\varepsilon_1-\varepsilon_{gap}}+\varepsilon_{gap}\sqrt{\varepsilon_1-\varepsilon_2}}\right)\right\}$$

Equation 21

$$\approx \frac{\lambda}{4\pi\sqrt{\varepsilon_1-\varepsilon_{gap}}}ln\left(\frac{\varepsilon_2'\sqrt{\varepsilon_1-\varepsilon_{gap}}-\varepsilon_{gap}\sqrt{\varepsilon_1-\varepsilon_2'}}{\varepsilon_2'\sqrt{\varepsilon_1-\varepsilon_{gap}}+\varepsilon_{gap}\sqrt{\varepsilon_1-\varepsilon_2'}}\right)$$

For,

$$\varepsilon_1 \ll |\varepsilon_2'|$$

Equation 22

The Equation 21 can be simplified to,

$$t_{min} \approx \frac{\lambda}{2\pi}\frac{\varepsilon_{gap}}{\sqrt{|\varepsilon_2'|}}\frac{1}{\varepsilon_1-\varepsilon_{gap}}$$

The real part of the effective refractive index $n_{MIS}$ is given by,

$$Re(n_{MIS}) \approx n_1 + \left[Re\left(\sqrt{\frac{\varepsilon_1\varepsilon_2}{\varepsilon_1+\varepsilon_2}}\right)-n_d\right]\left[1-\frac{t}{t_{min}}\right]^2$$

Equation 23

Using the approximation of Equation 22 and for the regime $0<t<t_{min}$ we can simplify Equation 23 to,

$$Re(n_{MIS}) \approx n_1\left[1+\frac{\varepsilon_1}{2}\left(\frac{\varepsilon_1-\varepsilon_{gap}}{\varepsilon_{gap}}\frac{2\pi}{\lambda}(t_{min}-t)\right)^2\right]$$

Equation 24

Under the same approximation of Equation 22 the losses $\alpha_{MIS}$ of the MIS plasmonic waveguide are giving by,

$$\alpha_{MIS} \approx \frac{4\pi}{\lambda}Im\left(\sqrt{\frac{\varepsilon_1\varepsilon_2}{\varepsilon_1+\varepsilon_2}}\right)\left(1-\frac{t}{t_{min}}\right)$$

Equation 25

Using the previous models we can approximate now the effective propagation losses of the MS waveguide $\alpha_{MS}$ (Equation 18) and the effective propagation losses of the MIS waveguide $\alpha_{MIS}$ (Equation 25). With this, using a stack of Ge as semiconductor and Cu as metal we obtain $\alpha_{MS}$=30119 cm$^{-1}$. On the other hand, using a MIS waveguide we obtain a value of $\alpha_{MIS}$=2410 cm$^{-1}$ using Equation 25. In this case the stack of the MIS is Cu as the metal, Si$_3$N$_4$ as the insulator and Ge as the semiconductor. The values used in the calculation are for Si$_3$N$_4$ $n_{Si3N4}$=1.9827 [119], for Cu $n_{Cu}$=0.3+11.1i [120] and for Ge $n_{Ge}$=4.23 [119] which corresponds to a wavelength of 1.647 μm.

In this section we only wanted to do a rought comparison between the effective propagation losses in the MS and in the MIS. The selection of the materials will be explained in the following section. Now, we want to conclude that the MIS waveguide has less propagation losses than the MS. Furthermore, the MIS is CMOS compatible.

# 3.6    Choice of the Materials of the MIS Waveguide

In this thesis we want to design a CMOS compatible plasmonic modulator. For this, it is worth noting to mention that it is important to carefully select the materials.

We design a plasmonic modulator that is composed by a MIS waveguide (Figure 35). As stated in the previous section, it is important to know which one is the metal and the insulator used to assure CMOS compatibility. Depending of those materials the design will be CMOS compatible or not.

In this section we discuss the material used to fabricate the CMOS compatible plasmonic modulator in terms of materials for both the metal and the insulator. We also want to obtain low optical loss for the plasmonic mode and good electrical reliability. The material for the semiconductor is already fixed since we want to use the FKE as the active principle of the modulator. Consequently, Ge is used.

## 3.6.1 Selection of the CMOS Compatible Metal

As explained in the previous section the losses of a plasmonic mode are mainly determined by the penetration of the optical electromagnetic field into the metal. Another critical issue that determines the losses of the plasmonic mode is the crystalline quality of the metal deposited over the insulator in the MIS waveguide.

One of the main disadvantages of plasmonic devices is that they use metals at optical frequencies. At those frequencies all metals (E.g.: Au, Ag, Al and Cu) possess a damping factor. This damping effect induces strong losses into the plasmonic modes that are much higher than the photonic ones.

Ttraditionally for several plasmonic devices [56], [121], [122] silver (Ag) and gold (Au) were used. It is worth noting that those metals are not compatible with a CMOS environment since they are killer contaminants of semiconductors like Si or Ge.

As a consequence, we discard them. On the other hand, in a CMOS environment, industry uses metals such as aluminum (Al) and Cu. Nevertheless, those metals are not much studied to design plasmonic devices [65], [120].

In the following Table 4 we summarize the refractive indices of the main metals at 1.55 μm and 1.647 μm. We present the refractive index of the materials at 1.55 μm because it is the telecommunication wavelength. Furthermore, we present it at 1.647 μm because at this wavelength the FKE is maximum [46]. When we note $Cu_{CEA-Leti}$ we refer to the Cu used in [120].

| Metal: | Ag [119] | Al [119] | Au [119] | Cu [119] | Cu$_{CEA\text{-}Leti}$ [120] |
|---|---|---|---|---|---|
| n (@1.55 μm): | 0.41+10.05i | 1.51+15.23i | 0.58+9.86i | 0.73+10.40i | - |
| n (@1.647 μm): | 0.44+10.69i | 1.62+16.24i | 0.63+10.49i | 0.79+11.04i | 0.3+11.1i |
| CMOS compatible: | | X | | X | X |

**Table 4: Efective refractive index of the metals for plasmonics**

In Figure 37 we plotted the propagation losses in a MS structure using Al and Cu. We do not plot Au and Ag because they are not CMOS compatible; consequently, we do not use them. The losses are plotted as a function of the wavelength of the light. For plotting the losses we used the following equation,

$$\alpha_{MS} = \frac{1}{4.3 L_{MS}} = \frac{2\pi}{4.3 \lambda_0} \sqrt{\frac{\varepsilon}{1+\varepsilon}}$$

Equation 26

Where, $\alpha_{MS}$ is the losses of the plasmonic mode in the MS waveguide in dB/μm, $\lambda_0$ is the wavelength of the light beam in free space and ε is the permittivity of the metal. Equation 26 describes the losses of a plasmonic mode that propagates in a MS. For calculating ε we used the special low loss Cu presented in [120] that was measured by ellipsometry ($n_{Cu}$=0.3+11.1i). For the value of Al ($n_{Al}$=1.62+16.24i) we took the value from Palik [119] both at $\lambda_0$=1.647 μm.

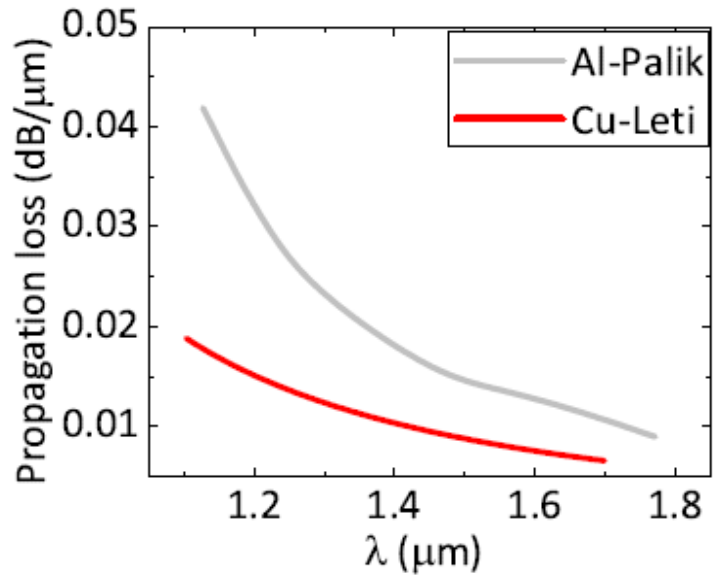

**Figure 37: Optical losses for Al and Cu in a MS plasmonic structure [123]**

From Figure 37 we see that the losses of the Cu (with the special process [120]) are lower than the losses of Al. According to this result, this Cu seems to be a better material than Al to develop a CMOS compatible plasmonic modulator. Furthermore, it is demonstrated in [120] that the Cu with the special process has lower optical losses than the one without the process reported on [119].

It is also known that in metallic interconnects the use of Al is being replaced by the use of Cu progressively [124]. Consequently Cu seems to be the best metal for the development of a CMOS compatible plasmonic modulator. It will be used as a metal in the MIS waveguide of the modulator that we present in this work.

In the following section we show why the use of the material $Si_3N_4$ is selected as the material for the insulator layer once we know that we will use $Cu_{CEA-Leti}$ [120] for the metal.

## 3.6.2 Selection of the Insulator Layer

Now we need to select the insulator layer between the metal and the semiconductor. Since we use the FKE as an active principle to perform the modulation we use Ge as the semiconductor.

In [123] a description of the selection of an insulator or diffusion barrier for a MIS waveguide is presented. The most common insulation and diffusion barriers for Cu in the electronic industry are: Ta, TaN, W, Pd, Ti and TiN [125], [126], [127].

Those materials can act both as a conductor and as a diffusion barrier. This is why they are widely used in the electronic industry. However the optical losses that they present are very high for an optical application like our plasmonic modulator.

In Table 5 we present the optical losses of the diffusion barriers that are employed between the metal and the semiconductor in the structures similar to the MIS waveguide. From this in [123] the optical losses of the mode in dB/$\mu$m are calculated. To obtain the values of Table 5 [123] the metal used is $Cu_{CEA-Leti}$ and the semiconductor used is Si. In [123] there is not information of the thickness of the diffusion barrier used.

From the results of Table 5 we can conclude that the use of a diffusion barrier done with Ti, TiN, Ta or TaN is higher than the ones without barrier case. This is due to the optical losses of such materials. We can also see that the best result is obtained for the $Si_3N_4$ barrier. A similar result is obtained by $SiO_2$ nevertheless its value is worser.

| Diffusion Barrier: | No barrier | Ti | TiN | Ta | TaN | $Si_3N_4$ | $SiO_2$ |
|---|---|---|---|---|---|---|---|
| Optical losses [dB/$\mu$m]: | 0.29 | 3 | 3.61 | 1 | 3.3 | 0.2 | 0.26 |

**Table 5: Optical losses in a MIS waveguide for the different diffusion barriers that can be used [123]**

In the metallic interconnects literature we can also see the use of non-conductive insulator materials like SiCO or SiCN [128]. Nevertheless both SiCO and SiCN have poor adhesion to Cu which difficult the integration into an integrated circuit. On the other hand, Cu has a good adhesion to $Si_3N_4$ and acts as a good diffusion barrier but it is difficult to obtain thin films.

As the discussion above we select $Si_3N_4$ as the insulator layer for the MIS waveguide of the plasmonic modulator that we present in this work. It will both reduce the losses of the guided mode with respect to the Cu-Ge MS structure and act as a barrier against the diffusion of Cu into Ge.

# 3.7    Conclusion

In this thesis the core of the plasmonic modulator will be formed by a MIS waveguide in which the stack is formed by cupper (Cu) as the metal, silicon nitride ($Si_3N_4$) as the insulator in the slot and germanium (Ge) as the semiconductor. We explain why we selected a $Si_3N_4$ material for the insulator in the previous section. We want to use a MIS waveguide to perform the modulation because with this waveguide you can achieve low propagation losses (it is linked to the insertion losses of a modulator) and the structure is easy changeable to induce a static electric field into the core of it (the semiconductor) to exploit the FKE in Ge. Thinking in an electro-absorption modulator, if you place a material in the core which changes the optical losses with a static electric field (E.g. Franzs-Keldysh effect in Ge, etc.) you can perform a modulation by using this effect. We selected the MIS waveguide instead of the MS one because the propagation losses of the MIS are smaller than in MS. Another point in favor of the MIS waveguide is that it is CMOS compatible.

Regarding other waveguides, the slot waveguide was not selected because although it is easy to induce a static electric field in the core it has significant higher propagation losses than the MIS. The other mentioned waveguides like coaxial, V-grooves, ridges, metallic wires, etc. were not selected because either they are difficult to integrate into a chip or it is difficult to induce a static electric field into the core.

As a first attempt we selected Ge in the core of the MIS waveguide to use the Franz-Keldysh effect to design an electro-absorption modulator. We also used $Cu_{CEA-Leti}$ as the metal to propagate the plasmon because it is the metal with less propagation losses [120]. The selection of the metal is further explained in chapter five.

This Cu-$Si_3N_4$-Ge stack will support a TM plasmonic mode. We want to work in the low-loss regime of the mode. For working in the low-loss regime of the MIS waveguide we need to select the thickness of the insulator layer t below the value of $t_{min}$ which is given by Equation 21. In Figure 36, in the graph below, we see that for small values of t (gap thickness) the losses of the plasmonic mode supported are smaller than for bigger values of t. The value $t_{min}$ is the one which assures that we work in the low loss regime of the MIS waveguide. Reducing the propagation losses of the MIS waveguide will reduce the insertion losses of the modulator.

For this we calculate the value of $t_{min}$ using Equation 21. We obtained a value of $t_{min}$=8.35 nm. So, our $Si_3N_4$ slot will have a thickness t<$t_{min}$= 8.35 nm. Since we want to work in the low loss regime of the MIS waveguide we selected the t of the thickness around 5 nm. This is thin enough for the $Si_3N_4$ layer to act as a diffusion barrier of the Cu into the Ge, furthermore, since it is below $t_{min}$=8.35 nm we assure that it is in the low loss regime of the MIS waveguide. In the case of using $SiO_2$ as an insulator in the MIS waveguide the value is $t_{min}$=4.42 nm. In this case we see that it is below 5 nm which is too small regarding the tolerance of the fabrication techniques used for the fabrication. Additionally, it is difficult to use as a diffusion barrier. In this case, the $Cu_{CEA-Leti}$ will diffuse into the Ge and will increase the propagation losses of the plasmon.

With this value ($t_{min}$=8.35 nm) we can obtain the effective refractive index $n_{MIS}$ using Equation 24 and the propagation losses $\alpha_{MIS}$ using Equation 25. The obtained values are $n_{MIS}$=4.31 and $\alpha_{MIS}$= 2410 $cm^{-1}$. With this we assure that we are working in the low loss regime of the plasmonic MIS waveguide. We compare those theoretical values with the optical simulation that we will perform in chapter four. This value ($\alpha_{MIS}$=2410 $cm^{-1}$) is smaller than the one given by the MS structure using Equation 18 which is around $\alpha_{MS}$=30119 $cm^{-1}$.

The values used in this section are for silicon nitride $n_{Si3N4}$= 1.9827 [119], for silicon oxide $n_{SiO2}$= 1.4476 [119], for cupper $n_{Cu}$=0.3+11.1i [120] and for Ge $n_{Ge}$=4.23 [119].

# 4 The Franz-Keldysh Effect

In this thesis we propose a plasmonic modulator that uses the Franz-Keldysh effect (FKE). In this section we explain in details the FKE. We present the basics of the FKE and the physical model used to simulate the effect. In our device we use the FKE in Ge.

## 4.1　　Introduction

The FKE is the change in the optical absorption coefficient of a material under a static electric field applied on it. The optical absorption changes the material losses near the band-edge of the material (at the bottom of it), around the direct bandgap. We have already presented the FKE in the introductory chapter of section 2.2.4.

The reason for designing an electro-absorption modulator using the FKE is to try to reduce the electrical power consumption of the modulator below 50 fJ/bit [33]. As it is well known, electro-refractive modulators that use a MZI have power consumption around pJ/bit. Using a RR the energy consumption can be reduced around 50-300 fJ/bit but it suffers from small bandwidth. Starting from an electro-absorption modulator and using plasmonic we will reduce the size of the modulator. It is expected that when reducing the size, the electrical power consumption is also reduced. Furthermore due to the large optical confinement associated with plasmons, there is more interaction of the field in the structure, leading in principle to a shorter modulator.

Now we are going to present more in details the FKE in this chapter. There are two models of the FKE. The first one is simpler and it is called the generalized theory of the FKE. This model is explained in the section 4.2 of this chapter. There is also a more complete model which also predicts the absorption of the material. It is explained in section 4.3. For our work we use this latter model. We do not use the generalized model in our simulations since it is difficult to find proper parameters used in the equations of the model to match with measurements present in the literature [46].

# 4.2    Generalized Model of the Franz-Keldysh Effect

In [129] the generalized theory of the FKE is introduced. This takes into account the presence of a static electric field in the material which produces the change in the absorption coefficient of it.

The change in the permittivity $\Delta\varepsilon$ of the material when a static electric field is present in the medium is defined as,

$$\Delta\varepsilon(E,F) = \varepsilon(E,F) - \varepsilon(E,0)$$
<div align="right">Equation 27</div>

Where E is the energy of the photon beam in the material and F is the static electric field in the medium.

It was shown in [129] that according to the FKE Equation 27 can be written as,

$$\Delta\varepsilon(E,F) = \frac{B}{E^2}\sqrt{\hbar\theta}[G(\eta) + iF(\eta)]$$
<div align="right">Equation 28</div>

Where B is a value related with the matrix element effects (it is related to transition probability by absorption of a photon). The other parameters $\theta$ and $\eta$ are defined as,

$$(\hbar\theta)^3 = \frac{e^2\hbar^2F^2}{2\mu_{||}}$$
<div align="right">Equation 29</div>

Where $\mu_{||}$ is the reduced effective mass of the electron-hole pair in the direction of the static electric field F and,

$$\eta = \frac{(E_0 - E + i\Gamma)}{\hbar\theta}$$
<div align="right">Equation 30</div>

Where $E_0$ is the direct energy gap of the material and $\Gamma$ is the broadening factor. Finally the parameters $G(\eta)$ and $F(\eta)$ are given by,

$$G(\eta) = \pi\left[A_i'(\eta)B_i'(\eta) - \eta A_i\ (\eta)B_i\ (\eta)\right]$$

And,

$$F(\eta) = \pi[A_i'^2(\eta) - A_i^2(\eta)]$$

Where $A_i(\eta)$ and $B_i(\eta)$ are the first and second order Airy functions and $A_i'(\eta)$ and $B_i'(\eta)$ are their derivatives. We plot both functions $G(\eta)$ and $F(\eta)$ in the following Figure 38,

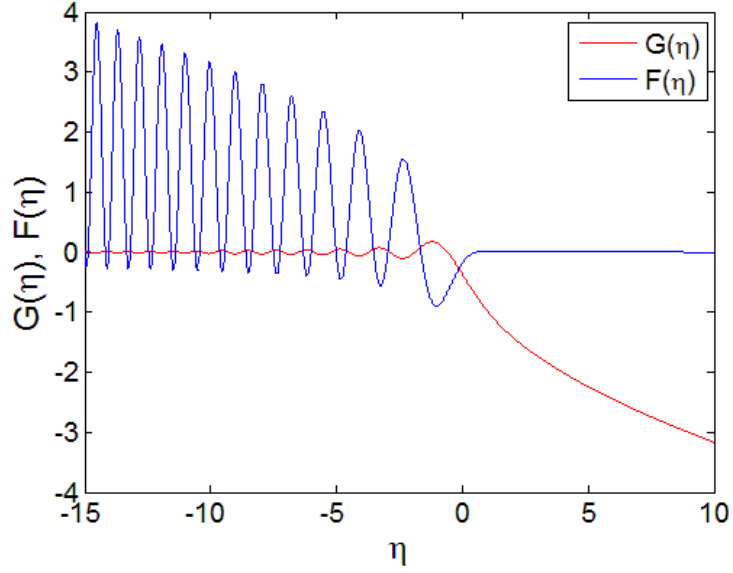

**Figure 38: Plot of the function G(η) and F(η)**

In the modulator that we propose in this thesis, the Ge will be deposited over Si. It was demonstrated in [47] that when Ge is deposited over Si, there is a biaxial tensile strain due to the process at which it is subdued. As a result of the biaxial tensile strain, the band-edge of the materials goes to longer wavelengths. Furthermore, it is observed that the light- and heavy-holes valence bands are non-degenerated. Due to this, the contributions from both the light- and heavy-holes needs to be taken into account (this is represented in Figure 6) and Equation 28 needs to be modified. The band gaps in the Γ point (defined in Figure 6) of the transition bands are defined as $E_g^\Gamma(lh)$ and $E_g^\Gamma(hh)$. Hence, the Equation 28 is modified as,

$$\Delta\varepsilon(E,F) = \frac{B_{lh}}{E^2}\sqrt{\hbar\theta_{lh}}[G(\eta_{lh}) + iF(\eta_{lh})]$$
$$+ \frac{B_{hh}}{E^2}\sqrt{\hbar\theta_{hh}}[G(\eta_{hh}) + iF(\eta_{hh})]$$

Equation 31

Where $B_{lh}$ and $B_{hh}$ are the constants of light- and heavy-holes transitions. G(η) and F(η) where described before (and represented in Figure 38) and,

$$\eta_{lh} = \frac{\left(E_g^\Gamma(lh) - E + i\Gamma_{lh}\right)}{\hbar\theta_{lh}}$$

$$\eta_{hh} = \frac{\left(E_g^\Gamma(hh) - E + i\Gamma_{hh}\right)}{\hbar\theta_{hh}}$$

Equation 32

Where $\Gamma_{lh}$ and $\Gamma_{hh}$ are the broadening factors of the light- and heavy-holes transition and $\hbar\theta_{lh}$ and $\hbar\theta_{hh}$ are the electro-optical energies of light- and heavy-holes transitions, they are given by,

$$(\hbar\theta_{lh})^3 = \frac{e^2\hbar^2 F^2}{2\mu_{lh,\parallel}}$$

$$(\hbar\theta_{hh})^3 = \frac{e^2\hbar^2 F^2}{2\mu_{hh,\parallel}}$$

Equation 33

Where e is the unity charge of the electrons, $\hbar$ is the Planck's constant, F is the static electric field and $\mu_{lh,\parallel}$ and $\mu_{hh,\parallel}$ are the reduced mass of electron-holes pairs of light and heavy holes.

The prediction of the values of $E_g^\Gamma(lh)$ and $E_g^\Gamma(hh)$ due to the biaxial tensile strain $\varepsilon_{\parallel}$ are important for the model. The deformation potential theory can predict this value and it is given by,

$$E_g^\Gamma(lh, \varepsilon_{\parallel}) = E_g^\Gamma(0) + a(\varepsilon_\perp + 2\varepsilon_{\parallel}) + \frac{\Delta_0}{2} - \frac{1}{4}\delta E_{100}$$
$$- \frac{1}{2}\sqrt{\Delta_0^2 + \Delta_0\delta E_{100} + \frac{9}{4}(\delta E_{100})^2}$$

Equation 34

$$E_g^\Gamma(hh, \varepsilon_{\parallel}) = E_g^\Gamma(0) + a(\varepsilon_\perp + 2\varepsilon_{\parallel}) + \frac{1}{2}\delta E_{100}$$

Equation 35

Where $E_g^\Gamma(lh)$ and $E_g^\Gamma(hh)$ are the direct bandgaps from the maximum of the light- and heavy-holes in the valence band to the bottom of the Γ valley, under tensile biaxial strain. a and b are the deformation potential constants and,

$$\delta E_{100} = 2b(\varepsilon_\perp - \varepsilon_{\parallel})$$

Equation 36

Finally, $\varepsilon_\perp$ and $\varepsilon_{\parallel}$ are the perpendicular and parallel (to F) biaxial tensile strains. We plotted Equation 35 and Equation 36 as a function of the biaxial tensile strain in the following Equation 23. The parameters of Equation 34 and Equation 35 were taken from [47].

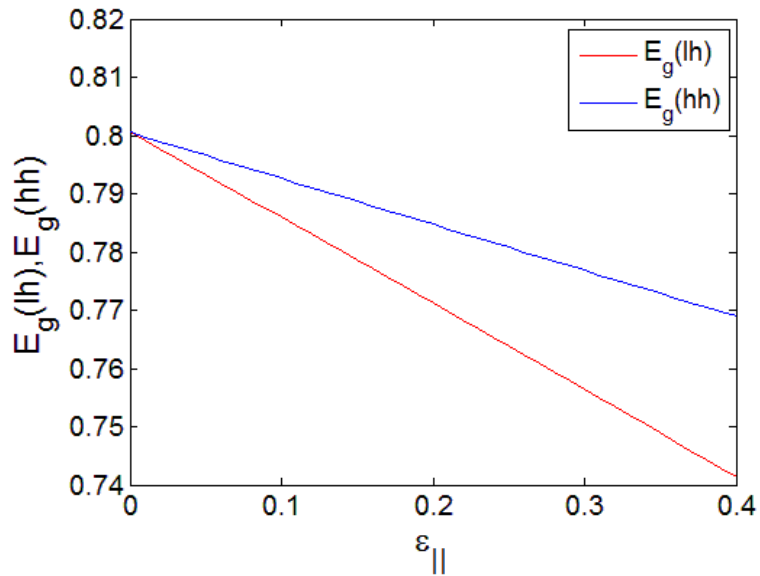

**Figure 39: Change in the energy bandgap of light and heavy holes due to biaxial tensile strain**

In this section we presented a simplified model of the FKE. This model cannot be used directly as it is difficult to find a proper value of the parameter B in the literature. However, we can use it to check the other model presented in the next section.

# 4.3    A Franz-Keldysh Effect Model

In this section we introduce a more complex model of the FKE. For this, we will explain the FKE in further details.

In Figure 5 we present an energy diagram of the FKE. In Figure 5 (a) there is the valence and conduction band of the material. When a photon of energy ℏω>$E_g$ excites an electron from the valence band to the conduction band, absorption of light occurs. Around $E_g$ there is the band-edge of the absorption as is represented in Figure 40.

When a static electric field is applied to the material, both the valence and the conduction bands are tilted as is represented in Figure 5 (b). In this case the wave-functions evanescently penetrate into the forbidden band-gap region. This means that the electrons can tunnel though the forbidden region. This allows a transition from the forbidden region close to the valence band to the conduction band. In this case, the energy needed by the electron is less than $E_g$. Due to the lower photon absorption energy ℏω<$E_g$ needed to excite the electron, there is absorption below the band-edge of Ge when a static electric field is applied. It is represented in Figure 40.

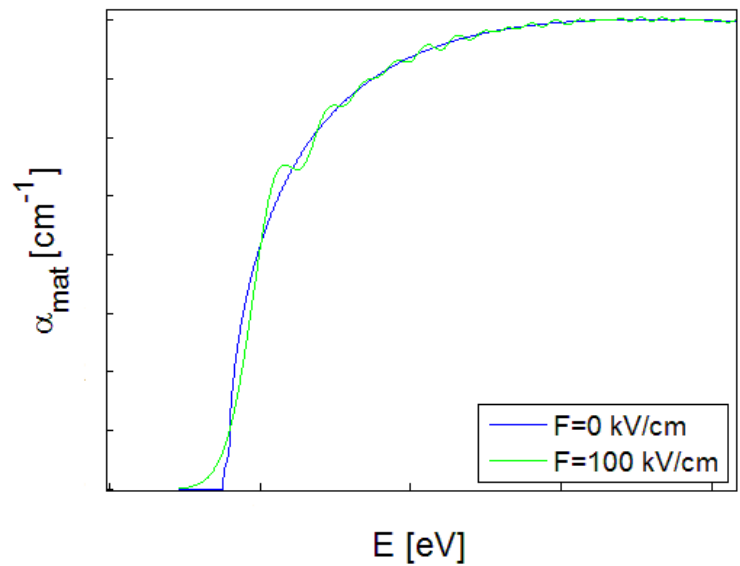

**Figure 40: Absorption spectrum of the material**

In the other case, when the energy of the photon beam is bigger than $E_g$ (ℏω>$E_g$) the transition of the electrons from the valence band to the conduction band corresponds to oscillatory part of the wave-function represented in Figure 5 (b). Due to the oscillatory properties of the wave-function in the conduction and valence bands the transitions are enhanced for some photon energies ℏω and reduced for some others.

Consequently, for photon energies larger than the band-gap energy ℏω>$E_g$ there are some oscillations in the absorption when a static electric field is applied.

Now we derive the absorption coefficient due to FKE under the static electric field F and the photon energy ℏω. For this, we solve the Schrödinger equation in the presence of a constant static electric field F. It is constant along the z-axis. It is a good approximation of the intrinsic part of a PIN structure, where the FKE normally takes place.

A PIN structure is a diode in which there is an intrinsic semiconductor region sandwiched between n-doped and p-doped semiconductors. The highly doped n and p regions are used to take ohmic contacts. It is like a PN diode but with an intrinsic semiconductor in the middle.

The PIN structure is normally used to induce an almost constant static electric field in the intrinsic part of the structure under a reverse bias.

The Coulomb potential of electron-hole interactions is neglected in the present derivation. Without this simplification, the Schrödinger equation used to develop the FKE model does not have any analytical solution. Furthermore, in the case of the FKE under a constant applied electric field, the electrons and holes near the band-edge are the ones that most contribute to the absorption due to band-to-band transitions.

In this section we solve the Schrödinger equation to develop the model of the FKE effect. The Schrödinger equation is given by,

$$\left(-\frac{\hbar^2}{2m_r}\frac{d^2}{dz^2} + eFz\right)\Phi(z) = E_z\Phi(z)$$

Equation 37

Where ℏ is the Planck constant, e is the electron charge, F is the applied static electric field in the material, $\Phi$ is the wave-function of the carriers and $m_r$ is the reduced effective mass of the electron-hole pairs given by,

$$m_r = \frac{m_e m_h}{m_e + m_h}$$

Equation 38

Where, $m_e$ is the mass of the electrons and $m_h$ is the mass of the holes. Continuing with Equation 37 the total energy E of the photons is given by,

$$E = \frac{\hbar^2}{2m_r}\left(k_x^2 + k_y^2\right) + E_z$$

Equation 39

Where $k_x$ and $k_y$ are the wave vector in the x and y-directions. The terms containing $k_x^2$ and $k_y^2$ represent the kinetic energy of the carriers while $E_z$ represents the potential of them. To solve the Equation 37 we need to perform a change of variable from z to Z. This change is given by,

$$Z = \left(\frac{2m_r eF}{\hbar^2}\right)^{1/3} \left(z - \frac{E_z}{eF}\right)$$

Equation 40

Consequently, the Equation 37 transforms to,

$$\frac{d^2\phi(Z)}{dZ^2} - Z\phi(Z) = 0$$

Equation 41

This is a well-known differential equation whose solutions are the Airy functions $A_i(Z)$ and $B_i(Z)$. The definitions of $A_i$ and $B_i$ are given by Equation 42 and Equation 43,

$$A_i(Z) = \frac{1}{\pi}\int_0^\infty cos\left(\frac{t^3}{3} + Zt\right)dt$$

Equation 42

$$B_i(Z) = \frac{1}{\pi}\int_0^\infty \left[exp\left(-\frac{t^3}{3} + Zt\right) + sin\left(\frac{t^3}{3} + Zt\right)\right]dt$$

Equation 43

In Equation 42 and Equation 43 the parameter Z is given by Equation 40 and t is the integration variable.

The representation of both $A_i(Z)$ and $B_i(Z)$ can be seen in Figure 41,

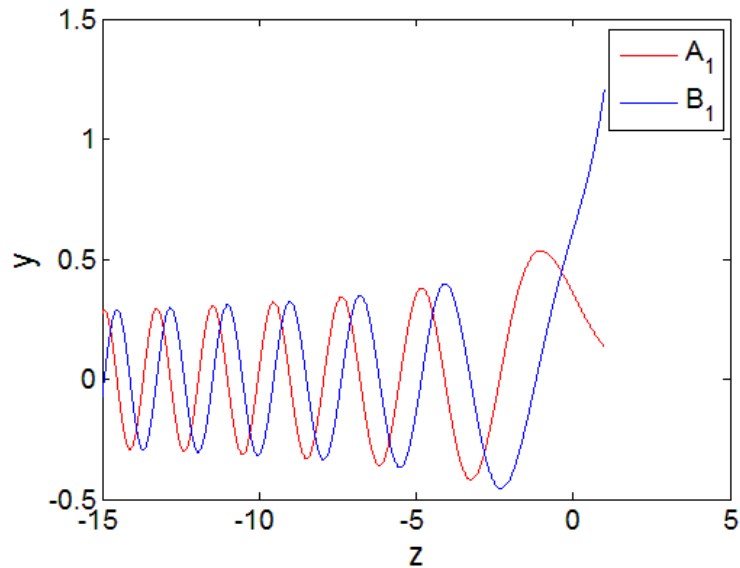

**Figure 41: Representation of the Airy functions A₁ and B₁**

Note that the function $B_i(Z)$ diverges when Z has values bigger than Z=0. Since the wave function must not diverge as Z tends to +∞ to have a physical meaning, then only the Airy function $A_i(Z)$ is selected.

To continue with the development of the FKE absorption, we applied the following normalization condition,

$$\int_{-\infty}^{+\infty} \phi_{E_{Z1}}(z)\phi_{E_{Z2}}(z)dz = \delta(E_{z1} - E_{z2})$$

<div align="right">Equation 44</div>

This normalization condition is further explained in [129]. Consequently, we have the following solution of the normalized wavefunction,

$$\phi_{E_z}(z) = \left(\frac{2m_r}{\hbar^2}\right)^{1/3} \frac{1}{(eF)^{1/6}} Ai\left[\left(\frac{2m_r eF}{\hbar^2}\right)^{1/3}\left(z - \frac{E_z}{eF}\right)\right]$$

<div align="right">Equation 45</div>

The total solution of Equation 37 is given by,

$$\psi_{k_x,k_y,E_z}(\boldsymbol{r}) = \frac{e^{i(k_x x + k_y y)}}{\sqrt{A}} \phi_{E_z}(z)$$

<div align="right">Equation 46</div>

Where A represents an area. This is the well-known solution of the Schrödinger equation. Now, using the wave-function derived in       Equation 46 we want to derive the optical absorption of the material. For this we use the Fermi's golden rule. In Fermi's Golden rule, the absorption coefficient can be calculated using the wave-function derived in       Equation 46. The Fermi's golden rule is given by,

$$\alpha(\varpi) = \frac{\pi e^2 E_p}{3n_r c\varepsilon_0 m_0 \varpi} \sum_n |\phi_n(\boldsymbol{r} = 0)|^2 \delta(E_n + E_g - \hbar\varpi)$$

<div align="right">Equation 47</div>

Where $E_p$ is a constant which is related to the transition matrix element of the material. In Ge the value is $E_p$=26.3 eV [130]. The parameter $n_r$ is the real part of the refractive index of the material (Ge in our case).

Now, to calculate the optical absorption we insert the wave-function derived in       Equation 46 into Equation 47. We obtain,

$$\alpha(\varpi) = \frac{\pi e^2 E_p}{3n_r c m_0 \varpi} \sum_{k_x}\sum_{k_y}\int \frac{dE_z}{A}\left|e^{i(k_x x + k_y y)}\phi_{E_z}(z)\right|^2_{x,y,z=0}\delta(E_t + E_z + E_g - \hbar\varpi)$$

<div align="right">Equation 48</div>

Where c is the velocity of light in vacuum and $k_x$ and $k_y$ are the wave-vectors of the carriers. The kinetic energy $E_t$ of the carriers in the transverse plane x-y is defined as,

$$E_t = \frac{\hbar^2(k_x^2 + k_y^2)}{2m_r} = \frac{\hbar^2 k_t^2}{2m_r}$$

<div align="right">Equation 49</div>

Using the following equation which is an identity between the operators (it is further explained in [129]),

$$\frac{1}{A}\sum_{k_x,k_y} = \int \frac{d^2 k_t}{(2\pi)^2} = \frac{m_r}{2\pi\hbar^2}\int dE_t$$

Into Equation 48 we have,

$$\alpha(\omega) = \frac{e^2 E_p}{6n_r c\varepsilon_0 \varpi \hbar^2}\left(\frac{m_r}{m_0}\right)\iint dE_t \, dE_z \, |\phi_{E_z}(z$$
$$= 0)|^2 \, \delta(E_t + E_z + E_g - \hbar\varpi)$$

<div style="text-align:right">Equation 50</div>

Operating in the integral using the properties of the delta function yields,

$$\alpha(\omega) = \frac{e^2 E_p}{6n_r c\varepsilon_0 \varpi \hbar^2}\left(\frac{m_r}{m_0}\right)\int_{-\infty}^{\hbar\omega - E_g} dE_z \, |\phi_{E_z}(z=0)|^2$$

And substituting the wave-function of     Equation 46 in the previous formula,

$$\alpha(\omega)$$
$$= \frac{e^2 E_p}{6n_r c\varepsilon_0 \varpi \hbar^2}\left(\frac{m_r}{m_0}\right)\int_{-\infty}^{\hbar\omega - E_g} dE_z \left(\frac{2m_r}{\hbar^2}\right)^{2/3}\frac{1}{(eF)^{1/3}}A_i^2\left[\left(\frac{2m_r eF}{\hbar^2}\right)^{1/3}\left(z\right.\right.$$
$$\left.\left. - \frac{E_z}{eF}\right)\right]$$

<div style="text-align:right">Equation 51</div>

Redefining the equation with the following formulas,

$$\hbar\theta_F = \left(\frac{\hbar^2 e^2 F^2}{2m_r}\right)^{\frac{1}{3}},$$

$$\tau = -\frac{E_z}{\hbar\theta_F},$$

$$\eta = \frac{E_g - \hbar\omega}{\hbar\theta_F}$$

<div style="text-align:right">Equation 52</div>

Rewriting Equation 51 and using Equation 52,

$$\alpha(\omega)$$
$$= \frac{e^2 E_p}{12n_r c\varepsilon_0 m_0 \omega}\left(\frac{2m_r}{\hbar^2}\right)^{3/2}\sqrt{\hbar\theta_F}[-\eta A_i^2(\eta)$$
$$+ A_i'^2(\eta)]$$

<div style="text-align:right">Equation 53</div>

In Equation 53 $A_i'$ is the derivative of $A_i$. Both functions are represented in the following Figure 42 to see how they look like and to confirm that they are finite and do not diverge to infinity.

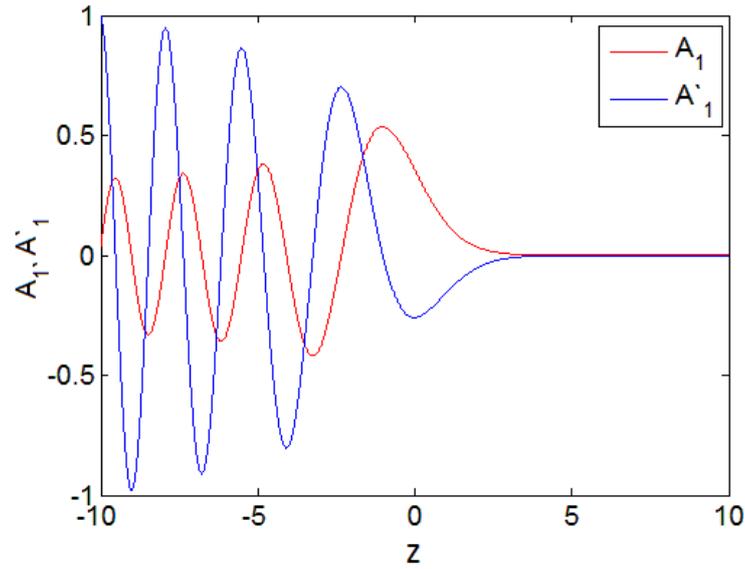

**Figure 42: Airy function $A_i$ and its derivative $A'_i$**

Equation 53 represents the optical absorption of a semiconductor material (in our case Ge) in which an electric field is applied. The mentioned formula quantifies the change of the absorption due to the static electric field.

We used Ge as a material to obtain the FKE. We are interested in the direct-bandgap around 1550-1610 nm. The exact wavelength of the FKE around the direct bandgap depends on the tensile strain present in the material. Since the FKE around the direct-bandgap is three times larger than the FKE around the indirect one [131], [132], the latter can be neglected.

Furthermore, the indirect band-gap of Ge is around 1900 nm and it is less sensitive to the strain than the direct one. Due to this small influence a small shift in the indirect bandgap due to the strain will not affect the absorption at the direct-bandgap.

Consequently, the equations derived in this section (where the indirect-bandgap is neglected) are valid for the Ge material.

Nevertheless, it is worth noting that we are interested in the weakly absorbing regime of the material (Ge or SiGe); E.g.: slightly over the direct band-gap where the optical absorption of the material is large absorption variation that we want to exploit for modulation due to the application of a static electric field (Figure 40). The reason of the change is the difference between the low absorption close to 0 V/cm and the larger absorption when a static electric field is applied due to the tunneling of the electrons through the forbidden region, doing transitions with $\hbar\omega < E_g$.

We can conclude that we can use Equation 53 to quantify the change in the optical absorption of Ge due to the FKE around the direct band-gap of Ge.

# 4.4    Franz-Keldysh Effect Model in Tensile Strained Ge

In this work we will place Ge over Si to make the structure of the modulator. It is biaxial tensile strained due to the process explained in [46]. In the previous section we derived a FKE model to quantify the optical absorption changes under the application of a static electric field for non-strained Ge. Due to the strain some modifications need to be done in the model to agree with the experimental data. These modifications are explained in this section.

The FKE around the direct band-gap has been studied in bulk Ge in [131], [132], [133], [134]. The change in the optical absorption below the direct band-gap under the application of a static electric field has been observed experimentally and the experimental results agree with the theory in most of the experiments [46].

The band structure of biaxial tensile strained Ge is presented in Figure 6.

The effect of the biaxial tensile strain in the FKE in Ge can be also modeled with Equation 53. Under tensile strain, both the light holes and the heavy holes become non-degenerated; consequently the absorption coefficient near the direct-gap can be predicted with a modification of Equation 53. It consists in separating the FKE due to heavy-holes and light holes. We remind that in this model we considered the absorption near the direct band-gap of Ge. The modified equation is given by,

$$
\begin{aligned}
\alpha(\omega) = \frac{e^2 E_p}{12 n_r c \varepsilon_0 m_0 \omega} \Bigg\{ & \left(\frac{2 m_{r,lh}}{\hbar^2}\right)^{3/2} \sqrt{\hbar \theta_{F,lh}} \, [-\eta_{lh} A_i^2(\eta_{lh}) \\
& + A_i'^2(\eta_{lh})] \\
& + \left(\frac{2 m_{r,hh}}{\hbar^2}\right)^{3/2} \sqrt{\hbar \theta_{F,hh}} \, [-\eta_{hh} A_i^2(\eta_{hh}) \\
& + A_i'^2(\eta_{hh})] \Bigg\}
\end{aligned}
$$

Equation 54

Where,

$$m_{r,lh} = \frac{m_e^{\Gamma} m_{lh}}{m_e + m_{lh}}$$

$$m_{r,hh} = \frac{m_e^{\Gamma} m_{hh}}{m_e + m_{hh}}$$

$$\hbar\theta_{F,lh} = \left(\frac{\hbar^2 e^2 F^2}{2m_{r,lh}}\right)^{1/3}$$

$$\eta = \frac{E_g^{lh} - \hbar\omega}{\hbar\theta_{F,lh}}$$

$$\hbar\theta_{F,hh} = \left(\frac{\hbar^2 e^2 F^2}{2m_{r,hh}}\right)^{1/3}$$

$$\eta = \frac{E_g^{hh} - \hbar\omega}{\hbar\theta_{F,hh}}$$

Equation 55

Where, $E_g^{lh}$ is the energy bandgap of the light-holes, $E_g^{hh}$ is the energy band-gap of the heavy-holes, $m_e^{\Gamma}$ is the effective mass of the electrons in the Γ valley (Figure 6), $m_{lh}$ is the effective mass of the electrons in the light-hole bands and $m_{hh}$ is the effective mass of the electrons in the heavy-hole bands.

Equation 53 (where there is no strain) is divided into a part with light-holes and another part with heavy-holes in

Equation 54 because the optical absorption of the material is a combination of both transitions from the valence band to the conduction band. As the valence band has now two non-degenerated bands (one corresponding to the heavy holes and another one corresponding to the light holes), there are two possible transitions contributing to the optical absorption: one from the light hole band to the conduction band and another one from the heavy hole band to the valence band. These two possible transitions are modeled with the two different terms of

Equation 54.

The values of the parameters in

Equation 54 and Equation 55 are taken from [46]. These values are: $E_P$=26.3 eV, $E_g^{lh}$=0.773 eV, $E_g^{hh}$=0.785 eV, $m_e^{\Gamma}$=0.038$m_0$ (electron effective mass in the Γ valley), $m_{lh}$=0.043$m_0$ and $m_{hh}$=0.33$m_0$ [46], where $m_0$ is the mass of the free electrons.

As we stated before the FKE due to the indirect band-gap is three times smaller than the one in the direct one [131], [132]. Consequently, it can be neglected in

Equation 54, in the same way as in Equation 53.Now with

Equation 54 we have a model to calculate the optical absorption of Ge material over Si due to the FKE. We can calculate the material absorption for any wavelength around the direct bandgap and any static electric field. It will be interesting to compare the model with the measurements done in the literature to know how accurate our model is. We compare the model with measurements taken in [46].

Figure 43 reports in dotted lines the spectral measurements of the Ge optical absorption under different static electric fields: 21, 46 and 30 kV/cm performed in [46]. In continuous line in Figure 43, we can see the prediction of the model for the same static electric field values using

Equation 54. We added the field of 0 kV/cm to know the optical absorption in the absence of a static electric field. From Figure 43 we can see that there is correct agreement between the model and the experimental results.

According to the study done in [46] the maximum change in absorption of the FKE in biaxial tensile strain Ge is around 1647 nm. Consequently, the proposed device in this thesis will work at this wavelength. This may be a disadvantaged related to devices that work in the telecommunication band (1550 nm), nevertheless adding a small amount of Si into the Ge ($Ge_{0.9925}Si_{0.0075}$) we can shift the band-edge of the Ge to 1550 nm and consequently the device can operate at the communication wavelength of 1550 nm.

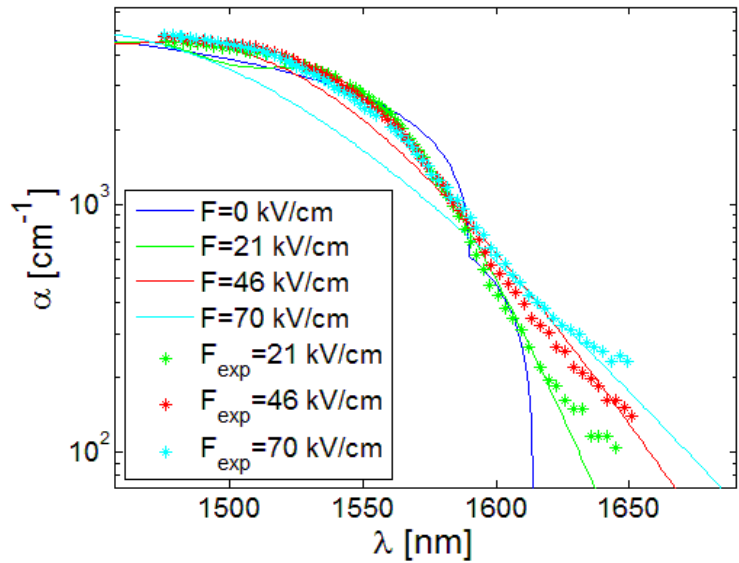

**Figure 43: Detail of the comparison of the model of the FKE (continuous line) and the measurements done in [46] (dotted line)**

## 4.5    Conclusion on the Franz-Keldysh Effect Models

As we stated before, the core of the modulator will be formed by Ge and we will exploit the FKE in the electro-absorption modulator. In this chapter we introduced the formulas to model the FKE in bulk Ge and in the induced biaxial tensile strained Ge. We are interested in the latter since our modulator will be fabricated on SOI substrates, meaning that Ge will be over Si. It will lead to a tensile strain of 0.2 %. The strainted Ge model will be used to calculate the change in optical absorption of the material when a static electric field is applied. In chapter five, we will use this

Equation 54 to simulate the behavior of the modulator that we propose.

According to the study done in [46], the maximum change in absorption due to the FKE in biaxial tensile strain Ge is around 1647 nm. Consequently, the proposed device in this thesis will work at this wavelength. This may be a disadvantage for the integration with other devices that work in the telecommunication band (1550 nm). Nevertheless, adding a small amount of Si into Ge to form $Si_xGe_{1-x}$ (typically an addition of 0.75 [35]-0.8% [51] of Si into the Ge will shift the FKE to 1550 nm), we can in principle shift the band-edge of Ge to 1550 nm [51] and consequently the device can operate at the communication wavelength of 1550 nm. The mentioned fact that the FKE around the direct band-gap is three times larger than the indirect one is also valid for SiGe [131], [132]. In the next section of this chapter, we will measure the FKE in both Ge and SiGe materials to shift the maximum effect around the telecommunication wavelength of 1550 nm.

# 4.6    Integrated Electro-Optic Simulator

In order to calculate the main benchmarks of the plasmonic modulator, we perform integrated opto-electronic simulations. The simulation flow is described in Figure 44. First, it is necessary to know the static electric field distribution in the Ge core to quantify how much static electric field is present in the Ge. With the FKE model described by

Equation 54 we calculated the change in the optical material absorption coefficient of strained Ge. Knowing the change in absorption we have to know the contribution in the effective propagation losses of the plasmonic mode. Knowing this value we can calculate the propagation losses and the extinction ratio of the device.

Using the commercial software ISE-dessis we calculate the static electric field distribution in a general structure as a function of the applied voltage between two electrodes. The main task is to design a structure that concentrates as much as possible static electric field in the Ge core. Knowing the static electric field distribution in Ge we calculate the change in the optical absorption of the material using the model described by

Equation 54. Knowing the change and the new distribution of the optical absorption coefficient of the material we import these values in a FDM mode solver to calculate the plasmonic mode characteristics. Using the losses of the material and the FDM mode solver we calculate the change in the effective absorption of the mode $\Delta\alpha_{eff}$ and the effective refractive index of the mode $n_{eff}$. Knowing $\Delta\alpha_{eff}$ and $\alpha_{eff}$ we calculate the extinction ratio and the insertion losses of the structure. The definition of $\Delta\alpha_{eff}$ is given by,

$$\Delta\alpha_{eff} = \alpha_{eff}(V = V_1) - \alpha_{eff}(V = V_2)$$

<div align="right">Equation 56</div>

Where $\alpha_{eff}$ is the effective losses of the mode when the voltages of the structure vary from $V=V_1$ to $V=V_2$. $V_1$ and $V_2$ are the voltages at which the modulator is driven.

In the next flow chart we see the steps done with the results of the simulator in order to perform and integrated opto-electronic simulations.

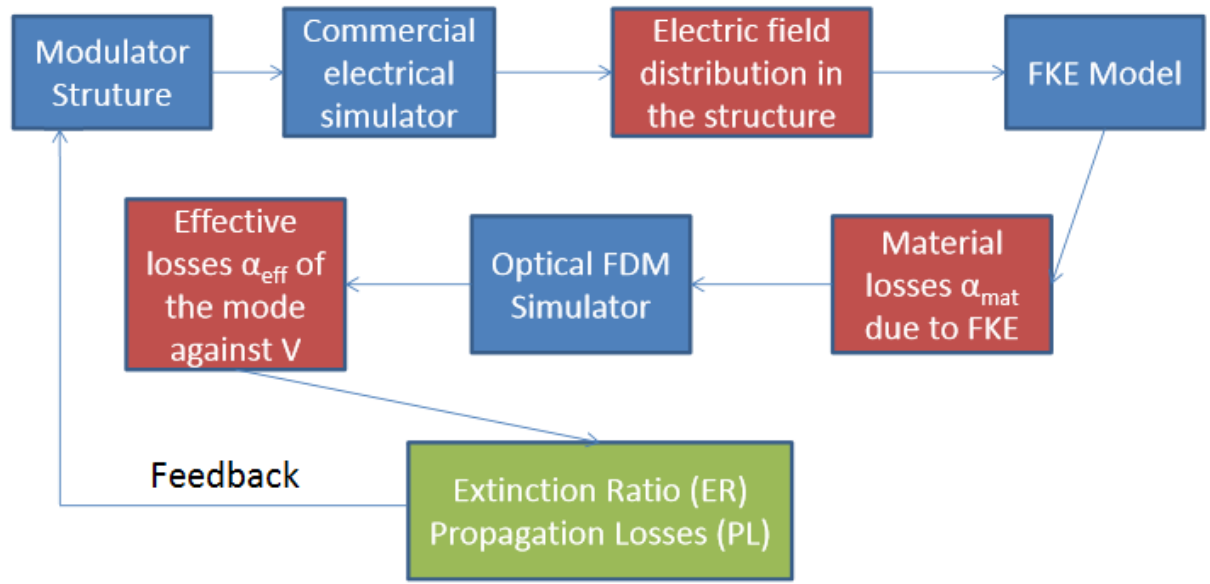

**Figure 44: Flow chart of an integrated opto-electronic simulation**

Using ISE-dessis we also calculate the intrinsic capacitance C of the modulator structure. Additionally, the access resistance R for the device is also theoretically calculated. The access resistance is the resistance that appears between the side contact and the Ge core of the modulator. Knowing this and the operational voltage used we are able to estimate the power consumption of the modulator. Furthermore, knowing C and R we can calculate the intrinsic bandwidth of the modulator.

As a result of the integrated opto-electronic simulation we know the extinction ratio, the propagation losses, the operational voltage, the electrical energy consumption and the bandwidth of the structure. Knowing these parameters we start the process again changing any parameter of the structure and we start the integrated electro-optical simulation again in the process of designing the device.

# 4.7 Validation of the Integrated Electro-Optic Simulator

For the implementation of the integrated electro-optic simulator is done it is interesting to contrast the result of the simulator with the results obtained in several publications. This will allow us to validate the implementation of our simulator. It is desirable to compare the simulator optically.

Both the optical and electrical simulations must be done separately since to the best of our knowledge there is not an integrated electro-optical simulation of a modulator. To calculate the static electrical field we use a commercial simulator called ISE-dessis. Since it is a widely used electrical simulator we do not validate it.

Now, we center our comparison in the FDM mode solver. Mainly, we compare if the effective refractive index $n_{eff}$ and the effective losses $\alpha_{eff}$ of the mode are similar with the published values in the literature. For this we analyze the same structure than in the published paper and we compare the results mainly in $n_{eff}$ and $\alpha_{eff}$. Sometimes it is difficult to reach a conclusion since in the publications not all parameters of the simulations (refractive index of the material, parameters of the mesh, etc.) are explained.

The FDM mode solver used in this thesis was previously done at University of Paris Sud 11. Nevertheless it was only used for photonics and it admitted only two materials which are Si and $SiO_2$. Furthermore, it was only tested for photonics waveguides using only these two materials. We improved this FDM mode solver by adding any kind of material including metals. Due to this modification it is necessary to test the code for plasmonics and the new materials.

We compare optically the FDM mode solver with the following papers: [121], [57], [111], [112], [117], [135]–[140].

As an example, we present in details the comparison between paper [57] and our integrated electro-optical simulator. The device described in [57] was previously introduced in the state-of-the-art of plasmonic modulators in chapter three. The structure of the device is presented in Figure 15. The tool used to perform the simulations in this paper was FullWAVE which is a FDM method. Our integrated opto-electronic simulator is based on FDM also.

The first parameter that we compare is the effective refractive index of the mode supported in a branch of the MZI of the structure of Figure 15 as a function of the refractive index of Si. The result is plotted in Figure 45. As is represented in the legend of the right figure in Figure 45, the cross-sections used are different. For this different materials like Ag, Si and $SiO_2$ are used. Different stacks are employed to form the plasmonic waveguide (like Ag-Si-Ag or Ag-$SiO_2$-Si-Ag). The effective refractive index of the paper is represented on left while the result of our integrated electro-optical modulator is presented on the right of Figure 45. It is possible to see that there biggest error is around 0.2-0.3 in the $n_{eff}$ for the case of $n_{Si}$=3.7 and Ag-Si(50 nm)-Ag. A maximum error smaller than 0.1 was achieved in the rest of the cases. This may be due to the fact that not all parameters of the simulation are described in the publication.

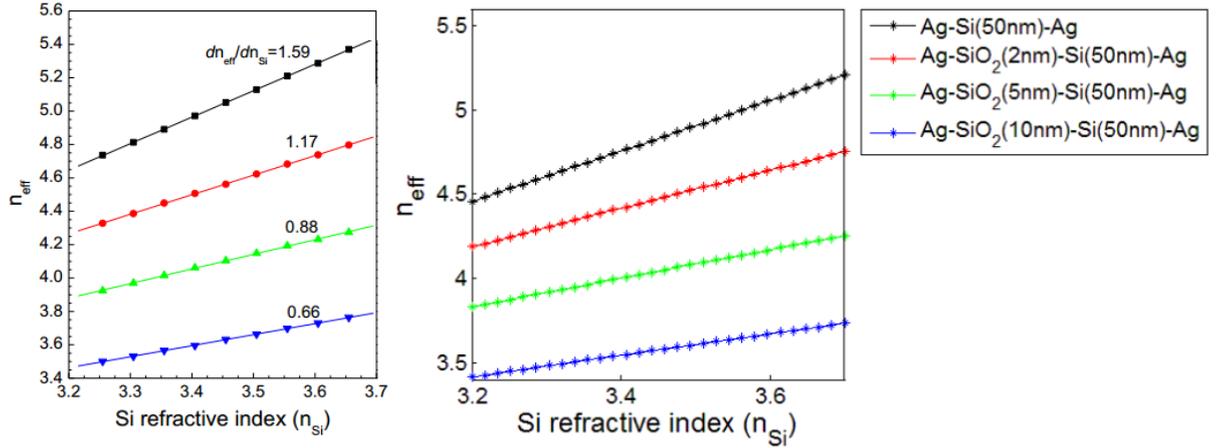

**Figure 45: Effective refractive index of the supported mode in one branch of the MZI as a function of the refractive index of Si published in [57] on the left and reproduced by our integrated electro-optical modulator on the right**

We further compare the propagation length of the plasmonic mode supported by the structure for different combinations of the materials Ag, Si and SiO₂. It is represented in Figure 46. On the left the variation of the effective propagation length $L_p$ as a function of the refractive index of Si $n_{Si}$ is represented. On the right we have the same result given by our integrated electro-optical simulator. As a conclusion, we can say that the biggest error is around 0.5 µm. This error may come from a different imaginary part in the refractive index of Ag. In general, we found it difficult to fully match the effective propagation loss coefficient since the refractive index of Ag is not showed in the publication.

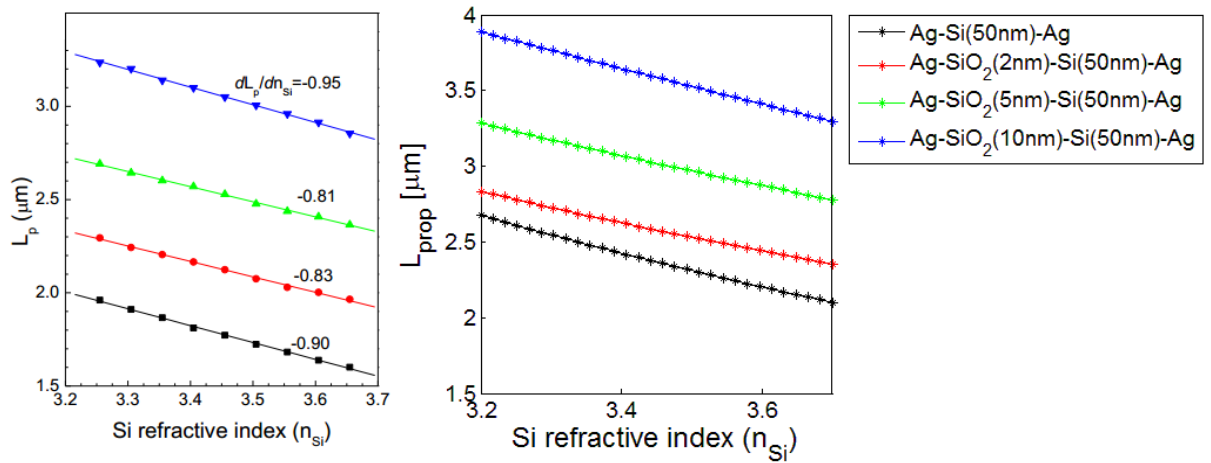

**Figure 46: Effective propagation length of the supported mode in one branch of the MZI as a function of the refractive index of Si published in [57] in the left and reproduced by our integrated electro-optical modulator in the right**

The next step is to compare the effective refractive index of the mode in the waveguide that forms the branch of the MZI as a function of the width of the waveguide $w_{Si}$. It is the width of the Si core that forms the core of the plasmonic waveguide. This width is scanned from 0 to 50 nm. The result is presented in Figure 47. In this case the maximum error in $n_{eff}$ is around 0.1. Note that for the biggining of the figure $w_{SiO2}$=5 nm which correspond to $w_{Si}$=45 nm since $w_{Si}$+$w_{SiO2}$=50 nm.

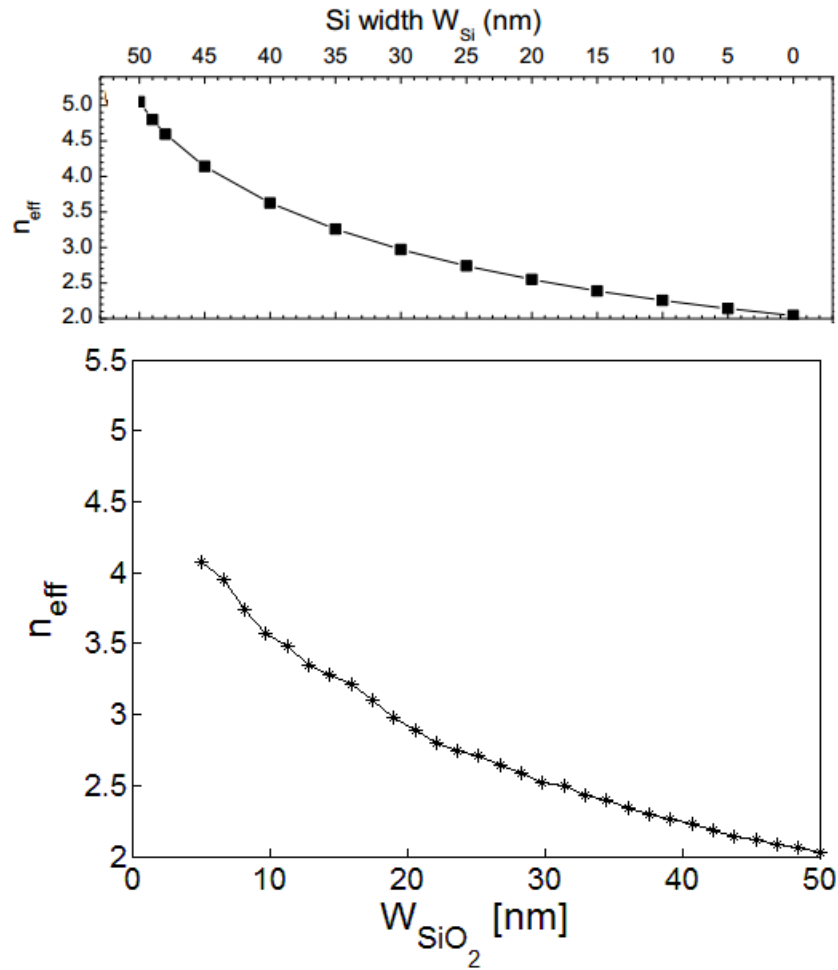

**Figure 47: Effective refractive index of the supported mode in one branch of the MZI as a function of the width wSi of the core of the waveguide. It is taken from reference [57] on top and it is reproduced by our integrated electro-optical modulator on the bottom**

The next step is to compare the propagation length of the mode $L_p$ as a function of the thickness of the $SiO_2$ layer. We vary the thickness of $SiO_2$ from $w_{SiO2}$=0 to 50 nm. $w_{SiO2}$ is the width of the verticals slots of $SiO_2$ which are between the core of the Si waveguide and the metal. It is represented in Figure 15. In Figure 15 we see the Ag metal in light blue, the $SiO_2$ layer in orange and the Si in grey. In this case $w_{Si}$ equals 50 nm. We vary the thickness of the $SiO_2$ from 0 nm to 50 nm. When the thickness of the $SiO_2$ layer is around 50 nm it covers all the silicon. It means, the core of the modulator has always a width of 50 nm. To clarify this case, if $w_{SiO2}$=25 nm (width of the $SiO_2$ layer) then the width of the Si core is also 25 nm.

Again, in Figure 48 we see that there is almost agreement between the results published in the paper and the outcome of the integrated electro-optical modulator. The maximum error is around 0.5 µm.

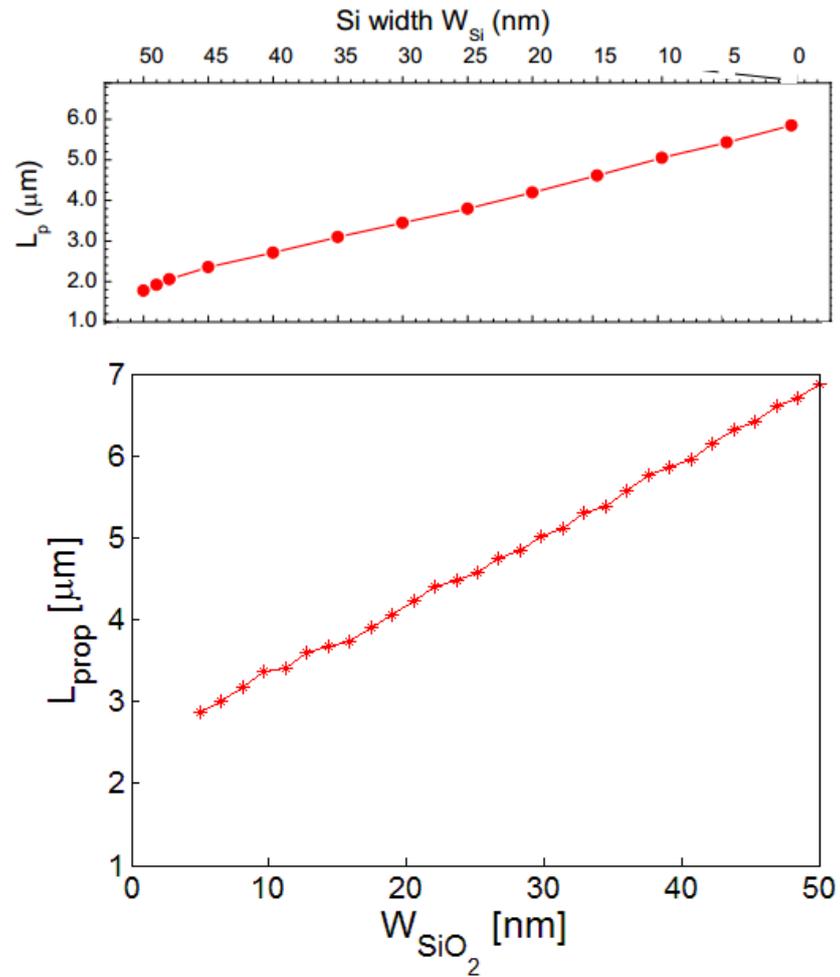

**Figure 48: Effective propagation length of the supported mode in one branch of the MZI as a function of the thickness of the SiO₂ layer. It was published in [57] on top and reproduced by our integrated electro-optical modulator at the bottom**

Next we compare the effective refractive index $n_{eff}$ of the mode with the width of the Si core $w_{Si}$. For this we will scan $w_{Si}$ from 10 to 100 nm for different stacks of the waveguide. The different stack appears in the legend of Figure 49. Again, we can see that there is agreement between the $n_{eff}$ of the paper and the result of our integrated electro-optical modulator. The error in the effective refractive index is around 0.2.

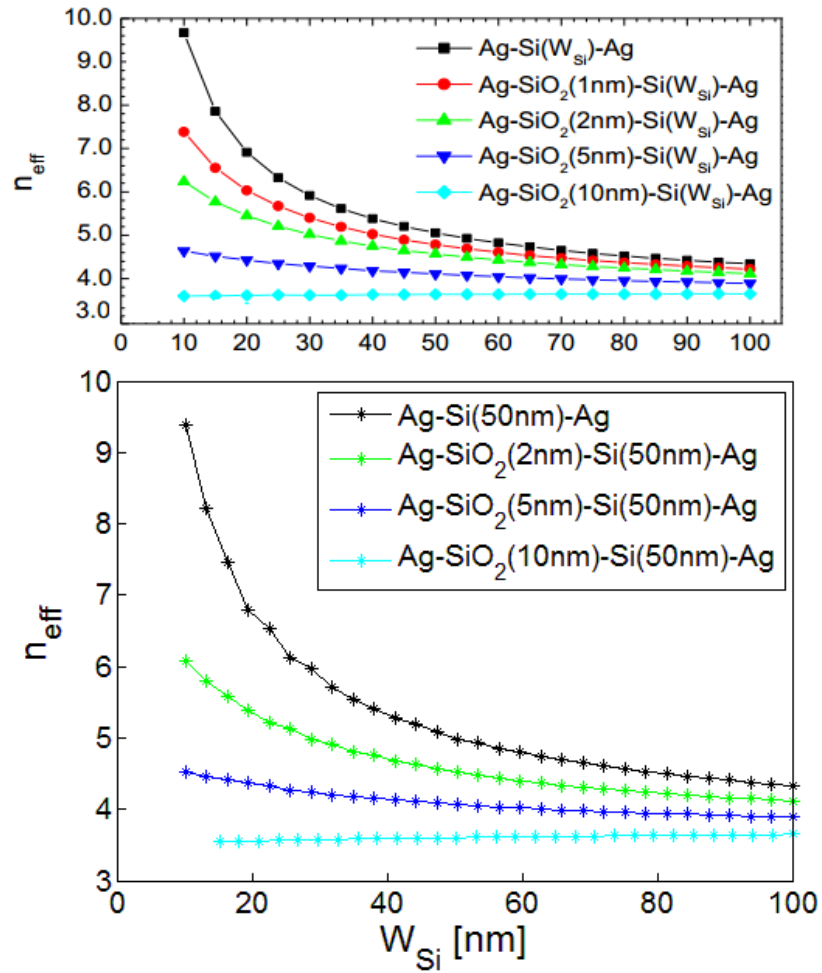

**Figure 49: Effective refractive index of the supported mode in one branch of the MZI as a function of the width w$_{Si}$ of the core of the waveguide. It is taken from reference [57] on top and it is reproduced by our integrated electro-optical modulator on the bottom**

Finally, we compare the effective propagation length L$_p$ of the waveguide of one branch of the modulator with the change in the width of the Si core w$_{Si}$. The results are represented in Figure 50. On the top is the calculations done in the paper and on the bottom the calculations done with our integrated opto-electrical simulator. There is also agreement between both calculations. Again, the maximum error in the propagation length is around 0.5 μm.

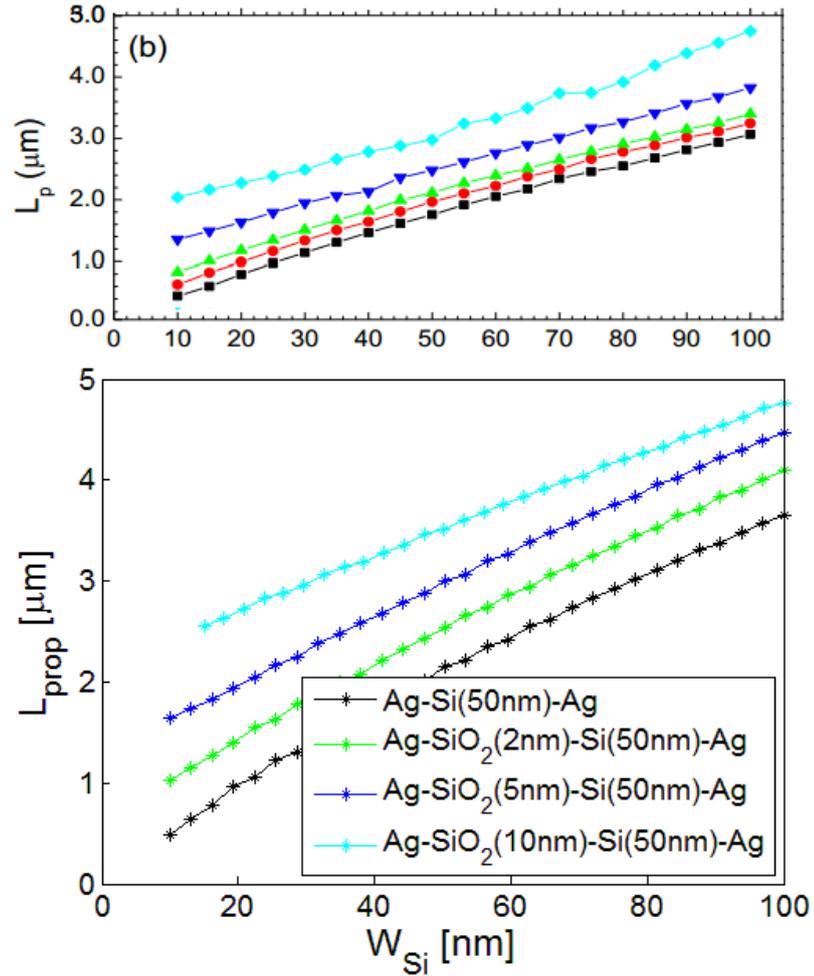

**Figure 50: Effective propagation length $L_p$ of the supported mode in one branch of the MZI as a function of the width $w_{Si}$ of the core of the waveguide. It is taken from reference [57] on top and it is reproduced by our integrated electro-optical modulator on the bottom**

As a conclusion we validated optically the evolutions of different parameters of the structure represented in Figure 15 like the effective refractive index and the propagation length. In all cases the agreement between the published and the reproduced results are satisfying.

With the agreements and maximum errors between our calculations and the ones done in [57] we performer additional comparisons with the following papers: [57], [111], [112], [117], [121], [135]–[140]. Using these new comparisons we saw that in most cases there is also agreement between the published results and the recalculations that we did using our simulator.

## 4.8     Measurements of the Franz-Keldysh Effect in SiGe

In this section we try to measure the FKE in SiGe material to evidence a shift of the absorption band edge from 1647 nm [46] to shorter wavelength (it is desirable to have it around 1550 nm).

## 4.8.1 Description of the Test Structure

For this purpose, we use a batch of standard photodetector structures represented in Figure 51,

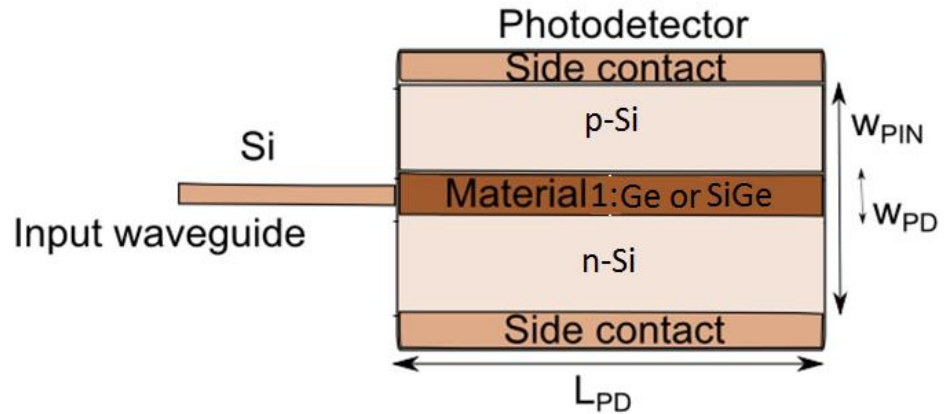

**Figure 51: Top view of photodetector butt-coupled to a standard Si rib waveguide. In this figure we have n-Si, p-Si and material 1 is Ge or SiGe.**

In Figure 51 we represent the top view of the photodetector and the input Si rib waveguide. The input Si rib waveguide arrives from the left. The photodetector is represented in the right. The photodetector has a total width of $w_{PIN}$=10 μm and the length is around $L_{PD}$=10 μm. Inside this cavity we build up a photodetector which is composed of n-Si, Material 1 and p-Si. The core of the photodetector (the place where the photodetector absorbs light to produce the photocurrent) is formed by Material 1, which can be either Ge or SiGe. The sides of the core consist of doped n- and p-doped Si on which electrical contacts are placed to drive the photodetector in the side of the structure of Figure 51. In this section we will compare the photodetector made of Ge with the one made of SiGe.

We want to compare the FKE in Ge and in SiGe. For measuring the FKE in Ge we use the following configuration: n-Si, intrinsic Ge is placed and in Material 1 and p-Si. In this configuration the electrical electrodes are placed in contact with n-Si and p-Si. First, in this configuration we will measure the FKE in Ge and later in SiGe.

To form SiGe in the cavity we anneal at 900°C to make Si diffuse into Ge to create SiGe and shift the absorption band edge to shorter wavelength. We do not control the percentage of Si diffused in Ge nevertheless to have the FKE around 1550 nm the percentage should be the following concentration: typically an addition of 0.75 [35]-0.8% [51] of Si into the Ge will shift the FKE to 1550 nm [51]. It means, by using SiGe instead of Ge in Material 1 we want to shift the maximum of the FKE to lower wavelengths.

Setting the structure of the Ge and SiGe photodetectors we place them in the following structure of Figure 52 to compare the absorption and FKE in Ge and in SiGe.

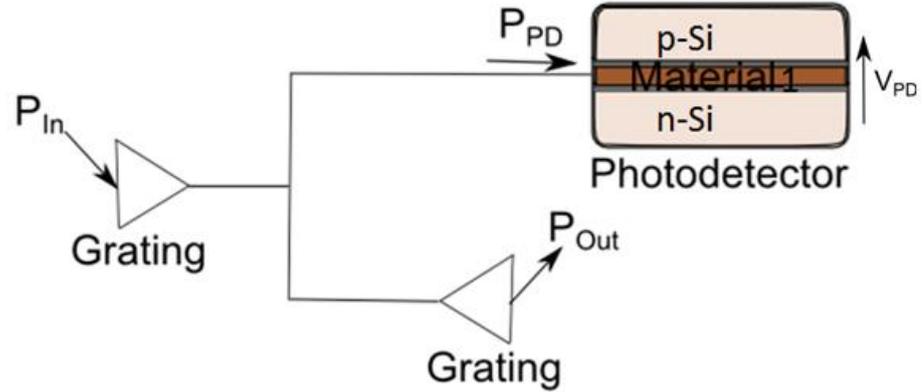

**Figure 52: Schematic of the structure to measure the FKE**

The test structure is presented in Figure 52. We have an input grating on the left for light injection from an optical fiber delivering the power $P_{in}$ to the planar test circuit. The light propagates through the Si waveguide to an MMI which equally divides the signal to the photodetector and to a reference arm containing an output grating to measure the light intensity received by the photodetector. The power measured at the ouput of the reference arm is named $P_{Out}$ and the power received onto the photodetector is called $P_{PD}$. We will measure two photodetectors one with Material 1 formed by Ge and another formed by SiGe (The concentration is not known since it is done by diffusing Si into Ge).

To evidence the FKE, we are interested in measuring the responsivity of the photodetector as a function of wavelength and reverse bias voltage. The reponsivity is defined as $R=I_{Photocurrent}/P_{PD}$. To measure the FKE using a photodetector we need to plot the responsivity R against the reverse bias voltage $V_{PD}$ which drives the photodetector. We expect that the responsivity R will increase with the reverse voltage for the wavelengths around 1647 nm due to the FKE. It means, when we increase the reverse voltage $V_{PD}$, the absorption of Ge close to the band-edge also increases due to the increase in the optical absorption due to a larger static electric field in Ge. The increase of absorption translates into an increase of photocurrent and consequently into an increase in the responsivity R of the photodetector. By using SiGe instead of Ge in Material 1 we want to shift the maximum of the FKE to lower wavelengths.To perform this measurement we use the following setup drawn in Figure 52.

To drive the photodetector we set in the multimeter the value of $V_{PD}$ and we measure the photocurrent $I_{PD}$ when we inject a continuous wave signal with a power of $P_{PD}$ onto the photodetetor. We measure the responsivity as $R=I_{PD}/P_{PD}$. To properly measure the photocurrent, we need to substract the dark current of the photodetector. For this we do not inject light onto the photodetector ($P_{PD}=0$ W) and we measure $I_{PD}=I_{Dark}$ when $V_{PD}$ is scanned from 0 to -5 V. To avoid burning the photodetector we set in the multimeter that the maximum allowed current is 1 mA.

To calculate the responsivity of the photodetector we use the following Equation 57,

$$R=I_{PD}/P_{PD}$$                                    Equation 57

## 4.8.2  Determination of the Optical Power Received by the Photodetector $P_{PD}$

We estimate the power $P_{PD}$ thanks to the reference arm, setting a set of equations which involve $P_{In}$, $P_{Out}$ and $P_{PD}$. We assume that the MMI is balanced (we will demonstrate it experimentally later). The set of equations are the following,

$$P_{Out}=P_{In}\text{-}IL_{GC}\text{-}3\text{-}IL_{GC} \text{ [dB]}$$
<div align="right">Equation 58</div>

$$P_{PD}=P_{In}\text{-}IL_{GC}\text{-}3 \text{ [dB]}$$

Where $P_{Out}$ and $P_{In}$ are defined in Figure 52, $IL_{GC}$ is the insertion loss of the grating couplers and the value of -3 dB is due to the MMI. Setting both equations we calculate $P_{PD}$ which is equal to,

$$P_{PD}=(P_{In}+P_{Out}\text{-}3)/2$$
<div align="right">Equation 59</div>

Using Equation 59 we estimate the value of $P_{PD}$ from the measurement of $P_{In}$ and $P_{Out}$. In the measurements $P_{In}$=0 dBm and we measure $P_{Out}$ using an external detector. Note that the estimation of the value $P_{PD}$ assumes a balanced MMI. It is also not necessary to know the value $IL_{GC}$ which is the losses of the grating couplers.

To verify that the MMI is well balanced and that we can apply Equation 59 we measured both paths o fthe MMI using a dedicated test stucture. The schematics of this measurement is the following Figure 53,

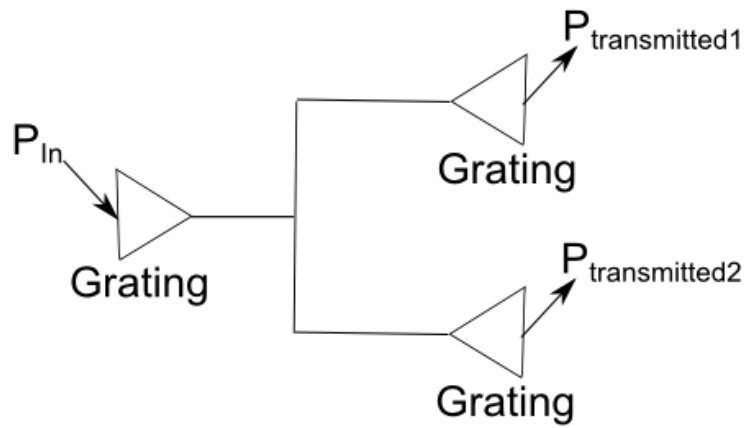

<div align="center">**Figure 53: Measurement setup of the MMI**</div>

We inject a monochromatic light that we scan from 1.5 μm until 1.64 μm and we measure $P_{transmitted1}$ and $P_{transmitted2}$ wich are the output powers at both arms of the MMI. The result is presented in Figure 54.

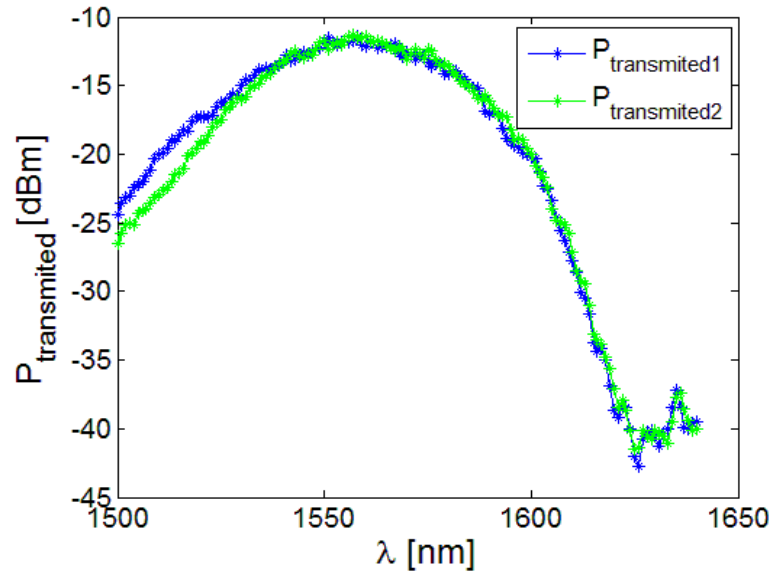

**Figure 54: Measurement of P$_{transmitted1}$ and P$_{transmitted2}$ of the MMI using the structure of Figure 53**

From Figure 54 we can observe that the MMI is well balanced since we obtain almost the same value of P$_{transmitted1}$ and P$_{transmitted2}$ powers at both arms. We can also observe a parabolic behaviour of both P$_{transmitted1}$ and P$_{transmitted2}$ when it is ploted versus wavelength. This is due to the bandwidth of the grating couplers. Furthermore, the grating couplers are centered at 1.55 μm by design which coincides with the maximum in Figure 54. With this we conclude that the MMI is well balanced.

Now, to estimate the value of P$_{PD}$ in order to calculate the responsivity R of the photodetectors, we measure the transmission from P$_{In}$ to P$_{Out}$ defined in Figure 52. The result is shown in Figure 55,

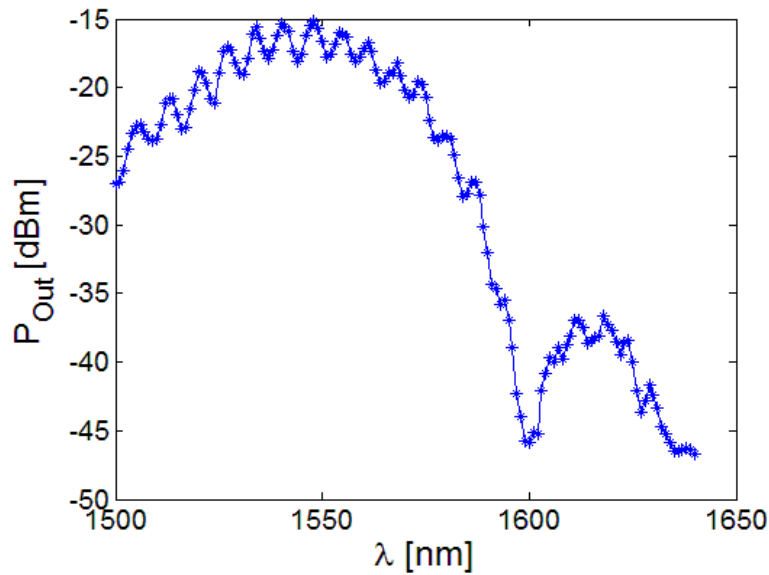

**Figure 55: P$_{Out}$ versus wavelength when at PIN we inject 0 dBm. The parameters are defined in Figure 53**

From Figure 55, we can see that the transmission in P$_{Out}$ has a similar shape that for P$_{transmitted1}$ and P$_{transmitted2}$ in Figure 54 which are the transmission in both arms of the MMI. The maximum of the principal lobe has the shape of the transmission of the gratings and from 1.60 μm to 1.64 μm there is a smaller second lobe due to the shape of the grating as well. The oscillations may come from residual reflection due to grating couplers.

### 4.8.3 Determination of the Photocurrent $I_{PD}$

First, we will measure the I-V curve of the photodetector with a Ge core and another one with SiGe core to asses a good photodiode behaviour.The I-V curve of the photodiode with the Ge core is given in Figure 56.

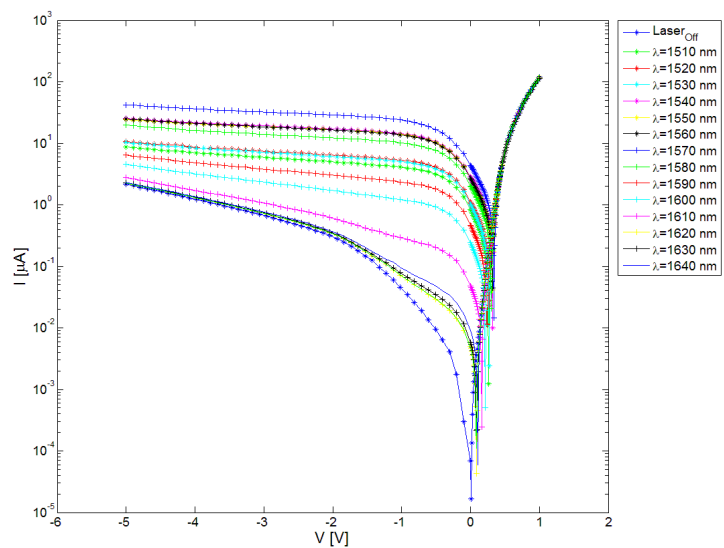

**Figure 56: I-V curve of the photodiode with the Ge core. It is represented the dark current Idark and the photocurrent for diferent wavelengths from 1510 nm to 1640 nm every 10 nm**

In Figure 56 we plot in blue (with the label Laser_Off) the dark current $I_{dark}$ of the photodiode for a reverse bias voltage which goes from $V_{PD}$=0 to -5 V. It is possible to observe that the dark current $I_{dark}$ is below the microampere regime. We also plot the photocurrent measured for $P_{In}$=0 dBm. We scan this from a starting wavelength of 1510 nm until 1640 nm every 10 nm. We observe that for those wavelengths the current in the I-V curve is larger than the dark current due to the existence of a photocurrent. We observe that the photocurrent is in the microampere regime. It depends on the wavelength (Figure 52).

In the next Figure 57 we measured the I-V curve of the SiGe photodetector.

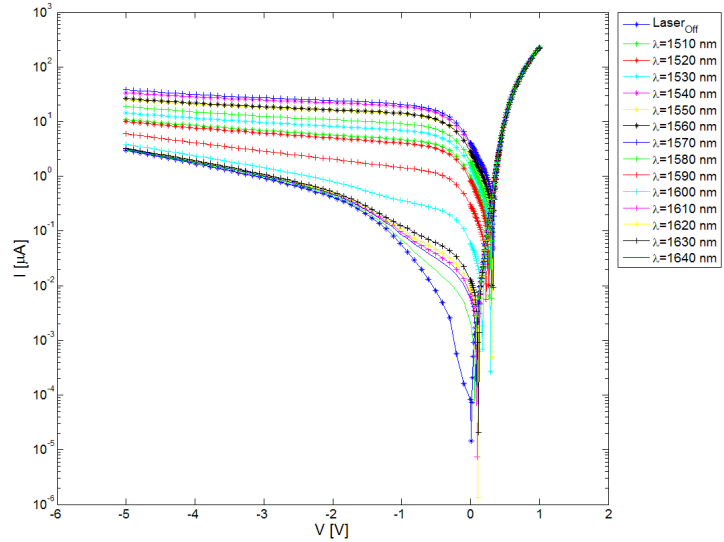

**Figure 57: I-V curve of the photodiode with the SiGe core. It is represented the dark current Idark and the photocurrent for diferent wavelengths from 1510 nm to 1640 nm every 10 nm**

As it was stated in the previous illustration, from Figure 57 we can see the dark current $I_{dark}$ (with the label Laser$_{Off}$) and the dark current with the photocurrent for other wavelengths.

From the measurements of Figure 56 and Figure 57 we can calculate the photocurrent $I_{PD}$. We only need to substract the dark current $I_{dark}$ to the values of current obtained in these figures. With this we have $I_{Tot}$-$I_{dark}$ which is the photocurrent $I_{PD}$. It will be calculated in the next section. We observe that the dark current is bigger in the case of SiGe. Probably, it is due to the diffusion of n-Si and p-Si in the annealing.

## 4.8.4 Responsivity Results

Having determined experimentally $P_{PD}$ and $I_{PD}$, we can derive the responsivity R. Responsivity results for the photodetector with the Ge core is presented in Figure 58,

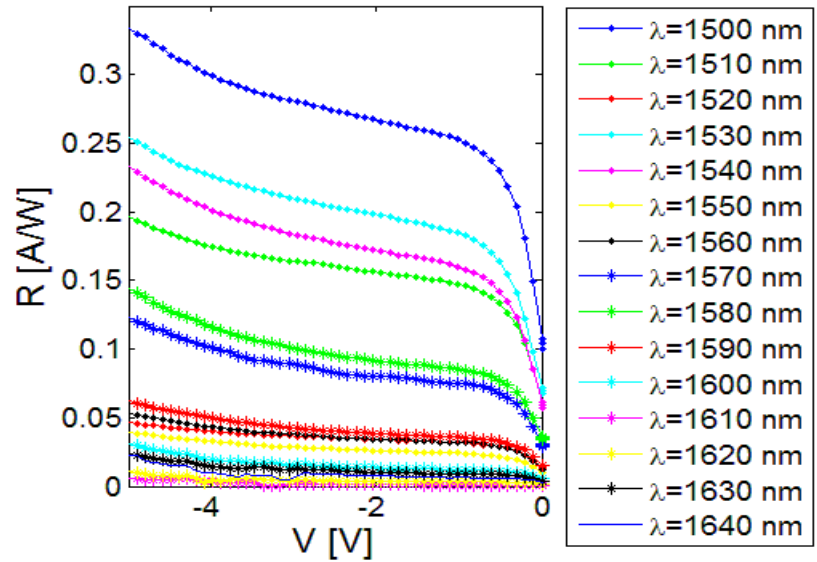

**Figure 58: Responsivity of the photodetector with Ge core versus the bias voltage V**

From Figure 58 we observe that the level of the responsivity R of the Ge photodetector is below 0.3 A/W. The results are similar to the values obtained in [141]. In [141] the responsivity measured in the Ge photodetectors are below 0.25 A/W.

From Figure 58 we observe that we have a slope in the responsivity R of the photodetector as the reverse bias voltage is increased from 0 V to -5 V. There are two kinds of regimes in the slope. The first one, between 0 and -0.7 volts in which the slope grows very fast. From -0.7 to -5 V we have a smaller slope. The photodetector we are using has a length of L=10 µm and a width of $w_{PD}$=1 µm.

Now, it is interesting to measure the responsivity R in the photodetector in which Si was diffused in the Ge to obtain SiGe. Such responsivity R versus the wavelength is presented in Figure 59,

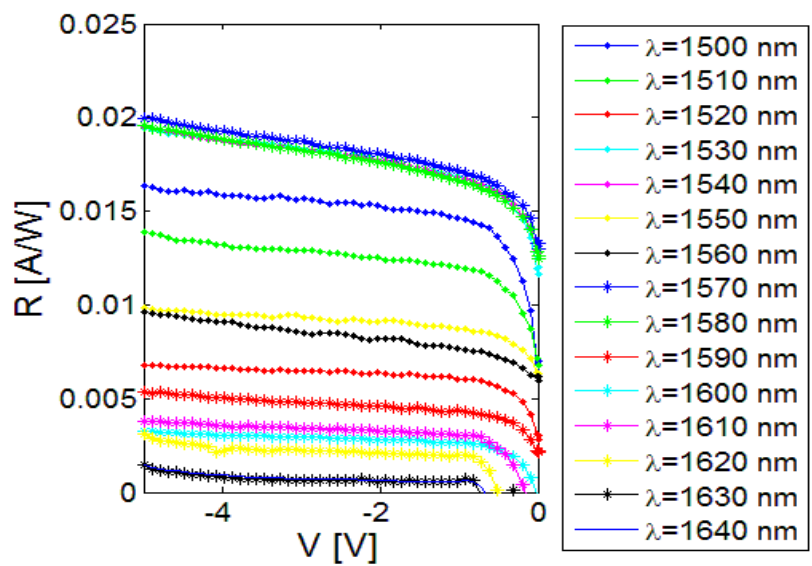

**Figure 59: Responsivity of the photodetector with SiGe core versus the bias voltage V**

The responsivity R of the SiGe photodetector in Figure 59 is significantly lower than in the Ge core photodetector shown in Figure 58. This is due to the fact that the diffusion of Si in the Ge core shifted the band edge of Ge to lower wavelengths. The direct bangap of the tensile strained Ge used is around 1.6 µm while the one of Si is around 1.2 µm. With the introduction of Si into the Ge the band-edge shifted to lower values less than 1.5 µm. this is confirmed by the low value of the responsivity R measured. We cannot measure lower wavelengths below 1.5 µm due to the bandwidth of the laser source. Additionally, the grating couplers used were designed for 1.55 µm.

To confirm that the band-edge of the SiGe core is shifted to values lower than 1.5 µm we change the laser. We use a fixed laser around 1.3 µm and we measured one point for this wavelength. It is represented in Figure 60,

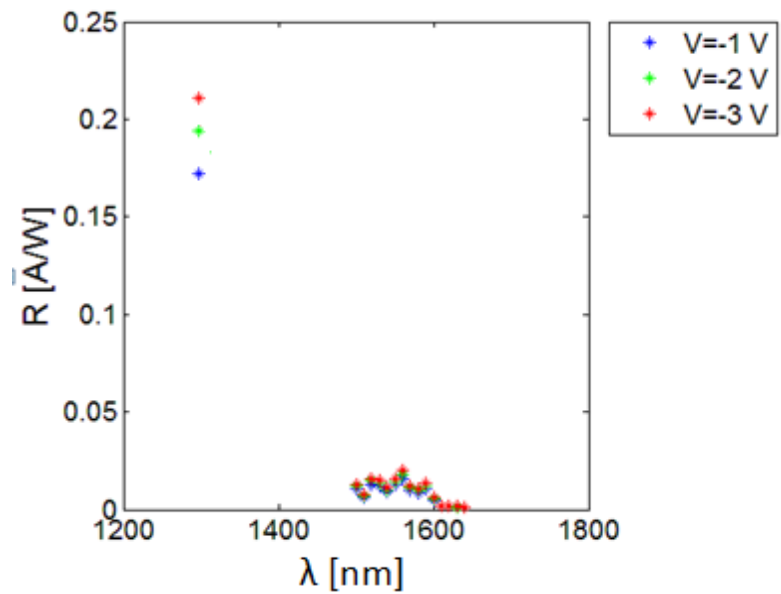

**Figure 60: Scanning of the responsivity R of a SiGe photodetector**

In Figure 60 we plotted the points already measured in Figure 59 but instead of representing it versus the bias voltage V we represented it versus the wavelength. We remind that in Figure 59 we scanned the wavelength from 1500 nm to 1640 nm every 10 nm. Additionally to these points, in Figure 60 we added three points measured with a laser at 1.3 µm. We can observe that the points around 1.3 µm have a larger responsibity R than the ones around 1.55 µm. This is due to the shift of the band-edge of Ge due to the diffusion of Si (obtaining SiGe). Additionally, the responsivity at 1.3 µm is around 0.15-0.2 A/W which is similar to the responsivity for wavelengths lower than 1.6 µm in [141]. It means, the Ge photodetector measured in [141] has a responsivity R=0.25 A/W for λ=1570-1580 nm which is below the band-edge of Ge . With this we confirm the fact that we diffused to much Si into Ge and the shift of the band-edge is at lower wavelengths than 1.55 µm.

We concluded that we were not able to measure the FKE in Ge and SiGe photodetectors but we can say that we diffused too much Si into Ge. The difficulty in this first set of experiments was that the Si content in Ge cannot be precisely controlled by annealing in order to diffuse the Si into the Ge to form SiGe. We launched a second set of wafers with a new approach. In this case we will grow SiGe epitaxially with a controlled fraction of Si into it. We will measure the same photodetectors but instead of annealing the Si will epitaxially grow SiGe with the following concentration (typically an addition of 0.75 [35]-0.8% [51] of Si into the Ge will shift the FKE to 1550 nm).

Although in our work we were not able to measure the FKE (there is not clear evidence), other researches at CEA-Leti measured the FKE using photodetectors. The results of their work is presented in Figure 61,

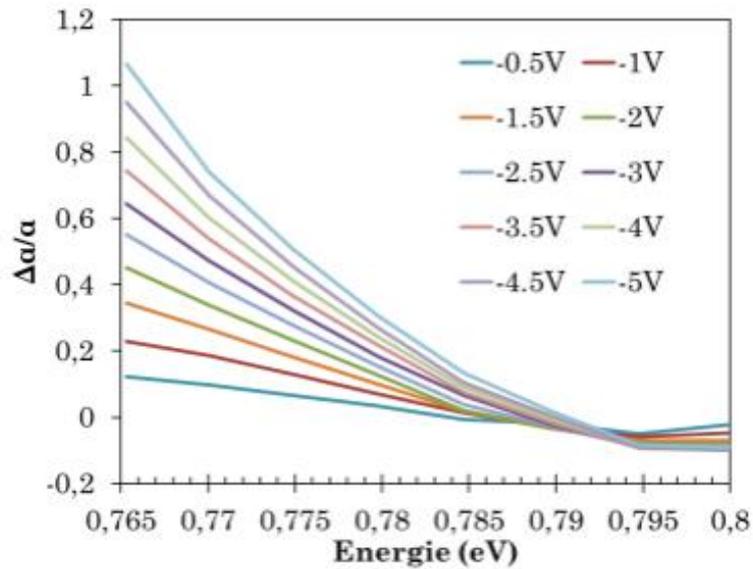

**Figure 61: Measurement of the FKE done by other researchers at CEA-Leti (Leopold Virot) using photodetectors**

In Figure 61 α is the absorption of the Ge when there is not applied bias and Δα is the difference in the absorption of the material when we apply a bias voltage to the photodetector.

# 5 Design of a Low-Power Consumption Plasmonic Modulator

Our main objective is to design a CMOS-compatible, low electrical power consumption optical modulator. We want that the electrical power consumption is below 50 fJ/bit as stated in the reference roadmap [33]. There are several FKE photonic modulators. We think that designing a FKE plasmonic modulator will reduce its dimension and consequently the electrical power consumption.

In this chapter we present a design of an electro-absorption plasmonic modulator based on FKE in Ge to perform the modulation. We propose several structures that use the FKE as an active principle to produce the modulation. Among these structures we selected the best one which is composed of a plasmonic MIS waveguide as introduced in the previous chapters.

One of the main tasks is to design a structure to concentrate the static electric field in the Ge core. It is important that the amount of static electric field into the Ge core varies with the voltage applied to the terminals of the structure. So, by changing the voltage it will be possible to change the absorption coefficient of Ge and produce intensity modulation.

Another important point of the device is to concentrate as much as propagating light into the Ge core, where the FKE takes place. It is due to the fact that the changes in electro-absorption of the material happen in the Ge core. The overlap between the static electric field and the electromagnetic field of the optical mode will determine the extinction ratio of the modulator. It means, by maximizing the overlap between the induced static electric field and the electromagnetic field of the light, we maximize the electro-absorption effect and thus increase the extinction ratio of the modulator.

We also study the reduction of the insertion losses of the device by reducing the propagation losses of the plasmon and the coupling losses from a standard Si input waveguide. A high bandwidth is also desirable. Furthermore, we want that the modulator is CMOS compatible.

As a summary we want to design a plasmonic modulator using a plasmonic waveguide to propagate a plasmonic mode and using the FKE to produce the modulation in Ge. For this we propose several structures that we analyze by simulations. We select the best one and we optimize it by using the integrated electro-optical simulator descrived in the previous chapter.

This simulator consists of a commercial electrical simulator to calculate the static electric field in the structure and calculate the modification of the absorption coefficient of Ge and a mode solver to evaluate the influence of the FKE on the effective refractive index and the effective losses of the plasmonic mode. A key point is to use the metal of the plasmonic waveguide as an electrical contact to induce the static electric field in the Ge core. The simulator was introduced in chapter four.

In this chapter, we explain the desired benchmarks for the device and we propose several CMOS compatible FKE-based plasmonic modulator structures. Using the integrated electro-optical simulator we optimize the structures to increase the extinction ratio, reduce the insertion losses, reduce the electrical power consumption, etc.

# 5.1    Specifications and Desired Benchmarks of the Device

The goal of the thesis is to develop a new kind of plasmonic modulator for performing an efficient electro-optical modulation in an integrated configuration. We will discuss the different benchmarks of the modulator like the CMOS compatibility, the electrical power consumption, the insertion losses and the extinction ratio.

## 5.1.1 CMOS Compatibility

As a first constraint we want that the proposed modulator is potentially able to be co-integrated with an electronic integrated circuit. For this, the dimensions of the device need to be small enough to match the dimensions of electronic elements. As shown in Figure 26, going to plasmonics will reduce the dimension mismatch between photonic and electronic devices. The plasmonic modulator must be therefore CMOS compatible. This will introduce constraints in both the materials and in the fabrication techniques used to build the device. The choice of materials and barriers were explained at the end of the third chapter. Furthermore, designing a CMOS compatible device will allow the use of the well-established microelectronics fabrication techniques and to benefit from the large scale price reduction of the devices.

## 5.1.2 Low Electrical Power Consumption

An important parameter of the design of a modulator related with the electrical energy consumption is the operational voltage of the device. As a general rule, the higher the operational voltage of a device the higher the power consumption uses to be. This also happens with the intrinsic capacitance of the device. It means, the higher the capacitance the bigger the electrical power consumption. One of the parameters which affect the capacitance is the length of the device. When the length of the device is increased then the capacitance is also increased and consequently the electrical energy consumption of the modulator. We can see that the photonic modulator presented in Table 3 have a higher operational voltage than the FKE photonic modulators presented in Table 2 due to the fact that the electro absorption needs less driving voltage to operate. We want to use a smaller driving voltage than the plasmonic modulators shown in Table 2.

One of the other constraints to impose to our design is that it has a low electrical power consumption in the order of a few fJ/bit. It is established in some roadmaps [33] that the required electrical energy consumption for a modulator in future optical links may be around 50 fJ/bit.

The power consumption for performing the modulation is given by Equation 17. It is key to reduce both operational voltage and capacitance to reduce electrical power consumption. This can be achieved by reducing the size of the device.

Setting the modulator as a plasmonic device will reduce the dimension of the modulator structure with respect to other photonic devices (both cross section and length). Reducing the size of the cross-section will increase the magnitude of the static electric field with respect to the same voltage; it means that with a lower voltage you induce the same static electric field as in a larger device. It is worth noting that modulators that use the carrier dispersion effect in a MZI configuration have an energy consumption around pJ/bit, which is well over the requirement of 50 fJ/bit reported by [33]. Photonic modulators based on FKE (Table 1) consume less energy than eletro-refractive modulators seen in Table 3 which use a MZI, a RR or a DR. In Table 2 we see key plasmonic modulator designs and it can be seen that they have less electrical power consumption than the photonic modulators that use the FKE reported in Table 1 and Table 3. There is a significant trend in which going from photonic to plasmonic reduces the electrical power consumption of the modulators.

## 5.1.3 Insertion Losses

Regarding the insertion losses of the device we want to obtain smaller insertion losses than the reported plasmonic devices presented in Table 2 to improve the performance of plasmonic devices.

In plasmonic devices, there is a trade-off between the confinement of light and the propagation losses of the optical mode as the higher the confinement of an optical mode, the higher the optical losses. Increasing the confinement of the mode will increase the electro-absorption effect and therefore lead to a compact device, limiting the impact of higher propagation losses on the total insertion losses of the modulator. The targeted value for the insertion losses of the device in this work will be less than 10 dB. We remind that the insertion losses are formed by the propagation losses of the plasmonic mode and the coupling losses which come from the coupling to the modulator form a standard Si waveguide.

### 5.1.4 Extinction Ratio

With respect to the extinction ratio of the modulators reported in the state-of-the-art of chapter two we can see that electro-absorption photonic modulators that use the FKE shown in Table 1 has a larger extinction ratio with respect to the key photonic modulators present in Table 3. Regarding the plasmonic modulators in Table 2 we can see that the extinction ratio is similar to the photonic FKE ones shown in Table 1. Concerning the extinction ratio we want to develop a modulator with an extinction ratio higher than 3 dB.

### 5.1.5 Summary of the Targeted Benchmarks

As a summary we want to design a plasmonic modulator that uses the FKE with electrical power consumption less than 50 fJ/bit, an operational voltage less than 3 V, an insertion losses in the order of 10 dB and an extinction ratio higher than 3 dB. We also want a high bandwidth (>40 GHz). Furthermore, we want that the device is CMOS compatible. For this, we will propose and simulate many structures, select the best one and optimize it using the integrated electro-optical simulator developed in this thesis.

## 5.2    Proposed Plasmonic Modulator Structures

### 5.2.1 Requirements and Figures of Merit

#### 5.2.1.1   Overlap of the Static Electric Field and the Electromagnetic Field in the Ge Core

As a general fact we want to use the FKE in Ge as the active principle of the modulator. So, the designed modulator needs to have a Ge core. In this core we need to find a structure to induce a static electric field using two or more terminals when a voltage is applied. Furthermore, we need to create a waveguide around the Ge in a way that the main part of the optical electromagnetic field of the plasmonic mode is in the Ge core. In this manner the effective losses of the plasmonic mode will change according to the application of the voltage to the structure. Changing the material absorption $\alpha_{mat}$, we change $\alpha_{eff}$ of the mode. $\alpha_{eff}$ is the effective losses of the plasmonic mode supported by the waveguide. We remind that the static electric field in the Ge induced by the voltage leads to variation of the optical absorption coefficient $\alpha_{mat}$ of the Ge. Hence, producing the modulation.

To summarize, we need to maximize the overlap between the static electric field induced by applying a voltage between terminals and the optical electromagnetic field. Both fields need to be present in the Ge core where the FKE is present to produce the modulation of the plasmonic mode.

### 5.2.1.2  Maximum Static Electric Field

Another requirement of the modulator is that the static electric field must vary from almost 0 kV/cm until a maximum value of 100 kV/cm to produce an effective change of $\alpha_{mat}$. The breakdown static electric field of Ge is around 100 kV/cm, so, this value cannot be exceeded. Furthermore, a low voltage to produce such a change has to be considered in order to reduce the electrical power consumption of the plasmonic modulator as was explained previously.

### 5.2.1.3  Definition of the Figure of Merit

At the beginning of this section we present a summary of the structures that we analyzed. We present a summary of the main idea and we say the cause of having a small performance. Later, we concentrate on the two best structures the MS and the MIS configurations and we explain the optimization of them in more detail.

Now, to compare the structures we need to introduce a Figure of Merit (FoM). We remind that the parameter $\alpha_{eff}$ is related with the propagation losses of the modulator and $\Delta\alpha_{eff}$ is related to the extinction ratio of it. It means, to obtain the propagation losses you need to multiply the parameter $\alpha_{eff}$ by L, where L is the length of the modulator. The same multiplication must be done with the parameter $\Delta\alpha_{eff}$ to obtain the extinction ratio. For the definition of $\Delta\alpha_{eff}$, it is given by Equation 56. We want to minimize the propagation losses and we want to increase the extinction ratio. Consequently, a good figure of merit will be FoM= $\Delta\alpha_{eff}/\alpha_{eff}$ =ER/PL. Maximizing the FoM we obtain the desired reduction of the propagation losses and the increase of the extinction ratio. This is the FoM that we are going to use throughout all the designs of the plasmonic modulator.

## 5.2.2 Vertical PIN Structures and Slot Plasmonic Waveguides

For inducing a static electric field into the Ge core first we thought about using a PIN structure in reverse bias. This structure is widely use in photonic modulators to induce a static electric field in the intrinsic region of the PIN. As a first approach we thought about a vertical PIN structure formed by p-doped Ge in the bottom below an intrinsic Ge core. An n-doped Ge is placed over the intrinsic Ge. The structure is presented in the following Figure 62. We placed a top contact over the n-doped Ge and two side contacts over the p-doped Ge at the bottom of the Ge PIN. Finally, the PIN structure is over an intrinsic Si layer over the buried oxide.

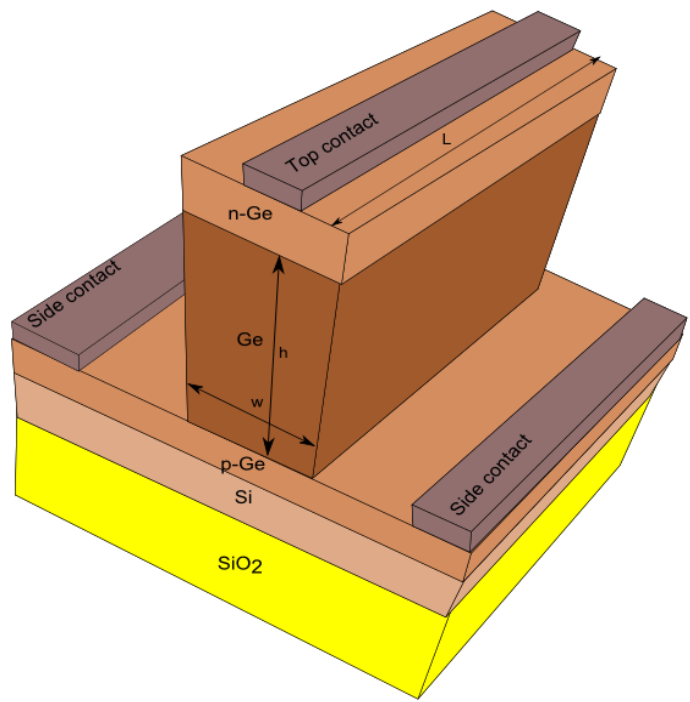

**Figure 62: Vertical PIN structure to induce a static electric field in the intrinsic core**

By applying a reverse bias voltage between the top contact and the two side contacts we are able to induce a static electric field in the intrinsic Ge core. As a first approach we selected the values w=50 nm and h=150 nm which are common values for PIN structures. We selected just those values to view as a first approach how the static electric field distribution is. We vary the reverse bias voltage from 0 V to 5 V. The doping levels used are $10^{19}$ cm$^{-3}$ for the n- and p-doped parts and $10^{16}$ cm$^{-3}$ for the intrinsic regions (those values are used all along this work). The result is represented in Figure 63.

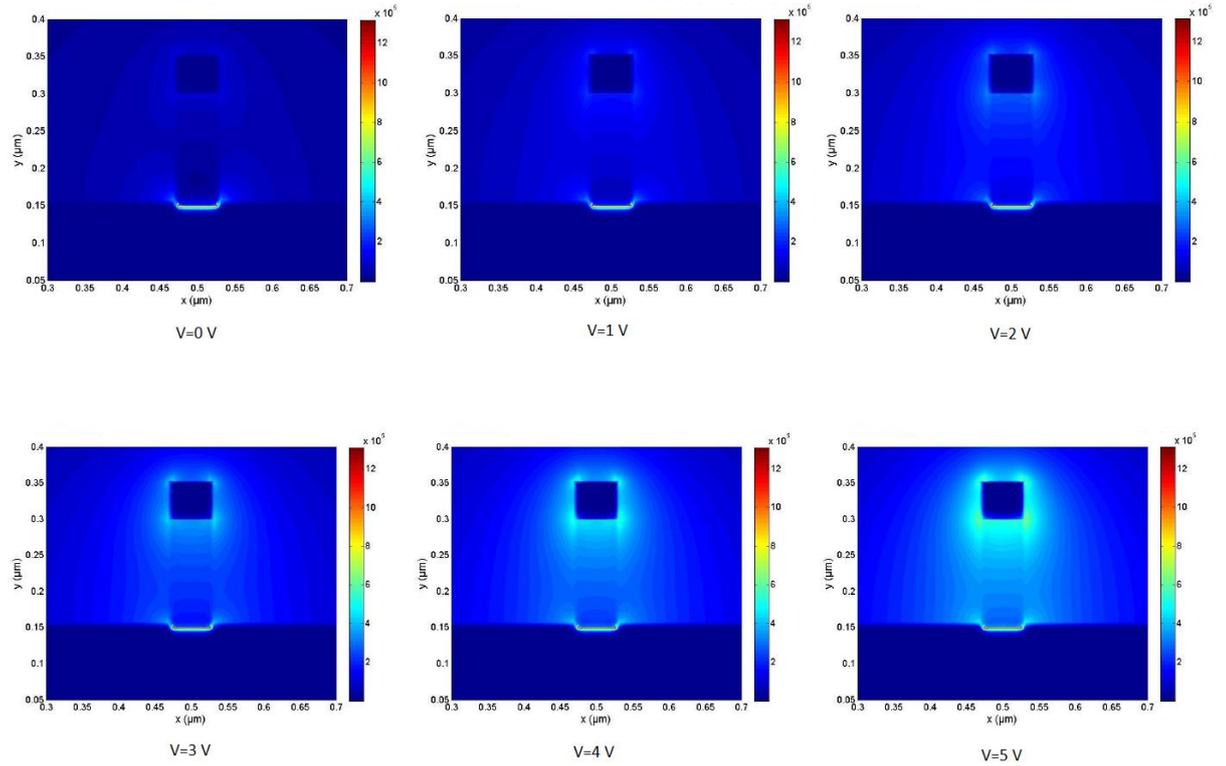

**Figure 63: Static electric field distribution induced in the intrinsic region of the vertical PIN of Figure 62 from 0 to 5 V in reverse bias**

From Figure 63 it is possible to see that a high static electric field can be present in the intrinsic Ge core. For V=5 V a field of $4.5 \times 10^5$ V/cm is induced in the Ge core. This value is considerably over the maximum static electric field that Ge can support ($1 \times 10^5$ V/cm). This value of $1 \times 10^5$ V/cm is achieved between 0 and 1 V. This gives us the idea that with a low driving voltage a high static electric field can be induced in the intrinsic region of the PIN structure.

Now that we have a structure to induce a high static electric field into the Ge core we need to add something that forms a plasmonic waveguide to induce the optical electromagnetic field in the Ge core. As a first approach we placed two metals around the intrinsic core to form a plasmonic slot waveguide. This structure is represented in Figure 64,

**Figure 64: Modulator structure with a vertical PIN and a slot plasmonic waveguide around the intrinsic Ge**

The structure of Figure 64 is basically a vertical PIN structure which has a plasmonic slot waveguide around the intrinsic Ge region of the PIN. The plasmonic slot waveguide is formed by the metal Cu and a small barrier of $Si_3N_4$. The use of $Si_3N_4$ is to avoid the diffusion of the Cu into the Ge core, this will also decrease the propagation losses of the supported plasmonic mode.

The plasmonic slot waveguide formed by the horizontal stack of Cu-$Si_3N_4$-Ge-$Si_3N_4$-Cu supports a plasmonic mode whose optical electromagnetic field is concentrated in the Ge core. The normalized optical intensity distribution of the plasmonic mode supported by the structure of Figure 64 is represented in Figure 65. It is possible to observe that the intensity of the optical mode is concentrated in the Ge core between the terminals in which the driving voltage is applied. It is also worth to mention that such a field is also concentrated into the $Si_3N_4$ slot.

The thin $Si_3N_4$ slot has the disadvantage that it is difficult to fabricate. Furthermore, the side metals of Cu in the plasmonic slot waveguide must be settled to V=0 V to avoid them to have a floating potential.

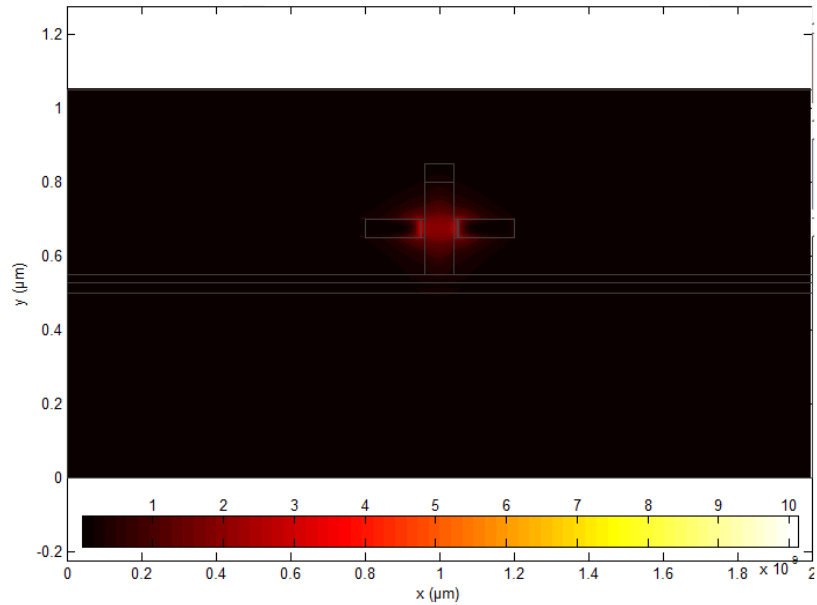

**Figure 65: Normalized optical intensity distribution of the structure of Figure 64**

Now, it is interesting to see how the presence of the metals that form the plasmonic slot waveguide affects the distribution of the static electric field in the structure of Figure 64 with respect to Figure 62. As a first approximation we settled w=50 nm, h=250 nm, $w_{Si3N4}$=5 nm, $h_{Cu}$=50 nm, $h_{Down}$=50 nm and $h_{Up}$=50 nm. We selected $h_{Cu}$=50 nm because it is a typical value in plasmonic slot waveguides, the rest of the parameters are optimized to minimize the propagation losses, for this, we scanned the parameters (it has been done optically). The parameters $w_{Si3N4}$, $h_{Down}$ and $h_{Up}$ are going to be scanned to know their influence on the static electric field distribution. With those parameters the static electric field distribution is presented in Figure 66,

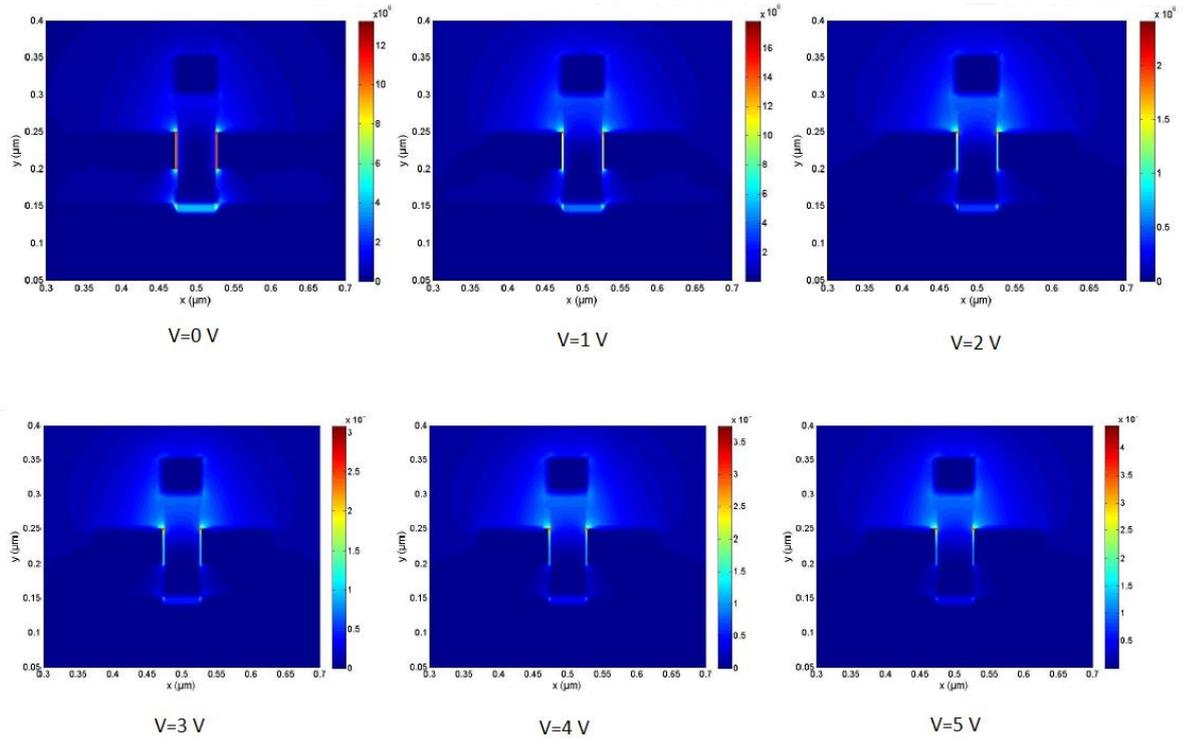

**Figure 66: Static electric field distribution induced in the intrinsic region of the structure of Figure 64 from 0 to 5 V in reverse bias**

From Figure 66 we observe that even for large bias voltages the static electric field is not confined in the core of the plasmonic slot waveguide. The static electric field is confined in parts of the structure where the electromagnetic field of the plasmonic mode is not present. It is mainly between the top metal and the side metals that form the slot. We remind that the optical electromagnetic field is present in the middle of the intrinsic Ge core and between the Cu metals (Figure 65). Furthermore, in this zone, the static electric field is confined in the $Si_3N_4$ slot and it is not present in the Ge where the FKE takes place. We varied the parameters $h_{Down}$ and $h_{Up}$ and we obtained the same problem, mainly low overlap between the static electric field and the optical electromagnetic field of the plasmonic mode which leads to a small performance of the device.

The next step we did is to settle $h_{Down}$ and $h_{Up}$ equal to 0 nm. In this case $h_{Cu}$ is equal to h (Figure 64). It means: all the plasmonic slot waveguide covers all the intrinsic region of the PIN diode. It is represented in Figure 67,

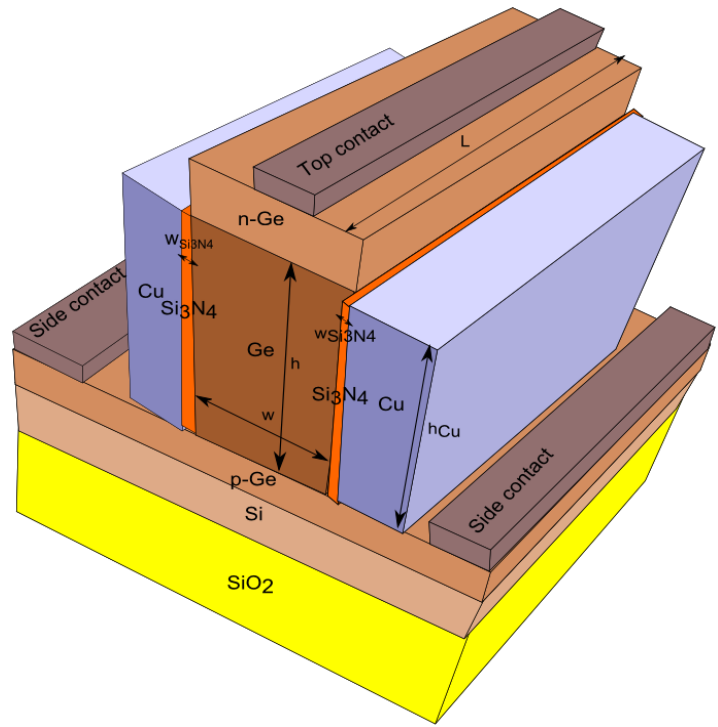

**Figure 67: Modulator structure with a vertical PIN and a slot plasmonic waveguide around the intrinsic Ge**

The normalized optical intensity of the plasmon supported by the structure of Figure 67 is represented in Figure 68. The optical field is concentrated in the low region of the intrinsic core of the vertical PIN structure.

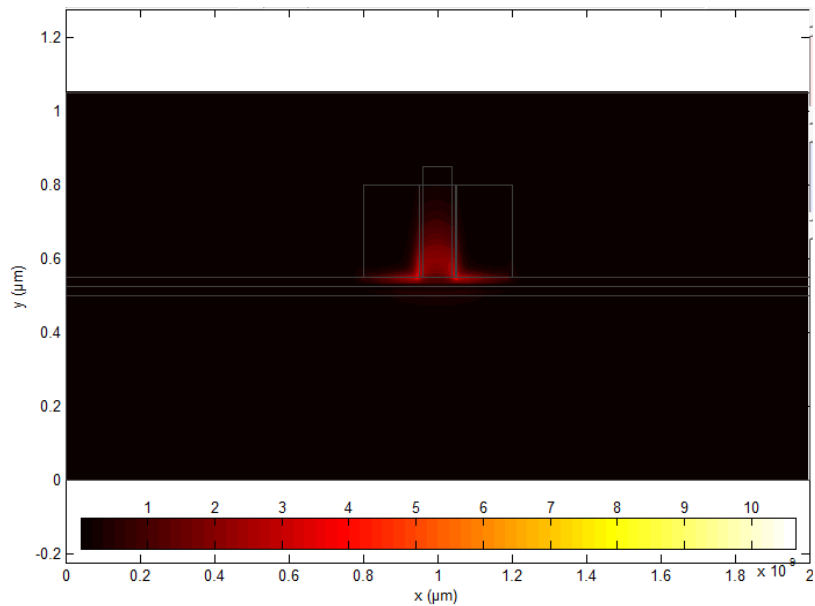

**Figure 68: Normalized optical intensity distribution of the optical field of the structure in Figure 67**

With this configuration we expect that the static electric field will be concentrated in the intrinsic Ge core of the vertical PIN structure. Again, we scanned the bias voltage V from 0 V to 3 V. We remind that the reverse bias is applied between the top contact and both side contacts. The pieces of metal of the plasmonic slot waveguide are set to ground. With this, we obtained the following static electric field distributions (Figure 69),

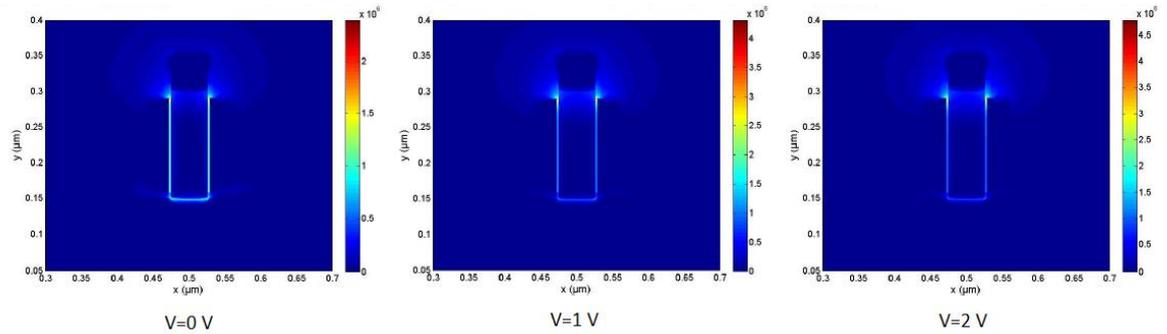

**Figure 69: Static electric field distribution for the structure of Figure 67**

It is possible to see that the static electric field is confined in the Si$_3$N$_4$ slots (where the static electric field is very high) and it is not present in the Ge core where the FKE takes place and where the optical electromagnetic field of the mode is present (Figure 68).

The best FoM found trying to optimize the structure of Figure 64 and Figure 67 with respect to the parameters w$_{Si3N4}$, w, h, h$_{Up}$, h$_{Down}$, and h$_{Cu}$ is in the order of $\Delta\alpha_{eff}/\alpha_{eff}$=0.11 applying a reverse bias of 5 V between the top contact and the side contacts. Furthermore, with this $\Delta\alpha_{eff}/\alpha_{eff}$ and 5 V there are some regions in Ge where the static electric field is larger than the breakdown static electric field of Ge (10$^5$ V/cm). For this performance the values used were w=120 nm, h=250 nm, w$_{Si3N4}$=5 nm, h$_{Down}$=60, h$_{Up}$=40 nm and h$_{Cu}$=150 nm (structure of Figure 64).

## 5.2.3 Horizontal PIN Structures

Now, we want to study a horizontal PIN structure. It is similar to the structure of Figure 62 but the new configuration of the PIN is horizontal. The stack is horizontal but the materials are the same namely from left to right it is formed by n-doped Ge, an intrinsic Ge and a p-doped Ge in the right. Everything is over a Si layer. Below the buried oxide is represented. This horizontal PIN is represented in Figure 70.

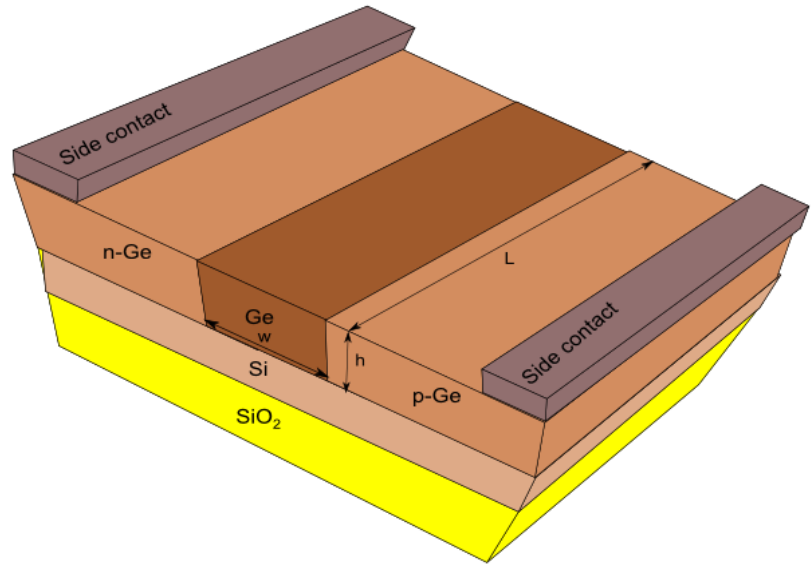

**Figure 70: Horizontal PIN structure to induce a static electric field in the intrinsic core**

Similarly with the vertical PIN structure (Figure 62) when there is a reverse bias voltage between the side contacts a static electric field is present in the intrinsic Ge core. Now, we need to add a plasmonic waveguide to try to induce also the optical electromagnetic field of the plasmonic mode in the Ge core.

For this, we add a $Si_3N_4$ and a Cu layers over the intrinsic Ge as shown in Figure 71. As explained before, the $Si_3N_4$ slot acts as a diffusion barrier. Furthermore, the stack of Ge-$Si_3N_4$-Cu forms a plasmonic MIS waveguide which, as explained before, is a waveguide that has low propagation losses.

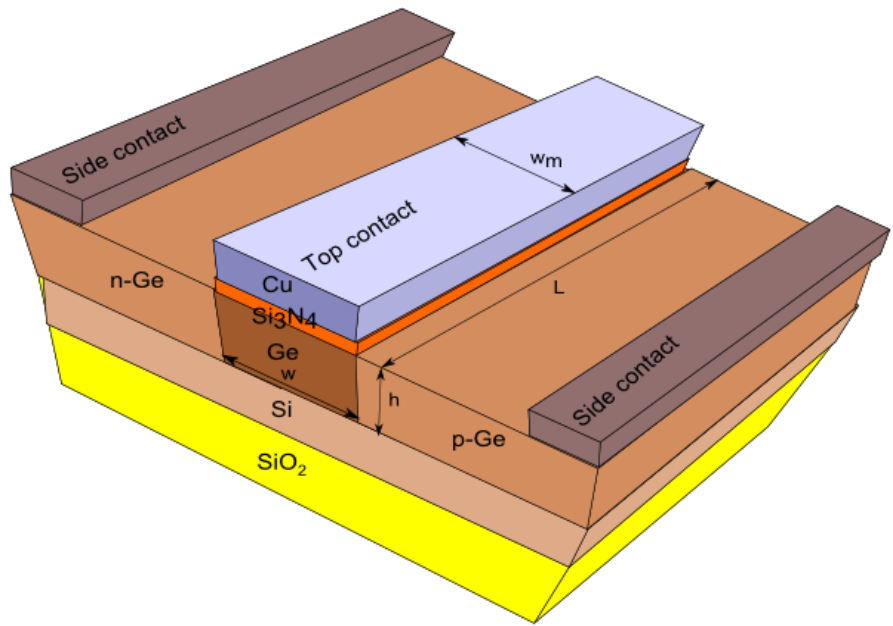

**Figure 71: Horizontal PIN structure with a plasmonic waveguide on top**

In the structure of Figure 71 the optical electromagnetic field is present in the $Si_3N_4$ slot and in the intrinsic Ge core. The field distribution is similar to a plasmonic MIS waveguide. The normalized optical intensity distribution is represented in Figure 72,

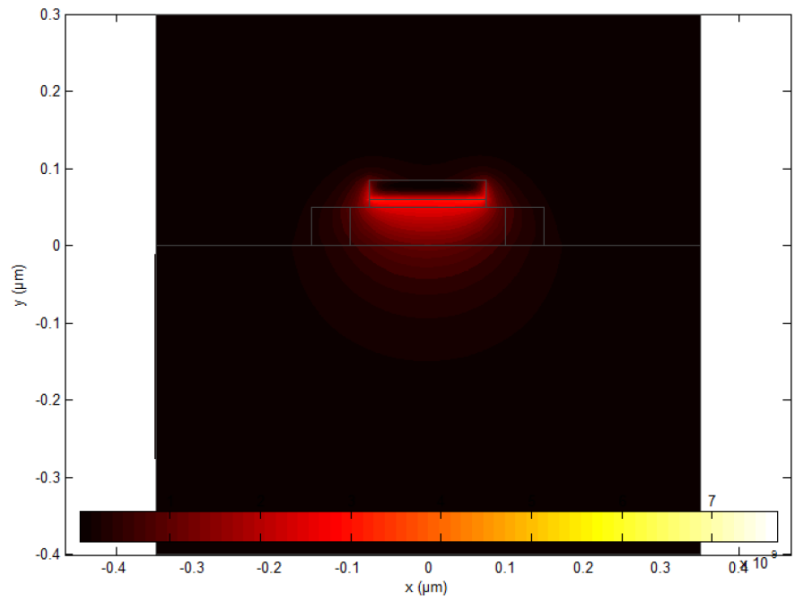

**Figure 72: Normalized optical intensity distribution in the horizontal PIN structure of Figure 71**

From Figure 72 we can see the intensity of the mode is concentrated in the intrinsic part of the Ge core. Nevertheless there is optical field outside this region like in the $Si_3N_4$ slot or at the sides of the top metal. The fact that not all the field is concentrated in the Ge core where the FKE takes place downgrades the performances of the structure.

Now that we studied the distribution of the optical field, it will be interesting to calculate the static electric field distribution in the structure. For this we applied 0 and 5 V reverse bias between the side contacts of the horizontal PIN (Figure 71). We set the top metal of the MIS waveguide to 0 V. With these considerations the static electric field distribution is in Figure 73,

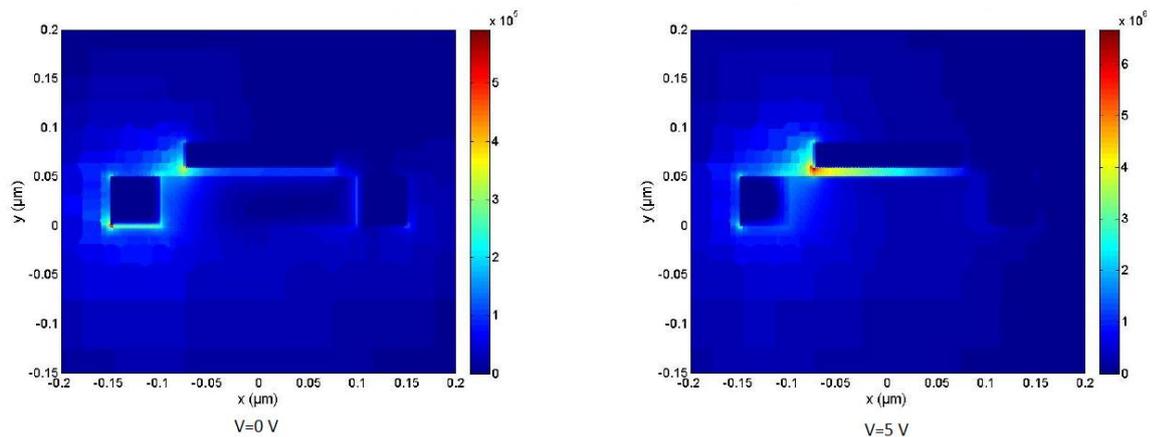

**Figure 73: Static electric field distribution for a reverse bias for V=0 V and V=5 V in the structure of Figure 71**

Form Figure 73 we can see that the static electric field distribution is concentrated between the left side of the top Cu and the n-doped Ge region of the horizontal PIN for both V=0 V and V=5 V. To try to improve the performance of this structure we tried to shift the $Si_3N_4$ barrier and the Cu to the left. The modified horizontal PIN structure is presented in Figure 74. With this we think that we can shift the entire optical field to the left where the biggest static electric field is concentrated giving the maximum change in the FKE. Although this also shift the maximum of the static electric field.

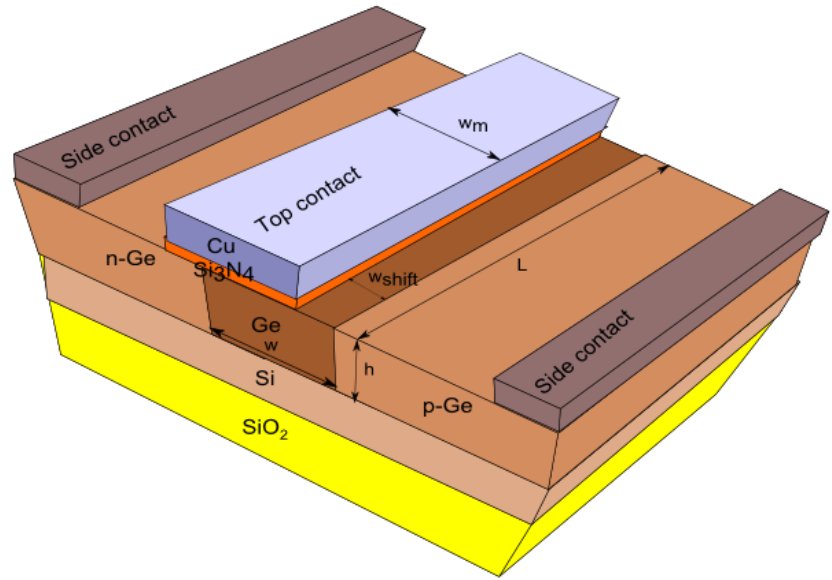

**Figure 74: Shifted horizontal PIN structure with a plasmonic waveguide on top**

Using the structure of Figure 74 and playing with the parameter $w_{shift}$ we did not obtain a very good performance of the device. The reason is low overlap between the static electric field and the optical electromagnetic field in the Ge core of the device. Regarding the structures of Figure 71 and Figure 74 the maximum $\Delta\alpha_{eff}/\alpha_{eff}$ is 0.14 for a bias voltage of 5 V. The parameters were w=200 nm and h=50 nm.

## 5.2.4 Horizontal Plasmonic Slot Waveguide

In the following structure we want to use a plasmonic slot waveguide in the horizontal direction. The structure is presented in Figure 75. The slot waveguide is formed by the horizontal stack of Cu-$Si_3N_4$-Ge-$Si_3N_4$-Cu. This structure supports a plasmonic mode whose optical electromagnetic field is mostly concentrated in the intrinsic Ge core. One of the advantages of this waveguide is that it can confine the light in a very small region in the Ge. Nevertheless some field is also present in the $SiO_2$ buffer and on the top of the Ge. Due to the trade-off of the confinement and the dimension this kind of waveguide suffers from big optical propagation losses.

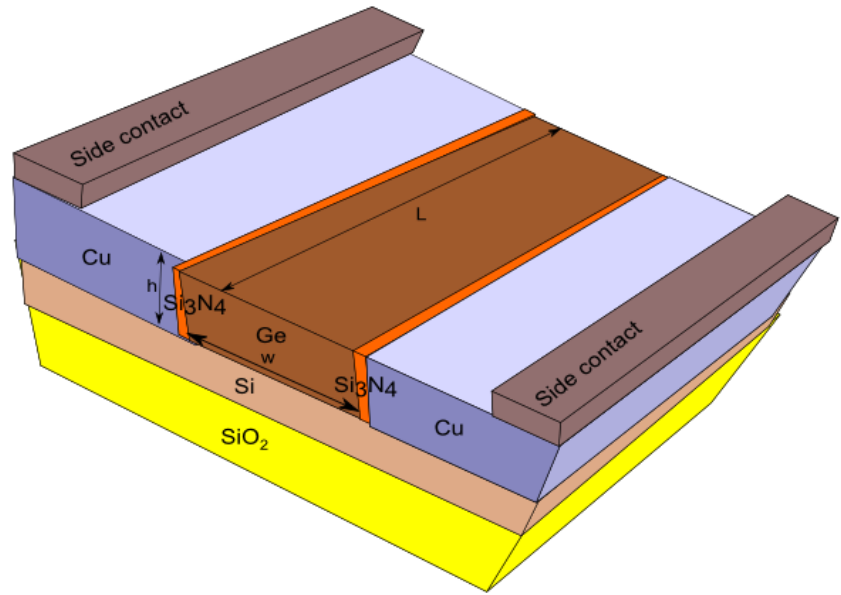

**Figure 75: Structure of the slot plasmonic modulator**

Optimizing the parameters of w, h and $w_{Si3N4}$ we obtained a maximum FoM $\Delta\alpha_{eff}/\alpha_{eff}$ in the order of 0.11 for 3 V. Nevertheless the voltage needs to be reduced because the static electric field in Ge exceeds the breakdown voltage of Ge ($10^5$ V/cm). If we reduce the voltage to 2 V then the FoM $\Delta\alpha_{eff}/\alpha_{eff}$ is around 0.075.

Now that we studied several structures we will summarize the FoM $\Delta\alpha_{eff}/\alpha_{eff}$ and driving voltage. It is also interesting to know if the breakdown static electric field is exceeded. We summarized everything in Table 6.

| Structure | Figure 64 and Figure 67 | Figure 71 and Figure 74 | Figure 75 |
|---|---|---|---|
| $\Delta\alpha_{eff}/\alpha_{eff}$: | 0.11 | 0.14 | 0.11 |
| Bias voltage: | 5 | 5 | 3 |
| Maximum static electric field in Ge: | Exceeded | Exceeded | Exceeded |

**Table 6: Summary of the plasmonic modulators structures studied until now**

According with Table 6 we obtained a maximum FoM $\Delta\alpha_{eff}/\alpha_{eff}$ around 0.11 for a bias voltage of 3-5 V. It is worth to mention that the breakdown static electric field in Ge is exceeded in all the structures. This means that if we reduced the bias voltage then the FoM $\Delta\alpha_{eff}/\alpha_{eff}$ should be smaller than 0.11.

## 5.2.5 Plasmonic MS Waveguide

It is widely known that the electric field in a gate of a MOSFET (Metal–Oxide–Semiconductor Field-Effect Transistor) transistor is very high. We remind that in the gate of transistors a capacitance is present where the core of it is the dielectric that acts as an insulator. Consequently, to satisfy our needs we think that using a capacitor structure, similar to the one used in the gate of transistors, will allow obtaining a high static electric field. The stack of the structure of the proposed modulator is represented in Figure 76. It is formed from the top to the bottom by a layer of metal, a layer of non-intentionally-doped (nid) Ge and a p doped Ge layer which is over the Si layer. Below, there is a SiO₂ substrate. One contact will be the top metal and the other one is taken on the p-doped Ge. Applying a voltage between the terminals it will induce a static electric field in the Ge core.

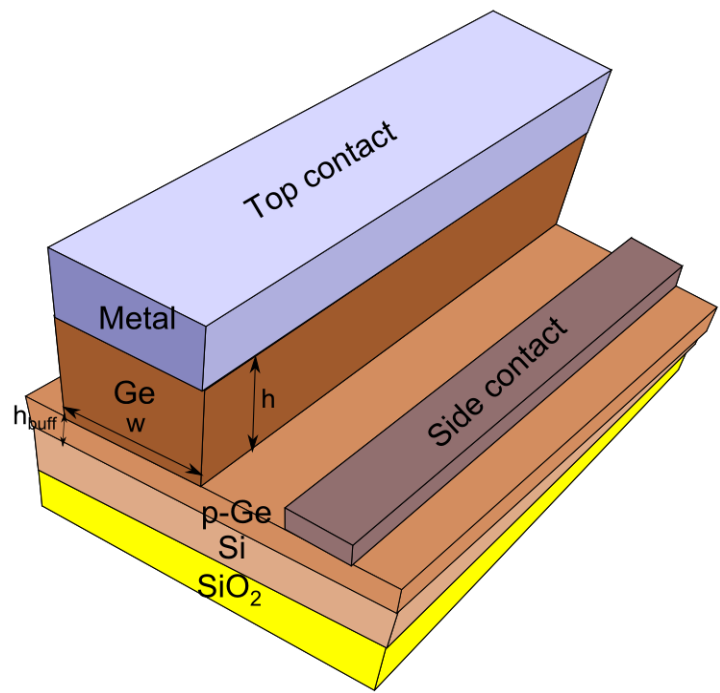

**Figure 76: Capacitor-like structure for the FKE plasmonic modulator**

Once we have defined the structure, we have to know which metal is the best suited for using it in our MS waveguide. The best metal is the one with the lower propagation losses of the plasmonic mode. For this we consider the following metals: Al, Ag, Au and Cu. Additionally we will use a Cu with a special process done at CEA-Leti [123] which reduces the propagation losses of the plasmonic mode with respect to standard Cu found in [119]. We will name this Cu as $Cu_{CEA-Leti}$. Although Ag and Au are not CMOS compatible we present them in order to do a complete comparison of all the metals. We take the refractive index of the metals from Palik [119]. The result is found in Table 7. With this we complete the study of the metal done at the end of chapter three.

| Material: | Ag [119] | Al [119] | Au [119] | Cu [119] | $Cu_{CEA-Leti}$ [123] |
|---|---|---|---|---|---|
| Refractive Index: | 0.590+11.05i | 1.73+16.6i | 0.62+10.41i | 0.657+9.61i | 0.325+11.4i |

**Table 7: Refractive index of the metals used to simulate the MS structure of Figure 76**

For the simulation we select a value of w=250 nm for the width of the Ge core and h=350 nm for the height of the Ge core. The thickness of the p-doped Ge layer is $h_{buff}$=400 nm. We apply a bias voltage of 3 V between the contacts. We selected the value of w=250 nm because it is typical for several plasmonic waveguides. Regarding the parameter h=350 nm we selected it because if we apply 3 V and we approximate the MS structure as a capacitance for h=350 nm we obtain a static electric field in the capacitance in the order of E=85x10$^5$ V/cm which is below the breakdown static electric field of Ge. To calculate the approximated static electric field in the MS structure we used the formula E=V/d where V is the applied voltage and d is the distance between the terminals of the capacitor (inclusing the doped layers). Regarding the choice of $h_{buff}$, we selected the value of 400 nm as a first attempt. As a first approximation we think that the influence of the parameter $h_{buff}$ on the extinction ratio, the insertion losses and the electrical power consumption is not important since both the static electric field and the optical electromagnetic field are not highly present in the buffer.

For the different metals, we calculate the figure of merit $\Delta\alpha_{eff}/\alpha_{eff}$ which accounts for the optimization of the modulator. It means, it is to increase the extinction ratio ($\Delta\alpha_{eff}$) and reduce the propagation losses ($\alpha_{eff}$) as stated before. We also present the maximum static electric field that appears into the Ge core in order to be sure that we do not exceed the breakdown electric field of Ge which is around 10$^6$ V/cm. The results are presented in the following Table 8,

| Material: | Ag | Au | Cu | $Cu_{CEA-Leti}$ |
|---|---|---|---|---|
| $\Delta\alpha_{eff}/\alpha_{eff}$ (3 V): | 0.13 | 0.10 | 0.07 | 0.23 |
| $E_{max}$ [kV/cm]: | 95 | 95 | 92 | 92 |

**Table 8: Performance of the plasmonic modulator of Figure 76 for different metals**

Where $E_{max}$ is the maximum static electric field found in the Ge core. From Table 8 we can see that the best metal in terms of efficiency for the MS waveguide of Figure 76 is given by $Cu_{CEA-Leti}$. It provides a FoM of 0.23. This is mainly due to the particular combination of the real and imaginary part of the $Cu_{CEA-Leti}$ material with respect to Ag, Au and normal Cu, which induce the lowest plasmonic losses. Furthermore all the metals provide a similar $\Delta\alpha_{eff}$ parameter since all induce in the Ge core a similar static electric field that varies between 92-95 kV/cm as can be seen in Table 8. According to these results although the best is $Cu_{CEA-Leti}$, when it is deposited over Ge there will be diffusion of the Cu into Ge, this will increase the propagation losses of the plasmonic mode. Consequently, we use the second best metal that is Ag with $\Delta\alpha_{eff}/\alpha_{eff}$ (3 V)=0.13 in the MS structure.

### 5.2.5.1   Influence of the Metal Width

In the next step we try to pattern the metal of the modulator to try to see its influence in FoM $\Delta\alpha_{eff}/\alpha_{eff}$ by reducing the metal in the lateral dimension of the modulator using a step on it. This modification is shown in the following Figure 77. As we can see, a step of thickness $h_{top}$ is patterned into the metal.

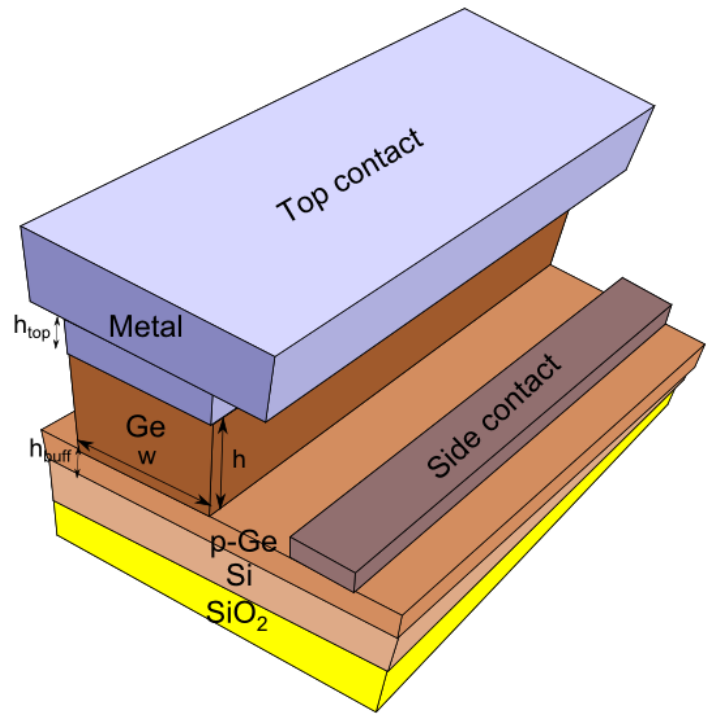

**Figure 77: Study of the influence of the patterned metal in the plasmonic modulator**

The influence of the height $h_{top}$ of the metal on the effective absorption $\alpha_{eff}$ of the plasmonic mode is summarized in the following Table 9. The dimensions used are w=250 nm, h=350 nm, $h_{buff}$=400 nm, V=3 V. We selected w, h and $h_{buff}$ with the same values as before for the reasons explained at the beginning of this section and we vary $h_{top}$ (0, 50 and ∞ nm). The results are presented in Table 9,

| $h_{top}$ [nm]: | 0 | 50 | ∞ |
|---|---|---|---|
| $n_{eff}$: | 3.98 | 4.19 | 4.12 |
| $\alpha_{eff}$ [cm$^{-1}$]: | 3167 | 3279 | 3835 |

**Table 9: Summary of the study of the patterned metal**

From Table 9 we can see that the larger the parameter $h_{top}$, the larger the effective optical losses of the plasmonic mode.

To study the influence of the parameter $h_{top}$ in the FoM= $\Delta\alpha_{eff}/\alpha_{eff}$, it is interesting to calculate $\Delta\alpha_{eff}/\alpha_{eff}$. For this, we represent the results for $h_{top}$=0 and ∞ nm in Table 10,

| $h_{top}$ [nm]: | 0 | $\infty$ |
|---|---|---|
| $\Delta\alpha_{eff}/\alpha_{eff}$ (3 V): | 0.11 | 0.10 |
| $E_{max}$ [kV/cm]: | 95 | 100 |

**Table 10: Summary of the study of the patterned metal**

From the results of Table 10 we can see that there is not a large influence of the parameter $h_{top}$ on the FoM $\Delta\alpha_{eff}/\alpha_{eff}$. We can see that the FoM is slightly higher with $h_{top}$=0 nm. Consequently, we set $h_{top}$=0 nm. The structure is similar to the one represented in Figure 76. Furthermore, this structure is easier to fabricate than the one with a finite $h_{top}$.

## 5.2.5.2   Influence of the p-Ge Layer Thickness

Next we studied the influence of the buffer layer of thickness $h_{buff}$ on the performance of the device. Since both the static electric field and the optical electromagnetic field are not highly present in the buffer layer, we expect a small influence in the FoM $\Delta\alpha_{eff}/\alpha_{eff}$ of the parameter $h_{buff}$. We remind that the parameters were set to the values w=250 nm and h=350 nm for the reasons explained before. We vary $h_{buff}$ from 50 to 400 nm. The results for $h_{buff}$=50, 200 and 400 nm are reported in the following Table 11,

| $h_{buff}$: | 50 | 200 | 400 |
|---|---|---|---|
| $\Delta\alpha_{eff}/\alpha_{eff}$ (1 V): | 0.02 | 0.03 | 0.03 |
| $\Delta\alpha_{eff}/\alpha_{eff}$ (2 V): | 0.06 | 0.07 | 0.08 |
| $\Delta\alpha_{eff}/\alpha_{eff}$ (3 V): | 0.12 | 0.12 | 0.13 |
| $E_{max}$ [kV/cm]: | 90 | 85 | 90 |

**Table 11: Summary of the study of the influence of the buffer layer**

According to Table 11 we can see as expected there is a little influence of $h_{buff}$ on the FoM $\Delta\alpha_{eff}/\alpha_{eff}$. Furthermore, if the buffer layer is too thick, the coupling of the device from a standard Si rib waveguide will be difficult when we want to integrate it into an integrated circuit. It also depends on the heigh h of the Ge core. The main reason is that the incoming light will be lost into the substrate. Consequently, the use of a thick buffer is a disadvantage compared to a thin one (although the FoM $\Delta\alpha_{eff}/\alpha_{eff}$ is slighty higher). The smaller the buffer, the better the coupling will be. We then select a buffer of $h_{buff}$=50 nm. Additionally, there is not too much difference between $h_{buff}$=400 nm where $\Delta\alpha_{eff}/\alpha_{eff}$=0.13 and when $h_{buff}$=50 nm where $\Delta\alpha_{eff}/\alpha_{eff}$=0.12. So, although we slightly reduce the FoM, we can obtain advantages regarding the coupling to the device.

### 5.2.5.3 Influence of the p-Ge Layer Width

It is also interesting to know how the pattern of the p-Ge layer of thickness $h_{buff}$ changes the FoM $\Delta\alpha_{eff}/\alpha_{eff}$ when it is confined laterally to a dimension equal to the width of the Ge core. It means, we reduce the infinite lateral dimension of the p-Ge layer shown in Figure 77 to a dimension of w. The result is shown in Figure 78. Now, to place the side contact, the Si layer below the p-Ge layer needs to be doped.

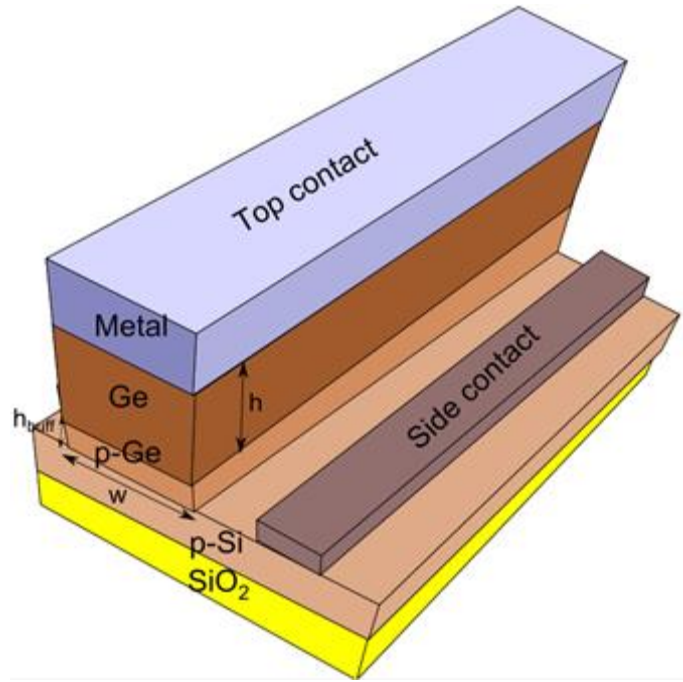

**Figure 78: Study of the influence of the patterned p-Ge layer in the plasmonic modulator**

The results with the laterally infinite p-Ge layer and with a p-Ge layer of w=250 nm is presented in the following Table 12,

| p-Ge layer width [nm]: | 250 | ∞ |
|---|---|---|
| $\Delta\alpha_{eff}/\alpha_{eff}$ (1 V): | 0.02 | 0.03 |
| $\Delta\alpha_{eff}/\alpha_{eff}$ (2 V): | 0.06 | 0.07 |
| $\Delta\alpha_{eff}/\alpha_{eff}$ (3 V): | 0.12 | 0.12 |
| $E_{max}$ [kV/cm]: | 90 | 90 |

**Table 12: Summary of the study of the influence of the buffer layer**

According to the Table 12 we can see that using an infinite or finite p-Ge layer is similar in both cases for a bias voltage of 3 V. Nevertheless, for 1 and 2 V, the FoM is sligthly better in the infinite case. So, regarding the FoM the infinite case is slightly better. Nevertheless, we can see that the width of the mentioned p-Ge layer is not a critical point. This is due to the fact that there is not a big portion of the electromagnetic field of the light in this layer, as the electromagnetic field of the plasmon in the MS structure located at in the interface between the metal and the Ge core.

Furthermore, using an infinite layer will allow placing the side contact on the p-Ge layer. In this case, the main function of the p-Ge layer is to act as a conductor material between the side contacts and the Ge core layer. Consequently, a capacitive structure is formed by the stack consisting of Metal/nid Ge/p-Ge. This will induce a trade-off in both the bandwidth of the modulator and the coupling efficiency. It means, when the thickness of the p-Ge layer increases it will increase the bandwidth of the modulator by reducing the access resistance of the p-Ge layer (with the parallel resistance due to the bottom p-Si layer). Another advantage of having both the p-Ge and the underlying p-Si layers infinite is that the access resistance is composed of two parallel resistances, one formed by the p-Ge layer and another by the p-Si layer. Since they are in parallel they will reduce the total resistance and consequently it will increase the bandwidth of the modulator. On the other hand, when $h_{buff}$ increases, more light will leak into the substrate and consequently the coupling efficiency from a standard Si rib waveguide will be lower.

Furthermore, from Table 11 and Table 12 we can observe an increase of the FoM $\Delta\alpha_{eff}/\alpha_{eff}$ when the driving voltage of the plasmonic modulator increases. This is due to the fact that for the same height h of the Ge core (h=250 nm in all cases), applying a higher voltage implies that there is a higher static electric field in the Ge core, this increases the change of the FKE which produces an increase in the parameter $\Delta\alpha_{eff}$. It is worth to mention that the increase of the FoM $\Delta\alpha_{eff}/\alpha_{eff}$ is due to an increase in $\Delta\alpha_{eff}$ rather than a decrease of the effective propagation losses $\alpha_{eff}$.

Until now, the best structure is the one which is shown in the following Figure 79,

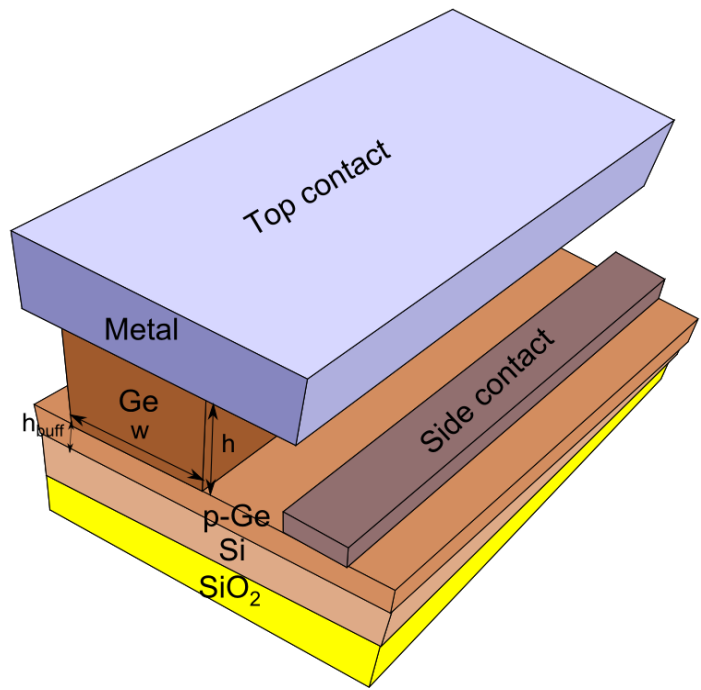

**Figure 79: Optimized structure until Table 8**

## 5.2.5.4   Influence of the Doping Level of the p-Si Layer

Now we study the influence of the doping level of the p-Si buffer layer which is between the $SiO_2$ substrate and the p-Ge layer. We obtain the FoM $\Delta\alpha_{eff}/\alpha_{eff}$ for the case where the Si layer is not intentionally doped with a concentration of holes of $10^{16}$ cm$^{-3}$ and for the case where it is p-doped with a concentration of holes of $10^{19}$ cm$^{-3}$. We study this change for two different configurations, one for w=150 nm and h=400 nm as used previously (Table 13) and the other for w=250 nm and h=350 nm (Table 14).

| p-Si Doping level [cm$^{-3}$]: | $10^{16}$ | $10^{19}$ |
|---|---|---|
| $\Delta\alpha_{eff}/\alpha_{eff}$ (3 V): | 0.11 | 0.11 |
| $E_{max}$ [kV/cm]: | 92 | 95 |

**Table 13: Study of the doping of the Si layer for w=150 nm and h=400 nm. Used to check the results of Table 14**

| p-Si Doping level [cm$^{-3}$]: | $10^{16}$ | $10^{19}$ |
|---|---|---|
| $\Delta\alpha_{eff}/\alpha_{eff}$ (3 V): | 0.11 | 0.11 |
| $E_{max}$ [kV/cm]: | 95 | 96 |

**Table 14: Study of the doping of the Si layer for w=250 nm and h=350 nm**

From these results, we can conclude that the FoM $\Delta\alpha_{eff}/\alpha_{eff}$ is not widely affected by the doping concentration of the p-Si layer. It is also independent on the parameters w and h. However, as it was already said in this section, it is better to have a highly doped p-Si layer (infinite in the lateral direction) below the p-Ge layer (infinite in the lateral direction) in order to reduce the access resistance from the side contact to the core of the modulator. This will increase the maximum operational frequency of the modulator. As we will explain later in this chapter the modulator can be reduced to an RC circuit in which we can calculate the operation bandwidth of the device. The resistance R is the resistance between the side contact and the p doped Si and Ge layers to the core of the modulator. The capacitance C is the capacitance of the non-intentionally doped Ge core of width w and heigh h which acts as a dielectric between the bottom and top contacts. Regarding the technology of fabrication, the use of a highly doped p-Si infinite layer also relaxes the Ge etching depth accuracy constraints down to the p-doped layers. Now, the best structure until now is the one represented in Figure 80.

## 5.2.5.5   Optimization of the Dimensions of the Ge Core

Knowing the structure of the modulator shown in Figure 80, we want to optimize the device with respect to the dimensions w and h. As we explained before we want to optimize the FoM=$\Delta\alpha_{eff}/\alpha_{eff}$ in order to maximize the extinction ratio of the device and reduce the propagation losses of the MS waveguide.

To do the optimization with respect to w and h of the structure of Figure 80 we select the values w=150 nm and w=250 nm and we vary h from 150 to 450 nm every 50 nm. The result for the width w=150 nm is presented in Table 15 while the result for w=250 nm is presented in Table 16. Regarding the selection of w=150 nm it is close to the 120 nm minimum feature that can be fabricated using 193 nm deep-UV lithography. To scan the influence of w in the FoM we selected w=250 nm just 100 nm bigger.

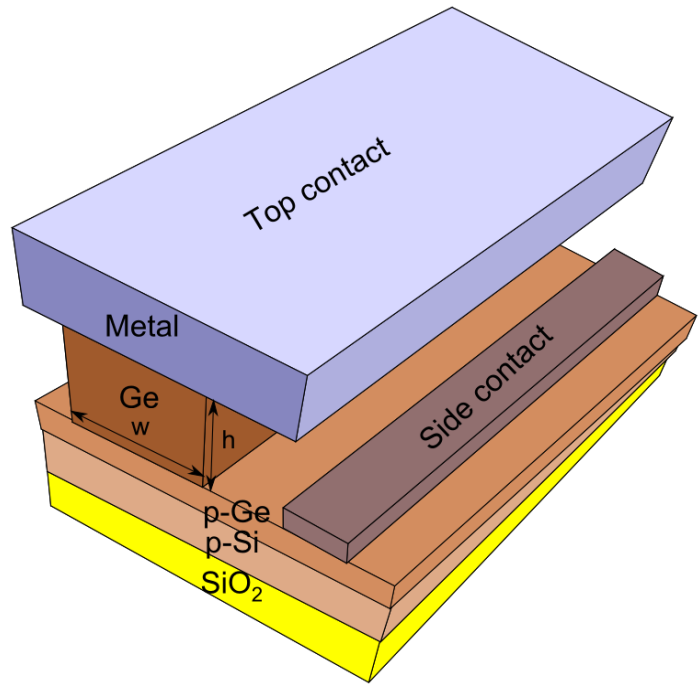

**Figure 80: Structure that is going to be optimized with respect to the parameters w and h of the NID Ge core**

| h [nm]: | 150 | 200 | 250 | 300 | 350 | 400 | 450 |
|---|---|---|---|---|---|---|---|
| V=1 V: | 0.07 | 0.06 | 0.05 | 0.04 | 0.03 | 0.03 | 0.02 |
| V=2 V: | Burnt | Burnt | 0.11 | 0.09 | 0.08 | 0.07 | 0.06 |
| V=3 V: | Burnt | Burnt | Burnt | Burnt | 0.13 | 0.11 | 0.10 |

**Table 15: Optimization of the modulator of Figure 80 with respect to h for a w=150 nm**

| h [nm]: | 150 | 200 | 250 | 300 | 350 | 400 | 450 |
|---|---|---|---|---|---|---|---|
| V=1 V: | 0.07 | 0.05 | 0.04 | 0.03 | 0.03 | 0.02 | 0.01 |
| V=2 V: | Burnt | Burnt | 0.1 | 0.08 | 0.07 | 0.06 | 0.05 |
| V=3 V: | Burnt | Burnt | Burnt | Burnt | 0.11 | 0.09 | 0.08 |

**Table 16: Optimization of the modulator of Figure 80 with respect to h for a w=250 nm**

Both Table 15 and Table 16 represent the FoM $\Delta\alpha_{eff}/\alpha_{eff}$ for different values of the driving voltages V and the height of the Ge core h. When we write "Burnt" in the Table it means that the static electric field in the Ge core is larger than the maximum static electric field of Ge. It means, the breakdown static electric field of the Ge core is exceeded. Regarding the rest of the materials the maximum static electric field in the p-Ge or p-Si is always below the breakdown electric field of the materials.

From Table 15 where the width of the Ge core is w=150 nm, we observe that for the same voltage V, when the parameter h is increased then the FoM $\Delta\alpha_{eff}/\alpha_{eff}$ is decreased. This is due to the fact that for the same voltage when the heigh is increased the static electric field in the core of Ge is also decreased. This reduces the parameter $\Delta\alpha_{eff}$ of the FoM. The propagation losses $\alpha_{eff}$ of the plasmonic mode are not highly affected by h. Furthermore, for the same h, we see that when we increased the voltage V the FoM $\Delta\alpha_{eff}/\alpha_{eff}$ also increases. This is due to the fact that when the voltage V is increased the maximum static electric field in the Ge core is also increased. This increases the parameter $\Delta\alpha_{eff}$ which increases the FoM $\Delta\alpha_{eff}/\alpha_{eff}$.

The best case of the FoM $\Delta\alpha_{eff}/\alpha_{eff}$ for w=150 nm is the value 0.13 for a driving voltage V=3 V and a height h=350 nm. If we continue increasing the voltage (E.g.: V=4, 5, 6 … V) the FoM is going to be bigger than 0.13 for a height h>350 nm. Nevertheless, we want to stop at V=3 V to limit the electrical power consumption of the device because when the driving voltage V is increased, the electrical power consumption of the device which is given by Equation 17 is also increased.

We can see a similar effect in Table 16. It means, for the same voltage V the FoM $\Delta\alpha_{eff}/\alpha_{eff}$ decreases with the increase of h. Furthermore, as in the previous case, we also see the increase in the FoM when we increase the driving voltage V for the same height h. The reasons for these behaviors are similar to the ones explained in the previous paragraph.

The best case of the FoM $\Delta\alpha_{eff}/\alpha_{eff}$ for w=250 nm is the value 0.11 for a driving voltage V=3 V and a height h=350 nm. In the same way as for w=150 nm we also see that if we increase the driving voltage above 3 V we are going to increase the FoM $\Delta\alpha_{eff}/\alpha_{eff}$. Nevertheless we stopped at V=3 V to limit the electrical power consumption of the device as explained before.

Comparing the results, we see that the best optimized device for w=150 nm (Table 15) is better than for the case of w=250 nm (Table 16). So, with respect to the width w of the Ge core we can observe that the smaller the width the better. So, our optimized modulator is for w=150 nm and h=350 nm yielding a value of the FoM $\Delta\alpha_{eff}/\alpha_{eff}$=0.13. We selected w=150 nm to have a well-defined plasmonic mode avoiding any tolerance and some margin for 193 nm UV-lithography definition.

## 5.2.5.6   Performance of the Optimized MS Modulator Structure

Once the modulator structure shown in Figure 80, it is interesting to calculate the extinction ratio (ER) and the propagation losses (PL) of the device. For the lengths L=10 μm and 20 μm, the results are shown in the following Table 17,

| L [μm]: | Extinction ratio [dB] | Propagation losses [dB] |
|---------|----------------------|-------------------------|
| 10      | 1.3                  | 10.6                    |
| 20      | 2.7                  | 21.3                    |

**Table 17: Summary of the performance of the device for the optimized modulator in Figure 43**

From Table 17 we see that for 10 μm long device the extinction ratio is not so large around 1.3 dB. Experimentally such an extinction ratio is impractical for optical communications. This is why we need to increase the length L of the device to 20 μm to increase the extinction ratio around 3 dB which is a better value. Nevertheless when the length of the device L is increased the propagation loss of the device is also increased. We selected L=10 μm and L=20 μm in Table 17 in order to have an extinction ratio around 1 dB and 3 dB respectively. The reached value of the propagation losses for 20 μm is around 21.3 dB which is too big to be practical. By now, we can conclude that we have a device with a practical extinction ratio but with too large propagation losses while maintaing or even increasing the extinction ratio to achieve the target performances defined at the beginning of this chapter.

## 5.2.5.7   Comparison of the Optimized MS Modulator with the State-of-the-art

Now, it is interesting to compare the performance of the structure of Figure 80 given in Table 17 with the state-of-the-art of photonic modulators that uses the FKE given in Table 1 (chapter two). This will give us information about the performance of the proposed modulator, what we can improve, etc. The extinction ratio of the proposed device (Figure 80) is below the extinction ratio of the photonic modulators that use the FKE (Table 1). The maximum extinction ratio of the proposed device is around 2.7 dB while the minimum extinction ratio in Table 1 is 4 dB for the device of Kotura Inc. presented in [53]. We take this into account to improve the extinction ratio in the designed device.

With respect to the propagation losses we see that the proposed modulator of Figure 80 has a high propagation loss (meanly also insertion losses) with respect to the photonic FKE modulators of Table 1. The maximum insertion loss of the photonic FKE modulators is 10 dB for the one presented by A*STAR in [52]. This higher loss of the designed device with respect to the photonic devices is due to the fact that we use a plasmonic instead of a photonic mode. This will allow us to achieve lower electrical power consumption for modulation. Using the plasmon and confining the optical mode in a small area increases the propagation losses. It is interesting to improve the trade-off between the confinement of light and the propagation losses. Furthermore, it will be interesting to reduce the propagation losses of our device. For this purpose, we will use a MIS waveguide instead of a MS one in the next section.

With respect to the driving voltage we see that it is similar to the photonic modulators that use the FKE. For most devices, it is around 3 V [35], [51], [54]. The lower driving voltage used in our structure is due to the fact that reducing the dimensions of the structure allows to use a smaller V to induce the same static electric field.

Comparing the performance of the device of Figure 80 with respect to the state-of-the-art of plasmonic modulators summarized in Table 2 (chapter 2) we see that the propagation losses of the proposed device are still large but the gap of the propagation losses between our modulator and the reported modulators is less than with the photonic FKE modulators discussed in the previous paragraphs. In Table 2 we see devices with 20 dB propagation losses like in [56] or with 18 dB insertion losses like in [63].

With respect to the extinction ratio of the proposed modulator of Figure 80 we see in Table 2 that it is smaller than the other reported plasmonic modulators. Nevertheless the gap in the extinction ratio between the proposed modulators and the state-of-the-art plasmonic devices is lower than with the photonics devices that uses the FKE. There are also plasmonic modulators with an extinction ratio around 1-2 dB like the one presented in [63] offering a lower performance that us. A comparison between the state-of-the-art photonic FKE photonic modulators summarized in Table 1 and the plasmonic modulators reported in Table 2 reveals that the plasmonic modulators have less extinction ratio than the photonic FKE ones.

With respect to the driving voltage the proposed modulator uses a smaller driving voltage that the summarized plasmonic modulators shown in Table 2. In this table there are devices that use up to 10 V of driving voltage like the one presented in [106].

To summarize the comparison of the proposed device with the photonic modulators that use the FKE and the plasmonic modulators we can conclude that although the proposed device in Figure 80 is compact and uses a low driving voltage we need to further improve the propagation losses. A good benchmark is to reduce them down to 10 dB. The extinction ratio is in the average of the modulators reported, so, in the best case it will be interesting to maintain it around 3 dB or increase it.

Depending on the metal used for the top contact in Figure 76 we need to add a diffusion barrier between the top metal and the Ge. Some metals like Cu diffuse into the semiconductor, in our case Ge, and this increases the propagation losses of the plasmonic mode. However, it is highly desirable that the modulator is CMOS compatible. This will restrict the materials that can be used. There are other metals, like silver (Ag) that can be deposited over Ge. Nevertheless, the structure is not CMOS compatible.

If we introduce a small slot (5-15 nm thick) of an insulator barrier between the metal and the semiconductor (Ge) we make the structure CMOS compatible. Furthermore, the formed MIS waveguide has less propagation losses $\alpha_{eff}$ than the MS waveguide. It will be studied in the next section.

## 5.2.6  Plasmonic MIS Modulator

The aim of this work from now on is to reduce the propagation losses of the plasmonic mode. In order to reduce the losses it is interesting to use a MIS waveguide instead of a MS structure. In this case, we not only reduce the propagation losses of the mode but also we can add an insulator diffusion barrier to the Cu metal to make it CMOS-compatible. Regarding the insulator barrier we have many options. It can be: Ti, TiN, Ta, TaN, $SiO_2$ and $Si_3N_4$ studied in chapter three. Table 5 of chapter three indicates that the propagation losses of the plasmonic mode of the MIS waveguide are smaller using Si₃N₄, consequently, we decided to use this barrier layer. The modified structure is represented in the following Figure 81,

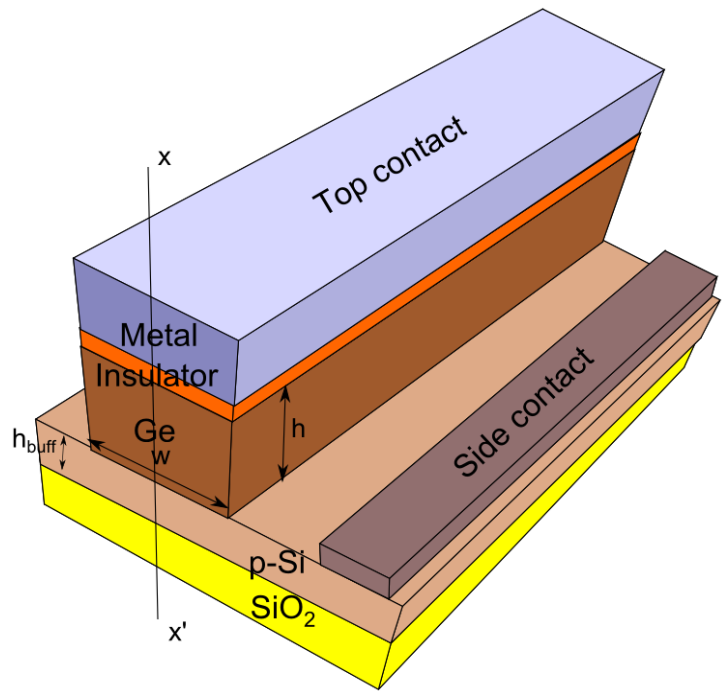

**Figure 81: Capacitor-like structure with a MIS waveguide**

Now we have one more parameter to optimize in the plasmonic modulator, namely: the height $h_{slot}$ of the Si₃N₄ insulator. Furthermore, we add the already studied parameters such as the height h of the Ge core and its width w. We already demonstrated that the parameters regarding the thickness of the layers below the Ge core do not have a high influence on the performances of the device (extinction ratio and propagation losses). This is due to the fact that both the static electric field and the electromagnetic field of the plasmonic mode are not highly present in the buffer.

First we need to evaluate how the presence of the insulator barrier affects both the distribution of the static electric field in the Ge core and the distribution and losses of the plasmonic mode.

## 5.2.6.1  Parameters of the Simulation

In this section we explain the parameters of the FDM simulation of the best structure which is the MIS. The mode solver used was implemented here: [142]. The simulation window used was 3 µm wide while the heigh was also 3 µm. With these dimensions we assured that the electromagnetic field of the guided mode is not present at the boundaries of the simulation window. The boundary conditions used were zero fields immediately outside of the boundary. In all the simulations we checked that all the fields of the guided mode where zero in the boundaries. Regarding the mesh we selected a uniform mesh in all the simulation windows with a horizontal and vertical resolution of 5 nm. When we set a small slot of 5 nm we assure that at least we have ten points inside the slot were the electric field of the optical mode changes fast. The resolution for the mesh around the slot (inside the slot and in the surroundings) is around 0.5 nm. To achieve those values we increased the mesh resolution until the effective index and effective absorptions of the modes do not change for two decimals. The maximum resolution of the mesh was limited by the RAM memory of 4 GB of the computer. With these parameters the simulation takes 25 minutes.

## 5.2.6.2   Influence of the Slot Thickness on the Static Electric Field Distribution

We perform a preliminary study of the static electric field distribution in the vertical cross-section x-x' of the structure of Figure 81. With respect to the previous optimized structure we do some simplifications since we do not take into account as a first study the p-Ge layer shown in Figure 80. Furthermore, as a difference with Figure 80 in Figure 81 we use a finite slot and metal in the lateral direction. As we will see the fact that the slot is not infinite in the lateral direction does not influence the performance of the device. Furhtermore, the aligment in the fabrication will be easier with a lateral infinite layer. Nevertheless, the infinite slot and metal will simplify the fabrication process since alignment constraints with the walls of the Ge core are suppresses.

In chapter three we calculated that the slot of the MIS waveguide needs to be smaller than $t_{min}$=8.35 nm to work in the low loss regime of the MIS waveguide. So, a thickness of 5 nm for the slot of the MIS waveguide will ensure that the static electric field is significantly present in the Ge core and it also ensures that we work in the low loss regime of the MIS structure. The former will increase the extinction ratio of the device while the latter will reduce the propagation losses of it. The extinction ratio of the structure (Figure 81) is improved for a thickness of 5 nm because if we increase this thickness the induced static electric field will be confined into the slot and is not going to be present into the Ge core.

As a first architecture to study the influence of the slot thickness on the static electric field we use the following dimensions for the structure of Figure 81: the width of the core is w=150 nm, the height of the core h=250 nm and the thickness of the slot $h_{Si3N4}$=15 nm. It is worth to mention that we want the static electric field present in the Ge core where the FKE is present. The applied bias voltage is equal to V=10 V. Although the bias voltage was settled to 3 V here we use 10 V to demonstrate that even for a very high voltage the static electric field is confined into the $Si_3N_4$ slot of heigh 15 nm. With these dimensions the static electric field of the cross section (Figure 81) is given in Figure 82.

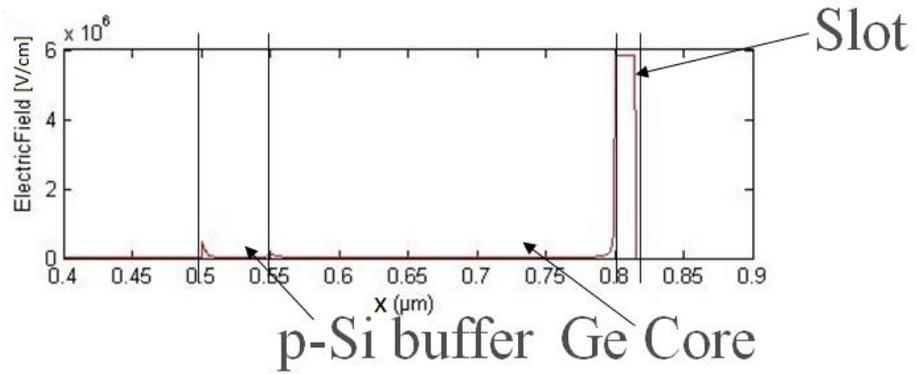

**Figure 82: Static electric field distribution in the middle of the device. In the cross-section x-x'. The applied bias voltage is around 10 V. It can be seen that the static electric field is confined in the Si₃N₄ barrier slot**

From Figure 82 we can see that the static electric field is concentrated in the Si₃N₄ slot under a driving voltage of V=10 V. It is observed that no static electric field is present in the Ge, core (for x between 0.55 and 0.8 μm). Due to the absence of static electric field in the core region of the Ge we do not expect to have a high extinction ratio using a slot of Si₃N₄ around 15 nm. We remind that to maximize the extinction ratio of the device it is important that as much as possible static electric field and electromagnetic field of the plasmonic mode need to be in the Ge core where the FKE takes place.

The performances of the structure for 10 V reverse bias voltage are summarized in the following Table 18,

| V [V]: | $n_{eff}$ | $\alpha_{eff}$ [cm⁻¹] | TE [%] | $\Delta\alpha_{eff}$ [cm⁻¹] | $E_{max}$(Ge) [V/cm] | $E_{max}$(Si₃N₄) [V/cm] |
|---|---|---|---|---|---|---|
| 0 | 3.47 | 252.68 | 0 | - | - | - |
| 10 | 3.47 | 278.96 | 0 | 26.28 | 2.5x10⁴ | 5.8x10⁶ |

**Table 18: Summary of the performance of the structure of Figure 81 with a slot thickness of 15 nm**

From Table 18 we can see that although we apply 10 V to the structure the modulator efficiency $\Delta\alpha_{eff}$ is around 26.28 cm⁻¹ since the variation of the effective losses of the mode $\alpha_{eff}$ is changing from 252.68 cm⁻¹ to 278.96 cm⁻¹. This value is very small to produce the modulation. Both from Figure 82 and Table 18 we can see that the maximum static electric field in Ge is 2.5x10⁴ V/cm and for Si₃N₄ is 5.8x10⁶ V/cm which is below the breakdown electric field of both materials.

On the other hand, for the same structure of Figure 81 with a width of the Ge core with w=150 nm, the height of the Ge core of h=250 nm and the thickness of the slot $h_{Si3N4}$=5 nm. In this case the driving voltage is V=3 V The static electric field distribution in the vertical cross section x-x' is presented below in Figure 83,

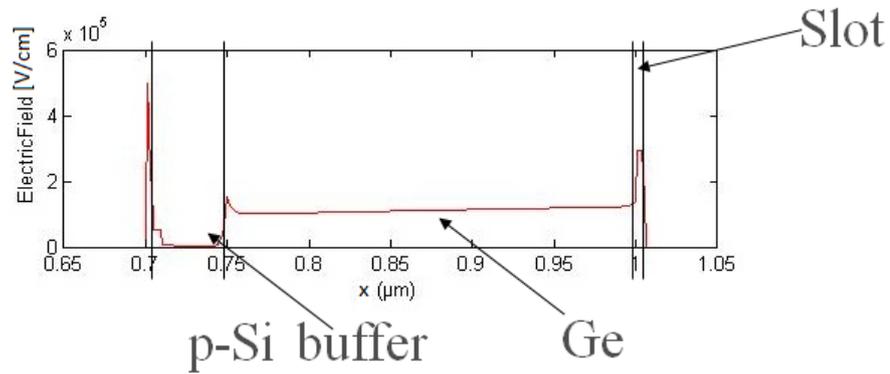

**Figure 83: Static electric field distribution in the middle of the device. In the cross-section x-x'. The applied bias voltage is around 3 V. It can be seen that the static electric field is desconfined in the Si₃N₄ barrier slot**

From Figure 83 we see now that with a thinner $Si_3N_4$ slot around 5 nm the static electric field is not all confined in the slot but also present in the Ge core. This will increase the extinction ratio of the device of Figure 81 with respect to the case in which the $Si_3N_4$ slot is around 15 nm. Nevertheless, it is worth noting that there is still a high electric field inside the slot but the value is below the breakdown electric field of the material $Si_3N_4$.

### 5.2.6.3   Influence of the Slot Thickness on the Propagation Losses of the Plasmonic Mode

In chapter three we calculated that the slot of the MIS waveguide needs to be smaller than $t_{min}$=8.35 nm to work in the low-loss regime of the MIS waveguide. So, a thickness of 5 nm for the slot of the MIS waveguide will ensure that the static electric field is significantly present in the Ge core and it also ensures that we work in the low-loss regime of the MIS structure. The former will increase the extinction ratio of the device while the latter will reduce the propagation losses of the device. The extinction ratio of the structure (Figure 81) is improved for a thickness of 5 nm because if we increase this thickness the induced static electric field will be confined into the slot and it is not going to be present into the Ge core.

### 5.2.6.4   Influence of the Thickness of the Slot on the Diffusion Barrier Properties

Regarding the thickness of 5 nm of $Si_3N_4$ it is still enough to act as an effective diffusion barrier for the Cu. The minimum value to act as an effective diffusion barrier for the $Si_3N_4$ is around 3 nm [123]. Using a thickness of 5 nm we take into account the tolerance of the fabrication process. It means, although having 3 nm [123] will maximize the extinction ratio with respect to the case of 5 nm, it is better to use 5 nm to take into account the fabrication tolerance and robustness. With a thickness of the Si₃N₄ slot of 5 nm we ensure that it acts as an effective diffusion barrier between the Cu and the Ge.

## 5.2.6.5  Optimization of the Ge Core of the Modulator

As a first consideration we used the structures of Figure 81 with a finite Cu and Si₃N₄ in the lateral direction to study the influence of the thickness of the slot. Now we consider the slot infinite as in the case of the structure shown in Figure 84. Having an infinite lateral metal and slot is better from an alignment and fabrication point of view. Regarding the substrate we now consider two p-Si and p-Ge layers (structure of Figure 80). Taking into accounts those two considerations, the structure that will be optimized is presented in the following Figure 84,

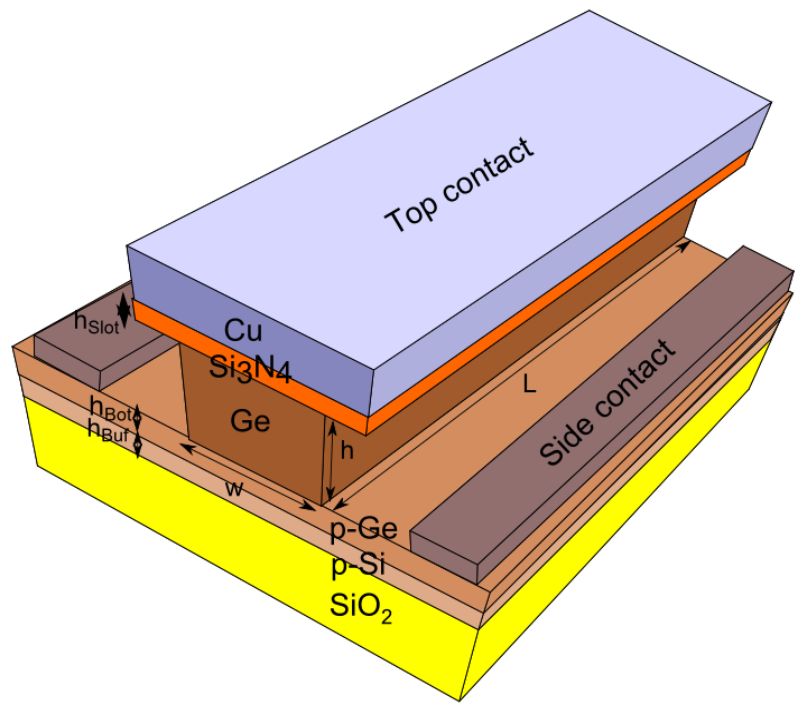

**Figure 84: Final structure of the FKE plasmonic modulator that is going to be optimized**

With respect to the structure of Figure 84 the main parameters to optimize are the width of the Ge core w and the height of the Ge core h.

First we simulate the distribution of the static electric field using similar dimensions as for the optimized modulator of Figure 80 to know what happens. Based on these dimensions, we plot the static electric field for w=150 nm, h=250 nm, h$_{Si3N4}$=5 nm, h$_{Buf}$=60 nm and h$_{Bot}$=40 nm. The static electric field distribution is represented in Figure 85 (a) for V=0 V and in Figure 85 (b) for a reverse bias of V=3 V.

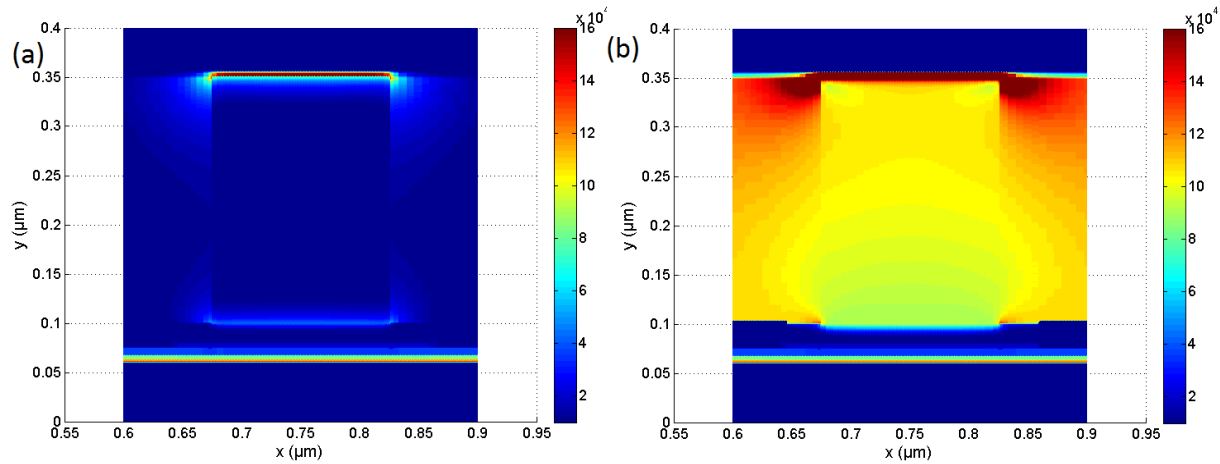

**Figure 85: Static electric field distribution for a driving voltage of (a) 0 V and (b) 3 V under the structure of Figure 48**

At 0 V, there is no static electric field in the Ge core. The static electric field in the Si$_3$N$_4$ slot is due to the residual charges present in the intrinsic Ge where the residual doping density is $10^{16}$ cm$^{-3}$ and to the presence of the Cu metal on the other side.

We can appreciate that according to Figure 85 (b) there is a static electric field induced in the Ge core when a reverse bias of V=3 V is applied. The static electric field in the Ge core is almost uniform. This changes the optical absorption of the material $\alpha_{mat}$ almost uniformly in the core which in turn leads to the modulation of the plasmonic mode by changing its effective damping constant (or loss constant) $\alpha_{eff}$.

When a voltage is applied to the structure, an accumulation of carriers at the interface between the Ge and the Si$_3$N$_4$ occurs. The refractive index of Ge will therefore also be modified close to the interface with the Si$_3$N$_4$. However the expected change is similar to the change in Si. According to reference [46] the maximum change of the real part of the refractive index of Ge due to the presence of a static electric field is around $8 \times 10^{-4}$ at 1647 nm for an static electric field of 60 kV/cm. Since we are doing an electro-absorption modulator, this change can be neglected since it is not going to affect the electromagnetic distribution of the plasmonic mode.

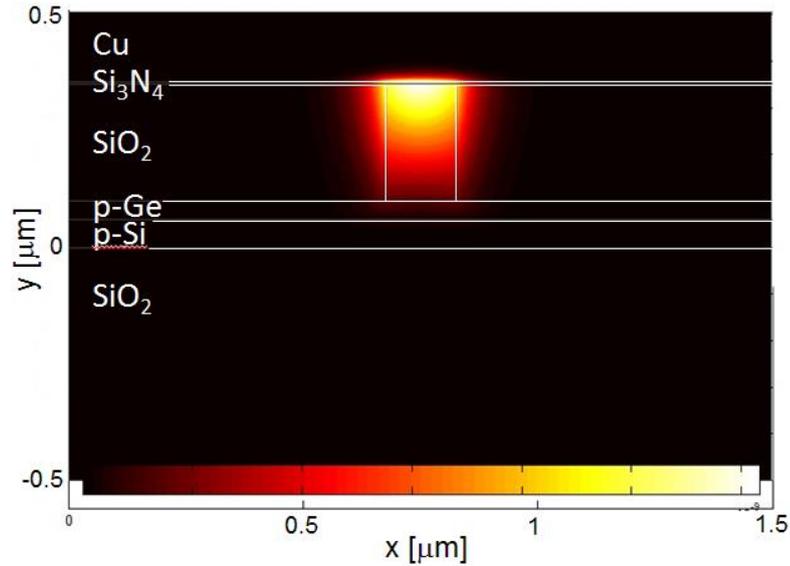

**Figure 86: Optical field distribution of the plasmon TM mode supported by the MIS waveguide in the structure of Figure 84**

Finally, in Figure 86 we represented the optical intensity of the plasmon mode supported by the MIS waveguide embedded in the device of Figure 84. We can see how the optical field is present into the Ge core where the FKE takes place.

In the following tables Table 19 and Table 20 we calculate the FoM $\Delta\alpha_{eff}/\alpha_{eff}$ in order to optimize the structure as we did for the MS structure in Table 15. Table 19 and Table 20 are also clarified using Figure 87 and Figure 88. We analyzed both Tables and Figures in order to facilitate the comprehension of the optimization. For this, we do a scan of the height of the Ge core h and the width of the Ge core w. The driving voltage V is also scanned.

| h [nm]: | 150 | 200 | 250 | 300 |
|---|---|---|---|---|
| V=1 V: | Burnt | 0.05 | 0.04 | 0.03 |
| V=2 V: | Burnt | 0.18 | 0.16 | 0.14 |
| V=3 V: | Burnt | Burnt | 0.34 | 0.26 |

**Table 19: Optimization of the modulator of Figure 84 with respect to h for a w=150 nm**

| h [nm]: | 150 | 200 | 250 | 300 | 350 |
|---|---|---|---|---|---|
| V=1 V: | Burnt | 0.04 | 0.03 | 0.03 | 0.2 |
| V=2 V: | Burnt | 0.18 | 0.16 | 0.15 | 0.13 |
| V=3 V: | Burnt | Burnt | 0.27 | 0.26 | 0.25 |

**Table 20: Optimization of the modulator of Figure 84 with respect to h for a w=250 nm**

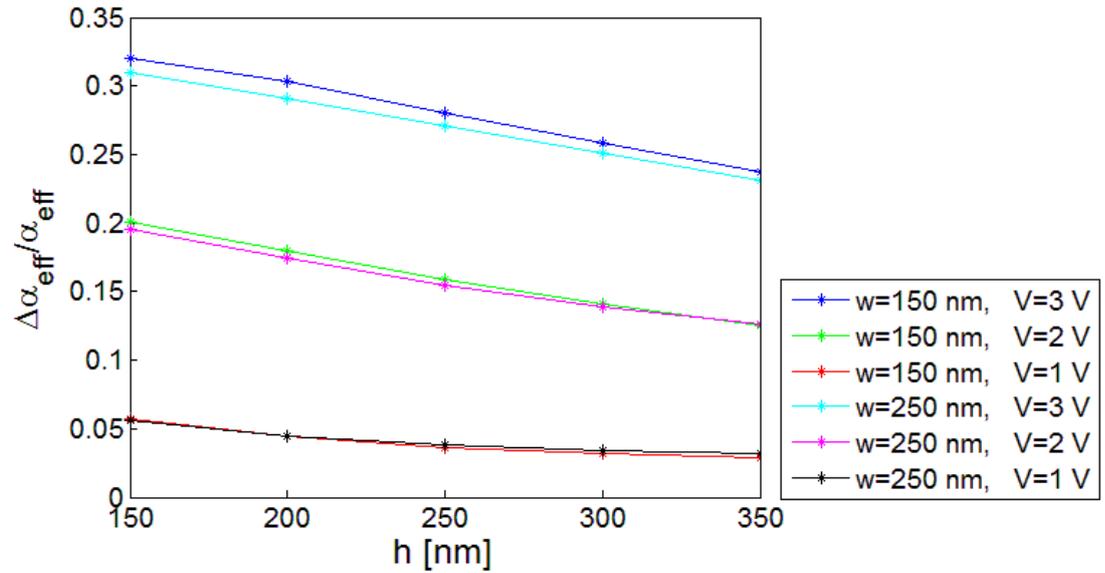

**Figure 87: $\Delta\alpha_{eff}/\alpha_{eff}$ against the height of the Ge core h for the modulator of Figure 84**

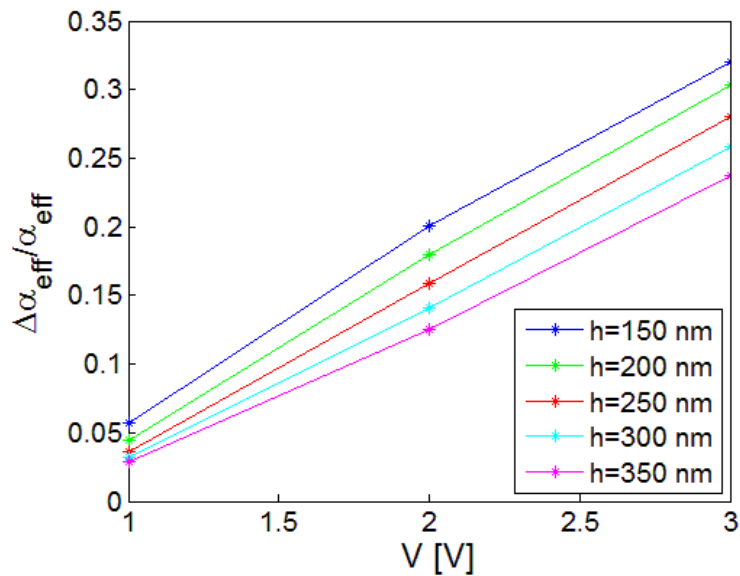

**Figure 88: $\Delta\alpha_{eff}/\alpha_{eff}$ against the driving voltage V for the modulator of Figure 84. The width used is w=150 nm**

In Table 19 the width of the Ge core is w=150 nm while in Table 20 the width of the Ge core is equal to w=250 nm. In some cells of the word 'Burnt' is written. It means that for these dimensions of h and w and for the selected driving voltage the static electric field reached in the Ge core is larger than the breakdown voltage of Ge. Consequently, the material is burnt. It is worth to mention that along the entire table the breakdown electric field of other materials like the $Si_3N_4$ slot are not exceeded. Nevertheless the static electric field in the slot is always below the breakdown electric field of the $Si_3N_4$ material.

In Table 19, the width of the Ge core is w=150 nm. In the same row we see that for the same voltage V, when we increase the height of the Ge core h, then the FoM $\Delta\alpha_{eff}/\alpha_{eff}$ is reduced (Figure 87). The reason of the decrease of the FoM $\Delta\alpha_{eff}/\alpha_{eff}$ is a decrease in the parameter $\Delta\alpha_{eff}$ due to the fact that for the same driving voltage V there is a weaker static electric field in the Ge core. The propagation losses $\alpha_{eff}$ of the plasmonic mode are not highly affected by the increase of h. Additionally, for the same height h (the same column of Table 19) we can see that when we increase the voltage V then the FoM $\Delta\alpha_{eff}/\alpha_{eff}$ is also increased (Figure 88). The reason is an increase in the parameter $\Delta\alpha_{eff}$. It is because when we maintain the height h of the Ge core and we increase the driving voltage V we also increase the static electric field in the Ge core.

If we look at the FoM $\Delta\alpha_{eff}/\alpha_{eff}$ in Table 19 we can see that the best FoM for a Ge core of w=150 nm is reached for a driving voltage of V=3 V and a height h=250 nm (Figure 87). FoM is 0.34 for those parameters. If the driving voltage of the structure V is increased above 3 V then the FoM will be larger than 0.34 for h>350 nm. Nevertheless we stop at 3 V in order not to increase the electrical power consumption of the device. We remind that according to Equation 17 the electrical power consumption of a modulator increases with the square of the driving voltage V.

On the other hand, from Table 20 we can see a similar effect in Table 19. It means, for the same voltage V the FoM $\Delta\alpha_{eff}/\alpha_{eff}$ decreases with the increase of h (Figure 87). Furthermore, as in the previous case, we also see the increase of the FoM when we increase the driving voltage V for the same height h (Figure 88).

The best FoM $\Delta\alpha_{eff}/\alpha_{eff}$ in the Table 20 is the value 0.27 for a driving voltage V=3 V and a height h=250 nm.

If we compare the influence of the Ge width w in the FoM $\Delta\alpha_{eff}/\alpha_{eff}$ we see that in (Table 19) for w=150 nm the maximum $\Delta\alpha_{eff}/\alpha_{eff}$=0.34 and for the case in which w=250 nm (Table 20) we can see that $\Delta\alpha_{eff}/\alpha_{eff}$=0.27.

So, our optimized modulator is for w=150 nm and h=250 nm leading to a value of the FoM $\Delta\alpha_{eff}/\alpha_{eff}$=0.34. We do not study the performance of the modulator below w=100 nm because this dimension is roughly the minimum feature that can be fabricated using 193 nm deep-UV lithography. We selected w=150 nm to have a well-defined plasmonic mode avoiding any tolerance.

### 5.2.6.6  Performances of the Optimized MIS Modulator Structure

Once the modulator structure of Figure 84, optimized, it is interesting to know the performances of the device in terms of the extinction ratio and the propagation losses. For the lengths L=10 μm and L=30 μm, the results are showed in the following Table 21 (the same information in form of graphs is presented in Figure 89 and Figure 90 for clarification),

| L [μm]: | ER [dB] | PL [dB] |
|---------|---------|---------|
| 10      | 1.1     | 3.7     |
| 30      | 3.3     | 11.2    |

**Table 21: Summary of the performance of the device for the optimized modulator in Figure 84**

In Table 21 we have a summary of the main characteristics of the device of Figure 84, mainly, the extinction ratio and the propagation losses. It is possible to see that for a length of L=10 μm the extinction ratio is around 1.1 dB, which is small for a practical optical communications. Due to this reason it is necessary to increase the length L of the device to 30 μm to reach 3 dB of extinction ratio, but the propagation losses are also increased.

In Figure 89 we represent the extinction ratio of the device for a length of 30 μm long. We can see that for a driving voltage of V=3 V and for a height of the Ge core of h=250 nm we can obtain a value of extinction ratio 3.3 dB. For a width of w=150 nm and V=3 V if we decreases the value h=250 nm then the material Ge is burnt due to a high static electric field present on it. This is why the minimum h we can take is 250 nm.

In Figure 90 we represent the propagation losses of the device for L=30 μm. We can see that for a driving voltage of V=3 V and for a height of the Ge core of h=250 nm we can obtain a value of propagation losses of 11.2 dB. This increases up to about 14 dB for h=150nm.

There is a trade-off between the extinction ratio and the propagation losses of the device with respect to h. If we decrease h the extinction ratio increases while the propagation loss increases. This reduces the FoM ER/PL.

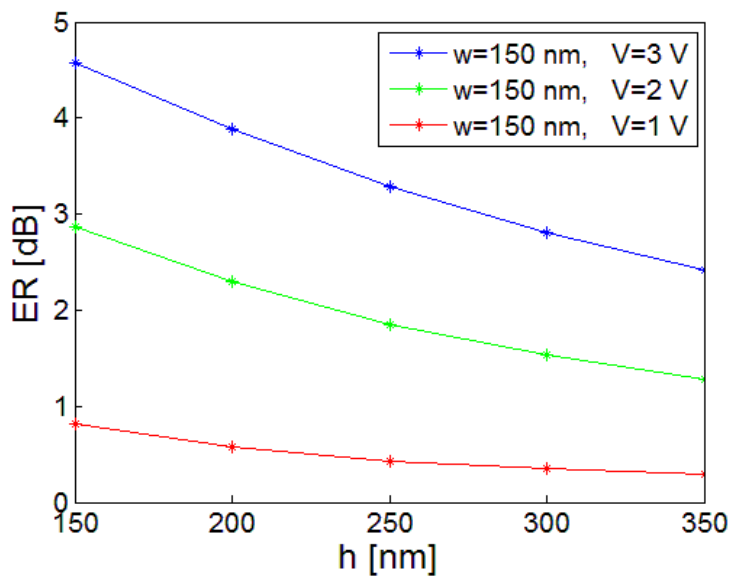

**Figure 89: Extinction ratio of the device of Figure 84 as a function of the height of the Ge core h for different driving voltages V. The length is 30 μm**

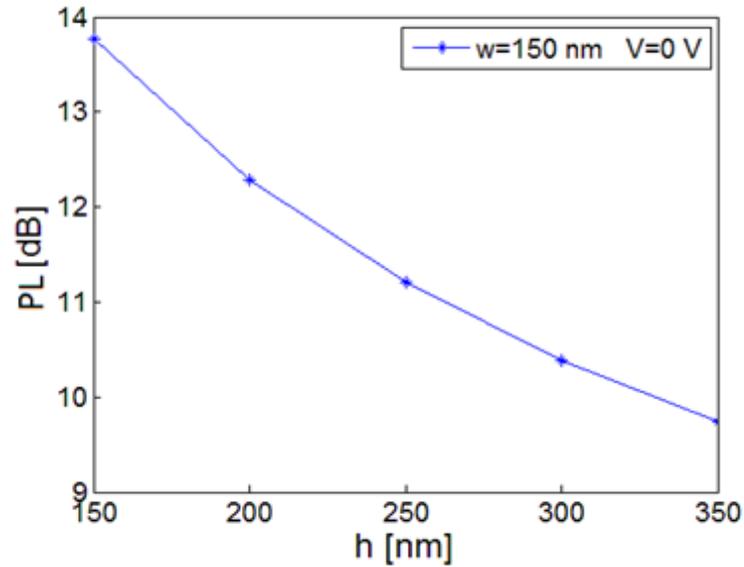

**Figure 90: Propagation losses of the device of Figure 84 as a function of the height of the Ge core h for different driving voltages V. The length is 30 µm**

For the optimized value w=150 nm, the computed extinction ratio and the propagation losses of the modulators are plotted in Figure 89 and Figure 90 respectively as a function of h. For the targeted value of h=250nm, an extinction ratio of 3.3 dB with propagation losses of 11.2 dB can be achieved for a device length of L=30 µm.

If the device length L is increased, then the extinction ratio is also increased at the cost of larger propagation losses. The extinction ratio per unit length is 0.11 dB/µm and the propagation losses is 0.37 dB/µm.

## 5.2.6.7 Operational Wavelength Range of the MIS Modulator

In order to evaluate the operating wavelength range of the modulator, integrated electro-optical simulations were performed at different wavelengths for the optimal modulator (w=150 nm, h=250 nm). The resulting $|\Delta\alpha_{eff}|/\alpha_{eff}$ FoM is plotted in Figure 91. The dispersion of the material was taken into account.

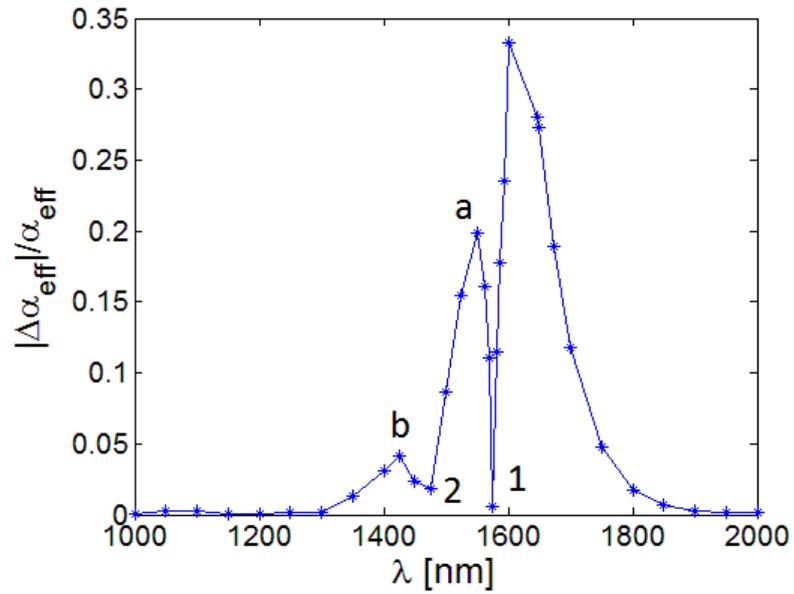

**Figure 91: Spectrum of the FoM $|\Delta\alpha_{eff}|/\alpha_{eff}$. The device is the optimized one. The maximums a and b and the minimums 1 and 2 below the bandgap wavelength of the material are due to FKE oscilations**

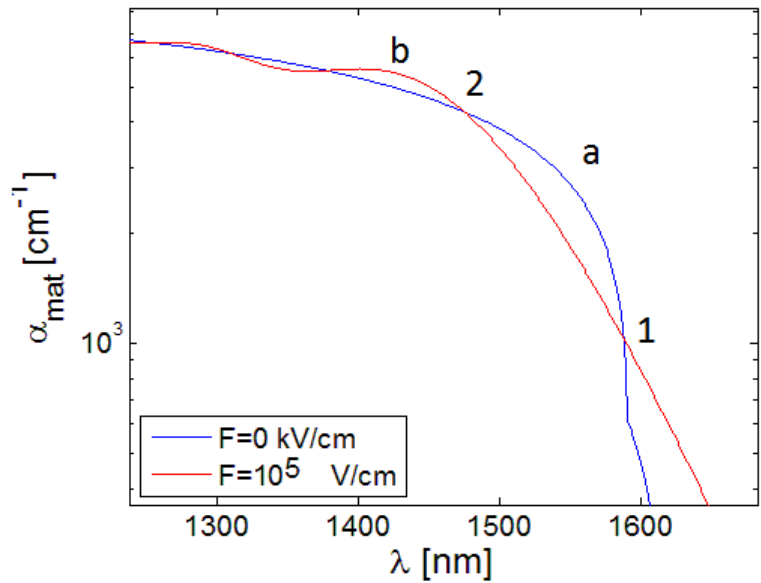

**Figure 92: Absorption of the material $\alpha_{mat}$ due to the FKE for a uniform static electric field in the material of 0 V/cm and $10^5$ V/cm**

As expected, the best operational wavelength of the device is around 1647 nm where the FKE is maximum, as we stated before. Above 1647 nm performances are degraded. Franz-Keldysh Oscillations (FKO) occurs below 1550 nm, E.g.: above the direct bandgap of Ge. To explain these oscillations, we plot the absorption of Ge against the operational wavelength in Figure 92 for zero static electric field and for $10^5$ V/cm, which is the field corresponding to 3 V applied voltage. At the points marked 1 and 2 in Figure 92 absorption values of the Ge core for 0 V (0 V/cm) and 3 V ($10^5$ kV/cm) are equal. Under such condition the parameter $|\Delta\alpha_{eff}|$ is close to zero, hence the drops in the modulator figure of merit in Figure 91 at the points also marked 1 and 2. The maximums a and b of Figure 91 also correspond with the points a and b in Figure 92 where the absorption difference between 0 and 3 V is maximum.

To further support the approximation of a constant static electric field that was done to plot Figure 92 just before, it can be seen in Figure 82 and Figure 83 that the electric field in the Ge is almost uniform when a driving voltage V is applied between the terminals. It is also represented in Figure 85 for V=0 V (a) and V=3 V (b). It is uniform especially in the upper part of Ge where the optical field of the mode is present (see Figure 86 for the representation of the optical field).

Furthermore this approximation is validated by the coincidence of points a and b in Figure 91 and Figure 92 and also the points 1 and 2 in Figure 91 and Figure 92. A constant static electric field in the Ge core was assumed to calculate the absorption in Figure 92 whereas the simulated static electric field distribution was used to calculate the relative absorption variation in Figure 91.

As a conclusion our modulator works optimally at a wavelength of 1647 nm. As we discussed in chapter four, it is theoretically possible to shift the operational wavelength to the communication band around 1550 nm. This can be done using SiGe instead of Ge in the core of the modulator. To shift the maximum peak of the FKE to 1550 nm we need to include an amount of Si (0.8%) into the Ge (99.2%) as was demonstrated in [51].

As can be seen in Figure 91, the predictive model of the FKE and the integrated electro-optical simulations say that the best operational wavelength for our device is 1647 nm where the FoM $\Delta\alpha_{eff}/\alpha_{eff}$ reaches a maximum value of 0.34.

## 5.2.6.8   Comparison of the Performances of MS and MIS Modulator Structures

Now we compare the different characteristics of the modulators like the extinction ratio, the propagation losses, etc. in the different structures proposed like the MS or the MIS modulators. Finally, we compare our best modulator structure with the state-of-the-art.

For the MS modulator structure, according to Table 17, when we increase the length from from L=10 μm to L=20 μm, we increase the extinction ratio from 1.3 dB (0.13 dB/μm) to 2.7 dB (0.135 dB/μm). But, the propagation losses are also increased from 10.6 dB (1.06 dB/μm) to 21.3 dB (1.065 dB/μm). It means, that with the structure of Figure 80 we go from an impractical extinction ratio (1.3 dB) to impractical propagation losses (21.3 dB) when we increase the length. The target values of ER > 3 dB and PL < 10 dB are not achieved to define a proper optical link.

Now, adding a small slot of $Si_3N_4$ to form the MIS structure, we reach 3.3 dB extinction ratio for a length of L=30 μm and propagation losses of only 11.2 dB for the same length. Now, with those values we have a more practical plasmonic modulator.

We can see that adding a slot, the efficiency of the extinction ratio is decreased from 0.135 dB/μm to 0.11 dB/μm. The reason of this change is that the static electric field is high in the slot and therefore less in the Ge core as can be seen in Figure 83.

Regarding the propagation losses of the structure, we can reduce the losses from 1.065 dB/μm to 0.363 dB/μm by adding a slot to the MS structure to form a MIS structure. The reason of the reduced losses by adding a slot is because the presence of the slot modifies the boundary conditions of the electromagnetic field and then it is taken out of the metals.

To summarize, adding a slot to the structure the extinction ratio is reduced from 0.135 dB/μm to 0.11 dB/μm. But for L=30 μm we still have a practical extinction ratio of 3.3 dB. The reason is that the static electric field also gets confined into the slot. Furthermore, we reduced drastically the losses from 1.065 dB/μm to 0.363 dB/μm. The reason of the reduced losses by adding a slot is because the presence of the slot modifies the boundary conditions of the electromagnetic field and then it is taken out of the metals. Hence, producing the reduction of the losses.

## 5.2.6.9   Comparison of the Performances MIS Modulator Structures with the State-of-the-art

The performances of the plasmonic modulator are now compared with the state-of-the-art of photonic modulator that uses the FKE summarized in Table 1. We compare the device with a length of L=30 μm since it is the one with the best benchmarks. We can see that the extinction ratio of 3.3 dB of the device is lower than the extinction ratio reported in Table 1. The minimum extinction ratio of Table 1 is 4 dB for the device of Kotura Inc. [53]. Comparing the propagation losses of the device for L=30 μm they are around 11.2 dB. These losses are higher than with respect to the losses presented in the photonic FKE modulator of Table 1. The maximum insertion losses of the photonic FKE modulators is 10 dB for the one presented by A*STAR in [52]. The fact that we use metals in the active region of the device to propagate the plasmon mode induces additional losses with respect to the photonic ones. Nevertheless, the plasmonic modulator proposed in this work has a smaller active region (the cross section is more compact) and is not longer than 30 μm. The average length of the modulator summarized in Table 1 is around 50 μm. The driving voltage used by the proposed modulator of Figure 84 is around 3 V which is similar to the driving voltage used by the photonic FKE modulators summarized in Table 1. For most devices is around 3 V as can be seen in the following references [35], [51], [54]. In our case, the driving voltage used in the structure is due to the fact that reducing the dimensions of the structure the induced static electric field for the same voltage V is bigger for a smaller V.

The FoM of 0.34 obtained as an optimization of the MIS plasmonic modulator is below the FoM of 2-3 reported in some photonic electro-absorption modulators [53], [66]. The reason to obtain a lower FoM in the plasmonic modulator is due to the higher effective losses $\alpha_{eff}$, due to the use of metal in the active region.

In comparison with the state-of-the-art of the reported plasmonic modulators summarized in Table 2, we obtain moderate propagation losses of 11.2 dB. There are devices with higher propagation losses around 20 dB like in [56] or the 18 dB insertion losses like in [63]. On the other hand, lower propagation losses around 8 dB are also reported. With respect to the extinction ratio of 3.3 dB for L=30 μm for our modulator, we can see that it is not so high in comparison with the plasmonic modulators summarized in Table 2. The average extinction ratio of such modulators is around 6.5 dB. Nevertheless there are also few modulators with less extinction ratio around 1-2 dB like the one reported in [63] or 3 dB for the one in [65]. Nevertheless the gap in the extinction ratio between the proposed modulators in this work and the state-of-the-art of the plasmonic devices is low.

A comparison between the state-of-the-art of photonic FKE modulators summarized in Table 1 and the plasmonic modulators reported in Table 2 reveals that our plasmonic modulator is more compact. Regarding the driving voltage of the proposed modulator, it is smaller than for the plasmonic modulators shown in Table 2.

Now we discuss the details of the plasmonic mode. The $\alpha_{eff}$ of the mode is in the order of 747.6 $cm^{-1}$ and the $\Delta\alpha_{eff}$ is about 186 $cm^{-1}$. This gives a FoM $\Delta\alpha_{eff}/\alpha_{eff}$ of 0.34 which is the maximum FoM that we reached doing the optimization of the device. Those parameters are in agreement with the values estimated in [46]. In [46] the tensile strained FKE is studied. For this they used a Ge-on-Si PIN structure to know if it is a good material for modulation, both in electro-refraction and electro-absorption. They estimate that for 80 kV/cm the parameter $\Delta\alpha_{eff}/\alpha_{eff}$ should be around 3 for photonic devices. This corresponds to a modulation of 7.5 dB for a device length of 100 μm (0.075 dB/μm). Extinction ratio is around 0.11 dB/μm (3.3 dB of extinction ratio for a length of 30 μm).

Regarding the propagation losses the reported value in [46] is 0.025 dB/μm. The efficiency in the propagation losses of the designed modulator is 0.363 dB/μm. The higher losses are due to the use of the metals in the active region. As a first conclusion, we increased the efficiency of the extinction ratio. As a trade-off, we degraded also the performance of the efficiency in the propagation losses.

At last, we want to remind that under 3 V bias voltage the static electric field present in the Ge core of the modulator is around $8 \times 10^4$ $Vcm^{-1}$. This value is below the $10^6$ V/cm breakdown static electric field of Ge.

## 5.2.6.10 Electrical Power Consumption

Now we need to calculate other benchmarks like the electrical power consumption and the operational frequency of the device.

To calculate the electrical power consumption we use Equation 17. We represented the equivalent electrical circuit of the modulator in Figure 93. There is a capacitance C in series with two parallel resistors $R_1$ and $R_2$. The side contact is separated from the core of the modulator by the distance $L_1$. A different voltage V is applied between the top contact and the side contact.

To calculate the electrical power consumption we need to know the driving voltage of the modulator (V) and the intrinsic capacitance of it (C). The driving voltage used is V=3 V as we explained early in this chapter. Regarding the intrinsic capacitance of the device we used ISE-dessis to calculate it. Doing the simulation we obtain a value of C=9 fF (for a device length around L=30 μm). It is interesting to know the change in the value of the capacitance C versus the driving voltage of the device V.

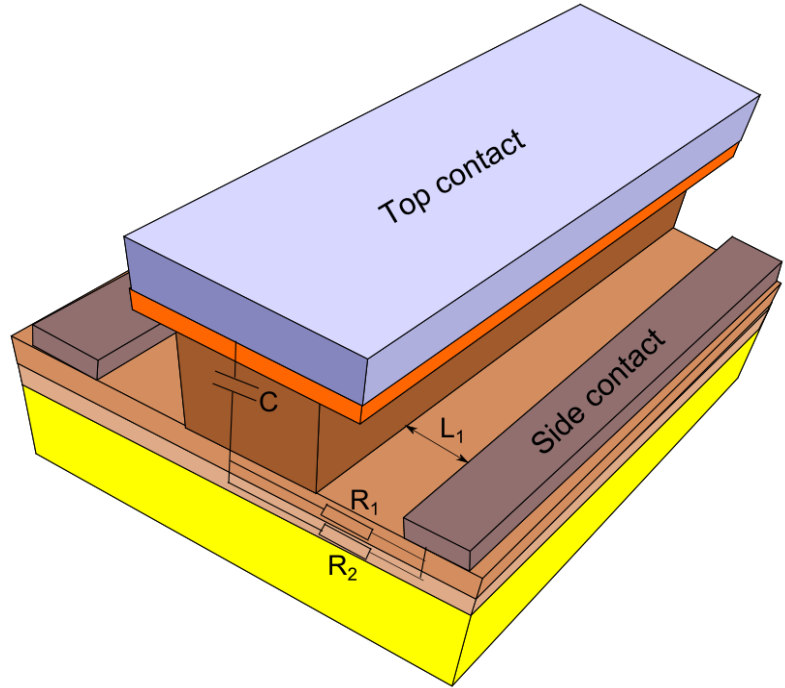

**Figure 93: Equivalent electrical circuit of the plasmonic modulator**

We want to know if the capacitance C changes with V. The main reason is that applying different V will change the charge distribution in the device and consequently the capacitance C may change. For a value of w=150 nm, h=250 nm, $h_{Si3N4}$=5 nm, $h_{Bot}$=40 nm and $h_{Buf}$=60, the capacitance study is presented in Table 22,

| V [V]: | 1 | 2 | 3 |
|---|---|---|---|
| C [fF]: | 8.52 | 8.73 | 9 |

**Table 22: Influence of the driving voltage V in the intrinsic capacitance of the modulator C**

According to Table 22 there is not a big influence of the voltage on the capacitance C, due to the redistribution of the charge when different driving voltages V are applied. Nevertheless, to calculate the electrical power consumption, we selected the highest capacitance C=9 fF to avoid underestimating the electrical power consumption.

To use the formula given in Equation 17 we need to ensure that the current (photocurrent and leakage current) can be neglected. It is worth to mention that in the derivation of Equation 17 the current that flows through the capacitance was neglected (both the photocurrent and the leakage current). Since the slot of $Si_3N_4$ has a thickness around 5 nm the electrons and holes can tunnel through them so the device of Figure 84 will have a current (photocurrent and leakage current). It is important to estimate the leakage current of the device to know if we can apply the formula given by Equation 17 to calculate the electrical power consumption of the device.

We simulated in ISE-dessis the tunneling of carriers through the thin $Si_3N_4$ barrier to estimate the current that passes through the device. We used two models called Fowler–Nordheim tunneling and non-local tunneling at interfaces and contacts. For the optimized device (w=150 nm, h=250 nm, $h_{Si3N4}$=5 nm, $h_{Bot}$=40 nm and $h_{Buf}$=60 nm, see Figure 84) operated at a bias of V=3 V, we obtain an electron current of $6.3 \times 10^{-12}$ A/cm and a hole current of $1.8 \times 10^{-12}$ A/cm. For a 30 μm long device, we have an electron current of around $1.89 \times 10^{-14}$ A and a hole current of $5.4 \times 10^{-15}$ A. Those current values are small, justifying our consideration of a capacitor without current in the calculation of the electrical power consumption using Equation 17.

Consequently, for a capacitance C=9 fF and a driving voltage V=3 V we obtained an electrical power consumption around 20 fJ/bit. We will compare this value with the state-of-the-art at the end of this section.

## 5.2.6.11 Bandwidth

To calculate the cut-off frequency of the device we use the following Equation 60,

$$f_c = \frac{1}{2\pi RC}$$

Equation 60

We already know the capacitance C of the device which is around C=9 fF.

R is the access resistance which was calculated using Equation 61 in the previous paragraph. This resistance is the resistance between the side contact and the core of the modulator, it is mainly given by the p-Si (thickness $h_{Buf}$) and p-Ge (thickness $h_{Bot}$) layers at the bottom of the modulator.

The distance between the contact and the modulator were taken between 1.5 μm and 2 μm. For the distance of 1.5 μm we obtain a lower resistance than with the separation of 2 μm. Consequently, when the distance is shorter we obtain a higher bandwidth that we call $F_{max}$ and with the separation of 2 μm we obtain a lower value that we call $F_{min}$. Consequently, if we place the side contact between 1.5 μm and 2 μm we will obtain an operational frequency between $F_{min}$ and $F_{max}$. To calculate the resistance of the device we used the Equation 61,

$$R = \rho \frac{L}{A}$$

Equation 61

Were $\rho$ is the resistance of the p-Si and the p-Ge, both values were taken equal to $10^{-4}$ Ω/cm, the distance L is 30 μm (the length of the device to have an extinction ratio of 3.3 dB according to Table 21). A, represents the area of the cross section of the materials p-Si and p-Ge. The height of the layer of p-Si $h_{Buf}$ is 60 nm while the $h_{Bot}$ of p-Ge is 40 nm. The length of both are L=30 μm. Consequently, for the p-Si layer A=$h_{Buf}$xL=60 nmx30 μm and for the p-Ge layer it is equal to A=$h_{Bot}$xL=40 nmx30 μm. The length between the contact and the Ge core $L_1$ is between $L_1$=1.5 μm and $L_1$=2 μm. This gives two results which correspond to $F_{min}$ and $F_{max}$. The maximum operational frequency $F_{min}$ correspond with $L_1$=2 μm and $F_{max}$ with $L_1$=1.5 μm.

Using the mentioned values for a distance of 1.5 μm we obtain $F_{min}$=237 GHZ and an $F_{max}$=475 GHz for a distance of 2 μm. It is summarized in Table 22. If we place the contact closer to the Ge core region we can increase the operational frequency of the device, it can be seen from $F_{min}$ and $F_{max}$, which increases when the contact is closer at 1.5 μm.

| L [µm]: | Electrical power consumption [fJ/bit] | $F_{min}$ [GHz] | $F_{max}$ [GHz] |
|---|---|---|---|
| 10 | 6.7 | 237 | 475 |
| 30 | 20.3 | 237 | 475 |

**Table 23: Power consumption, minimum and maximum operational frequency ($F_{min}$ and $F_{max}$) of the modulator**

As it was mentioned before the operational frequency of the device is not limited by the FKE which is ultrafast but it is RC limited. However, the cut-off frequency of our device is much larger than the application requirements of 25 to 50 GHz. In conclusion, we are not limited by the bandwidth of our device.

If we compare the operational frequency of the device with the photonic modulators that use the FKE (Table 1) we can see that  there are modulators with a much lower value of 1.2 GHz like the one presented in [51]. The rest of the modulators [35], [53], [54] are between 30-40 GHz.

Regarding the comparison with the plasmonic modulators (Table 2) most of the reported devices are limited by the bandwidth below 15 GHz [56], [59], [71]. The main limitation factors of such devices are either the RC limitation or the intrinsic limitation of the physical effect used to perform the modulation. E.g.: Electro-refraction in Si based on carrier injection is limited by the carrier velocity. Nevertheless, there are also other devices reported with a bandwidth is over 500 GHz like in [57]. This value is over the modulators presented here. There is even a modulator reported in the THz regime in [63].

Regarding the electrical power consumption reported in Table 23 we want to compare it with the electrical power consumption summarized in the photonics FKE modulators summarized in Table 1 and with the plasmonic modulators summarized in Table 2. Since it is an important parameter of the device we present all mentioned modulators in the following Table 24,

| Team: | MIT [51] (2008) | A*STAR [65] (2011) | Kotura Inc. [53] (2011) | Leeds [71](2011) | Berkeley [66] (2012) | Stanford [143] (2013) | Our work (2013) |
|---|---|---|---|---|---|---|---|
| Electrical power consumption [fJ/bit]: | 50 | 200 | 100 | 360 | 56 | 100 | 20 |
| Type: | Photonic | Plasmonic | Photonic | Photonic | Plasmonic | Photonic | Plasmonic |
| Structure: | Absorption | Absorption | Absorption | MZI | Absorption | Absorption | Absorption |
| Effect: | FKE | Plasma disperssion | FKE | Plasma disperssion | ITO | ITO | FKE |

**Table 24: Summary of the photonic FKE modulators and plasmonic modulators which reports the electrical power consumption**

From Table 24 we see that the proposed modulator in this thesis has the lowest electrical power consumption to the best of our knowledge. With this we confirmed the initial idea of reducing the electrical power consumption by going from a photonic FKE modulator to a plasmonic one.

# 5.3 Conclusion

In this chapter, we proposed a new design of a low electrical power consumption Ge based FKE electro-absorption plasmonic modulator. We have performed an integrated electro-optical simulation in order to determine the performances of the device. Using the static electric field distribution we calculated the change in absorption $\Delta\alpha_{mat}$ occurring in Ge due to the FKE. FDM optical simulations were performed in order to determine the extinction ratio and the propagation losses of the modulator. The operational bandwidth, the maximum frequency and the electrical power consumption were also calculated.

Using this method, we optimized the structure in order to maximize the FoM $\Delta\alpha_{eff}/\alpha_{eff}$. The optimized device has a compact active region of w=150 nm and h=250 nm and is fully compatible with the CMOS fabrication technology. For a length of L=30 µm the device has an extinction ratio of 3.3 dB and propagation losses of 11.2 dB, while working at a bias voltage of V=3 V. High-speed operation can be achieved with a cut-off frequency beyond 300 GHz.

The energy consumption of such a device is around 20 fJ/bit which is well below the energy consumption of previous reported modulators [46], [35], [51], [52], [53] and [54].

The final objective of the work is to integrate the plasmonic modulator device into a photonic integrated circuit. For this it is important to study and optimize the coupling to the device from a standard Si rib waveguide. This study is the objective of next chapter.

In this chapter we introduced the propagation losses of the device but the insertion losses of the device include both the propagation losses and the coupling losses to conventional standard Si waveguides. The coupling losses depend on the coupling scheme used to excite the modulator. In the next chapter we present different coupling structures. We select the best one and we calculate the coupling losses. With this information, we will derive the total insertion losses of the device.

# 6 Coupling to the FKE Plasmonic Modulator

To fully integrate our plasmonic modulator into a photonic integrated circuit it is important to study its coupling with a standard Si rib waveguide. The core of the modulator has been optimized into the following dimensions: w=150 nm, h=250 nm, $h_{Slot}$=5 nm, $h_{Bot}$=40 nm and $h_{Buf}$=60 nm. The parameters are represented in Figure 84.  Both schematic views of the optical modulator and the Si rib waveguide are reported in Figure 94. The methodology of the coupling study is the following.

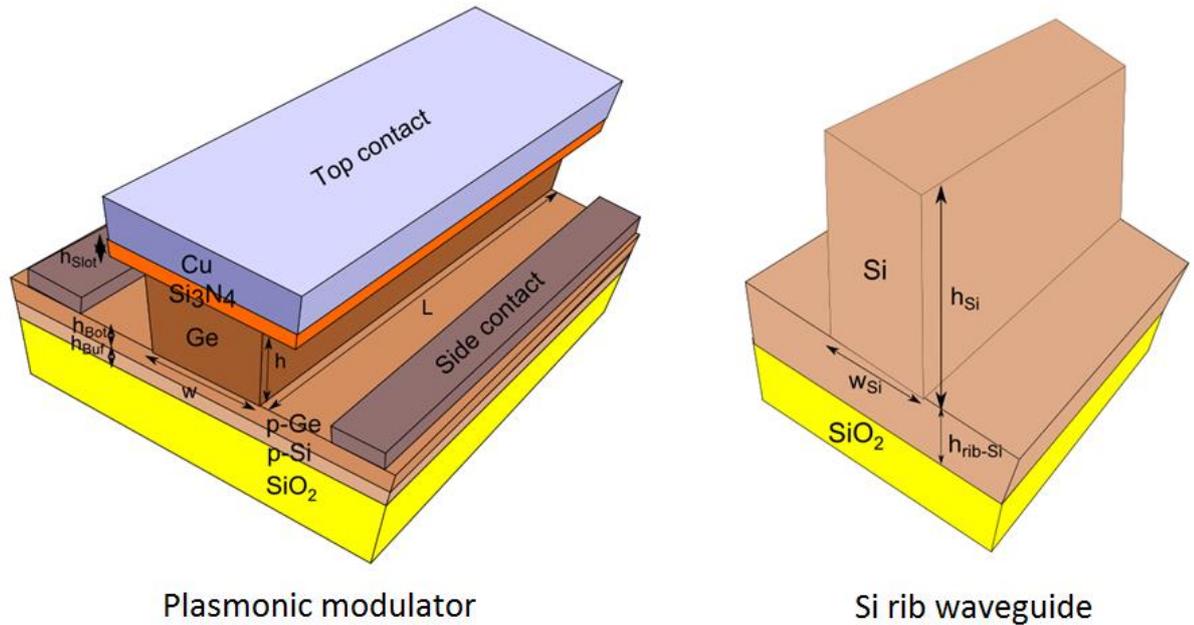

**Figure 94: Structures to couple. Optimized plasmonic modulator (left) and Si rib waveguide (right)**

First we study the modes of the MIS waveguide that forms the core of the plasmonic modulator. It means, we scan the effective refractive index $n_{eff}$ and the effective optical absorption $\alpha_{eff}$ that the plasmon has (the plasmonic mode supported by the modulator is TM) for different parameters of the structure and the mode profile. We also study the mode of a Si rib photonic waveguide in the same way. It means, we scan the parameters of the waveguide to know the evolution of $n_{eff}$ and $\alpha_{eff}$.

Knowing the parameters $n_{eff}$ and $\alpha_{eff}$ in both the MIS waveguide of the modulator and the Si rib waveguide we studied a butt-coupling approach to excite the modulator. In the butt-coupling structure the parameters of the modulator are fixed. Consequently, we select the proper dimension of the Si rib waveguide to match the effective refractive index $n_{eff}$ between the Si rib waveguide and the plasmonic one into the modulator. This allows reducing the reflection from the interface between the Si rib waveguide and the plasmonic waveguide. In this butt-coupling structure we also analyzed the influence of the overlap between the modes in the Si rib waveguide and the MIS structure.

Knowing the performance of the butt-coupling structure we try to engineer the coupling to try to improve the coupling efficiency. We introduced a taper between the Si rib waveguide and the plasmonic modulator. In this case, the structure is formed by the Si rib waveguide, followed by a taper and finally the plasmonic modulator is reached. It means, we have three different waveguides. The objective is that the effective refractive index in all the structures that are coupled varies in a continuous way and hence, minimize the reflection. First, we analyzed the performance of a Ge taper and finally we propose a Si-Ge taper.

# 6.1    Tools Used in the Coupling Study

In the coupling study presented in this section we performed 3D Finite-difference time-domain (FDTD) simulations using the commercial software Lumerical® to study the propagation of the light from the Si rib waveguide to the optimized plasmonic modulator. Using this tool we obtain the value of the coupling efficiency. Furthermore, using Lumerical® we perform 2D simulations to calculate the overlap integral between the mode that we excite in the plasmonic modulator from the standard Si rib waveguide and the fundamental mode that we want to excite. This parameter gives the information of the fraction of electromagnetic field coupled in the fundamental TM plasmonic mode that we want to excite. We remind that we only want to excite the fundamental TM mode of the optimized plasmonic modulator.

Regarding the FDTD parameters we used a square simulation window of 12 µm length, 5 µm wide and 4 µm high. Regarding the boundary conditions we used a perfectly matched layer (PML) in all the corners. We observed that there is not field in the boundary when performing the simulation. We used an automatic non-uniform mesh with a minimum mesh step of 2 nm. The maximum mesh step is around 10 nm. In the sensible areas like the slot of the modulator we used a uniform refinement of the mesh. In this case the maximum step size is around 0.5 nm (if the thickness of the slot is 5 nm then we have 10 mesh points inside the slot). The same is done for the abrupt interface between the Si rib waveguide and the optimized plasmonic modulator. Regarding the injection of light we place a modal source in the Si rib waveguide which excites a continuous wave temporal excitation. This source was configured to excite the fundamental TM mode of the Si rib waveguide in direction to the optimized plasmonic modulator (x-axis). To validate all the parameters set in the 3D FDTD simulation we reproduced the results obtained in [144]. This paper studies the coupling between a Si strip waveguides and a plasmonic MIS waveguide. We selected to reproduce the results of this publication because it is similar to the one we want to simulate. We only simulated the first 10 µm of the plasmonic modulator because with this length the simulation takes 24 hours in the cluster at CEA-Leti. We employed 16 cores.

# 6.2    Properties of the Plasmonic Mode in the Plasmonic Modulator

First we analyze the effective refractive index $n_{eff}$ and the effective optical absorption $\alpha_{eff}$. This modulator is presented in Figure 84 and the optimized values are w=150 nm, h=250 nm, $h_{Slot}$=5 nm, $h_{Bot}$=40 nm and $h_{Buf}$=60 nm. The mode profile of the optimized plasmonic modulator is presented in Figure 95. Since it is a fundamental TM mode we only represent the magnitude of the magnetic field H.

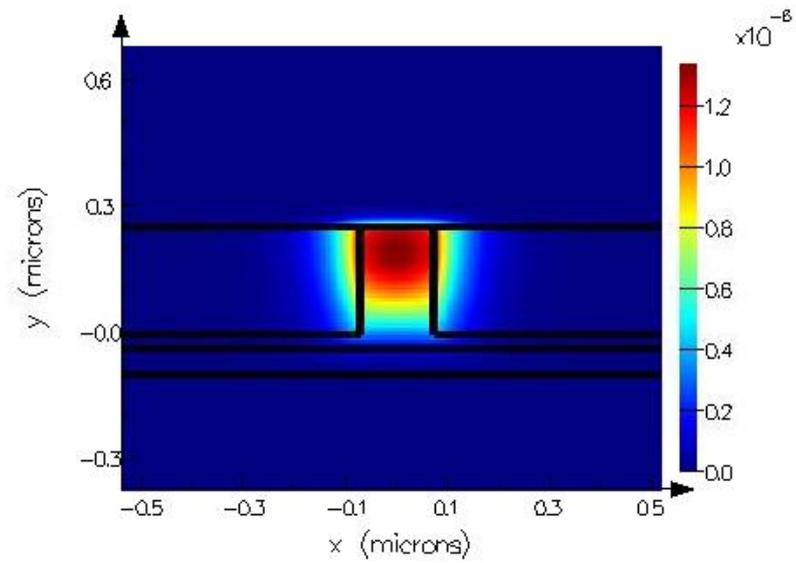

**Figure 95: Magnitude of the magnetic field H distribution of the fundamental TM mode of the optimized plasmonic modulator**

In the following study we change w to know the evolution of the parameters ($n_{eff}$ and $\alpha_{eff}$) of the plasmonic mode supported by the modulator. The effective refractive index $n_{eff}$ is presented in Figure 96 and the effective absorption $\alpha_{eff}$ is presented in Figure 97.

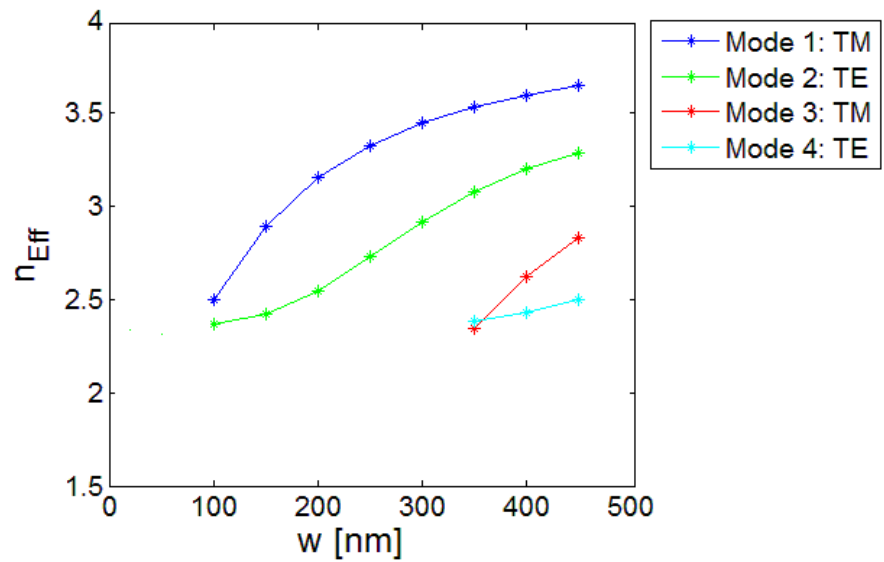

**Figure 96: Effective refractive index $n_{eff}$ of the plasmonic modulator of Figure 84. It is difficult to calculate the effective index below 100 nm due to the amount of evanescent field**

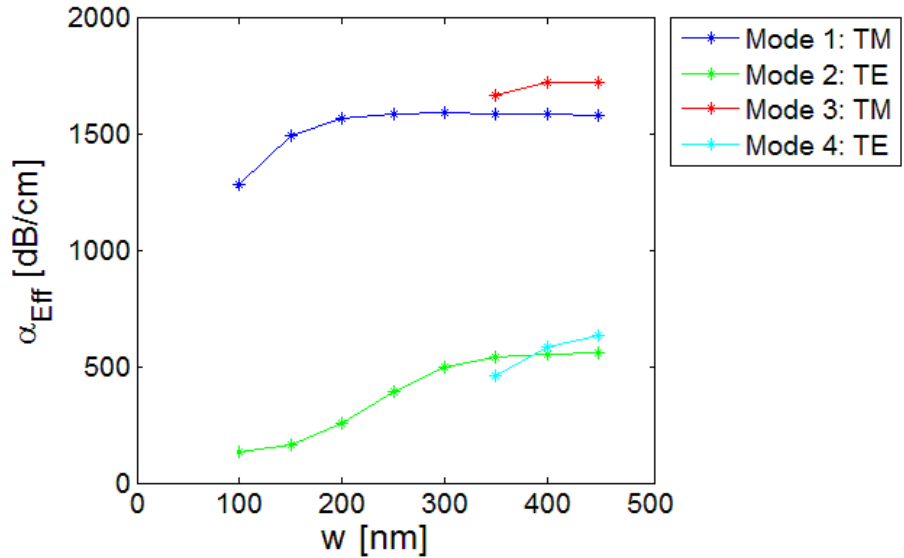

**Figure 97: Effective absorption α_eff of the plasmonic modulator of Figure 84. It is difficult to calculate the effective index below 100 nm due to the amount of evanescent field**

The plasmon that we excite in the plasmonic modulator is the one represented in blue in the previous figures (Figure 96 and Figure 97). The effective refractive index increases from $n_{eff}$=2.5 to 3.5 when w increases from w=100 nm to 450 nm. The value for the optimized modulator is around 2.89 (w=150 nm). From Figure 97 it is possible to see that the effective losses of the plasmon goes from 1250 dB/cm (0.125 dB/μm) to 1500 dB/cm (0.15 dB/μm) when w goes from w=100 nm to w=450 nm. For the optimized values of the modulator, the effective losses are around 1491 dB/cm (0.149 dB/μm).

As a conclusion, in the optimized modulator we want to excite the fundamental TM mode whose effective refractive index is $n_{eff}$=2.89 and the effective propagation losses are $\alpha_{eff}$=1491 dB/cm (0.1491 dB/μm). Consequently, in the Si rib waveguide which will excite the optimized modulator a TM mode must be present.

The plasmonic modulator also supports other modes like a fundamental TE mode (green), a second order TM mode (red) and another second order TE mode (cyan). It is worth to mention that both TE modes (green and cyan) are photonic ones. Such modes have low losses (Figure 97). The other two TM modes are plasmonic and have higher losses. Note also that the second order TE and TM modes appear for a width around w=350 nm. We want to excite the fundamental TM mode (blue) nevertheless, it is worth to mention that if w>350 nm two TM modes exist so we may excite the incorrect one and they may also beat.

## 6.3 Properties of the Photonic Mode in a Silicon Rib Waveguide

In this section we study the properties of the Si rib waveguide which is used to excite the plasmonic modulator defined in the previous chapter. The structure of the Si rib waveguide is represented in the following Figure 98. We study the influence of $w_{Si}$ in the effective refractive index $n_{eff}$. Since we need to couple to the optimized plasmonic modulator we select the parameter $h_{Si}$ of the Si rib waveguide to be the same as the plasmonic modulators, so, in this study, $h_{Si}$=255 nm in the Si rib waveguide. We remind that the height of the optimized plasmonic modulator is h=250 nm for the height of the Ge core and $h_{slot}$=5 nm for the thickness of the $Si_3N_4$ layer (see Figure 84). Regarding the parameter of $h_{rib-Si}$ for the Si rib waveguide we selected it to be 100 nm, so, it equals the sum of $h_{Bot}$=40 nm and $h_{Buf}$=60 nm in the optimized plasmonic modulator (see Figure 84). It means, by selecting the parameter $h_{Si}$=255 nm and $h_{rib-Si}$=100 nm we obtain the same height for the Ge core of the optimized modulator and also the same slab (which is $h_{Bot}$+$h_{Buff}$=100 nm).

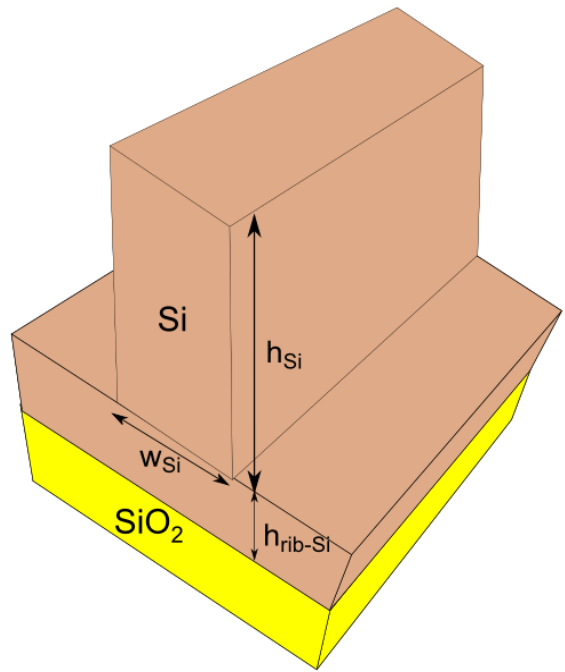

**Figure 98: Silicon rib waveguide and its parameters**

Setting $h_{Si}$=255 nm and $h_{rib-Si}$=100 nm we vary $w_{Si}$ in the Si rib waveguide (Figure 98) to know the influence of w in the effective refractive index $n_{eff}$. It is represented in Figure 99.

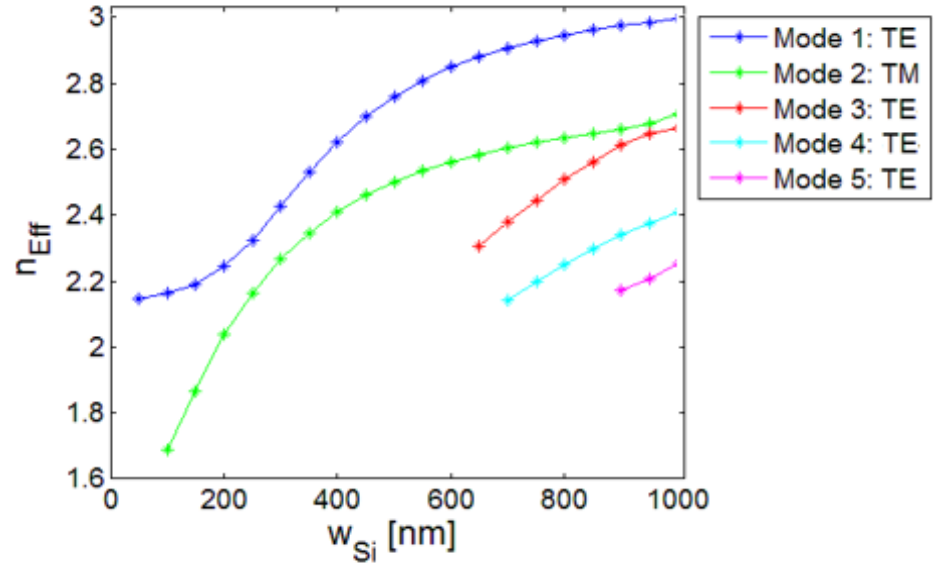

**Figure 99: n_eff as a function of w for the Si rib waveguide. It is difficult to calculate the effective refractive index below w_Si=150 nm due to the deconfinement of the mode**

From Figure 99 it is possible to see that there is a fundamental TE mode (blue) and a fundamental TM mode (green). For values of w larger than 650 nm a second TE modes appear (red). Around $w_{Si}$=700 nm a TE mode appears (cyan). Finally, around $w_{Si}$=900 nm another TE mode appears (pink).

From the Si rib waveguide we need to excite the fundamental plasmonic TM mode in the optimized plasmonic modulator. For this we need to have a photonic TM mode in the input Si rib waveguide (green).

From Figure 99 we see that for the values 100 nm<w<800 nm the effective refractive index $n_{eff}$ increases from 1.6 to 2.65 for the TM mode (green). We remind that the effective refractive index of the plasmon in the optimized plasmonic modulator is around 2.89. So, there is a mismatch in the effective refractive index between 1.29 and 0.24. This difference will induce reflections between the two waveguides in the butt-coupling configuration. We can increase $w_{Si}$ to increase $n_{eff}$ around 2.89. Nevertheless, we do not want to increase the width $w_{Si}$ for values bigger than 800 nm since it is too big. Furthermore, the mode mismatch may also increase for wider waveguides. It means, there is a trade off between reflection and profile mismatch.

The magnitude of the electrical field E and the magnitude of the magnetic field H of the Si rib waveguide for $w_{Si}$=450 nm is represented in Figure 100. Furthermore, the same fields E and H for $w_{Si}$=800 nm is also represented in Figure 101. The rest of the parameters are: $h_{Si}$=250 nm and $h_{Si\text{-}rib}$=100 nm.

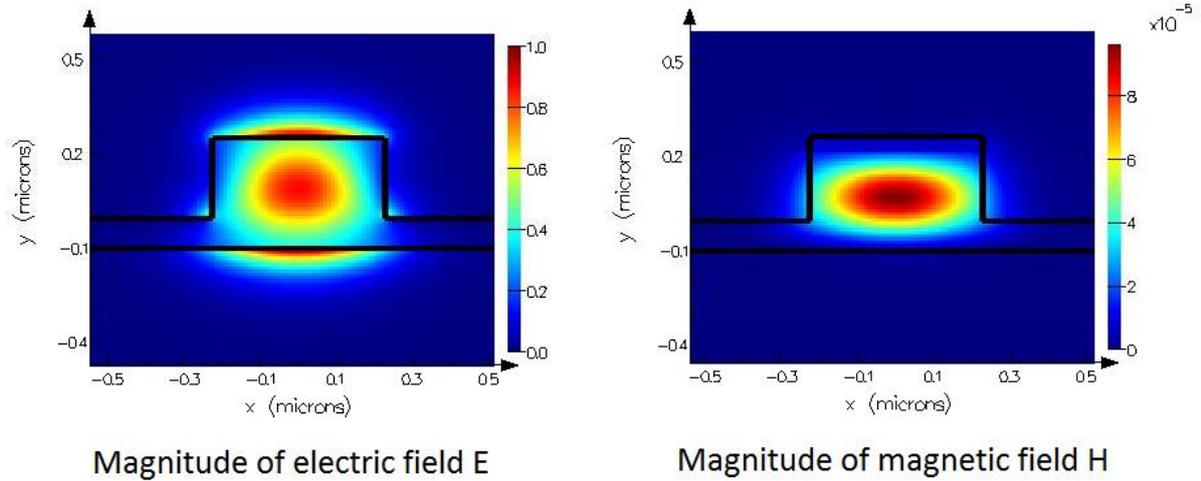

Figure 100: Magnitude of the electric field E and the magnetic field H in a Si rib waveguide with w_Si=450 nm, h_Si=255 nm and h_Si-rib=100 nm

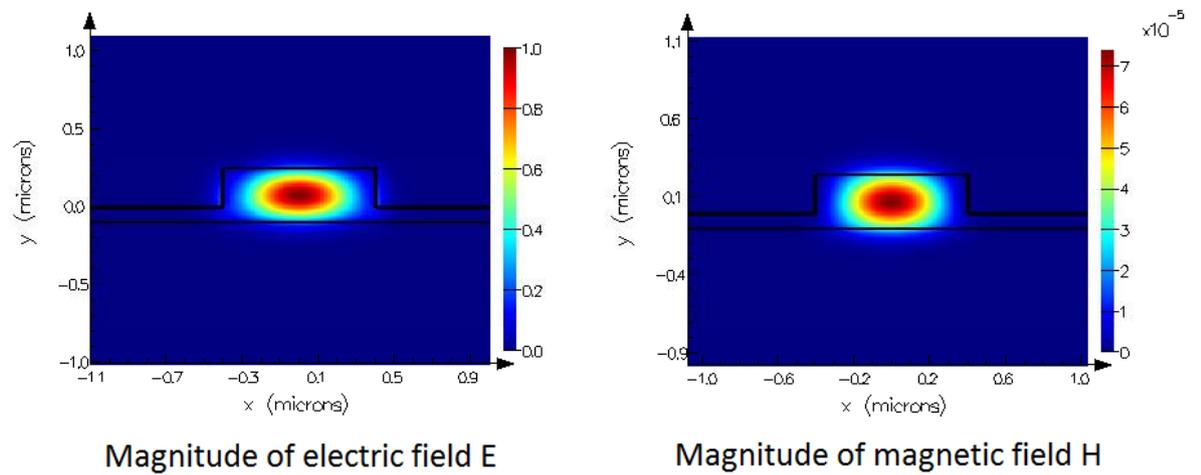

Figure 101 : Magnitude of the electric field E and the magnetic field H in a Si rib waveguide with w_Si=800 nm, h_Si=255 nm and h_Si-rib=100 nm

Using this Si rib waveguide we study a butt-coupling configuration with the optimized plasmonic modulator.

## 6.4    Butt-Coupling Structure

In this section we study the simplest coupling configuration between the Si rib waveguide and the optimized plasmonic modulator. This configuration is called butt-coupling. It is represented in Figure 102,

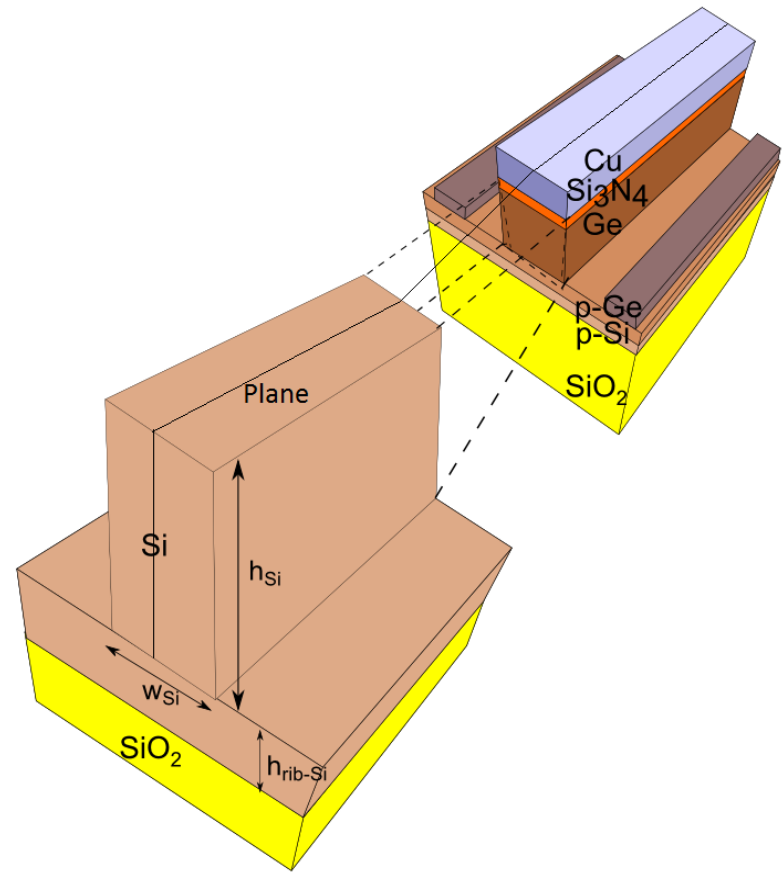

**Figure 102: Representation of the butt-coupling structure between the Si rib waveguide and the optimized plasmonic modulator. The fixed parameters are $h_{Si}$=255 nm and $h_{rib-Si}$=100 nm**

In this configuration, the Si rib waveguide is placed just in front of the optimized plasmonic modulator with an abrupt transition. We align the top of the Si rib waveguide with the top of the $Si_3N_4$ slot. We fixed h and $h_{Si}$ in this way since h has a big influence in the performance of the modulator while w has less influence as explained in chapter five. Consequently, we will study both w and $w_{Si}$. Regarding the horizontal dimension the middle of the Si rib waveguide coincides with the middle of the optimized plasmonic modulator. It means, the two waveguides are centered.

# 6.4.1　　Overall Transmission and Reflection

We study the structure of Figure 102 fixing the parameter of the Si rib waveguide to $w_{Si}$=450 nm (we will show later that this value is the best one. Furthermore, with this dimension we assure that we have a fundamental TM mode in the Si rib waveguide), $h_{Si}$=255 nm and $h_{rib-Si}$=100 nm. We varied $w_{Ge}$ which is the width of the Ge core in the plasmonic modulator (we remind that the optimized plasmonic modulator the value is w=150 nm). We will study the parameter w=150, 250, 350 and 450 nm. For this we simulated 10 µm of the plasmonic modulator to know which mode we excite. For w=150 nm it is represented in Figure 103, for w=250 nm it is represented in Figure 104, for w=350 nm it is represented in Figure 105. Finally, for w=450 nm it is represented in Figure 106. In the left of the vertical white line there is the Si rib waveguide while in the right of the vertical white line there is the plasmonic modulator. The view is the magnitude of the magnetic field along the propagation direction of the light. The plane is in the middle of the waveguides and it is named as 'Plane' in Figure 102. The light propagates in the x-axis.

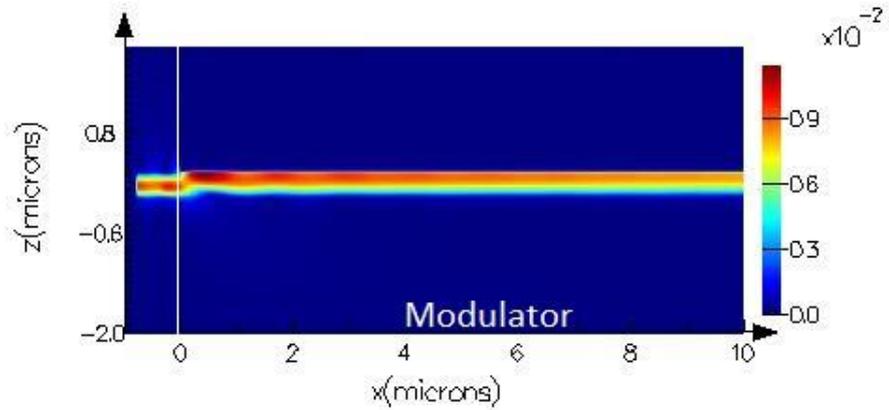

**Figure 103: Simulation of the magnitude of the magnetic field in the plasmonic modulator for w=150 nm. In the left of the vertical withe line the Si rib waveguide is present while in the left of the vertical with line the plasmonic modulator is present. It is a side view of the structure of Figure 102**

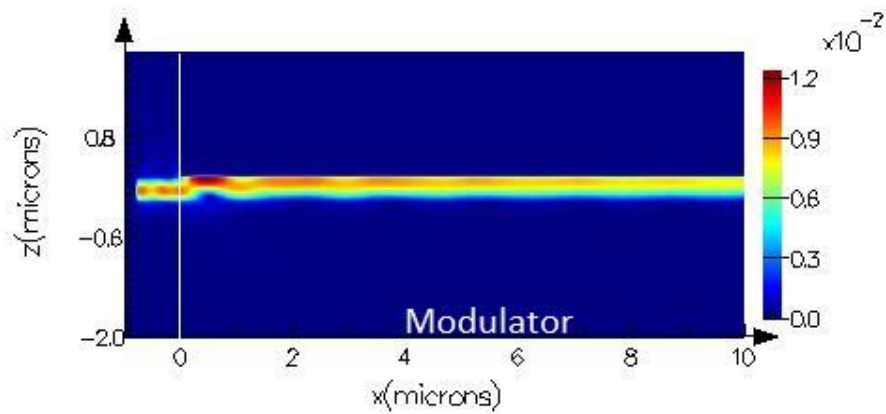

**Figure 104: Simulation of the magnitude of the magnetic field in the plasmonic modulator for w=250 nm. In the left of the vertical withe line the Si rib waveguide is present while in the left of the vertical with line the plasmonic modulator is present. It is a side view of the structure of Figure 102**

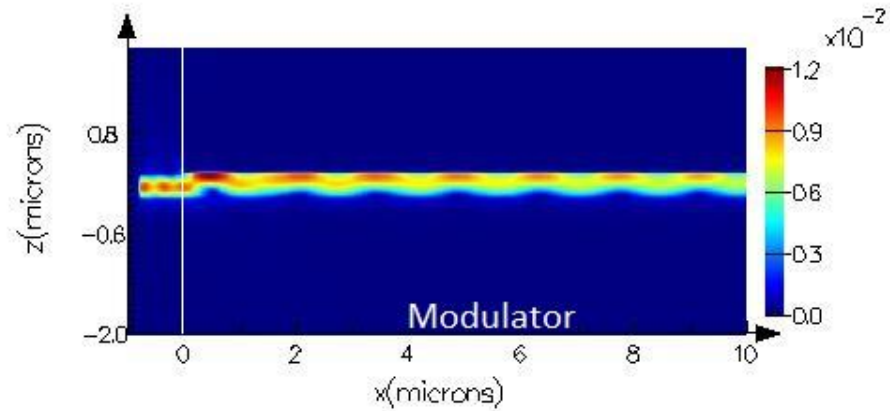

**Figure 105: Simulation of the magnitude of the magnetic field in the plasmonic modulator for w=350 nm. In the left of the vertical withe line the Si rib waveguide is present while in the left of the vertical with line the plasmonic modulator is present. It is a side view of the structure of Figure 102**

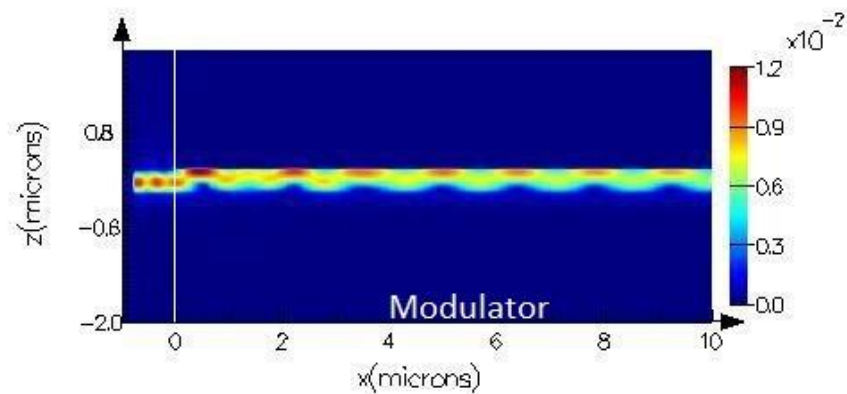

**Figure 106: Simulation of the magnitude of the magnetic field in the plasmonic modulator for w=450 nm. In the left of the vertical withe line the Si rib waveguide is present while in the left of the vertical with line the plasmonic modulator is present. It is a side view of the structure of Figure 102**

From Figure 103 (w=150 nm) and Figure 104 (w=250 nm) we observed that a mode is excited in the modulator. In the first 5 μm along the x-axis for both w=150 nm and w=250 nm we see that there are some oscillations of the magnitude of the magnetic field. We will see later that such oscillations are due to the stabilization of the mode. As they disappear after approximately 5 μm of propagation. The transmission is better in the optimized plasmonic modulator (where w=150 nm) than in the case where w=250 nm. It also has fewer oscillations.

From Figure 105 (w=350 nm) and Figure 106 (w=450 nm) we see that the oscillations do not disappear after several μm. We will show later that it is due to the beating of two modes supported by the structure of the plasmonic modulator. We observe in both cases that after 5 μm the spatial frequency of the oscillations are settled in a periodic way. Nevertheless between 0 and 5 μm the spatial frequency is not constant, we attributed this fact to the stabilization of the mode.

Now, it is interesting to calculate the total transmission against the propagation direction in the (x-axis) along the modulator. The length of the modulator is around 30 μm (to have a 3.3 dB value). We cannot simulate the complete length, so, we only simulate the propagation in the first 10 μm of the modulator. To measure the total transmission for a given propagation length we place a monitor that covers the entire cross section of the plasmonic modulator at a distance $x=x_0$ in the x-axis. With this, we can measure the total transmission $T_{Total}$ which includes the propagating mode $T_{In\ the\ mode}$, the total scattering $T_{Scattered}$ and the energy coupled to other modes $T_{Other\ modes}$ which is the energy coupled to other modes which are not the TM mode that we want to excite. It means, we have,

$T_{Total}=T_{In\ the\ mode}+ T_{Other\ modes}+ T_{Scattered}$                          Equation 62

The parameter $T_{Total}$ as a function of x for different values of w=150, 250, 350 and 450 nm is represented in Figure 107,

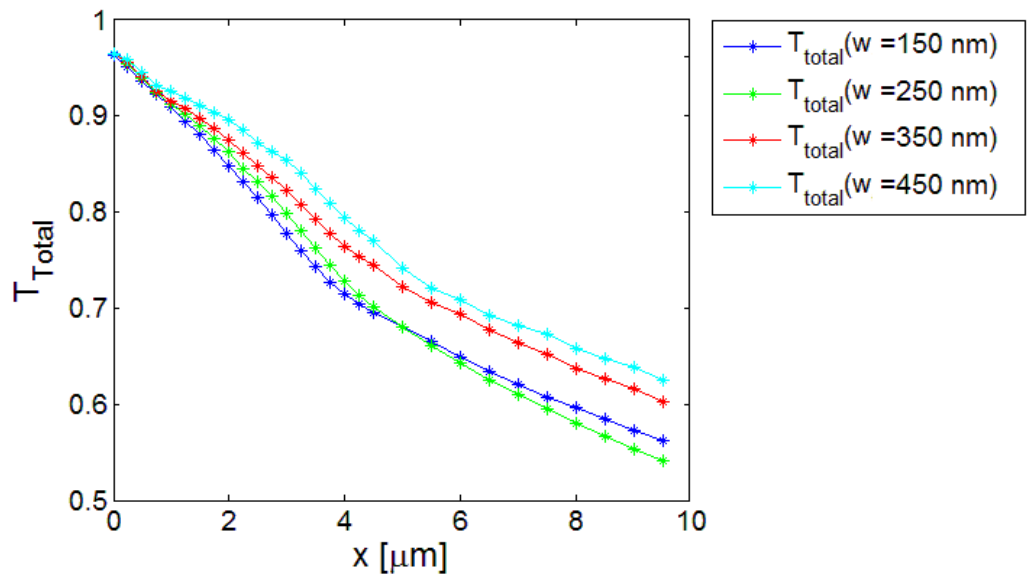

**Figure 107: $T_{Total}$ as a function of the propagation length x for different values of w. The selected width of the Si rib waveguide is 450 nm.**

In Figure 107 it is possible to observe that at x=0 $T_{Total}$=0.96 and it is almost independent of the parameter $w_{Ge}$. If 96% of the light is propagating into the positive direction of x it means that at x=0, 4% of light is reflected back. In the reflected light we count the light into the reflected mode and the scattering due to the interface between the Si rib waveguide and the plasmonic modulator. Obtaining only 4% reflection is a very low value compared with other coupling structures.

## Validation

In [144] a similar structure like the one presented here is simulated. This structure is a Si strip waveguide of width 450 nm and height 250 nm which is butt-coupled to a MIS waveguide formed by the stack of Si-Al$_2$O$_3$-Ag. It is represented in Figure 108. The width of the MIS is 150 nm and the height is 250 nm; the thickness of the Al$_2$O$_3$ slot is 50 nm. In this paper they say that they measure a reflection below 4% and that this value is quite independent of the parameters of the MIS waveguide. With this comparison we validate our result of obtaining a very low reflection from the structure of Figure 102.

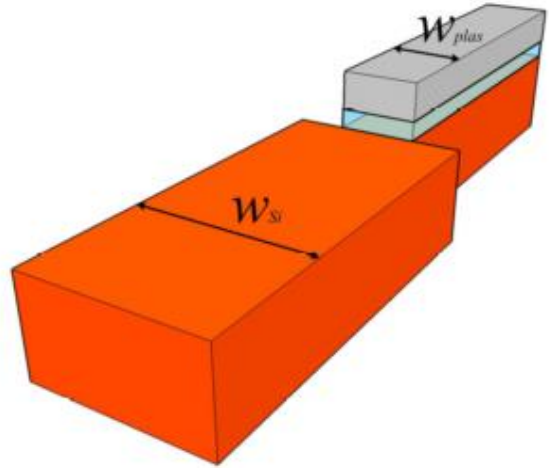

**Figure 108: Structure simulated in [144]. The red color is Si, the light blue is Al₂O₃ and the grey is Ag**

# 6.4.2 Coupling into the Plasmonic Mode, into Other Modes and Scattering

## 6.4.2.1 Mode Analysis

Regarding Figure 107, it is also possible to see that T_Total is smaller when w=150 nm and that it increases when w=450 nm. It is necessary to analyze which one is the scattering and the coupling to other modes to find the cause of this. To extract the power coupled to the plasmonic mode we need to calculate the overlap integral between the electromagnetic field distribution present in the modulator section (calculated in the 3D FDTD simulation) and the fundamental TM mode that the plasmonic modulator supports (calculated in the 2D FDM mode solver). It means, we want to know how much field is coupled to the plasmonic mode of interest. We remind that the plasmonic modulator supports 2 modes for w<350 nm which are the fundamental TE and the fundamental TM modes. On the other hand, when w>350 nm second order TM and TE modes appear. In Figure 109 we calculate the overlap integral between these modes and the mode extracted from the 3D FDTD simulation. We represent the overlap integral between the modes as a function of x.

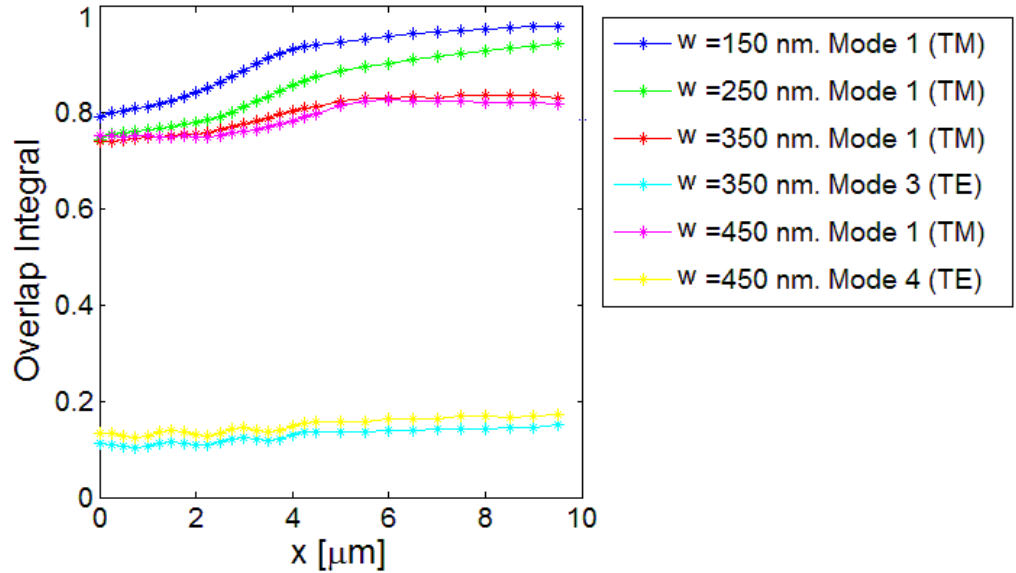

**Figure 109: Overlap integral between the modes supported by the plasmonic modulator and the mode excited by the Si rib waveguide**

From Figure 109 we see that for the case of the optimized modulator when w=150 nm we excite only the fundamental TM mode of the modulator. This overlap integral goes from 80% at the beginning of the modulator until 98% after 10 µm. We conclude that until 5 µm the mode stabilizes in the modulator as was explained before regarding the oscillations of Figure 103 which also disappear after 5 µm. The overlap integral with respect to the other TE fundamental mode supported by the optimized plasmonic modulator is zero and consequently it is not represented in Figure 109. We also observe that in all cases the overlap integral increases as the light propagates in the x direction. This is due to the fact that the modes are settled down in the plasmonic modulator. After 5 µm it stabilizes.

Regarding the overlap integral when w=250 nm we see that only the fundamental TM mode is excited. Again, the overlap integral with the other fundamental TE mode is zero. We see that the overlap integral in the first 5 µm is increasing due to the stabilization of the mode. After x=5 µm the increase is not so rapid. It is worth noting to mention that the oscillations of Figure 104 also disappeared after 5-6 µm.

For both cases: w=150 nm and w=250 nm there is no beating with other mode since the rest of the overlap integrals are zero.

In the case in which w=350 nm in Figure 105 we see that the oscillations do not disappear after 10 µm in contrast to what happens in w=150 or 250 nm where the oscillations disappear after 5 µm. These oscillations may come from the beating of two modes. Calculating the overlap integral of the field in the plasmonic modulator when excited from the Si rib waveguide and the modes supported by the plasmonic modulator we observe in Figure 109 that now there is overlap with two modes. The red overlap integral is with the fundamental TM mode of the plasmonic modulator while the cyan one is with the third order mode, it means, the second order TE mode. For w=350 nm the TE percentage of the fundamental TM mode is around 10% and the TE fraction of the TE mode is around 82%. Around x=5 µm again we can identify a change of regime in the overlap integral since the stabilization oscillations disappear. It is represented by the red line in Figure 109. Regarding the overlap integral we see that we excite around 82% the fundamental TM mode of the plasmonic modulator and only around 10% the second order TE mode.

When w=450 nm we have a similar behavior as with w=350 nm. Now, we also may have a beating between modes. These modes are the fundamental TM mode (represented in pink in Figure 109) and the second order TE mode (represented in yellow in Figure 109). Again, in Figure 106 we see that there are some oscillations due to the stabilization of the mode at the beginning of the modulator which disappear after a few µm. Some oscillations continue that may come from the beating between the fundamental TM mode and the second order TE mode of the plasmonic modulator. This will be analyzed later in this section. Oscillations will be seen in Figure 111.

As a conclusion of the analysis done in Figure 107 and Figure 109 we observe that although the optimized plasmonic modulator with w=150 nm has the lowest $T_{Total}$ (Figure 107) it is the one with the maximum overlap integral (Figure 109) meaning that w=150 nm is the configuration that best excites the fundamental TM mode that we want to use to modulate. For w=350 nm and w=450 nm they have a similar performance in the overlap integral and they are below w=150 nm and w=250 nm.

In the next section we use the overlap integral calculated in Figure 109 to calculate the coupling efficiency and the amount of light scattered.

## 6.4.2.2    Transmission into the Plasmonic Mode

Now, we calculate the percentage of power of the source that couples to the fundamental plasmonic TM mode $T_{In\ the\ mode}$ which is the one used to perform the modulation. This value is given by the multiplication of the total transmission $T_{Total}$ and the overlap integral OI (it was calculated in Figure 109). It means,

$$T_{In\ the\ mode}=P_{In\ the\ mode}/P_{Source}=T_{Total}xOI \qquad\qquad \text{Equation 63}$$

Where $T_{In\ the\ mode}$ is the percentage of power of the source that couples to the fundamental TM mode of the plasmonic modulator which is the one used to modulate (the coupling loss is given by 1-$T_{In\ the\ mode}$). $P_{In\ the\ mode}$ is the power present in the fundamental TM mode of the plasmonic modulator and $P_{Source}$ is the power injected by the source into the Si rib waveguide that later excites the plasmonic modulator. So, to calculate $T_{In\ the\ mode}$ we need to multiply the data given in Figure 107 by the data given in Figure 109. $T_{In\ the\ mode}$ is represented in Figure 110,

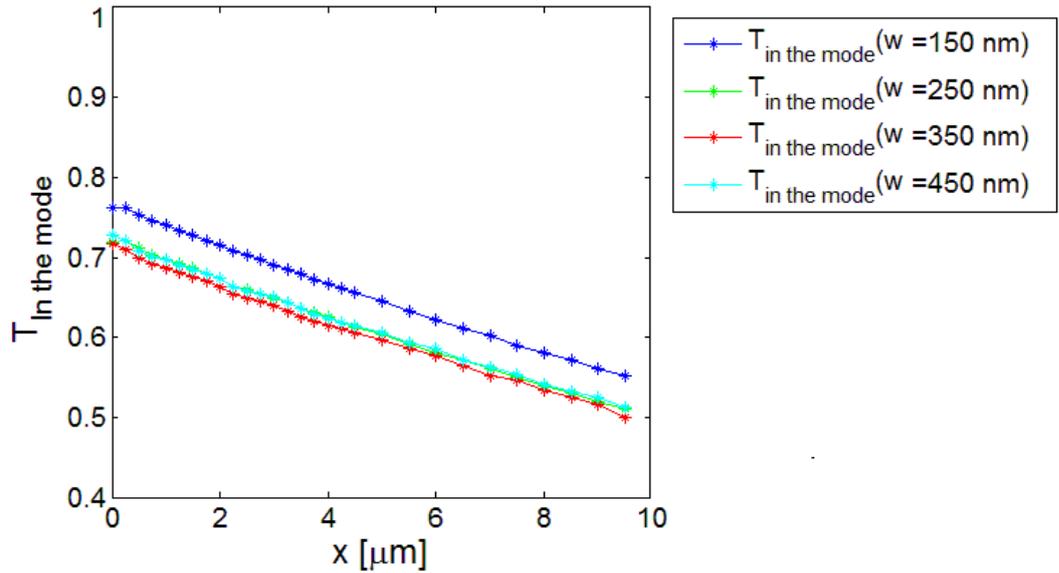

**Figure 110: T$_{\text{in the mode}}$ inside the plasmonic modulator in the first 10 μm**

In Figure 110 the best option to couple to the fundamental TM mode of the plasmonic modulator is when the width of the Ge core is around w=150 nm. This parameter is the optimum for coupling and coincides with the value of the optimized plasmonic modulator in which we maximized the FoM Δα$_{\text{eff}}$/α$_{\text{eff}}$. Now, we can say that for w=150 nm then at least 76% of the power of the source in the Si rib waveguide couples to the fundamental TM mode of the plasmonic modulator which is the mode we want to modulate. Furthermore, in Figure 110 we see how the transmission coupled to the mode decreases, this is due to the propagation losses of the plasmonic mode.

# 6.4.2.3     Coupling to Other Modes

For the case of w=350 nm and 450 nm, part of the power of the source is coupled to other modes. They are the second order TE mode that appears for w>350 nm. The transmission to other modes T$_{\text{Other modes}}$ is given by,

T$_{\text{Other modes}}$=P$_{\text{Other modes}}$/P$_{\text{Source}}$=T$_{\text{Total}}$xOI$_{\text{Other modes}}$          **Equation 64**

Where T$_{\text{Other modes}}$ is the percentage of power of the source in the Si rib waveguide that couples to the second order TE mode of the plasmonic modulators for the cases in which w=350 nm and w=450 nm. P$_{\text{Other modes}}$ is the amount of power present in other modes which are not the TM fundamental one. OI$_{\text{Other modes}}$ is the overlap integral between the excited fields of the other modes presented in the optimized plasmonic modulator. The parameter T$_{\text{Other modes}}$ is represented in Figure 111,

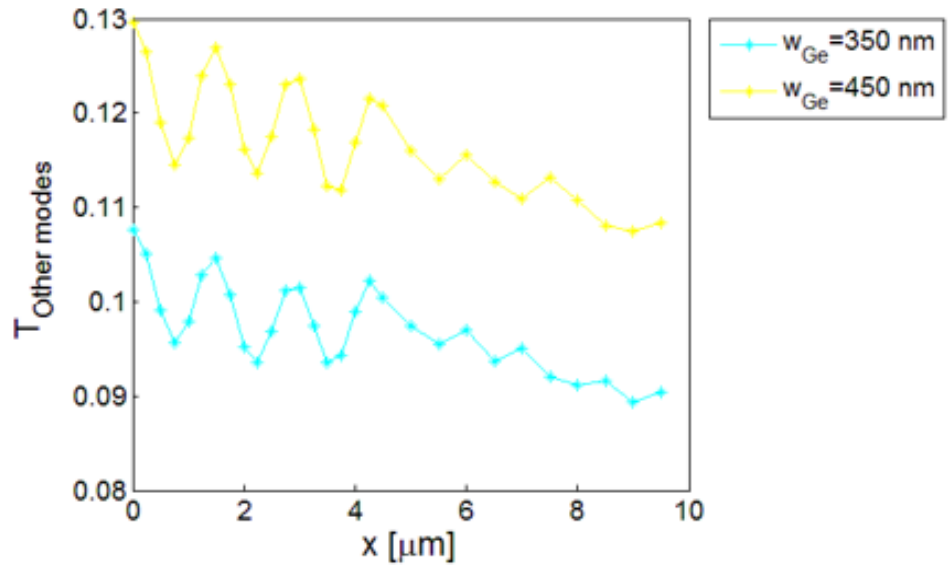

**Figure 111: T$_{Other\ modes}$ inside the plasmonic modulator in the first 10 μm**

In Figure 111 the coupling to the second TE mode of the plasmonic modulator is larger for w=450 nm, in agreement with Figure 109. The oscillations in the first 4-6 μm are due to the stabilization of the mode and also may have beating between modes (the second order TE mode and the fundamental TM of the plasmonic modulator). These oscillations are also present in Figure 105 (w=350 nm) and Figure 106 (w=450 nm).

## 6.4.2.4    Scattering

Now that we know how much power is coupled to the fundamental TM mode that we want to modulate and to the other modes (the second order TE mode for w=350 nm and 450 nm) we can calculate how much light is scattered due to the interface between the Si rib waveguide and the plasmonic modulator. For this, we define the parameter T$_{Scattering}$ which defines how much energy of the power source scatters. T$_{Scattering}$ is defined by,

$\quad$ T$_{Scattering}$=P$_{Scattered}$/P$_{Source}$=T$_{Total}$−(T$_{In\ the\ mode}$+ T$_{Other\ modes}$) $\qquad\qquad$ Equation 65

Where P$_{Scattered}$ is the power that scatters in the structure. We represent the parameter T$_{Scattering}$ in Figure 112. In Equation 65 the reflection is also taken into account since it is already substracted from T$_{Total}$.

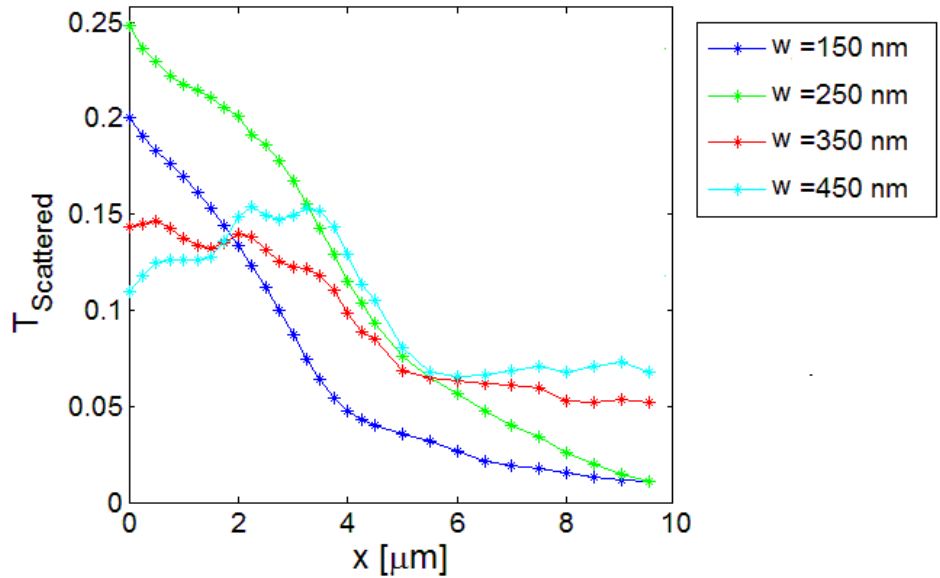

**Figure 112: T<sub>Scattered</sub> inside the plasmonic modulator in the first 10 µm**

From Figure 112 the largest scattering occurs for w=250 nm. For both w=150 nm and w=250 nm the maximum of the scattering occurs at the interface between the Si rib waveguide and the plasmonic modulator (x=0 µm) In Figure 122 we calculate the scattering from the interface between the Si rib waveguide and the optimized modulator (x=0 µm) until 10 µm inside the optimized modulator. The scattering is reduced as the light propagates through the modulator as the scattered light is progressibly absorved at the boundaries. For w=350 nm and w=450 nm, in the first 4 µm the scattering is maximum and almost constant, after 5 µm it starts to decrease. It is difficult to predict why the scattering is bigger for w=250 nm rather than for w=150 nm. The same happen with the evolution when w=350 nm and w=450 nm. In the case of w=350 nm and w=450 nm there are also beating between modes.

Until now we have analyzed the influence of the parameter w (width of the Ge core of the plasmonic modulator) on the coupling efficiency. We concluded that the best dimension is w=150 nm which coincides with the width for the optimized plasmonic modulator obtained in chapter five. Now, it will be interesting to study the influence of the Si rib waveguide w$_{Si}$. Consequently, in the future study we will consider w=150 nm and we call it the optimized plasmonic modulator.

# 6.4.3    Optimization of the Input Si Rib Waveguide

To study the influence of the width of the Si rib waveguide w$_{Si}$, we first calculate the parameter T$_{Total}$ in the plasmonic modulator for different widths w$_{Si}$. As was explained before we need a TM mode in the Si rib waveguide to excite the fundamental TM mode of the plasmonic modulator. We selected w$_{Si}$=250, 350, 450, 625 and 800 nm to study the influence. T$_{Total}$ is represented in Figure 113.

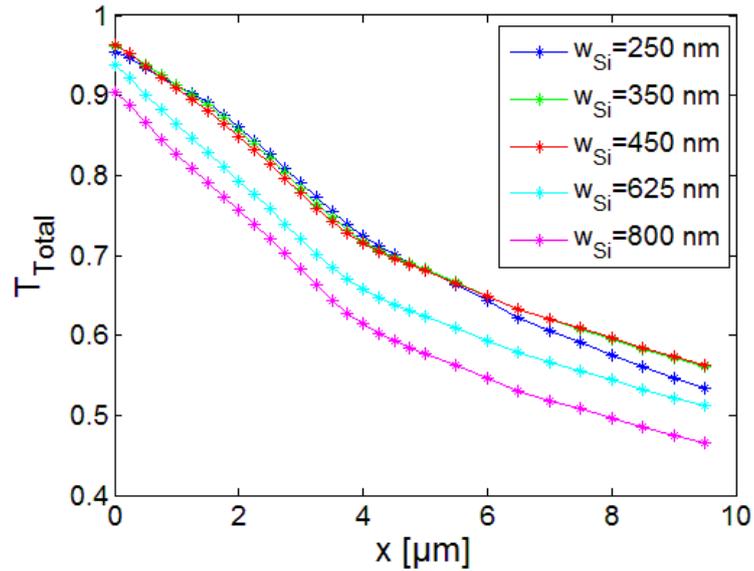

**Figure 113: T$_{Total}$ as a function of the propagation length x for different values of w$_{Si}$**

From Figure 113 it is possible to observe that the parameter T$_{Total}$ is larger for a narrow Si rib waveguide presenting the maximum for w$_{Si}$=250, 350 and 450 nm. The lowest T$_{Total}$ is for w$_{Si}$=800 nm. It is possible to observe that the parameter T$_{Total}$ decreases as the light propagates in the x direction due to the fact that some light is lost by scattering and also due the propagation losses of the fundamental TM mode. From T$_{Total}$ we cannot say which one is the best configuration so it is needed to calculate the overlap integral and T$_{In the mode}$.

Now, it is interesting to calculate the overlap integral between the optical electromagnetic field excited from the Si rib waveguide in the plasmonic modulator (using a 3D FDTD simulator) and the optical mode that the optimized plasmonic modulator supports (using a 2D mode solver). This overlap integral is presented in Figure 114,

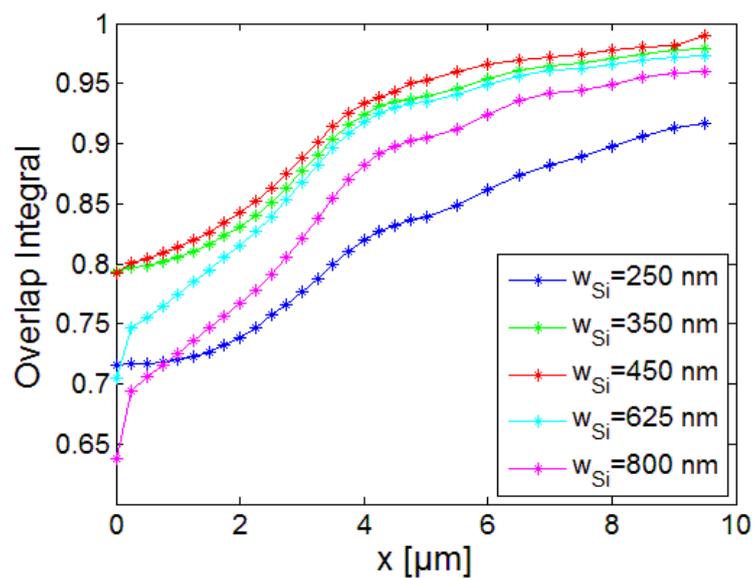

**Figure 114: Overlap integral between the modes supported by the plasmonic modulator and the mode excited by the Si rib waveguide**

In Figure 114 it is possible to observe that the width of the Si rib waveguide $w_{Si}$ which most excites the fundamental TM mode in the plasmonic modulator is $w_{Si}$=450 nm. It is also possible to observe that for the three cases the overlap integral is increased for increasing values of x. This is due to the fact that the mode is settled down into the plasmonic modulator. From Figure 113 and Figure 114 we can conclude that the best width to excite the fundamental TM mode of the plasmonic modulator is $w_{Si}$=450 nm since both the $T_{Total}$ and the overlap integral are the best ones. We use Equation 63 to calculate $T_{in\ the\ mode}$ for the three cases of $w_{Si}$. It is represented inFigure 115,

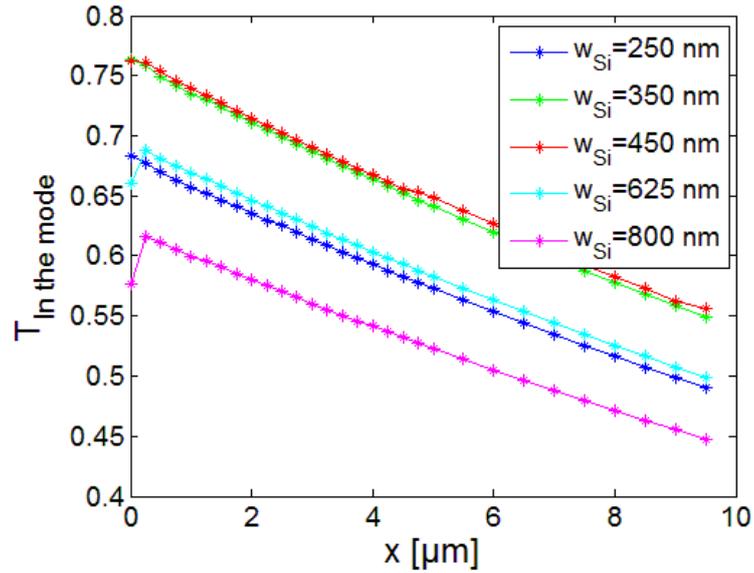

Figure 115: $T_{in\ the\ mode}$ **inside the plasmonic modulator in the first 10 µm**

With Figure 115 we confirm that the best parameter is $w_{Si}$=450 nm. With this 76% of the source in the Si rib waveguide couples to the fundamental TM mode of the optimized plasmonic modulator that we want to modulate. This is due to both more overall transmission (Figure 113) and higher overlap integral (Figure 114) for the case of $w_{Si}$=450 nm. On the other hand, the effective index matching is not better between the two guide's modes. It means $n_{eff}$ in the Si rib waveguide with $w_{Si}$=450 nm is $n_{eff}$=2.4 while the fundamental TM mode of the plasmonic modulator is $n_{eff}$=2.89. For the case of $w_{Si}$=800 nm the effective refractive index $n_{eff}$=2.65. Nevertheless the profile mismatch of the mode also plays a role.

## 6.4.4    Conclusion

As a conclusion of this section a butt-coupling structure of Figure 102 was analysed to maximize the transmission into the TM mode of the plasmonic modulator from a Si rib waveguide. The optimized parameter for the Si rib waveguide are $w_{Si}$=450 nm, $h_{Si}$=255 nm and $h_{Slab}$=100 nm. With this waveguide we excite the optimized plasmonic modulator that we designed in chapter five. The maximum coupling efficiency achieved with the butt-coupling configuration is around $T_{in\ the\ mode}$=76%. Additionally the reflection R=4% and the scattering transmission is around $T_{Scattering}$=20%. In the case of the optimized plasmonic modulator with w=150 nm there is not coupling to other modes. In the following sections we will do some engineering proposing several structures to increase this value.

In the following Table 25 we summarize the main parameters in the best case of the butt-coupling scheme. We remind that the parameters of the optimized plasmonic modulator are: w=150 nm, h=250 nm, $h_{Slot}$=5 nm, $h_{Buf}$=60 nm and $h_{Bot}$=40 nm. On the other hand, the parameters of the input Si rib waveguide are: $w_{Si}$=450 nm, $h_{Si}$=255 nm and $h_{Slab}$=100 nm.

| Parameter: | R | $T_{Total}$ | $T_{In\ the\ mode}$ | $T_{Scattered}$ | $OI_{Fundamlental\ mode}$ | $OI_{Other\ modes}$ |
|---|---|---|---|---|---|---|
| Value [%]: | 4 | 96 | 76 | 20 | 66 | 0 |

**Table 25: Summary table of the main parameter of the butt-coupling structure**

In Table 25 R stands for the reflection, $OI_{Fundamental\ mode}$ is the overlap integral between the excited mode from the Si rib waveguide and the fundamental TM mode that we want to excite in the optimized modulator. Furthermore, $OI_{Other\ modes}$ is the overlap integral between the mode excited in the optimized modulator from the Si rib waveguide and the other TM modes supported by the optimized plasmonic modulator.The parameters of Table 25 are summarized in Figure 116.

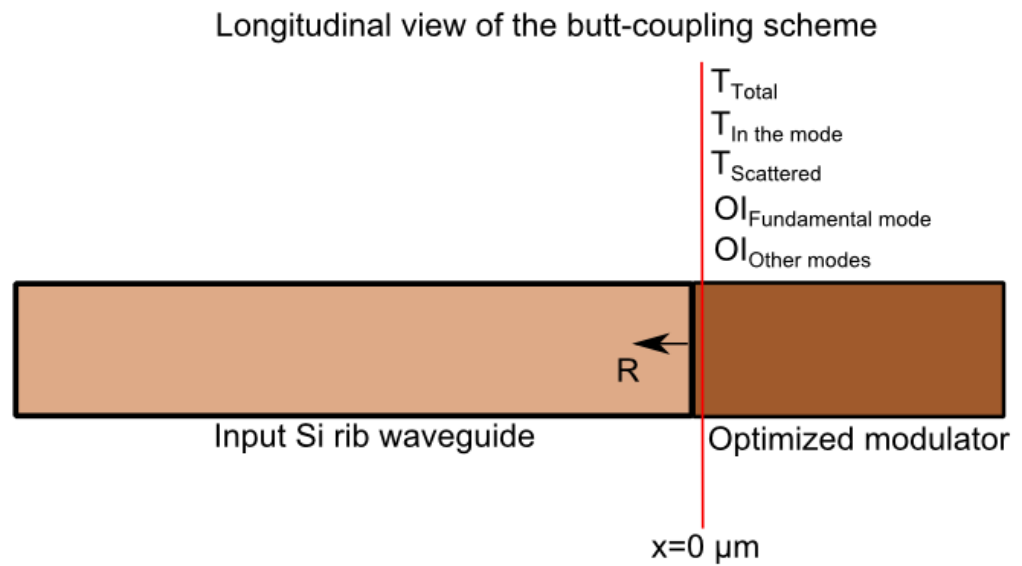

**Figure 116: Summary table of the main characteristics of the Ge taper summarized in Table 25**

## 6.5     Germanium Taper

In this section we will add a Ge taper between the Si rib waveguide and the optimized plasmonic modulator. There are several main factors that prevent a good coupling between the structures: the first is due to a different field distribution in the cross section of the waveguides that we want to couple. It is due to a profile mismatch of the modes in the cross section of the waveguides also leading to light diffraction.  The second is the reflection that we have at the interface between the two waveguides. This is mainly due to a different effective index $n_{eff}$ in both waveguides. It means, due to the difference of the effective indices of the modes in the waveguides there is reflection.

In this section we will minimize the reflection between the waveguides. Although the reflection in the butt-coupling scheme was only 4% we will see how the scattering, the profile missmatch and the overlap integral behave. This may lead to an improvement of the coupling. For this, we will insert between the Si rib waveguide and the optimized plasmonic modulator a taper that matches both effective indexes of the Si rib waveguide and the optimized modulator. The structure that we want to use is presented in Figure 117,

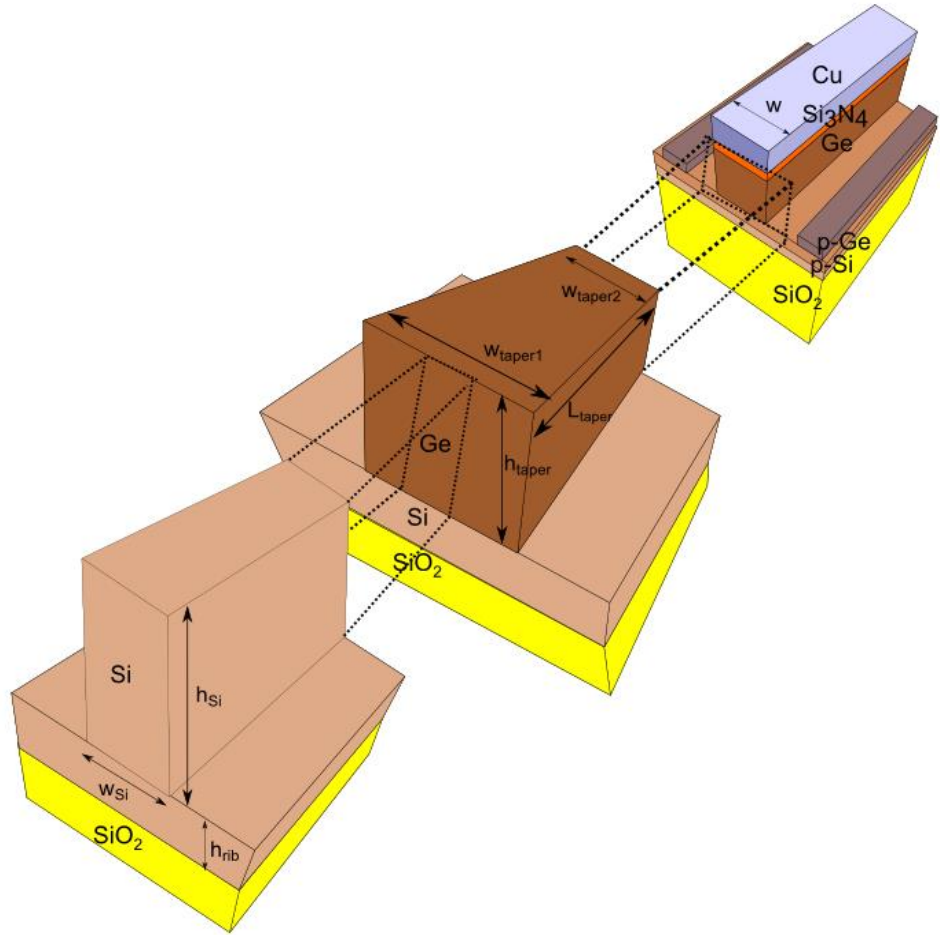

**Figure 117: Ge taper approach by inserting an intermediate Ge rib taper to match the effective refractive index of the Si rib waveguide and the optimized plasmonic modulator**

Now, we want to gradually vary the refractive index of the modes between the Si rib waveguide and the optimized plasmonic modulator. The effective index of the fundamental TM plasmon in the optimized plasmonic modulator is around $n_{eff}=2.89$ we recall this parameter $n_{Eff,waveguide}$ in order to present it in Figure 119. The objective is to achieve phase matching at the interface modulator/waveguide and to evaluate the mode mismatch, scattering and reflection.

# 6.5.1    Study of Several Intermediate Tapers

To match such effective indices we want to place a taper between them. We analyzed several tapers for this purpose. They are summarized in Figure 118. The first one is to introduce a Si rib taper between them. The second one is a Ge rib taper. These two tapers are photonic like. Since we want to excite a plasmonic mode we also though about a plasmonic taper. In this case the third taper in Figure 118 is a metal taper over Si and the last case is a metal taper over Ge.

To know which case is the better we want to plot the parameter $n_{Eff, waveguide} - n_{Eff, mode\ to\ couple}$. Where $n_{Eff, mode\ to\ couple}$ is the effective index of the mode in the respective tapers of Figure 118. If we want to match the effective indices of both waveguides the parameter $n_{Eff, waveguide} - n_{Eff, mode\ to\ couple}$ has to be close to zero. It means that the effective index at the end of the taper is the same as the mode in the optimized plasmonic modulator. This parameter is represented in Figure 119 for all the tapers of Figure 118.

In the case of the Si rib waveguide we vary the parameter $w_{Si}$ to know the evolution of $n_{Eff, waveguide} - n_{Eff, mode\ to\ couple}$. In the same way in the Ge rib waveguide we vary $w_{Ge}$. Finally, in the metal taper over Si and Ge we vary $w_{Cu}$. The rest of the parameters are fixed and follow the same dimensions as the optimized plasmonic modulator for the reasons explained before. We summarize them: $h_{Si}$=255 nm, $h_{rib-Si}$=100 nm, $h_{Ge}$=255 nm, $h_{rib-Ge}$=100 nm, w=150 nm, h=250 nm and $h_{Slot}$=5 nm.

In Figure 119 we represent the evolution of these parameters ($w_{Si}$, $w_{Ge}$ and $w_{Cu}$) versus $n_{Eff, waveguide} - n_{Eff, mode\ to\ couple}$.

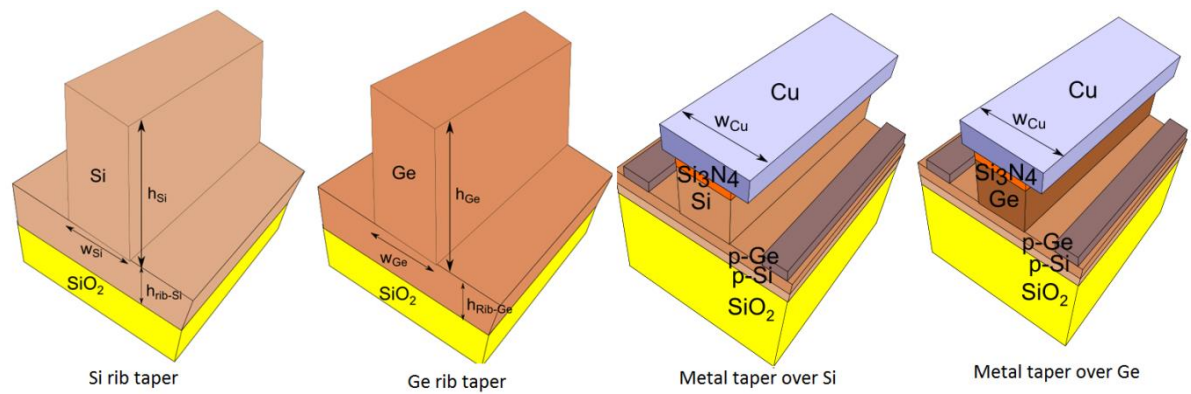

**Figure 118: Summary of the waveguides that are study to match the effective index between the standard Si rib waveguide and the optimized plasmonic modulator**

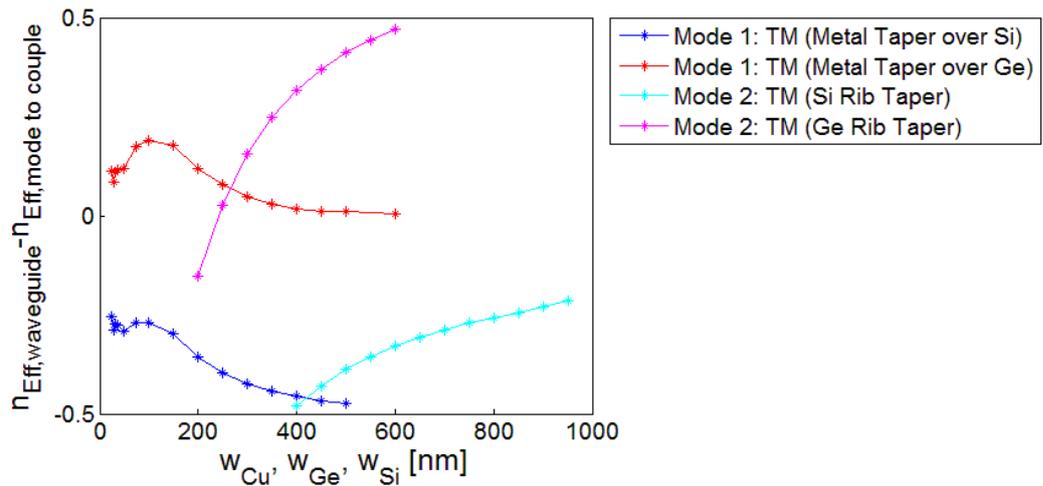

**Figure 119: All tapers comparison for a metal taper over Si, metal taper over Ge, Si rib waveguide taper and Ge rib waveguide taper**

From Figure 119 we observe that the only taper that crosses through zero is the Ge rib taper (pink). Consequently, we will center in the study of this waveguide. This assures a matching between the intermediate taper and the plasmonic modulator. The interface between the input Si rib waveguide and the intermediate taper can also be matched when the pink line crosses the light blue one. We will select the proper values later in this section.

It is interesting to mention that the metal taper over Ge also approaches zero when the parameter $w_{Cu}$ is big. This is due to the fact that the optimized plasmonic modulator structure is reached when $w_{Cu}$ tends to infinity. Furthermore, the two plasmonic tapers (red and blue in Figure 119) have the same evolution versus $w_{Cu}$. This is because they have the same structure but only changing Si by Ge. Since Ge has a refractive index bigger than Si then the red curve (Ge in the core) is over the blue one (Si in the core).

## 6.5.2    Study of the Ge Rib Intermediate Taper

Now we will analyze the intermediate Ge rib waveguide that we want to use as a taper between the Si rib waveguide and the optimized plasmonic modulator. A representation of this waveguide is illustrated in Figure 120,

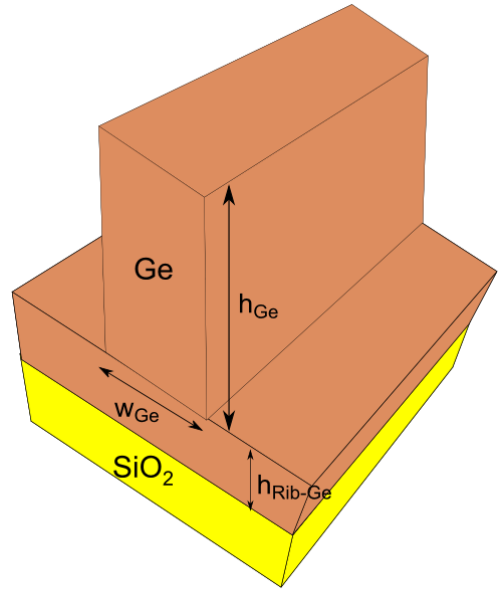

**Figure 120: Germanium rib waveguide and its parameters**

We will select $h_{Ge}$=255 nm and $h_{Rib-Ge}$=100 nm for the same reasons we chose $h_{Si}$ and $h_{Rib-Si}$ in the Si rib waveguide. It means, to align the top of the Ge rib waveguide with the top of the $Si_3N_4$ layer of the optimized plasmonic modulator. Regarding the parameter $h_{Rib-Ge}$ we selected it to be 100 nm so it is equal to the sum of $h_{Bot}$=40 nm and $h_{Buf}$=60 nm of the optimized plasmonic modulator of Figure 84. Setting $h_{Ge}$ and $h_{Rib-Ge}$ we will vary the parameter $w_{Ge}$ to know the variation of $n_{eff}$ in the Ge rib waveguide. It is presented in Figure 121. In Figure 122 present the effective propagation losses $\alpha_{eff}$.

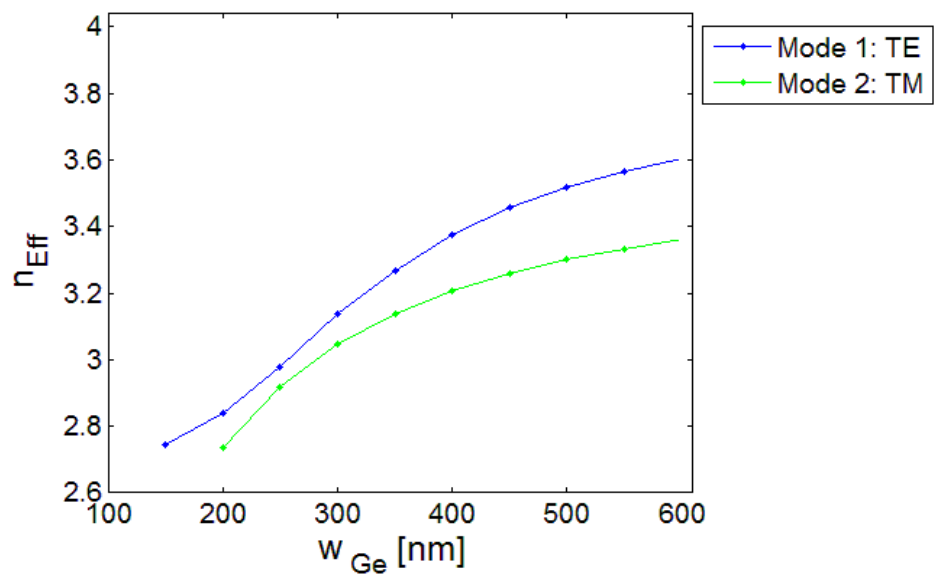

**Figure 121: $n_{eff}$ as a function of w for the Ge rib waveguide. It is difficult to calculate the effective refractive index below 150 nm due to the deconfinement of the mode**

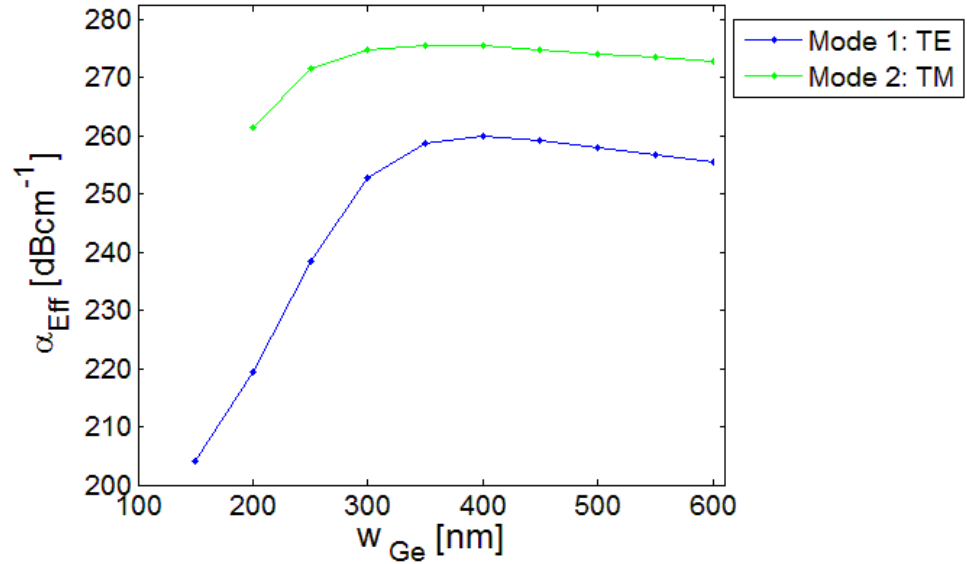

**Figure 122: α$_{eff}$ as a function of w for the Ge rib waveguide. It is difficult to calculate the effective absorption coefficient below 150 nm due to the deconfinement of the mode**

Since the mode that we need to excite in the optimized plasmonic modulator is the fundamental TM mode, we need to have a TM mode in the Ge rib taper that we insert between the modulator and the Si rib waveguide. In Figure 121 the TM mode is represented in green. We see that the effective refractive index n$_{eff}$ varies from 2.74 when w$_{Ge}$=200 nm to 3.36 when w$_{Ge}$=800 nm. As explained before, we need to match at the input of the taper the effective index of the Si rib waveguide which is around n$_{eff}$=2.65. For this, we selected w$_{taper1}$=175 nm (Figure 117) where n$_{eff}$ is around 2.6. At the end of the taper we need to match the effective refractive index of the optimized plasmonic modulator which is around 2.89, for this we select w$_{taper2}$=250 nm (Figure 117). Like this, the intermediate waveguide tries to gradually vary the effective index of the mode. It means, there is no mismatch in n$_{eff}$.

Regarding the propagation losses α$_{eff}$ of the mode in the Ge taper we remind that Ge absorbs light at 1.647 μm (the operational wavelength of the optimized plasmonic modulator). The propagation losses of the Ge rib waveguide are presented in Figure 122. From this it is possible to observe that the losses of the TM mode are over the propagation losses of the TE mode. Due to the losses of the TM mode we expect a trade-off between the length of the intermediate Ge rib taper and the coupling efficiency. If the taper is too long the losses may decrease the transmission into the fundamental TM mode of the optimized plasmonic modulator but if it is too short the effective index will not vary smoothly enough to achieve adiabatic transition of the mode.

# 6.5.2.1    Optimization of the Length of the Taper

Once the widths of the taper are chosen $w_{taper1}$=175 nm and $w_{taper2}$=250 nm. We need to optimize the length for the taper $L_{taper}$. For this, we scan the length $L_{taper}$ between 100 nm and 4 μm to find the maximum of the transmission into the plasmonic mode $T_{In\ the\ mode}$ given by **Equation 63**. It is represented in Figure 123,

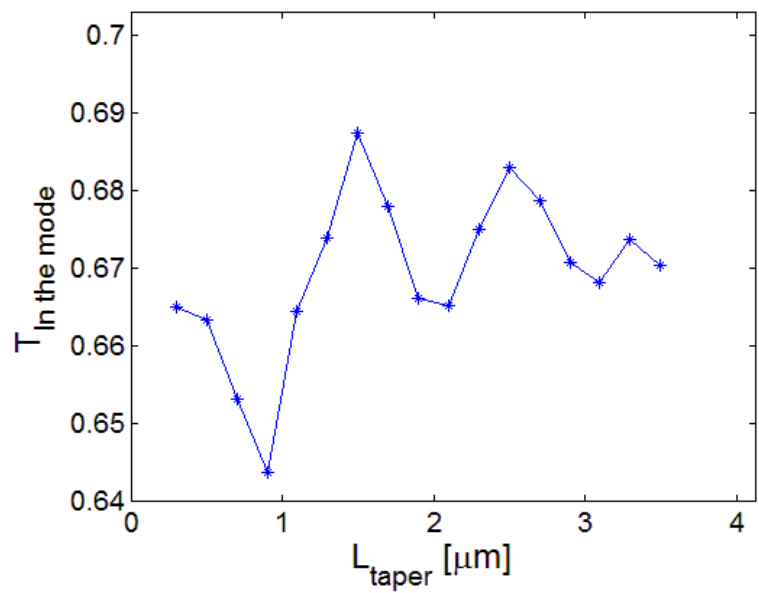

**Figure 123: Scanning the parameter $L_{taper}$ to achieve the maximum of transmission into the fundamental TM mode of the optimized plasmonic modulator $T_{In\ the\ mode}$**

In Figure 123 we observe that the maximum transmission into the fundamental plasmonic TM mode $T_{In\ the\ mode}$ is achieved for $L_{taper}$=1.5 μm. It is around $T_{In\ the\ mode}$=0.69 for $L_{taper}$ equal to 0 μm and 4 μm the transmission is significantly smaller.

From Figure 123 it is possible to see that using butt-coupling we obtained $T_{In\ the\ mode}$=0.75 which is over $T_{In\ the\ mode}$=0.69 obtaining with a taper. Consequently, although we optimized the effective refractive index to reduce the reflection we did not increase the $T_{In\ the\ mode}$ parameter. We remind that already in the butt-coupling structure we observed a small reflection from the interface of the Si rib waveguide and the optimized plasmonic modulator. It is around 4%. Now, we analyze why this configuration does not offer a better performance. We will settle the parameter $L_{taper}$ to $L_{taper}$=1.5 μm since the maximum performance was obtained with this value.

In Figure 123 we see that there is oscillations in the parameter $T_{In\ the\ mode}$ which comes from a Fabry-Perot cavity formed by the Ge taper between the Si rib waveguide and the optimized plasmonic modulator. This is due to the fact that small reflections still remind at the interfaces, we discuss it in further details at the end of this section when we calculate the reflections from the interfaces. The length of the Fabry-Perot cavity is $L_{taper}$.

## 6.5.2.2     Overal Transmission

First, we plot the parameter $T_{Total}$ as a function of the distance traveled by light inside the optimized plasmonic modulator x. We did a similar plot with the butt-coupling configuration done in Figure 107. Now, for the optimized taper it is represented in Figure 124. We remind that the definition of $T_{Total}$ is given in Equation 62. In Figure 124 we simulated the whole structure of Figure 117 nevertheles we present only the $T_{Total}$ inside the modulator. x=0 µm represents the interface between the intermediate taper and the plasmonic modulator. On the other hand, x=10 µm represents the first 10 µm inside the plasmonic modulator.

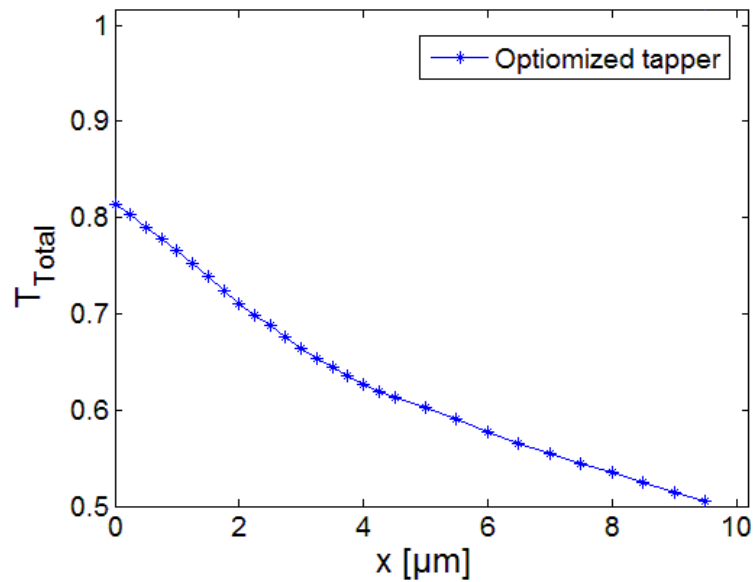

**Figure 124: $T_{Total}$ as a function of the propagation length x for the Ge taper**

If we compare the parameter $T_{Total}$ in Figure 124 for the optimized coupler and in Figure 107 for the butt-coupling structure we observe that the total transmission into the optimized plasmonic modulator is larger in the butt-coupling configuration. Now, it will be interesting to discriminate which portion of the total transmission goes to the fundamental TM mode of the optimized plasmonic modulator, the scattering and the reflection.

## 6.5.2.3     Coupling into the Plasmonic Mode

We start by calculating the parameter $T_{In\ the\ mode}$ inside the optimized plasmonic modulator. It is represented in Figure 125. We remind that the definition is given by Equation 63.

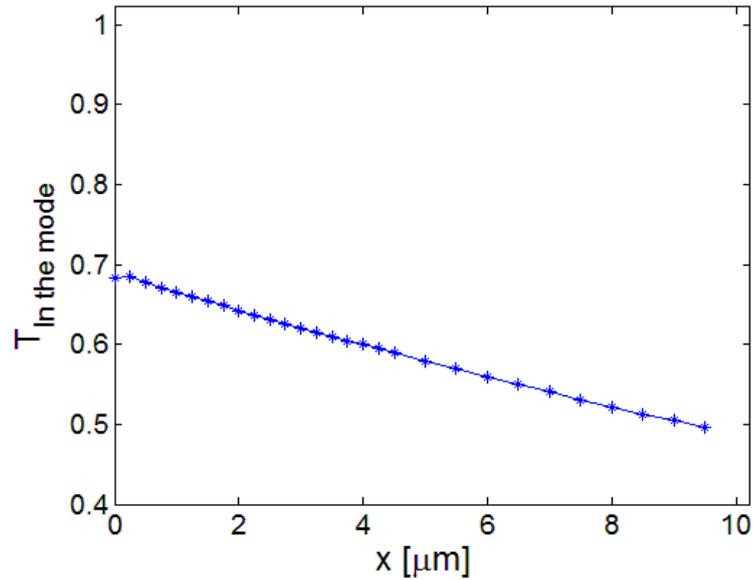

**Figure 125: T$_{In the mode}$ inside the plasmonic modulator in the first 10 μm**

According to Figure 125 the parameter T$_{In the mode}$ is worse with the Ge taper with respect to the butt-coupling structure. It starts from T$_{In mode}$=0.69 and decreases due to the scattering and the propagation losses of the plasmon. This confirms what we obtained in Figure 123. In the following study we will determine which one is the cause of obtaining a T$_{In the mode}$ less than using the butt-coupling structure.

# 6.5.2.4    Overlap Integral

The next step is to calculate the ovelap integral between the electric field distribution excited in the optimized plasmonic modulator and the fundamental TM mode of the optimized plasmonic modulator. In is given in Figure 126,

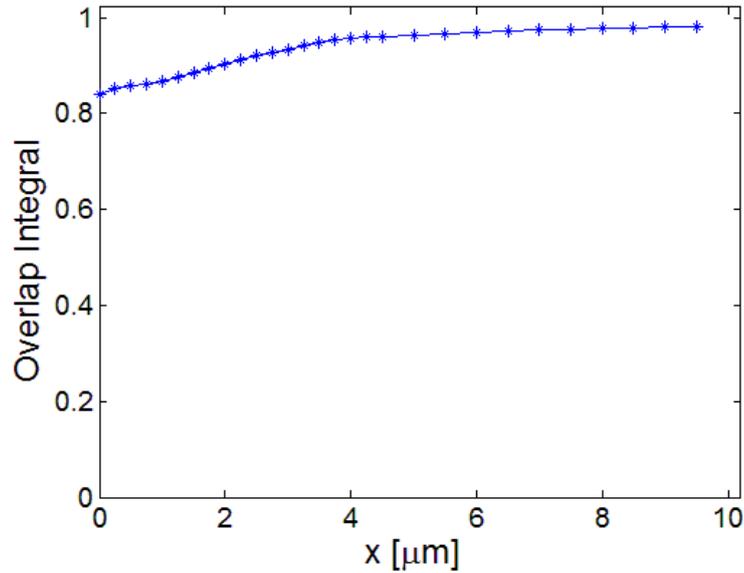

**Figure 126: Overlap integral between coupled electromagnetic field in the modulator and the mode excited by the Ge taper**

Comparing the overlap integral of the Ge taper in Figure 126 and the overlap integral in the butt-coupling structure represented in Figure 109 we observed that the overlap integral is slightly larger in the Ge taper rather than in the butt-coupling structure. This means that with the Ge taper the excited field inside the optimized modulator is slightly more similar to the fundamental TM mode rather than in butt-coupling.

# 6.5.2.5      Scattering, Reflection and Absorption

Now, it will be interesting to calculate the scattering present into the optimized plasmonic modulator. For this we use Equation 65.

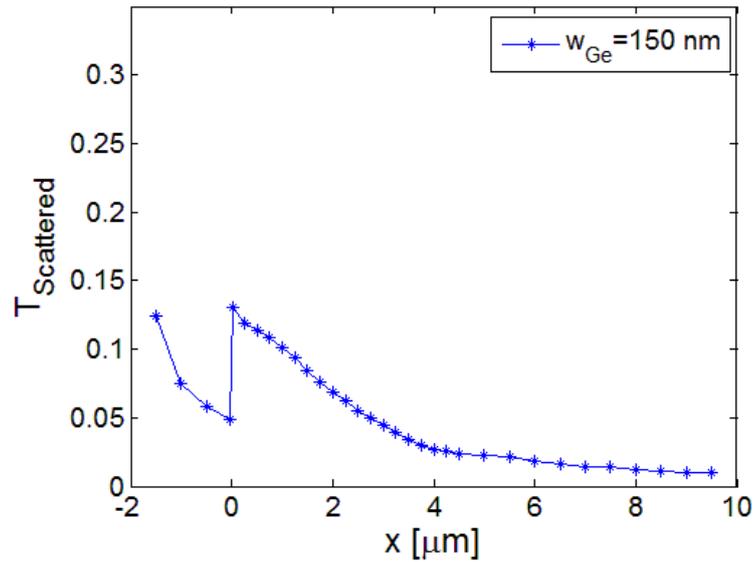

**Figure 127: T$_{Scattered}$ inside the Ge taper (-1.5 μm<x<0 μm) and the optimized plasmonic modulator in the first 10 μm (0 μm<x<10 μm)**

In Figure 127 we represent the scattering T$_{Scattered}$ in the Ge taper, it is presented versus the propagation distance x [μm]. The Ge taper is presented between -1.5 μm<x<0 μm (since L$_{taper}$=1.5 μm in the best case). The T$_{Scattered}$ is presented inside the first 10 μm of the optimized plasmonic modulator for 0 μm<x<10 μm. The two maximums of the function T$_{Scattered}$ are due to scattering in the first interface between the Si rib waveguide and the Ge taper and the second one is between the Ge taper and the optimized plasmonic modulator.

Comparing Figure 127 with Figure 112 we observed that with the Ge taper we obtained less scattering in each interface than using the butt-coupling structure. The maximum of the butt-coupling was T$_{Scattered, 2}$=0.2 for w=150 nm (the best case). On the other hand, the maximum of the Ge taper is around T$_{Scattered,1}$=0.12 in the first interface and T$_{Scattered,2}$=0.13 in the second interface (between the Ge taper and the optimized plasmonic modulator). It means, we named T$_{Scattered, 1}$ the scattering in the first interface and T$_{Scattered, 2}$ the maximum scattering in the second interface. Nevertheless, in comparison with the butt-coupling structure there is now scattering in both interfaces.

We also studied the reflection at both interfaces of the Ge taper. The reflection in the first interface is R$_{First}$=6.5% and in the second interface it is around R$_{Second}$=4.5%. Note that the second interface is a similar structure than the butt-coupling one, except that the butt-coupling interface is a Si rib waveguide with the optimized plasmonic modulator and the second interface of the Ge taper is between a Ge rib waveguide and the optimized plasmonic modulator. It is important to note that it is difficult to fully match the effective indices of the different waveguides by finding the proper dimensions of the waveguides. It is because of this small size mismatch in dimensions.

Furthermore, the light absorption by the Ge in the taper also plays a role in the coupling efficiency. For this we measured the total transmission T$_{Total}$ with loss and not loss in the material Ge due to the Ge absorbing. The transmission with loss T$_{Total, absorving}$=0.879821 while the Ge without loss gives a transmission of T$_{Total, non-absorving}$=0.887921. The amount of light of the source absorved by Ge is T$_{Total, non-absorving}$- T$_{Total, absorving}$=0.007. It means that 0.7% of light is absorved by the Ge in the Ge taper. Another factor that may influence the coupling efficiency is the amount of scattered light, which in this case is less than in the butt-coupling configuration.

# 6.5.3    Conclusion

As we stated before, there are several factors that influence the good coupling from one waveguide to the other one. One is the difference in the effective refractive index $n_{eff}$ that leads to reflection in the interfaces. Other is the mode profile mismatch. It means the mismatch between the field distribution in the cross section at the end of one waveguide and at the beginning of the other (it is represented in Figure 128). There are others factors like scattering or absorption by the materials also.

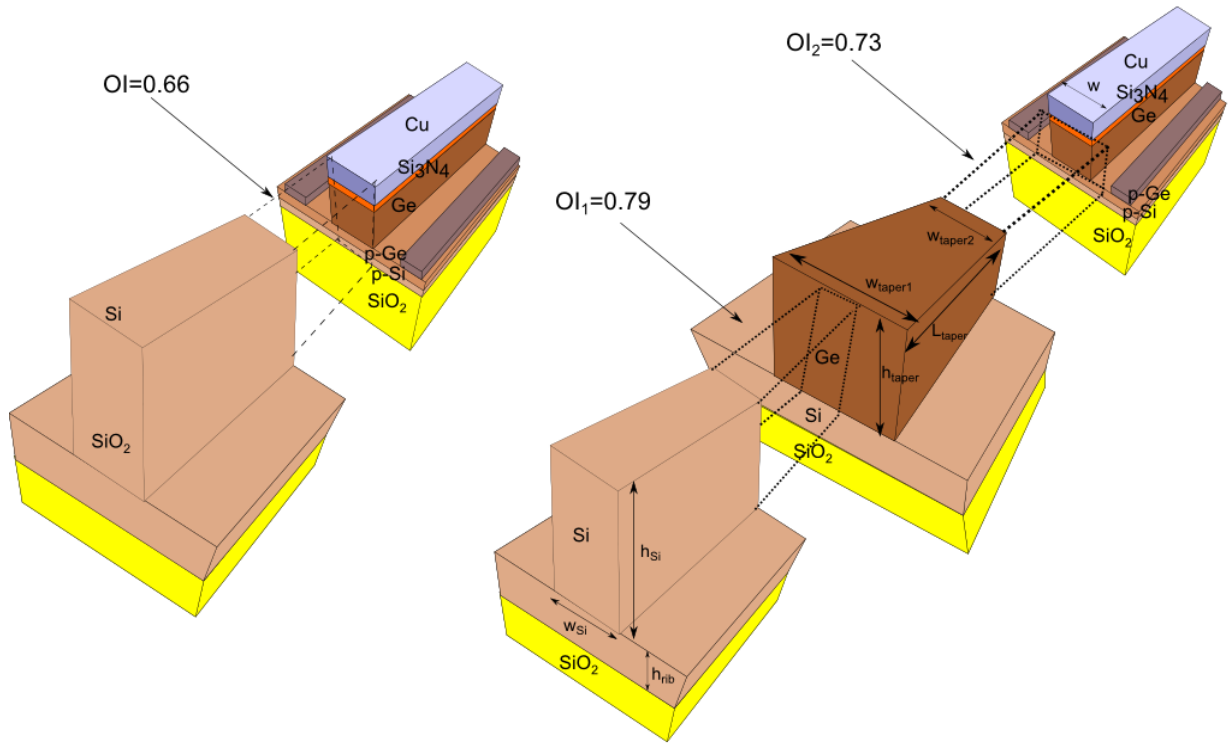

**Figure 128: Overlap integral to measure the difference in the field profile between the two adjacent waveguides for the butt-coupling structure (left) and the Ge taper (right)**

According to Figure 128 the overlap integral between the photonic mode of the input Si rib waveguide with the fundamental TM mode of the optimized plasmonic modulator is OI=0.66. In the case of the Ge taper, the first interface (between the Si rib waveguide and the Ge rib taper) leads to an overlap integral of $OI_1$=0.79. In the case of the second interface of the Ge taper (between the Ge rib taper and the optimized plasmonic modulator) it is around $OI_2$=0.73.

As a conclusion we observed that $T_{Total}$ and $T_{in\ the\ mode}$ are smaller in the Ge taper than in the butt-coupling structure. Consequently, the butt-coupling approach is better than the Ge taper. This is the main conclusion to say that the butt-coupling structure is better. There are several factors that influence the worser result of the Ge taper. The first is that you have more reflection in the total structure of the Ge taper ($R_{First}$=6.5% in the first interface of the Ge taper $R_{Second}$=4.5% in the second one). We obtain such a reflection because it is difficult to match the proper interface. In the butt-coupling structure R=4%. Additionally, the Ge taper has scattering light at both interfaces around 12% and 13% respectively. While in the butt-coupling it is around 20%. Finally, the Ge taper has material absorption which is around 0.72%.

In the following Table 26 we summarize the main parameters in the best case of the Ge taper scheme. In this Table 26 we used the optimized plasmonic modulator. The parameters of the Ge taper are: $w_{Si}$=800 nm, $h_{Si}$=255 nm and $h_{Slab}$=100 nm for the input Si rib waveguide. The parameters of the Ge taper are: $w_{taper1}$=175 nm, $w_{taper2}$=250 nm, $h_{taper}$=255 nm and $L_{taper}$=1.5 µm.

| Parameter: | $R_{First}$ | $R_{Second}$ | $T_{Total}$ | $T_{In the mode}$ | $T_{Scattered, 1}$ | $T_{Scattered, 2}$ | $OI_{Fundamental mode (first interface)}$ | $OI_{Fundamental mode (second interface)}$ | $OI_{Other modes}$ | Absorption by Ge |
|---|---|---|---|---|---|---|---|---|---|---|
| Value [%]: | 6.5 | 4.5 | 81 | 68 | 12 | 13 | 79 | 73 | 0 | 0.75 |

**Table 26: Summary table of the main parameter of the Ge taper structure structure**

In Table 26 we use two new parameters: $OI_{Fundamental mode (first interface)}$ and $OI_{Fundamental mode (second interface)}$. In the case of $OI_{Fundamental mode (first interface)}$ refers to the overlap integral calculated between the fundamental TM mode at the end of the input Si rib waveguide and the beginning of the Ge taper with width $w_{taper1}$=175 nm. On the other hand $OI_{Fundamental mode (second interface)}$ refers to the overlap integral between the mode at the end of the taper (the width is $w_{taper2}$=250 nm) and the mode at the beginning of the optimized plasmonic modulator. The parameters of Table 26 are summarized in Figure 129.

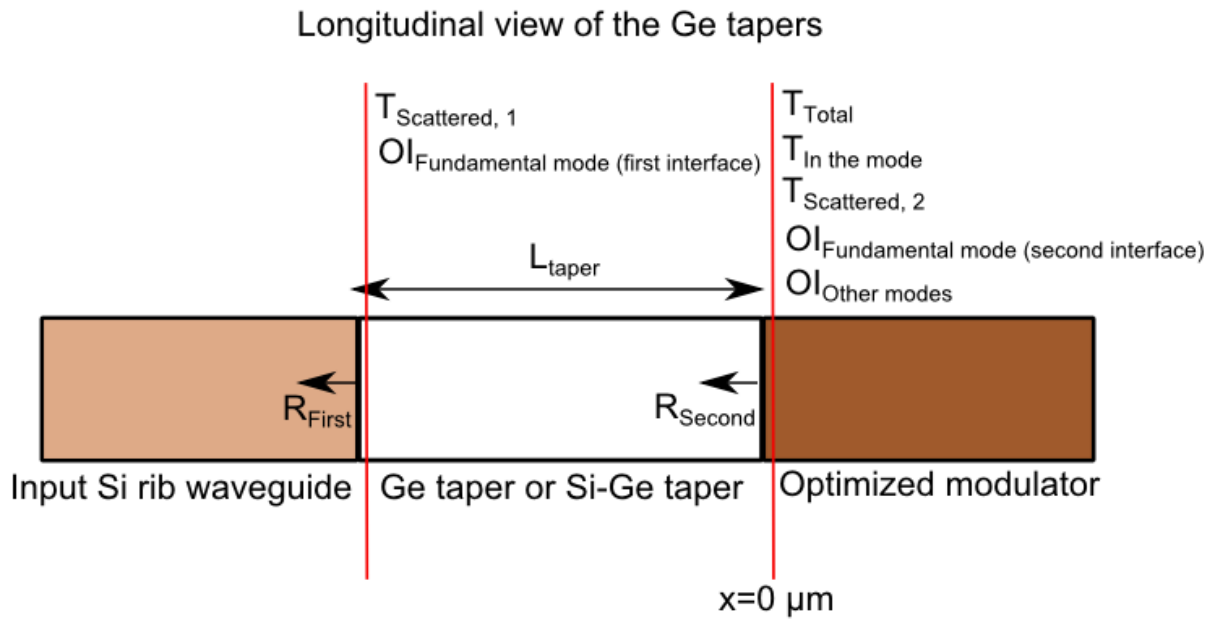

**Figure 129: Summary Figure of the main characteristics of the Ge taper summarized in Table 26**

In the next section, instead of trying to reduce the reflection by matching the effective indices, we will try to reduce the mode profile mismatch in the coupler design by using another kind Si-Ge taper.

# 6.6    Silicon-Germanium Taper

In the previous section we tried to reduce the reflection and scattering at the interface by matching the different effective indices between the different waveguides that appear in the Ge taper configuration. Furthermore, we studied the reflections, scattering and mode missmatch. Now we will try to optimize the matching in the profile of the modes in different waveguides. It means, we want to maximize the overlap integral between the modes that are going to be coupled. For this, we will represent the mode profiles of the optimized plasmonic modulator and of the Si rib waveguide. The profile of the optimized plasmonic modulator is represented in Figure 130. The mode profile of the Si rib waveguide is presented in Figure 131,

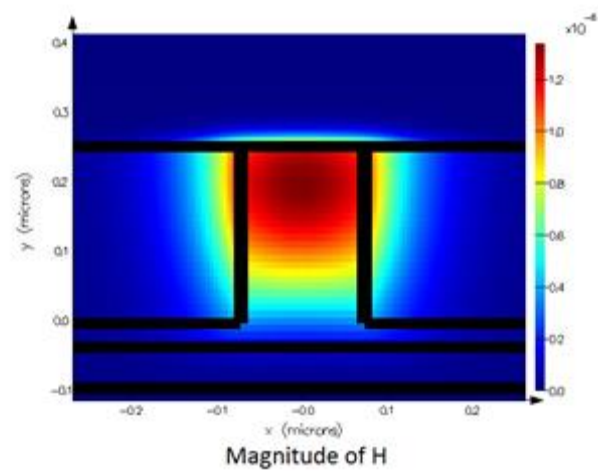

Magnitude of H

**Figure 130: Magnitude of E and H of the fundamental plasmonic TM mode of the optimized modulator**

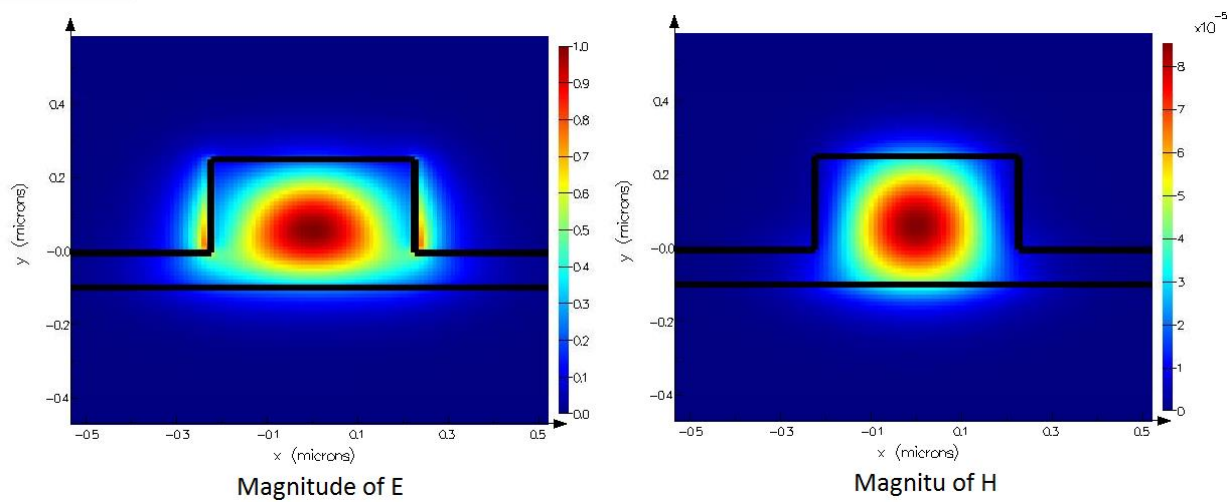

Magnitude of E          Magnitu of H

**Figure 131: Magnitude of E and H of the fundamental Si rib waveguide**

The modes of the Figure 130 and Figure 131 are TM mode. Consequently, the magnetic field seems to be more important that the electrical one. Regarding the field distribution of the MIS plasmonic waveguide (Figure 130) we see that the field is confined in the Si$_3$N$_4$ slot. It means, up in the waveguide.

Furthermore, in the lateral direction the field is more confined in the Ge. The plasmonic waveguide has a width of $w_{Ge}$=150 nm.

Regarding the field in the Si rib waveguide (Figure 131), we see that the magnitude of the magnetic field H is concentrated in the middle of the waveguide, it means, at a height of $h_{Si}$/2=125 nm. Furthermore, in the lateral direction it is present in all the width of Si which is around $w_{Si}$=450 nm.

As described in the two last paragraph in the optimized plasmonic modulator the field distribution in the horizontal direction is around w=150 nm while in the Si rib waveguide the field in the horizontal direction is distributed around $w_{Si}$=450 nm. In the vertical direction the spatial difference is smaller and it is around 50-70 nm. The fact that the main difference is in the lateral direction leads to a use of a horizontal taper. Consequently, we will study the use of a new taper in the form of the one showed in Figure 132,

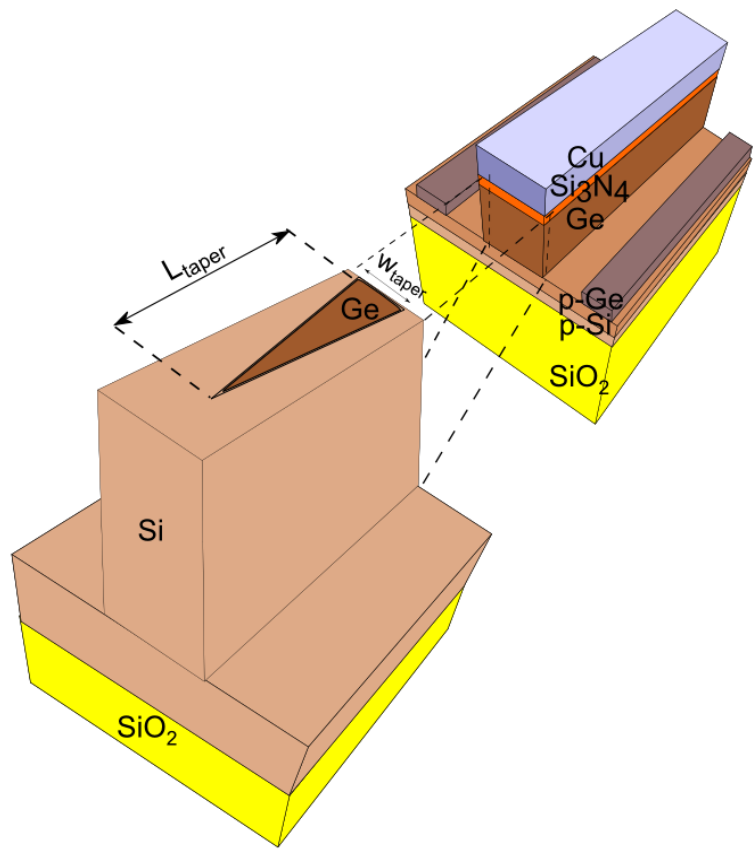

**Figure 132: Si-Ge taper into the Si rib waveguide to couple the optimized plasmonic modulator**

# 6.6.1    Mode Analysis of the Silicon-Germanium Taper

First, we study the effective index $n_{eff}$ and the effective propagation losses $\alpha_{eff}$ of the taper shown in Figure 132. The cross section of such a taper is represented in Figure 133,

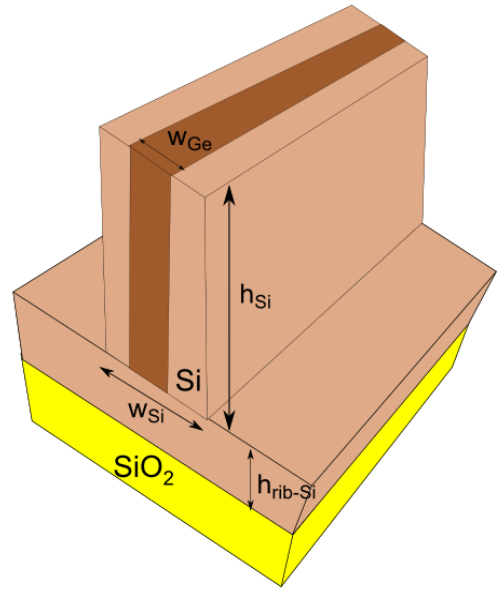

**Figure 133: Cross section of the Si-Ge taper of Figure 132**

We study the evolution of the effective index $n_{eff}$ and the effective absorption $\alpha_{eff}$ of the mode supported by the Si-Ge taper proposed in Figure 132. The cross section of the Si-Ge taper is presented in Figure 133. We studied the evolution of these parameters versus the parameter $w_{Ge}$ that we vary from 0 to 450 nm. We selected $w_{Ge}$=0 nm to have completely Si (like at the beginning of the waveguide). At the end of the scan we selected $w_{Ge}$=450 nm, so the waveguide is completely made of Ge. We selected the parameter $w_{Si}$=450 nm to have a marging to scan the parameter $w_{Ge}$ and because it is the best width in the butt-coupling structure. The other two parameters are $h_{Si}$=255 nm and $h_{rib-Si}$=100 nm for the reasons explained at the beginning of this chapter. The influence of $w_{Ge}$ in $n_{eff}$ and $\alpha_{eff}$ is represented in Figure 134 and Figure 135.

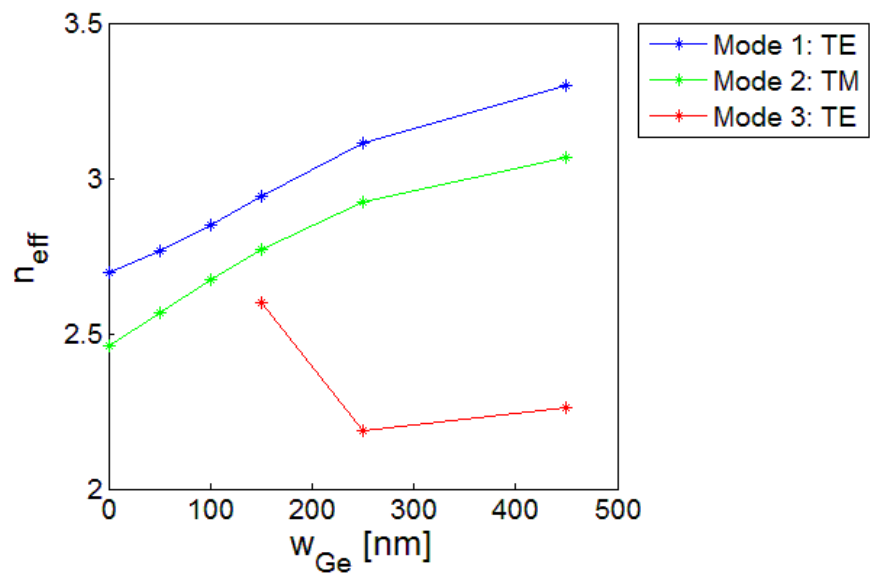

**Figure 134: $n_{eff}$ as a function of $w_{Ge}$ for the cross-section of the Si-Ge taper**

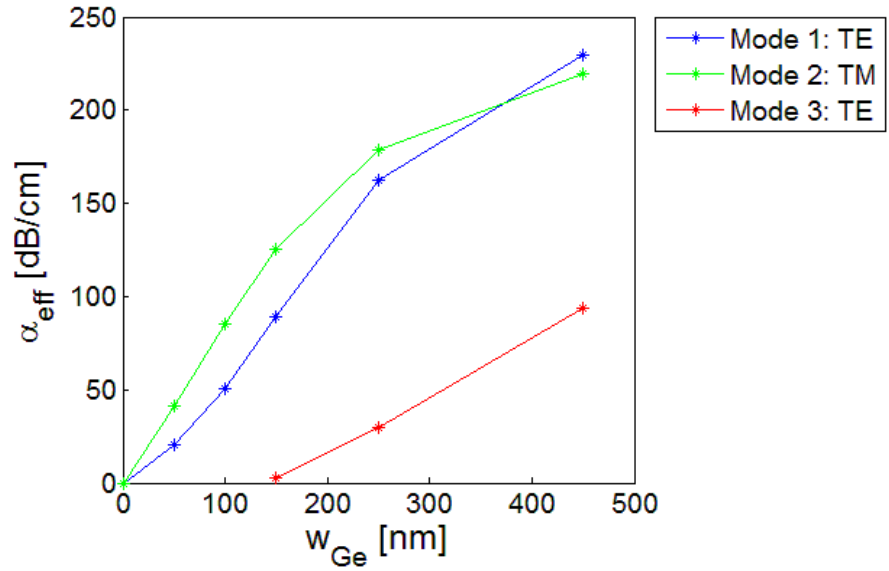

**Figure 135: $\alpha_{eff}$ as a function of $w_{Ge}$ for the cross-section of the Si-Ge taper**

From Figure 134 the effective index of the waveguide of Figure 133 has two fundamental modes starting at $w_{Ge}=0$. One is TE (blue) and the second TM (green). Around $w_{Ge}=150$ a second order TE mode appears. All modes are photonic as we can see the low losses they have in Figure 135.

In the input Si rib waveguide we want to excite the fundamental TM mode in order to excite the fundamental TM mode in the optimized plasmonic modulator. Consequently, we want to excite the fundamental TM mode of the waveguide of Figure 133. It corresponds to the green curve in Figure 134. The effective index of the plasmonic mode in the optimized modulator is $n_{eff}=2.89$. For this, the value of $w_{Ge}$ that most matches this effective refractive index is $w_{Ge}=250$ nm. The parameter $w_{Si}=450$ nm and $h_{Si}=255$ nm.

## 6.6.2    Total Transmission

In Figure 132 the structure to be studied is represented. One interesting parameter to scan is $w_{taper}$ (width of the taper in the interface with the optimized plasmonic modulator). For this value we selected the parameters $w_{taper}=150$ nm, 250 nm and 450 nm. We selected $w_{taper}=150$ nm because it is the same width as the optimized plasmonic modulator $w=150$ nm. On the other hand, we selected $w_{taper}=450$ nm because it has the same width as the input Si rib waveguide $w_{Si}=450$ nm. We selecte a third value around $w_{taper}=250$ nm in the middle of this range. With these values of $w_{taper}$ we scanned the parameter $L_{taper}$ from 1 μm to 5 μm. We selected those values because they are similar to the values used in a plasmonic taper presented in [144]. We scanned $L_{taper}$ every 1 μm. The result for the parameter $T_{Total}$ is presented in Figure 136. We remind that $h_{Si}=255$ nm and h=255 nm with $h_{slot}=5$ nm (Figure 84). The monitor to measure the parameter $T_{Total}$ is 20 nm inside the optimized plasmonic modulator. It means 20 nm from the interface between the taper and the modulator and inside the modulator. The input Si rib waveguide is $w_{Si}=450$ nm, $h_{Si}=250$ nm and $h_{rib-Si}=100$ nm (Figure 102).

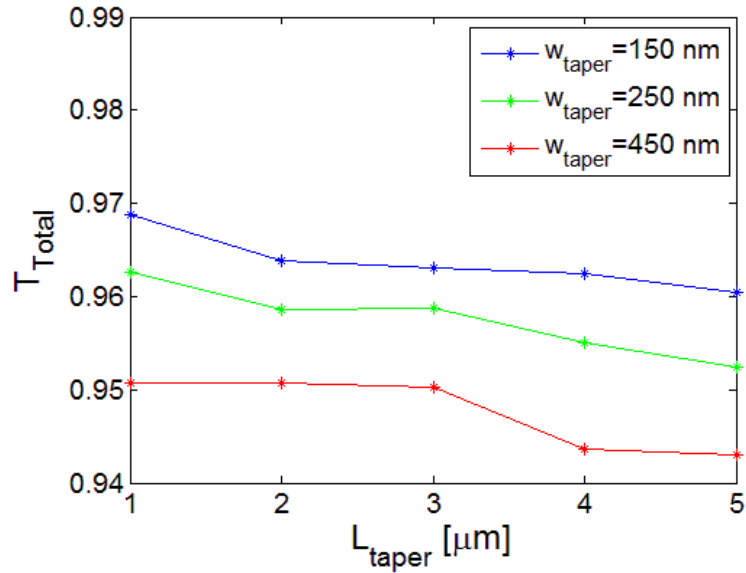

**Figure 136: Study of the parameter w_taper in the total transmission through the modulator T_Total for different length of the length of the Si-Ge taper L_taper**

From Figure 136 we observe that the maximum transmission is for the parameter $w_{taper}$=150 nm and the maximum transmission into the modulator is around $L_{taper}$=1 μm. For these parameters the transmission is around $T_{Total}$=0.97.

We remind that $T_{Total}$ captures all the light that passes the cross section of the modulators just 20 nm away from the interface between the Si-Ge taper and the modulator.

# 6.6.3    Overlap Integral and Transmission into the Mode

The next interesting parameter is to study the influence of $w_{taper}$ and $L_{taper}$ on the overlap integral between the mode at the end of the Si-Ge taper and the mode at the beginning of the modulator. For this, we selected the same values $w_{taper}$ and $L_{taper}$ as in the previous section. The result is presented in Figure 137.

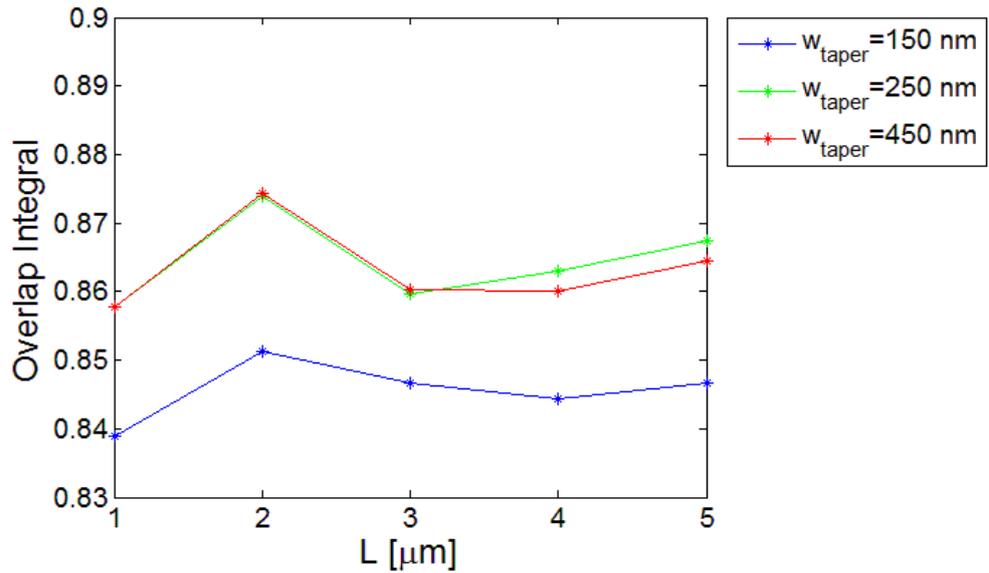

**Figure 137: Study of the parameter w_taper in the overlap integral OI through the modulator T_Total for different length of the length of the Si-Ge taper L_taper**

In Figure 137 we can see that the overlap integral is higer in $w_{taper}$=450 nm for $L_{taper}$<2 μm. For L>2 μm is is similar to $w_{taper}$=250 nm. On the other hand, the case in which $w_{taper}$=150 nm it is always the worst case. We see that around $w_{taper}$=250 nm the effective index of the optimized plasmonic modulator and the waveguide of Figure 133 match. The effective index to match is $n_{eff}$=2.89 and the one at the end of the taper is around $n_{eff}$=2.9. As we can see the reflection is again similar to the butt-coupling structure around 4%.

Comparing Figure 136 and Figure 137 we observe a trade-off since the parameter $T_{Total}$ is better for the case in which $w_{taper}$=150 nm and on the other hand, the overlap integral is better for $w_{taper}$=450 nm. To know which effect is dominan we need to calculate the parameter $T_{In\ the\ mode}$ to know which one is the best case. It is illustrated in Figure 138.

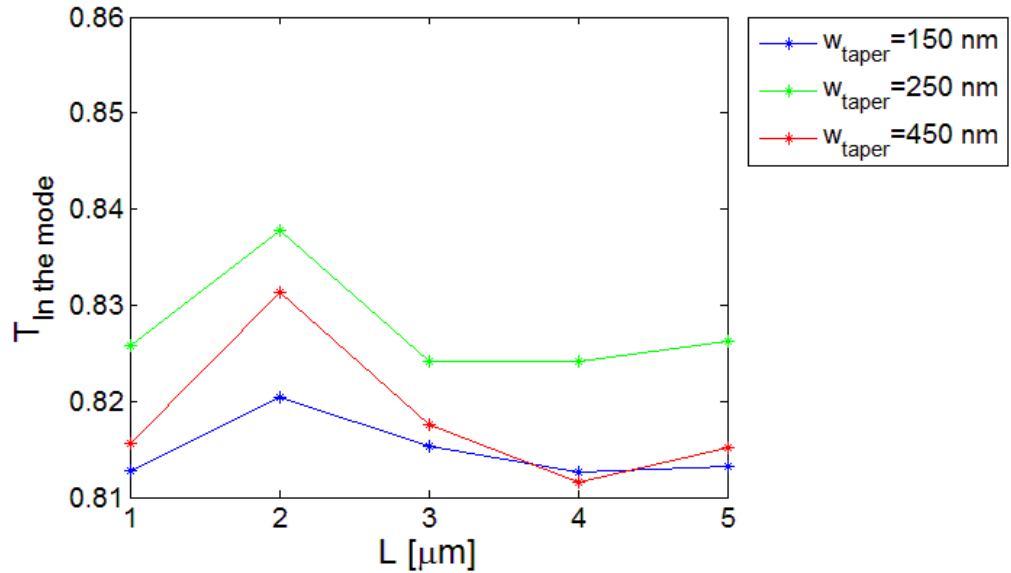

**Figure 138: Study of the parameter w_taper in the transmission in the mode TIn the mode through the modulatorfor different length of the length of the Si-Ge taper Ltaper**

In Figure 138 the parameter $T_{In\ the\ mode}$ is calculated versus the length of the Si-Ge taper $L_{taper}$. From Figure 138 we can clearly see the the parameter $T_{In\ the\ mode}$ is better for $w_{taper}$=250 nm and $L_{taper}$=2 μm. The maximum value is $T_{In\ the\ mode}$=0.84.

Additionally, the Si-Ge taper is done by Ge which absorbs light at the operational wavelength of the modulator which is around 1.647 μm. It is used in the Ge taper. This is an advantage of the butt-coupling structure (where Ge is not used) since the Si is transparent and consequently there is not absorption of light. To calculate the amount of light absorved in the Si-Ge taper of the Ge taper we measured the parameter $T_{Total}$ at the interface at x=0 μm (the interface of width $w_{taper2}$) for absorving and non-absorving Ge. For non-absorving Ge we set zero for the imaginary part of the refractive index of Ge in the tpaer. With this we obtained $T_{Total,\ absorving}$=0.996 for the absorving case and $T_{Total,\ non-absorving}$=0.999 for the non-absorving case. The amount of light absorved by the Si-Ge taper is then $T_{Total,\ non-absorving}$- $T_{Total,\ absorving}$=0.0034. It means that 0.34 % of the light pasing through the taper is absorbed by the Ge.

## 6.6.4    Conclusion

In the following Table 27 we summarize the main parameters in the best case of the Ge taper scheme. In this we used the optimized plasmonic modulator. We remind that the parameters of the optimized plasmonic modulator are: w=150 nm, h=250 nm, $h_{Slot}$=5 nm, $h_{Buf}$=60 nm and $h_{Bot}$=40 nm. The parameters of the Ge taper are: $w_{taper}$=250 nm, $w_{Si}$=450 nm, $h_{Ge}$=255 nm and $L_{taper}$=2 μm.

| Parameter: | R | $T_{Total}$ | $T_{In\ the\ mode}$ | $T_{Scattered}$ | $OI_{Fundamental\ mode}$ | $OI_{Other\ modes}$ | Absorption by Ge |
|---|---|---|---|---|---|---|---|
| Value [%]: | 4 | 96 | 84 | 12 | 87 | 0 | 0.34 |

**Table 27: Summary table of the main parameter of the Si-Ge taper structure structure Figure 132**

The best coupling efficiency was using butt-coupling with the following parameter: $w_{Si}$=450 nm, $h_{Si}$=255 nm and $h_{Slab}$=100 nm (Figure 84) for the Si rib input waveguide. The maximum is $T_{In\ the\ mode}$=0.76. Now, we demonstrated that using a Si-Ge taper within the input Si rib waveguide (Figure 132) we can obtain a better value of $T_{In\ the\ mode}$=0.84. This means that using the Si-Ge taper we increased 8% the coupling efficiency from the input Si rib waveguide to the optimized plasmonic modulator. The parameters are: $w_{Si}$=450 nm, $h_{Si}$=255 nm and $h_{Slab}$=100 nm, $L_{taper}$=2 μm and $w_{taper}$=250 nm (Figure 132). In Figure 139 there is a schematic with the definition of the parameters of Table 27.

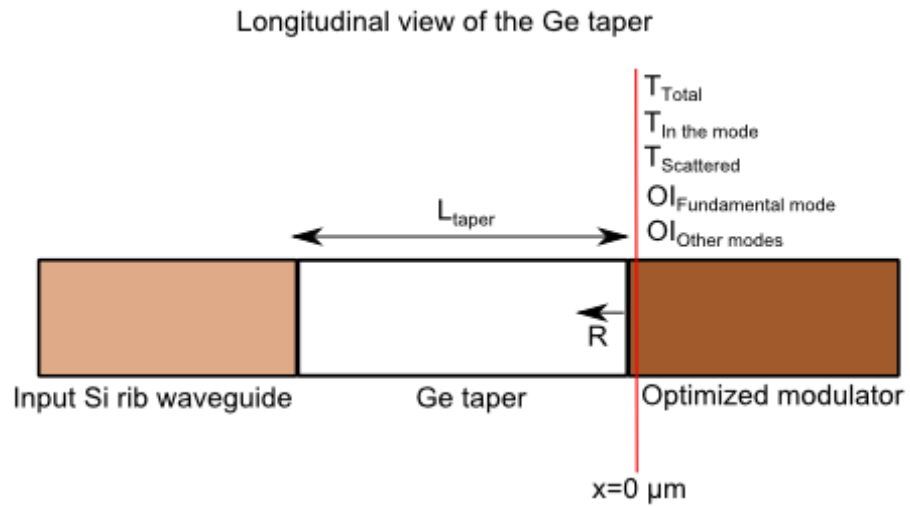

**Figure 139: Summary table of the main characteristics of the Ge taper summarized in Table 27**

Comparing the Si-Ge taper with the butt coupling structure: The better value in $T_{In\ the\ mode}$ of the Si-Ge taper is due to a better overlap integral (we excite more the fundamental TM mode of the optimized plasmonic modulator) and a less scattering, given by the parameter $T_{Scattered}$. We also have a similar reflection for both cases around R=4% but we have a material absorption around 0.34% in the Ge taper.

# 6.7    Comparison between the Coupling Schemes

In this section we compare the three coupled schemes proposed in this chapter. First, we summarize the main results of each individual study. It is represented in,

| Parameter: | $R_{First}$ | $R_{Second}$ | $T_{Total}$ | $T_{In the mode}$ | $T_{Scattered, 1}$ | $T_{Scattered,2}$ | $OI_{Fundamental mode, 1}$ | $OI_{Fundamental mode, 2}$ | $OI_{Other modes}$ | Absorption by Ge |
|---|---|---|---|---|---|---|---|---|---|---|
| Butt-Coupling [%]: | 0 | 4 | 96 | 76 | 0 | 20 | - | 66 | 0 | 0 |
| Ge taper [%]: | 6.5 | 4.5 | 81 | 68 | 12 | 13 | 79 | 73 | 0 | 0.75 |
| Si-Ge taper [%]: | 0 | 4 | 96 | 84 | 0 | 12 | - | 87 | 0 | 0.34 |

**Table 28: Summary table of the main characteristics of the coupling schemes studied. It is the butt-coupling, Ge taper and the Si-Ge taper**

Where $R_{First}$ is the reflection of the first interface and $R_{Second}$ is the reflection in the second interface. $T_{Scattered, 1}$ is the maximum scattering in the first interface and $T_{Scattered,2}$ is the maximum of the scattering of the second interface. $OI_{Fundamental mode, 1}$ is the overlap integral in the first interface and $OI_{Fundamental mode, 2}$ is the overlap integral in the second interface. In the case of the butt-coupling structure there is only one interface, in this case we consider it as a second interface. This is why $R_{First}=0$ and $T_{Scattered, 1}=0$ in the butt-coupling structure. The rest of the parameters were defined along the chapter. A schematic view of each one is represented in Figure 140.

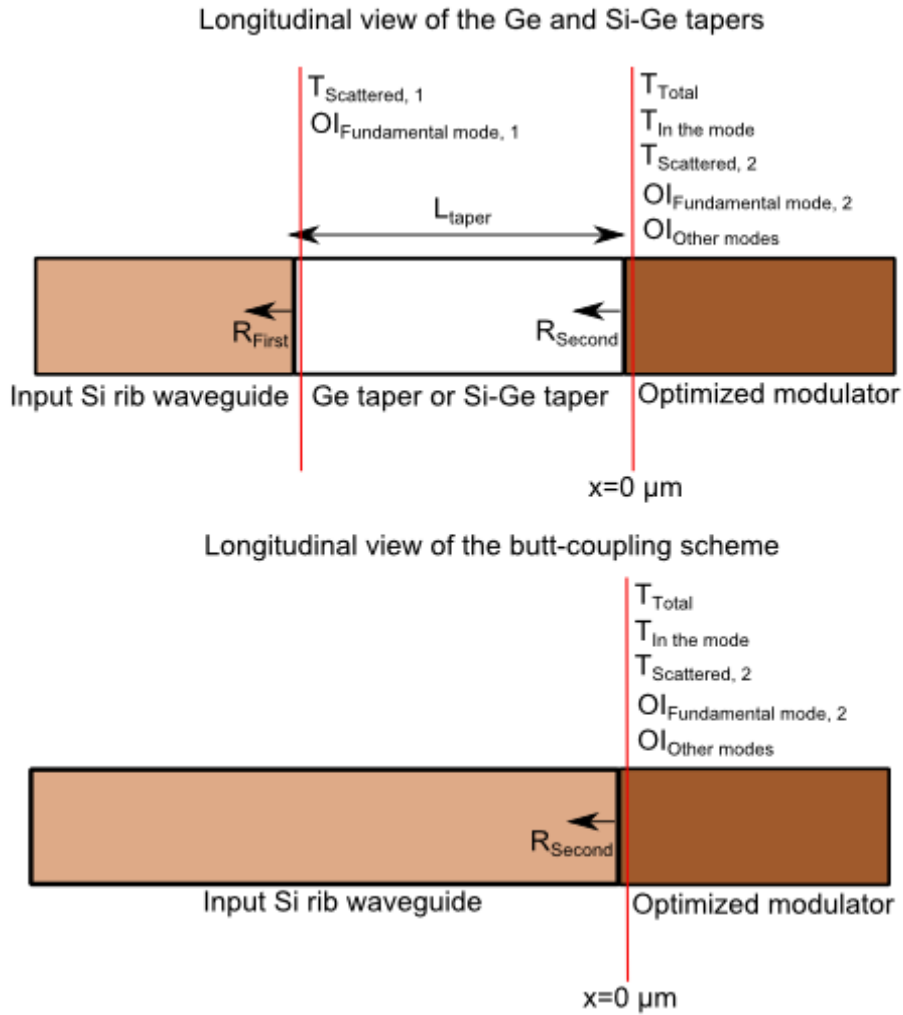

**Figure 140: Diagram of the parameters meassured in Table 28 for the Ge and Si-Ge tapers and the butt-coupling scheme**

To conclude we obtained a good result using the butt-coupling structure. We improved the coupling efficiency of this scheme by using a Si-Ge taper. The main factor for improving this was by having a better overlap integral of the mode we want to excite in the optimized plasmonic modulator and the mode at the end of the Si-Ge taper. Furthermore, it is observed less scattering in the Si-Ge taper.

In this chapter we studied the coupling scheme to the optimized plasmonic modulator from a standard Si rib waveguide. We started studying a direct butt-coupling scheme which give us a coupling efficiency around $T_{In the mode}$=0.76. We proposed several coupling schemes like the Ge taper and the Si-Ge taper. Among all these schemes the best one was the Si-Ge taper with a coupling efficiency around $T_{In the mode}$=0.84. This provides coupling efficiency around 8% better than the butt-coupling scheme. We realized that the increase in the complexity of the taper does not significantly increase the coupling efficiency.

# 7 Conclusion and Future Work

## 7.1   Conclusions

The objective of the thesis was to design a CMOS compatible low electrical power consumption modulator based on the Franz-Keldysh effect in germanium. We selected a plasmonic waveguide configuration to reduce the size of the optical modulator and to increase the optical confinement leading to the reduction of the electrical energy consumption of the device.

The first task was to design the plasmonic modulator. For this we developed an integrated electro-optical simulator which consists of a commercial electrical simulator (ISE-dessis) integrated with a finite difference method (FDM) mode solver. The simulator can model the Franz-Keldysh effect in germanium. With this integrated electro-optical modulator we can know the influence on the optical properties of the mode (mainly $n_{eff}$, $\alpha_{eff}$ and TE fraction) induced by the presence of a static electric field. For example, we can calculate $n_{eff}$, $\alpha_{eff}$ and TE fraction of guided mode versus the driving voltage V applied to the structure. I started from a FDM mode solver developed at IEF and I modified it to support plasmonic modes, metals and effects in germanium including the Franz-Keldysh effect.

In the case of the Franz-Keldysh effect, the absorption coefficient of the material at the band-edge of the gap changes under a static electric field. To model the Franz-Keldysh effect, we used an explicit formula which predicts the absorption. It is integrated with the electro-optical simulator. First, we calculate the static electric field distribution in a desired structure with the commercial simulator ISE-dessis. Knowing the static electric field distribution in germanium, we use the Franz-Keldysh effect model to calculate the new distribution of the germanium absorption. Afterwards, we import the structure to the optical FDM mode solver and we calculate the properties of the modes supported by the structure. Using this tool we can determine the effective index $n_{eff}$, and the effective losses $\alpha_{eff}$ of the mode versus the voltage applied to the structure. Once this tool was developed we tested it with 10 publications to validate it.

In the design chapter we tested several structures to induce electro-absorption effect in Ge based on plasmonic concepts. We studied several plasmonic waveguides for the optical structure of the modulator including slot waveguides, metal-semiconductor, metal-insulator-semiconductor, etc. We also tested several electrical structures to induce the static electric field like lateral and vertical PIN structures and capacitive structures. To compare all the structures we studied the extinction ratio and the propagation losses. For this, we defined the figure of merit $\Delta\alpha_{eff}/\alpha_{eff}$ where $\Delta\alpha_{eff}=\alpha_{eff}(V=V_{Ref})-\alpha_{eff}(V=0\ V)$. As a result of the whole process of comparing all the proposed structures, the modulator with best performances was composed of a metal-insulator-semiconductor waveguide. The best material to place as a metal was a special copper with a process developed at CEA-Leti. Furthermore, the metal of the metal-insulator-semiconductor waveguide was used to guide the plasmon and as an electrode to induce a static electric field into the semiconductor (germanium). It was also used as a capacitive structure to induce a static electric field to perform the modulation. Once we identified the best structure we optimized it with respect to all the parameters and dimensions of it. The best performance of the device is an extinction ratio of 3.3 dB, an insertion loss of 13.2 dB for a length of 30 μm, a bandwidth larger than 40 GHz, CMOS compatibility and a dynamic electrical power consumption of 20 fJ/bit. Regarding the electrical power consumption we compared it with all the Franz-Keldysh effect modulators and all the plasmonic modulators that report the energy consumption. The result of this comparison was that the device proposed in this work has the lowest electrical power consumption reported in the literature.

Once we designed and optimized the device we study the integration of this device in a standard integrated circuit. For this, we study by the means of mode analysis and 3D FDTD simulations three possible ways of coupling the plasmonic modulator with a silicon rib waveguide. The first approach is a butt-coupling configuration between the silicon rib waveguide and the optimized plasmonic modulator. The simulated coupling efficiency was around 76%. In a second approach we added a germanium taper between the silicon rib waveguide and the optimized germanium-based plasmonic modulator. The dimensions of the waveguides were settled to match the effective index at all the interfaces. The performance showed by this approach was not as good as in the butt-coupling scheme. The reason was the increase of the reflection at waveguide–taper interface and consequently reduced the power into the modulator. Furthermore, the Ge taper also absorbs light. Finally, in order to try to improve the overlap integral between the silicon rib waveguide and the optimized plasmonic modulator we investigated a Si-Ge taper. This taper showed an efficiency of 84%, overcoming the performances of the butt-coupling scheme.

With the design of the optimized plasmonic modulator and the coupling scheme, we prepared the process flow for the whole fabrication and the layout of several configurations of modulators (Appendix C). The fabrication was launched at CEA-Leti but unfortunately it will finish after the end of this PhD.

Regarding the operational wavelength of the designed plasmonic modulator, it is supposed to optimally work at 1647 nm where the Franz-Keldysh effect is maximum. It is desired to shift the maximum of the Franz-Keldysh effect to 1550 nm for telecom applications. It can be done by including some silicon into the germanium. For this, we fabricated test structures consisting of a lateral PIN diode formed by p-doped silicon, intrinsic germanium and n-doped silicon. Once fabricated, the structure was heated at high temperature to make silicon diffuse laterally to the germanium core to obtain silicon-germanium in the intrinsic core. We characterized the devices and we found some indication that the absorption edge of the SiGe material was shifted to wavelengths lower than 1550 nm. We faced the difficulty of controlling the amount of diffused silicon into the germanium using this technique. Knowing this, we designed a second batch of silicon-germanium PIN structures to try to shift the maximum of the Franz-Keldysh effect to 1550 nm. In this case, a mixture of silicon and germanium is directly epitaxially grown with a desired composition. The batch is still under fabrication at the time of writing and we expect to have the first characterization results at the end of the PhD.

# 7.2    Future Work

Unfortunately, due to finite time, several tasks could not be finalized.

Firstly the first generation of CMOS-compatible, low-energy plasmonic modulator designed and fabricated in this work will have to be characterized once the fabrication will be completed. Secondly, it is necessary to complete the demonstration of the Franz-Keldysh effect at 1550 nm in a silicon-germanium waveguide photonic waveguide. Once the Franz-Keldysh effect is demonstrated at 1550 nm a second generation of plasmonic modulator may be designed with a core of intrinsic silicon germanium instead of pure germanium. With this, we want to shift the operational wavelength of the device at 1550 nm.

One of the main disadvantages of plasmonic modulators in general is the high propagation losses. Consequently, a reduction in the loss of metal-insulator-semiconductor waveguides is still necessary. To try to reduce the propagation losses a nanostructuration of the metal of the metal-insulator-semiconductor waveguide could be investigated. It is well known that when the light travels in a medium which consists of a subwavelength periodic alternation of materials with refractive indices $n_1$ and $n_2$ the dielectric constant felt by the light is the mean of the refractive index constants present in the medium [145] . We think that by performing a subwavelength periodic nanostructuration of the metal of a metal-insulator-semiconductor waveguide, we may reduce the propagation losses due to a lower content of metal.

Another approach could be also to study the plasma dispersion effect in germanium for a germanium based electro-refraction modulator. To the best of my knowledge there is not a deep study of electro-refraction modulators in germanium. The structure may be photonic or plasmonic.

# 8 Appendix A

In this appendix the permittivity of a metal is presented. For this we will use the Maxwell's equations and the Drude model for metals. It is important to calculate the so called plasma frequency of the metals. Below this frequency (in the infra-red and visible regime) the metals behave as a dielectric. Is at those frequencies when the metals are used in a waveguide to break the diffraction limit of light.

The physical model of metals at telecommunication frequencies are described by Maxwell's equations and the Drude model. In the Drude model there is a set of carriers with density N and mass m that freely moves into a periodic array of cores. Metals possess a high density of carriers in the material. Consequently, there is not any discretization of the electron energy levels with respect to the thermal excitation $k_BT$. When an electromagnetic field is present in the material the electrons oscillate around the positive ion cores. The collision frequency is given by Equation 66,

$$\gamma = \frac{1}{\tau},$$

Equation 66

Where γ is the collision frequency and τ is the free electron relaxation time. The dispersion relation ω(k) relates the frequency ω of the oscillations with the wave-vector k. The plasma of electrons under the presence of an electric field **E** is described by Equation 67,

$$m\frac{d^2x}{dt^2} + m\gamma\frac{dx}{dt} = -e\boldsymbol{E}$$

Equation 67

Where x is the electron displacement between the electron and the ion core, m is the mass of the electron, e is the unitary charge of the electron, γ is the damping factor and **E** is the electric field present in the material.

If the material is excited by a monochromatic electric field described by Equation 68,

$$\boldsymbol{E}(t) = \boldsymbol{E_0}e^{-i\omega t}$$

Equation 68

We can essay a solution of the form given by,

$$x(t) = \frac{e}{m(\omega^2 + i\gamma\omega)} E(t)$$

Equation 69

When the electrons are displaced around the positive ion core a polarization vector P appears. It is described by Equation 70,

$$P = -nex$$

Equation 70

Where n is the refractive index of the material. Substituting Equation 69 into Equation 70 we obtain,

$$P = -\frac{ne^2}{m(\omega^2 + i\gamma\omega)} E$$

Equation 71

Using the relation between the displacement vector, the electric field and the polarization vector given by Equation 72,

$$D = \epsilon_0 E + P$$

Equation 72

Consequently, the displacement vector is,

$$D = \varepsilon_0 \left(1 - \frac{\omega_p^2}{\omega^2 + i\gamma\omega}\right) E$$

Equation 73

Where $\omega_p$ is the damping frequency given by,

$$\omega_p^2 = \frac{ne^2}{\varepsilon_0 m}$$

Equation 74

Using Equation 72 and Equation 73 the complex dielectric function of a metal is given by,

$$\varepsilon(\omega) = 1 - \frac{\omega_p^2}{\omega^2 + i\gamma\omega}$$

Equation 75

Expressing the real and imaginary part of Equation 75 in the form ε= ε′+ iε′′ we obtain

Equation 76,

$$\varepsilon'(\omega) = 1 - \frac{\omega_p^2\tau^2}{1 + \omega^2\tau^2}$$

$$\varepsilon''(\omega) = \frac{\omega_p^2\tau}{\omega(1 + \omega^2\tau^2)}$$

Equation 76

The plasma frequency $\omega_p$ is the frequency over which the dielectric constant of a metal is positive. For the optical electric field it behaves like a dielectric.

The Drude model is accurate and it predict well the behavior of a metal in the infra-red and visible regime. Over this frequency the predictions are not so accurate. Then, another model called Lorentz-Drude needs to be used.

# 9 Appendix B

In this section we are going to obtain the main characteristics of a metal-semiconductor (MS) plasmonic waveguide. We start from Maxwell's equations which we apply to the MS structure. We also impose the boundary conditions in the interface between the metal and the semiconductor.

The objective of this study is to determine the penetration depth of the plasmonic mode supported by the MS structure into the metal. The losses of the plasmonic mode are also calculated. We will deduct that those losses are smaller than the losses of the plasmon in a metal-insulator-semiconductor (MIS) waveguide. This will be done at the end of chapter three.

Maxwell's equations for a linear, homogeneous and isotropic medium are,

$$\nabla \cdot \boldsymbol{D} = \rho \qquad\qquad \text{Equation 77}$$

$$\nabla \cdot \boldsymbol{B} = 0 \qquad\qquad \text{Equation 78}$$

$$\nabla \, x \, \mathbf{E} = -\frac{\partial \boldsymbol{B}}{\partial t} \qquad\qquad \text{Equation 79}$$

$$\nabla \, x \, \mathbf{H} = \mathbf{J} + \frac{\partial \boldsymbol{D}}{\partial t} \qquad\qquad \text{Equation 80}$$

Where, $\boldsymbol{D}$ is the electric displacement field, $\rho$ is the charge density, $\boldsymbol{B}$ is the magnetic field, $\boldsymbol{E}$ is the electric field, $\boldsymbol{H}$ is magnetic field strength and $\boldsymbol{J}$ is the current density. For a medium in which there is neither current $\boldsymbol{J} = 0$ nor charge $\rho = 0$ we can obtain the wave equation from previous ones,

$$\nabla \, x \, (\nabla \, x \, \mathbf{E}) = -\mu_0 \frac{\partial^2 \boldsymbol{D}}{\partial t^2} \qquad\qquad \text{Equation 81}$$

Using the vector identity $\nabla \, x \, (\nabla \, x \, \boldsymbol{V}) = \nabla(\nabla \cdot \boldsymbol{V}) - \nabla^2 \boldsymbol{V}$ where $\boldsymbol{V}$ is a generic vector. Applying this equation we obtain,

$$\nabla^2 \boldsymbol{E} - \frac{\varepsilon}{c^2} \frac{\partial^2 \boldsymbol{E}}{\partial t^2} = 0 \qquad\qquad \text{Equation 82}$$

Applying the Fourier transform for the Equation 82 $\left(\frac{\partial}{\partial t} = -i\omega\right)$ and using a monochromatic electric field of the form,

$$\boldsymbol{E}(t) = \boldsymbol{E_0}e^{-i\omega t}$$                    Equation 83

We obtained,

$$\nabla^2\boldsymbol{E} + \mathrm{k}_0^2\varepsilon\boldsymbol{E} = 0$$                    Equation 84

Where $\mathrm{k}_0$ is the free space wave-vector.

In Figure 141 there is the geometry that is going to be solved. It consists on a metal embedded in a dielectric. This is one-dimensional problem in which the metal is infinite in the xy-axis. The interface between the metal and the dielectric is around z=0. The propagation direction of the plasmonic mode will be around the x-axis. In this case the permittivity of the structure depends only on z, it means: ε=ε(z).

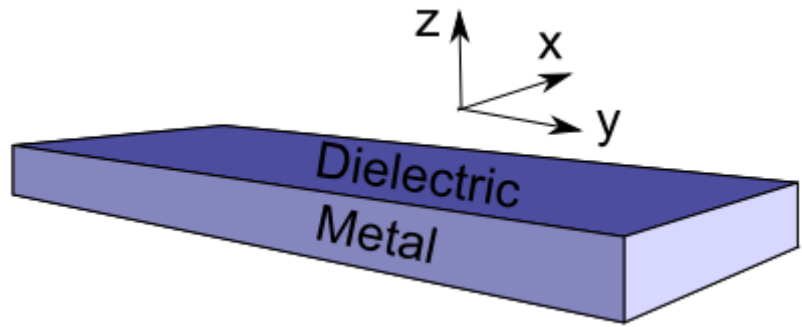

**Figure 141: Metal-dielectric interface to solve the propagation of SPP**

The expected electric field will have the form,

$$\boldsymbol{E}(x,y,z) = \boldsymbol{E}(z)e^{i\beta x}$$                    Equation 85

Where, $\beta$=$\mathrm{k}_x$ is the propagation constant, which is the wave-vector in the direction of the propagating plasmon. Substituting the desired solution Equation 85 into Equation 84, we obtain,

$$\frac{\partial\boldsymbol{E}(z)}{\partial z^2} + (k_0^2\varepsilon - \beta^2)\boldsymbol{E} = 0$$                    Equation 86

There is another equation similar to Equation 86 but for the magnetic field strength **H**. The spatial field profile and the propagation vector will be calculated using Equation 86. The expressions for **E** and **H** will be calculated. Using harmonic time dependence (Equation 85), we obtain,

$$\frac{\partial E_z}{\partial y} - \frac{\partial E_y}{\partial z} = i\omega\mu_0 H_x$$

$$\frac{\partial E_x}{\partial z} - \frac{\partial E_z}{\partial x} = i\omega\mu_0 H_y$$

$$\frac{\partial E_y}{\partial x} - \frac{\partial E_x}{\partial y} = i\omega\mu_0 H_z$$

$$\frac{\partial H_z}{\partial y} - \frac{\partial H_y}{\partial z} = -i\omega\mu_0 E_x$$

$$\frac{\partial H_x}{\partial z} - \frac{\partial H_z}{\partial x} = -i\omega\mu_0 E_y$$

$$\frac{\partial H_y}{\partial x} - \frac{\partial H_x}{\partial y} = -i\omega\mu_0 E_z$$

Equation 87

When the propagation is in the x-direction we have,

$$\frac{\partial}{\partial x} = i\beta$$

And in the y-direction,

$$\frac{\partial}{\partial y} = 0$$

The system simplifies to,

$$\frac{\partial E_y}{\partial z} = -i\omega\mu_0 H_x$$

$$\frac{\partial E_x}{\partial z} - i\beta E_z = i\omega\mu_0 H_y$$

$$i\beta E_y = i\omega\mu_0 H_z$$

$$\frac{\partial H_y}{\partial z} = -i\omega\varepsilon_{0\varepsilon} E_x$$

$$\frac{\partial H_x}{\partial z} - i\beta H_z = -i\omega\varepsilon_0 \varepsilon E$$

$$i\beta H_y = -i\omega\varepsilon_0 \varepsilon E_z$$

Equation 88

The system of Equation 88 allows two possible solutions that are called transverse magnetic (TM) or transvers electric (TE). In the case of TM only the components $E_x$, $E_z$ and $H_y$ are different from zero. And in the case of TE only the values of $H_x$, $H_z$ and $E_y$ are nonzero.

Taking into account the TM mode the system of Equation 88 simplifies to,

$$E_x = -i \frac{1}{\omega \varepsilon_0 \varepsilon} \frac{\partial H_y}{\partial z}$$

<div align="right">Equation 89</div>

$$E_z = -\frac{\beta}{\omega \varepsilon_0 \varepsilon} H_y$$

With the wave equation giving for TM modes equal to,

$$\frac{\partial H_y}{\partial z^2} + (k_0^2 \varepsilon - \beta^2) H_y = 0$$

<div align="right">Equation 90</div>

In the same way the systems of Equation 88 simplifies for TE modes to,

$$H_x = -i \frac{1}{\omega \mu_0} \frac{\partial E_y}{\partial z}$$

<div align="right">Equation 91</div>

$$H_z = -\frac{\beta}{\omega \mu_0} E_y$$

And the wave equation,

$$\frac{\partial E_y}{\partial z^2} + (k_0^2 \varepsilon - \beta^2) E_y = 0$$

<div align="right">Equation 92</div>

In Figure 141 the infinite interface between the metal and the dielectric is at z=0. For z<0 the metal is described by a permittivity given by $\varepsilon_2(\omega)$ and for z>0 the permittivity of the dielectric is given by $\varepsilon_1$. We assume that the frequency of the beam of light is below the plasma dispersion frequency of the metal $\omega_p$. It means, $\omega < \omega_p$. Under this condition we obtain that Re($\varepsilon_2$)<0. This condition is important for the existence of a SPP in the surface between the metal and the dielectric.

We will analyze the TM polarization plasmon first. Using Equation 89 and Equation 90 for the upper half of Figure 141 (z>0) we obtain,

$$H_y(z) = A_2 e^{i\beta x} e^{-k_2 z}$$

$$E_x(z) = iA_2 \frac{1}{\omega \varepsilon_0 \varepsilon_2} k_2 e^{i\beta x} e^{-k_2 z}$$

<div align="right">Equation 93</div>

$$E_z(z) = iA_1 \frac{\beta}{\omega \varepsilon_0 \varepsilon_2} k_2 e^{i\beta x} e^{-k_2 z}$$

And for the other half (z<0) we obtain,

$$H_y(z) = A_1 e^{i\beta x} e^{-k_1 z}$$

$$E_x(z) = iA_1 \frac{1}{\omega \varepsilon_0 \varepsilon_1} k_1 e^{i\beta x} e^{-k_1 z}$$

<div align="right">Equation 94</div>

$$E_z(z) = iA_1 \frac{\beta}{\omega \varepsilon_0 \varepsilon_2} k_2 e^{i\beta x} e^{-k_1 z}$$

Where $k_i$ which $i \in \{1, 2\}$ is the component of the wave-vector which is perpendicular to the interface in the dielectric (z>0) and in the metal (z<0). And the constants $A_i$ which $i \in \{1, 2\}$ depends on the input power.

The continuity of $H_y$ and $\varepsilon_i E_z$ on Equation 93 and Equation 94 in the metal-dielectric interface gives,

$$A_1 = A_2$$

$$\frac{k_2}{k_1} = -\frac{\varepsilon_2}{\varepsilon_1} \qquad \qquad \text{Equation 95}$$

Applying the wave equation for TM polarization Equation 90 we obtain,

$$k_1^2 = \beta^2 - k_0^2 \varepsilon_1 \qquad \qquad \text{Equation 96}$$

$$k_2^2 = \beta^2 - k_0^2 \varepsilon_2$$

Combining Equation 95 and Equation 96 we obtain the following dispersion relation,

$$\beta = \sqrt{\left( \frac{\varepsilon_1 \varepsilon_2}{\varepsilon_1 + \varepsilon_2} \right)} \qquad \qquad \text{Equation 97}$$

A similar analysis can be done for a TE polarized plasmon. Nevertheless the continuity of $E_y$ and $H_x$ leads to the following relation,

$$A_1(k_1 + k_1) = 0 \qquad \qquad \text{Equation 98}$$

As we mentioned at the beginning of this discussion to have a plasmon we need to fulfill Re{$k_1$}>0 and Re{$k_2$}>0. In this case Equation 98 has only solutions for $A_1$=0 and $A_2$=0. It means that we cannot have a TE polarization in the configuration of Figure 141. SPP modes only exist in TM polarization.

When the plasmonic mode is bounded, the perpendicular propagation constants $k_i$, $i \in \{1, 2\}$ in both mediums (the dielectric and the metal) is imaginary, since,

$$k_i = \frac{\omega^2}{c} - \beta^2 < 0$$

And,

$$\beta^2 > \frac{\omega^2}{c}$$

Taking into account Equation 97 we can see that the wave-vector of the SPP will be bigger than the wave-vector of the free propagating light into the dielectric. Furthermore, the dielectric constant is complex in metals since they have absorption. It means that the component $\beta$ will also be complex through Figure 141. We can divide it by,

$$\beta = \beta' + i\beta''$$

<div align="right">Equation 99</div>

Using the fact that the imaginary part of the dielectric constants in metals is smaller than the real part ($\varepsilon_2'' < \varepsilon_2'$) we can obtain a simplified format for the real and imaginary part of the propagation wave-vector of the SPP,

$$\beta' = \frac{\omega}{c} \sqrt{\left(\frac{\varepsilon_2' \varepsilon_1}{\varepsilon_2' + \varepsilon_1}\right)}$$

<div align="right">Equation 100</div>

$$\beta'' = \frac{\omega}{c} \sqrt{\left(\frac{\varepsilon_2' \varepsilon_1}{\varepsilon_2' + \varepsilon_1}\right)^3} \frac{\varepsilon_2''}{2\varepsilon_2'^2}$$

<div align="right">Equation 101</div>

As stated before, the real part of the propagation wave-vector of the SPP (Equation 100) must be real for the mode to be bounded. This implies that,

$$Re\{\varepsilon_2' \varepsilon_1\} < 0$$

And,

$$Re\{\varepsilon_2' + \varepsilon_1\} < 0$$

Metals can fulfill these two conditions at telecommunication wavelengths (also in the visible and in the infrared regions).

The dispersion relation of plasmonic mode for different wavelengths is represented in Figure 142.

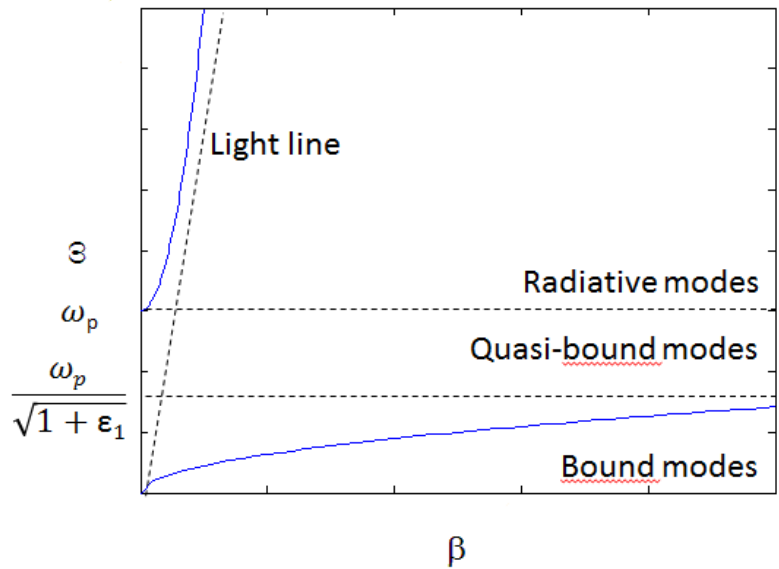

**Figure 142: Dispersion relation of SPP**

The wavelengths of the plasmons approaches the light line for the case in which $\varepsilon_2' \gg \varepsilon_1$. For the case in which $\varepsilon_2'$ approaches $\varepsilon_1$ we can see that the wave-vector of the plasmon is closer to the surface plasmon frequency. This surface plasmon frequency is given by,

$$\omega_{sp} = \frac{\omega_p}{\sqrt{1 + \varepsilon_1}}$$

Equation 102

As is represented in Figure 142 between $\omega_{sp}$ and $\omega_p$ the wave-vector of the plasmon is imaginary. In this region there is not propagation of the mode. Over the frequency $\omega_p$ the plasmons behaves as a dielectric.

There are still two more characteristics of the plasmonic mode. They are: the penetration depth and the propagation length.

The penetration depth is the penetration of the optical electromagnetic field into the dielectric and into the metal. It is defined as the length from the metal-dielectric interface to the length at which the optical electromagnetic intensity drops until 1/e. The penetration depth into the dielectric is given by,

$$e_1 = \frac{1}{2k_1'}$$

Equation 103

The penetration depth into the metal is given by,

$$e_2 = \frac{1}{2k_2'}$$

Equation 104

The propagation length is the length in which the intensity of the propagating SPP drops by 1/e with respect of the initial intensity. It is given by,

$$L_p = \frac{1}{2\beta''}$$

Equation 105

In [118] there is a derivation of the main parameters of a plasmon in the MS structure. We are only to present the results. In the approximation of a perfect conductor,

$$\varepsilon_2' < 0, \varepsilon_2'' \ll |\varepsilon_2'| \ and \ \varepsilon_1 \ll |\varepsilon_2'|$$

Equation 106

Doing a Taylor expansion of Equation 97 we can obtain an approximation for the difference $\Delta n_{MIS} = n_{MIS} - n_1$. It is given by,

$$\Delta n_{MS} = Re(n_{MS}) - n_1 \approx \frac{n_1^3}{2|\varepsilon_2'|}$$

Equation 107

Where $n_{MS}$ is the effective refractive index of the SPP mode in the MS waveguide. It depends on the third power of the dielectric cladding $n_1$. This shows a high value for dielectric with high refractive indexes. The definition of the propagation losses are given by,

$$\alpha_{MS} = \frac{4\pi}{\lambda} Im(n_{MS})$$

Equation 108

Using a derivation similar to Equation 107, the losses $\alpha_{MS}$ of an SPP in an MS waveguide is given by,

$$\alpha_{MS} \approx \frac{2\pi}{\lambda} \frac{n_1^3 \varepsilon_2'}{\varepsilon_2'^{\,2}}$$

<div align="right">Equation 109</div>

Now we calculated the approximated losses of the plasmonic mode supported by the MS plasmonic waveguide. At the end of chapter three we are going to introduce a similar waveguide called metal-insulator-semiconductor (MIS) waveguide. At the end of chapter three we demonstrate that the MIS waveguide has less propagation losses than the MS waveguide. For this, we will use Equation 109.

# 10  Appendix C

As a result of the study done in chapters five and six, the plasmonic modulator designed will be fabricated. Before fabrication can be launched, it is necessary to elaborate a process flow and do the layout of the maskset. This chapter describes first the process flow which was established at CEA-Leti. Then we present the layout which was done partly at IEF and at CEA-Leti. This work was done in the frame of the ANR project MASSTOR (plasMon ASSisTed electro-optical modulatOR).

## 10.1   Process Flow

The process flow is composed of seven lithographic levels. We start with SOI wafers of 440 nm thickness. The main steps are the following:

- Definition of the input and output silicon waveguides and the grating couplers (Figure 143) and partial etch for the TM fiber grating couplers to launch light at thie input of the circuit and and collect it at the output.
- Protection of the grating couplers and further etch of silicon down to the slab for the silicon waveguides (final result in Figure 144)
- Definition and etching of a cavity in Si in which Ge will be grown epitaxially (Figure 145)
- Epitaxial growth of a small layer of p-doped Germanium, followed by undoped Germanium and chemical-mechanical planarization (Figure 146)
- Definition and implantation of two regions of Ge for the future side contacts of the modulator (Figure 147)
- Definition of the Ge modulator waveguide and the lateral regions for the future contacts (Figure 148) and encapsulation

- Definition of the plasmonic stack / MOS structure of the modulator : etching of a cavity, deposition of a thin layer of Si₃N₄ followed by Cu deposition and chemical-mechanical palanrization (Figure 149)
- Definition of the AlCu sides contacts of the plasmonic modulator (Figure 150)

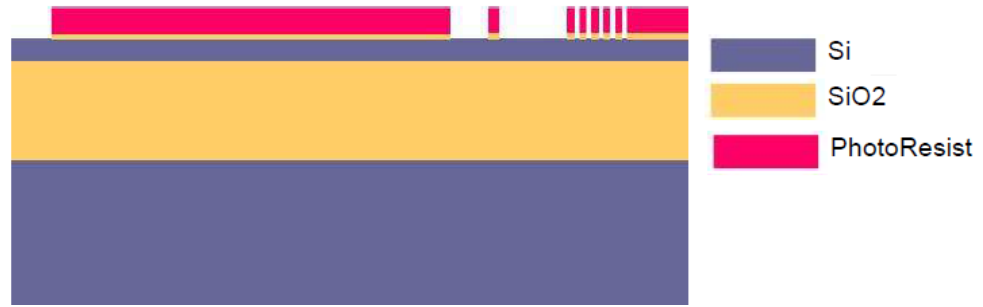

**Figure 143: First step of the layout fior the definition of the Si rib input/output waveguide and the grating**

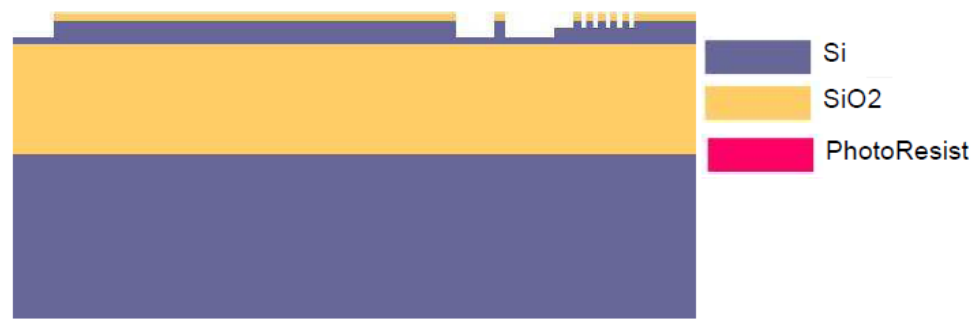

**Figure 144: Definition of the input/output Si rib waveguide and the grating**

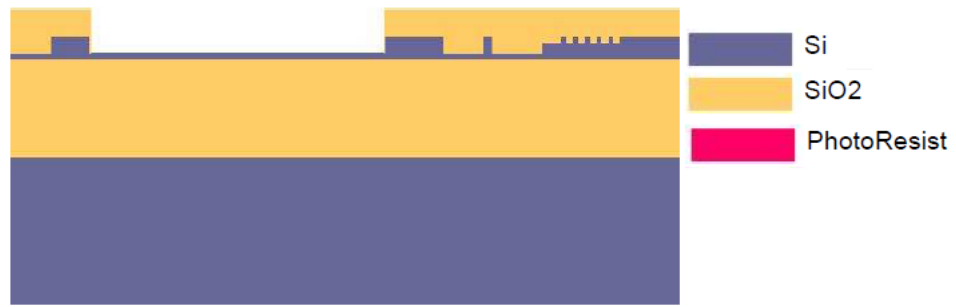

**Figure 145: Etching of a cavity in Si to place the core of the plasmonic modulator**

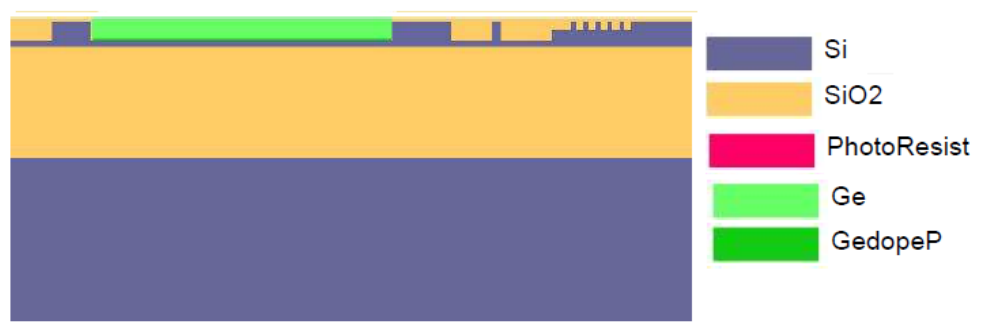

**Figure 146: Cavity filled with Ge and p-doped Ge**

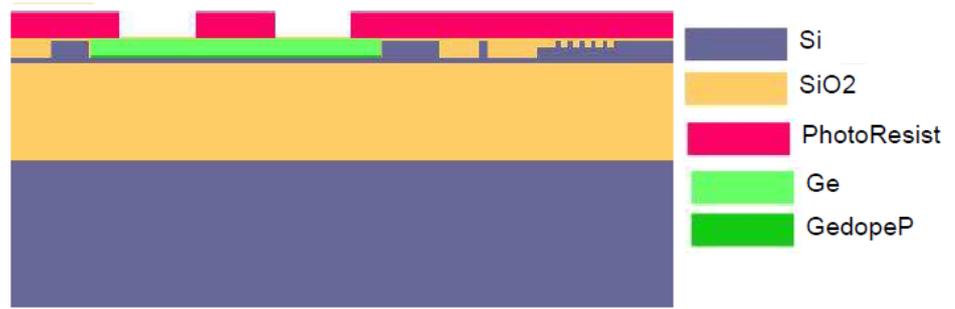

**Figure 147: Definition and doping of the side contacts**

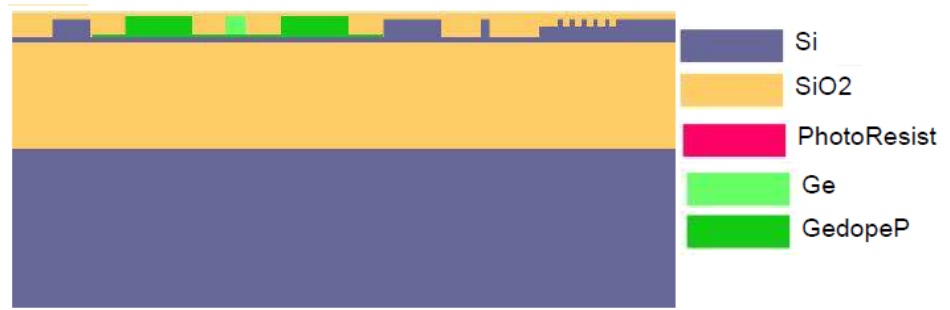

**Figure 148: Definition of the intrinsic Ge core of the modulator with side contacts**

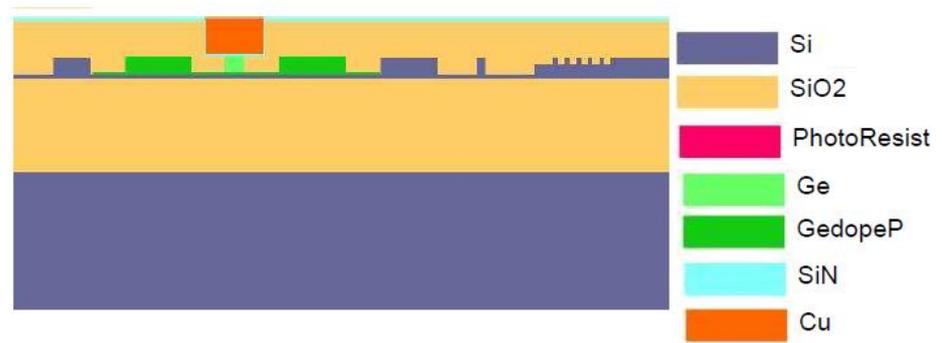

**Figure 149: Deposition of the Si$_3$N$_4$ and the Cu layer**

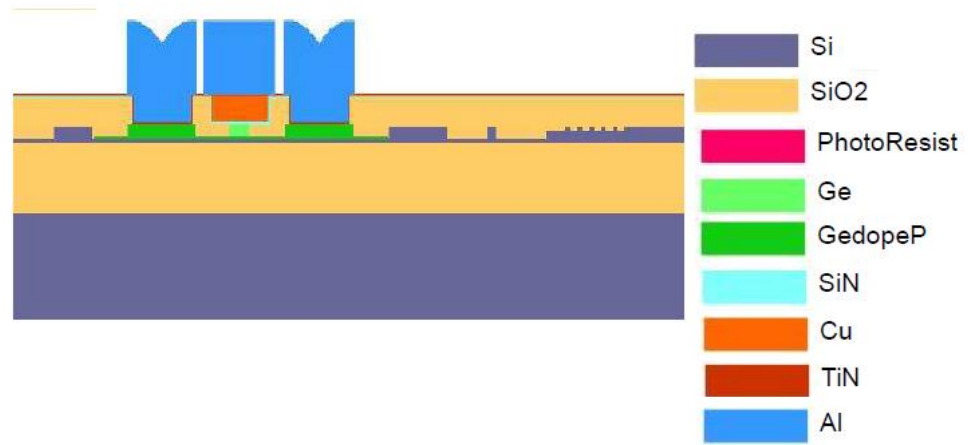

**Figure 150: Final layout with the plasmonic modulator with side contacts and the input/output Si rib waveguide with the grating**

Figure 150 shows a cut-view the SOI substrate at the end of the process flow. From left to right, it is composed by the core of the plasmonic modulator with the side contacts (cross-section), the input/output Si rib waveguide (cross-section view) and the TM grating couplers (longitundinal view).

## 10.2   Layout Design

An overview of the final layout is presented in Figure 151. The overall layout cell size is around 10 mmx10 mm. In this global cell, we placed 24 unit cells containing 16 modulator devices each, i.e. 384 plasmonic modulators in total. All the modulators are different from each other. The unit cell of 16 modulators is detailed afterwards on Figure 152.

In addition, there are several test structures for the elementary building blocks like waveguides, grating couplers, electrodes to check their performances independently of the Ge plasmonic modulator circuit. These structures are also detailed later in this section.

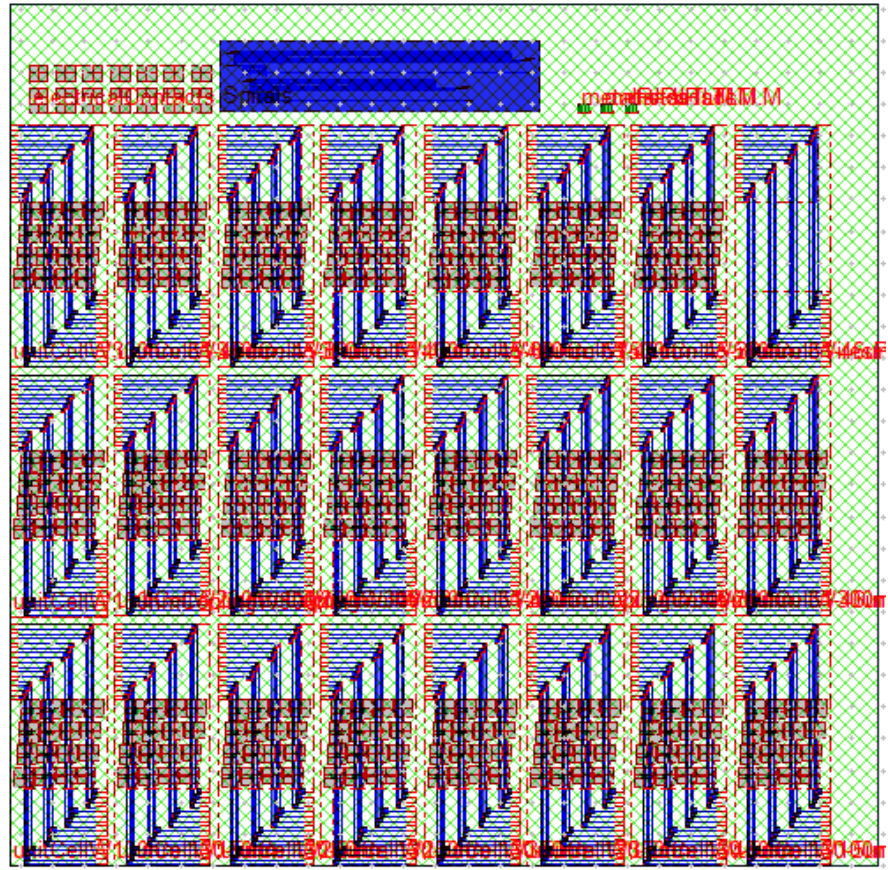

**Figure 151: Final layout of the plasmonic modulators. We can observe 24 sets of 16 modulators each**

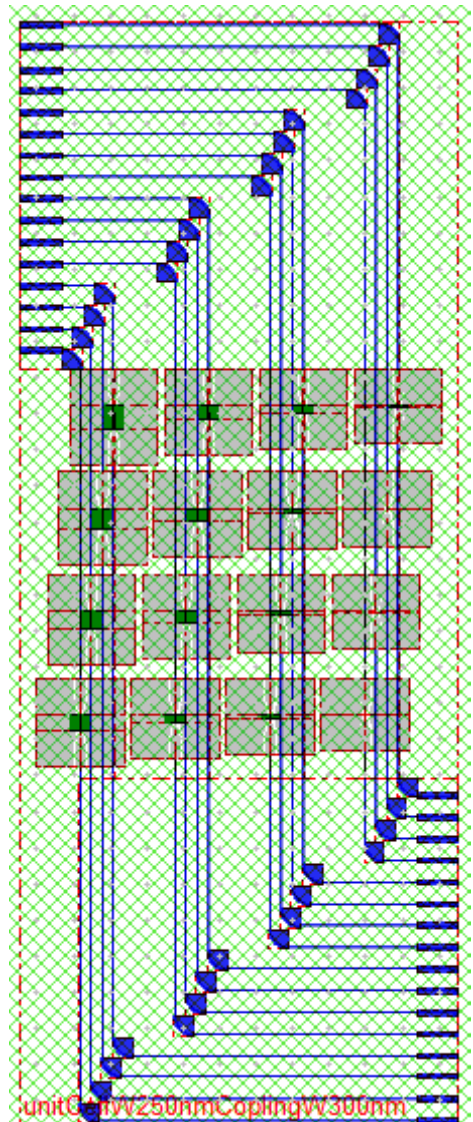

**Figure 152: Detail of a set of 16 Ge plasmonic modulators**

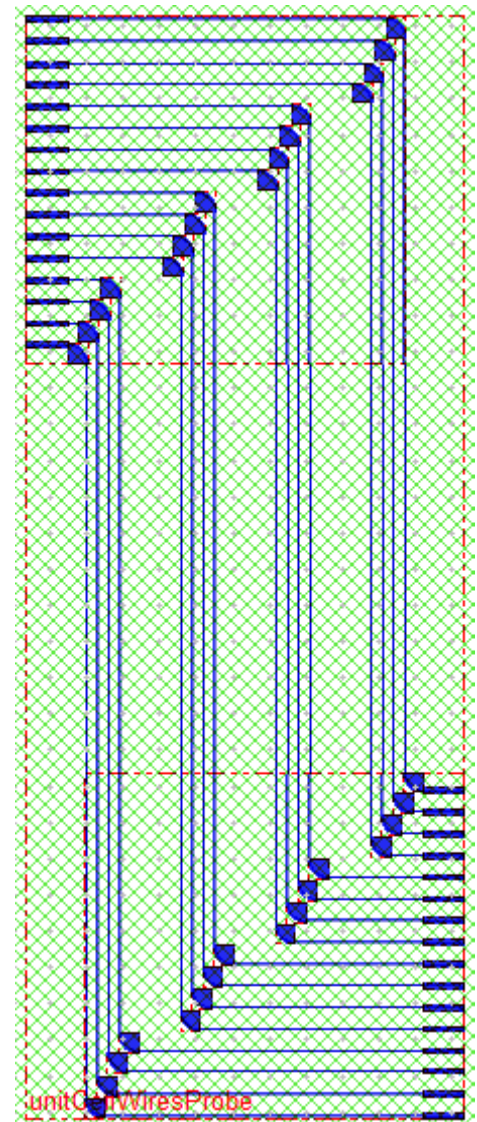

**Figure 153: Empty waveguides to meassure the losses of the input/output Si rib waveguides**

In Figure 152, we can observe one set of 16 modulators. The core of the Ge plasmonic modulators is in green and the AlCu electrodes are in grey. In blue, we can see the input and output silicon waveguides with the fiber grating couplers at both outputs.

We implemented 384 plasmonic modulators in total. Although in chapter 5, we optimized the plasmonic modulator with respect to the parameters w, h, $h_{Slot}$ and L, in the layout we scanned both the width w and the length L of the Ge core to verify experimentally the optimal configuration. The dimensions of the optimized modulators is around L=30 µm and w=150 nm.

We scan w and L around the optimimum, knowing that the smallest value allowed for 193 nm lithography is 100 nm :

- The width w has the following values w ∈ {100,150,200,250,300,350,400,450,500} in nanometers.
- The length L has the following values L ∈ {5,7,10,15,20,25,30,40,50,60,70,80,90,100,115,130} in micrometers.

Futhermore, we did a finer scan around the optimal dimensions w=150 nm and L=30 μm. For this, we did the following sets of parameters

- $w_1 \in \{200,250,300,350,400,450,500\}$ in nanometers and
- $L_1 \in \{30,31,32,33,34,35,36,37,38,39,40,41,42,43,44,45\}$ in micrometers

and a third set of parameters :

- $w_2 \in \{200,250,300,350,400,450,500\}$ in nanometers
- $L_2 \in \{46,47,48,49,50,51,52,53,54,55,56,57,58,59,60,61\}$ in micrometers.

To extract the performances of the Ge plasmonic modulator waveguides alone, we also added a set of reference structures without modulators, arranged in the same configuration, as shown in Figure 153. These reference structures consist of an input fiber grating coupler, a Si rib waveguide and an output fiber grating coupler.

The layout of the fiber grating couplers is shown in Figure 154. Since the plasmonic Ge modulator supports a TM mode, a fiber grating coupler was design specifically for this maskset to couple light to the TM photonic mode of the silicon input and output rib waveguides, at the wavelength of 1647 nm, optimal for the Franz-Keldysh effect.

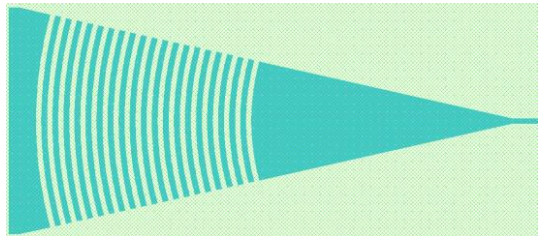

**Figure 154: Layout of the fiber grating coupler designed to excite the fundamental TM mode at 1.647 μm of the Si rib waveguide**

Test structures of the propagation losses of the TM photonic mode of the input and output silicon waveguides were added. Five spirals like the one represented in Figure 155 of length $L_{spiral}$=10000, 25000, 50000, 75000 and 100000 μm were designed. We designed them to have around less than 30 dB of total losses from the input grating to the output grating. The expected propagation loss of the Si rib waveguide is around 0.5-5 dB/cm.

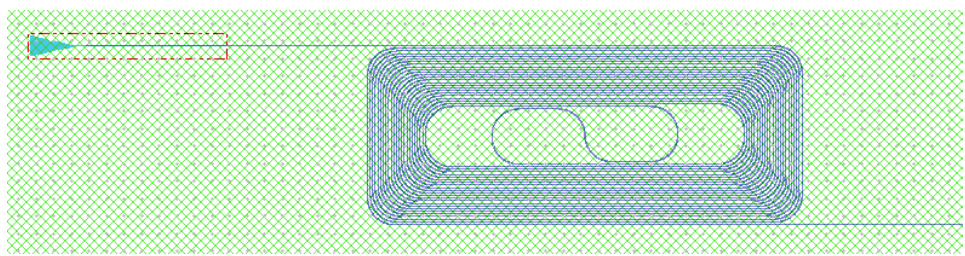

**Figure 155: Example of a test spiral to meaasure the propagation losses of the TM photonic mode of the Si input/output rib waveguide**

Finally, we designed electrical and RF test structures (not represented here):

- TLM (Transmission Line Method) are useful to measure the resistivity of the doped layers like p-doped Si and p-doped Ge in our case (and therefore check the doping level) and to check the contact resistance.
- Several electrode structures (in a short and open configuration) are also useful to perform the deembedding of the future RF measurements.

# Bibliography

[1]　G. E. Moore, "Cramming more components on to integrated circuits." Electronics, pp. 114–117, 1965.

[2]　I. Ross, "The invention of the transistor," *Proc. IEEE*, 1998.

[3]　W. Brinkman, "A history of the invention of the transistor and where it will lead us," *Solid-State Circuits, ...*, 1997.

[4]　D. Miller, "Physical reasons for optical interconnection," *Int. J. ...*, 1997.

[5]　W. Paper, "4004 single chip 4 bit p-channel microprocessor," no. 01, pp. 1–9.

[6]　W. Paper, "2nd generation Intel core vPro processor family."

[7]　H. B. Bakoglu, *Circuits, interconnections, and packaging for VLSI*. Addison-Wesley, 1990.

[8]　G. E. Moore, "No exponential is forever : but ' forever ' can be delayed !," 2003.

[9]　A. Jain, S. Rogojevic, S. Ponoth, N. Agarwal, I. Matthew, W. N. Gill, and P. Persans, "Porous silica materials as low- k dielectrics for electronic and optical interconnects," vol. 399, pp. 513–522, 2001.

[10]　A. a. Istratov and E. R. Weber, "Physics of copper in silicon," *J. Electrochem. Soc.*, vol. 149, no. 1, p. G21, 2002.

[11]　L. Pavesi and D. Lockwood, *Silicon photonics*. 2004.

[12]　R. Soref, "Silicon photonics technology: past, present and future," *Proc. SPIE*, 2005.

[13]　"Simply silicon." Nature Photonics, p. 491, 2010.

[14]　G. T. Reed, "The optical age of silicon," *Nature*, vol. 427, pp. 595–596, 2004.

[15]　B. Jalali and S. Fathpour, "Silicon photonics," vol. 24, no. 12, pp. 4600–4615, 2006.

[16]　M. Lipson, "Guiding, modulating, and emitting light on Silicon-challenges and opportunities," *J. Light. Technol.*, vol. 23, no. 12, pp. 4222–4238, Dec. 2005.

[17]　C. Gunn, "CMOS Photonics for High Speed Interconnects," pp. 58–66, 2006.

[18]　Intel Labs, "The 50G silicon photonics Link," *White Pap.*, 2010.

[19]　M. H. and T. Baehr-Jones, "Towards fabless silicon," *Nat. Photonics*, vol. 4, pp. 492–494, 2010.

[20]　K. Wada, "Si CMOS Photonics," *5th Int. Symp. Adv. Sci. Technol. Silicon Mater.*

[21]　B. Jalali, S. Member, S. Yegnanarayanan, T. Yoon, T. Yoshimoto, I. Rendina, and F. Coppinger, "Advances in silicon-on-insulator optoelectronics," vol. 4, no. 6, pp. 938–947, 1998.

[22]　R. A. Soref and S. Member, "Silicon-based optoelectronics," vol. 81, no. 12, 1993.

[23]　B. G. Celler, C. Scientist, and M. Wolf, "Smart Cut," *SOITEC's White Pap.*, no. 2003, 2004.

[24]　S. O. Rib and M. Characteristics, "Silicon-on-insulator optical rib waveguide loss and mode characteristics," *J. Light. Technol.*, vol. 12, no. October, pp. 0–5, 1994.

[25]　R. A. Soref, J. Schmidtchen, and K. Petermann, "Large single-mode rib waveguides in GeSi-Si and Si-on-SiO2," *IEEE J. Quantum Electron.*, vol. 27, no. 8, pp. 1971–1974, 1991.

[26]　N. Novgorod, S. I. Ll, and C. O. N. N. Ns, "Low loss singlemode optical waveguides with large cross-section in silicon-on-insulator," *Electron. Lett.*, vol. 27, no. 1, pp. 4–6, 1991.

[27]　D. Liang and J. E. Bowers, "Recent progress in lasers on silicon," *Nat. Photonics*, vol. 4, no. 8, pp. 511–517, Jul. 2010.

[28]　G. Reed, G. Mashanovich, F. Gardes, and D. Thomson, "Silicon optical modulators," *Nat. Photonics*, 2010.

[29]　J. Michel, J. Liu, and L. C. Kimerling, "High-performance Ge-on-Si photodetectors," *Nat. Photonics*, vol. 4, no. 8, pp. 527–534, Jul. 2010.

[30]　A. Alduino, L. Liao, R. Jones, M. Morse, B. Kim, W.-Z. Lo, J. Basak, B. Koch, H.-F. Liu, H. Rong, M. Sysak, C. Krause, R. Saba, D. Lazar, L. Horwitz, R. Bar, S. Litski, A. Liu, K. Sullivan, O. Dosunmu, N. Na, T. Yin, F. Haubensack, I. Hsieh, J. Heck, R. Beatty, H. Park, J. Bovington, S. Lee, H. Nguyen, H. Au, K. Nguyen, P. Merani, M. Hakami, and M. Paniccia, "Demonstration of a high speed 4-channel integrated silicon photonics WDM link with hybrid silicon lasers," *Integr. Photonics Res. Silicon Nanophotonics Photonics Switch.*, p. PDIWI5, 2010.

[31]　F. Tooley, "Challenges in optically interconnecting electronics," *... Top. Quantum Electron. IEEE J.*, 1996.

[32]    H. Cho, P. Kapur, and K. Saraswat, "Power comparison between high-speed electrical and optical interconnects for interchip communication," *J. Light. Technol.*, 2004.

[33]    D. Miller, "Device requirements for optical interconnects to silicon chips," *Proc. IEEE*, vol. 97, no. 7, pp. 1166–1185, Jul. 2009.

[34]    "Integrated optics an introduction," *Bell Syst. Tech. J.*, vol. 48, no. 7, pp. 2059–2069, 1969.

[35]    J. Liu, D. Pan, S. Jongthammanurak, K. Wada, L. C. Kimerling, and J. Michel, "Design of monolithically integrated GeSi electro-absorption modulators and photodetectors on a SOI platform.," *Opt. Express*, vol. 15, no. 2, pp. 623–8, Jan. 2007.

[36]    S. A. Clarka, B. Cuishawa, E. J. C. Dawnayb, and I. E. Days, "Thermo-optic phase modulators in SIMOX material I," *Proc. SPIE*, vol. 3936, 2000.

[37]    S. Manipatruni, R. K. Dokania, B. Schmidt, N. Sherwood-droz, C. B. Poitras, A. B. Apsel, and M. Lipson, "Wide temperature range operation of micrometer- scale silicon electro-optic modulators," vol. 33, no. 19, pp. 2185–2187, 2008.

[38]    K. Hassan, L. Markey, A. Dereux, A. Pitilakis, O. Tsilipakos, and E. E. Kriezis, "Thermo-optic plasmo-photonic mode interference switches based on dielectric loaded waveguides," vol. 241110, no. 2011, pp. 10–13, 2012.

[39]    R. Goldstein, "Electro-optic devices in review," *Lasers Appl. April*, 1986.

[40]    P. Weinberger, "John Kerr and his effects found in 1877 and 1878," *Philos. Mag. Lett.*, vol. 88, no. 12, pp. 897–907, Dec. 2008.

[41]    R. S. Jacobsen, K. N. Andersen, P. I. Borel, J. Fage-Pedersen, L. H. Frandsen, O. Hansen, M. Kristensen, A. V Lavrinenko, G. Moulin, H. Ou, C. Peucheret, B. Zsigri, and A. Bjarklev, "Strained silicon as a new electro-optic material," *Nature*, vol. 441, no. 7090, pp. 199–202, May 2006.

[42]    B. Chmielak, M. Waldow, C. Matheisen, C. Ripperda, J. Bolten, T. Wahlbrink, M. Nagel, F. Merget, and H. Kurz, "Pockels effect based fully integrated, strained silicon electro-optic modulator.," *Opt. Express*, vol. 19, no. 18, pp. 17212–9, Aug. 2011.

[43]    R. A. Soref and B. R. Bennett, "Electrooptical effects in silicon," *IEE J. Quantum Electron.*, no. 1, pp. 123–129, 1987.

[44]    W. Franz, "Einflußeines elektrischen Feldes auf eine optische Absorptionskante," *Zeitschrift Naturforsch. Tl. A*, vol. 13, p. 484, 1958.

[45]    L. ~V. Keldysh, "Behavior of non-metallic crystals in strong electric fields," *Sov. J. Exp. Theor. Phys.*, vol. 6, p. 763, 1958.

[46]    S. Jongthammanurak, J. Liu, K. Wada, D. D. Cannon, D. T. Danielson, D. Pan, L. C. Kimerling, and J. Michel, "Large electro-optic effect in tensile strained Ge-on-Si films," *Appl. Phys. Lett.*, vol. 89, no. 16, p. 161115, 2006.

[47]    J. Liu, D. Cannon, K. Wada, Y. Ishikawa, D. Danielson, S. Jongthammanurak, J. Michel, and L. Kimerling, "Deformation potential constants of biaxially tensile stressed Ge epitaxial films on Si(100)," *Phys. Rev. B*, vol. 70, no. 15, p. 155309, Oct. 2004.

[48]    D. Miller, D. Chemla, and T. Damen, "Band-edge electroabsorption in quantum well structures: The quantum-confined Stark effect," *Phys. Rev. Lett.*, 1984.

[49]    Y. Kuo, S. Member, Y. K. Lee, Y. Ge, S. Ren, J. E. Roth, T. I. Kamins, D. A. B. Miller, and J. S. Harris, "Quantum-Confined Stark Effect in Ge / SiGe Quantum Wells on Si for Optical Modulators," *IEEE J. Sel. Top. Quantum Electron.*, vol. 12, no. 6, pp. 1503–1513, 2006.

[50]    C. K. and W. F. J. Leuthold, "Non linear silicon photonics," *Nat. Photonics*, pp. 535–544, 2010.

[51]    J. Liu, M. Beals, A. Pomerene, S. Bernardis, R. Sun, J. Cheng, L. C. Kimerling, and J. Michel, "Waveguide-integrated, ultralow-energy GeSi electro-absorption modulators," *Nat. Photonics*, vol. 2, no. 7, pp. 433–437, May 2008.

[52]    A. E.-J. Lim, T.-Y. Liow, F. Qing, N. Duan, L. Ding, M. Yu, G.-Q. Lo, and D.-L. Kwong, "Novel evanescent-coupled germanium electro-absorption modulator featuring monolithic integration with germanium p-i-n photodetector.," *Opt. Express*, vol. 19, no. 6, pp. 5040–6, Mar. 2011.

[53]    N.-N. Feng, D. Feng, S. Liao, X. Wang, P. Dong, H. Liang, C.-C. Kung, W. Qian, J. Fong, R. Shafiiha, Y. Luo, J. Cunningham, A. V Krishnamoorthy, and M. Asghari, "30 GHz Ge electro-absorption modulator integrated with 3μm silicon -on-insulator waveguide," *Opt. Express*, vol. 19, no. 8, pp. 7062–7, Apr. 2011.

[54]    D. Feng, S. Liao, H. Liang, J. Fong, B. Bijlani, R. Shafiiha, B. J. Luff, Y. Luo, J. Cunningham, A. V Krishnamoorthy, and M. Asghari, "High speed GeSi electro-absorption modulator at 1550 nm wavelength on SOI waveguide.," *Opt. Express*, vol. 20, no. 20, pp. 22224–32, Sep. 2012.

[55]    R. a. Soref, G. Sun, and H. H. Cheng, "Franz-Keldysh electro-absorption modulation in germanium-tin alloys," *J. Appl. Phys.*, vol. 111, no. 12, p. 123113, 2012.

[56]    J. a Dionne, K. Diest, L. a Sweatlock, and H. a Atwater, "PlasMOStor: a metal-oxide-Si field effect plasmonic modulator.," *Nano Lett.*, vol. 9, no. 2, pp. 897–902, Feb. 2009.

[57]    S. Zhu, G. Q. Lo, and D. L. Kwong, "Theoretical investigation of silicon MOS-type plasmonic slot waveguide based MZI modulators.," *Opt. Express*, vol. 18, no. 26, pp. 27802–19, Dec. 2010.

[58]    a. Melikyan, L. Alloatti, a. Muslija, D. Hillerkuss, P. C. Schindler, J. Li, R. Palmer, D. Korn, S. Muehlbrandt, D. Van Thourhout, B. Chen, R. Dinu, M. Sommer, C. Koos, M. Kohl, W. Freude, and J. Leuthold, "High-speed plasmonic phase modulators," *Nat. Photonics*, vol. 8, no. 3, pp. 229–233, Feb. 2014.

[59]    R. Thomas, Z. Ikonic, and R. W. Kelsall, "Plasmonic enhanced electro-optic stub modulator on a SOI platform," *Photonics Nanostructures - Fundam. Appl.*, vol. 9, no. 1, pp. 101–107, Feb. 2011.

[60]    A. V Krasavin, T. P. Vo, W. Dickson, M. Bolger, and A. V Zayats, "All-plasmonic modulation via stimulated emission of copropagating," *Nano Lett.*, pp. 2231–2235, 2011.

[61]   M. P. Nezhad, K. Tetz, and Y. Fainman, "Gain assisted propagation of surface plasmon polaritons on planar metallic waveguides," *Opt. Express*, vol. 12, no. 17, pp. 4072–4079, 2004.

[62]   a Marini, a V Gorbach, D. V Skryabin, and a V Zayats, "Amplification of surface plasmon polaritons in the presence of nonlinearity and spectral signatures of threshold crossover.," *Opt. Lett.*, vol. 34, no. 18, pp. 2864–6, Sep. 2009.

[63]   a Melikyan, N. Lindenmann, S. Walheim, P. M. Leufke, S. Ulrich, J. Ye, P. Vincze, H. Hahn, T. Schimmel, C. Koos, W. Freude, and J. Leuthold, "Surface plasmon polariton absorption modulator.," *Opt. Express*, vol. 19, no. 9, pp. 8855–69, Apr. 2011.

[64]   E. Feigenbaum, K. Diest, and H. a Atwater, "Unity-order index change in transparent conducting oxides at visible frequencies.," *Nano Lett.*, vol. 10, no. 6, pp. 2111–6, Jun. 2010.

[65]   S. Zhu, G. Q. Lo, and D. L. Kwong, "Electro-absorption modulation in horizontal metal-insulator-silicon-insulator-metal nanoplasmonic slot waveguides," *Appl. Phys. Lett.*, vol. 99, no. 15, p. 151114, 2011.

[66]   V. J. Sorger, N. D. Lanzillotti-Kimura, R.-M. Ma, and X. Zhang, "Ultra-compact silicon nanophotonic modulator with broadband response," *Nanophotonics*, vol. 1, no. 1, pp. 17–22, Jan. 2012.

[67]   B. a Kruger, A. Joushaghani, and J. K. S. Poon, "Design of electrically driven hybrid vanadium dioxide (VO2) plasmonic switches.," *Opt. Express*, vol. 20, no. 21, pp. 23598–609, Oct. 2012.

[68]   V. E. Babicheva, N. Kinsey, G. V Naik, M. Ferrera, A. V Lavrinenko, V. M. Shalaev, and A. Boltasseva, "Towards CMOS-compatible nanophotonics : Ultra-compact modulators using alternative plasmonic materials," *Opt. Express*, vol. 21, no. 22, pp. 5833–5835, 2013.

[69]   A. Boltasseva and H. a Atwater, "Low-loss plasmonic metamaterials," *Science*, vol. 331, no. 6015, pp. 290–1, Jan. 2011.

[70]   P. R. West, S. Ishii, G. Naik, N. Emani, V. M. Shalaev, and A. Boltasseva, "Searching for Better Plasmonic Materials," *Laser Photon. Rev.*, pp. 1–28.

[71]   R. Thomas, Z. Ikonic, and R. W. Kelsall, "Electro-optic metal–insulator–semiconductor–insulator–metal Mach-Zehnder plasmonic modulator," *Photonics Nanostructures - Fundam. Appl.*, vol. 10, no. 1, pp. 183–189, Jan. 2012.

[72]   A. Liu, R. Jones, L. Liao, and D. Samara-rubio, "A high-speed silicon optical modulator based on a metal – oxide – semiconductor capacitor," *Nature*, vol. 427, no. February, pp. 615–618, 2004.

[73]   P. Dong, S. Liao, D. Feng, H. Liang, and D. Zheng, "Low Vpp, ultralow-energy, compact, high-speed silicon electro-optic modulator," *Opt. Express*, vol. 17, no. 25, pp. 22484–22490, 2009.

[74]   L. Liao, D. Samara-Rubio, M. Morse, A. Liu, D. Hodge, D. Rubin, U. Keil, and T. Franck, "High speed silicon Mach-Zehnder modulator.," *Opt. Express*, vol. 13, no. 8, pp. 3129–35, Apr. 2005.

[75]  W. M. Green, M. J. Rooks, L. Sekaric, and Y. a Vlasov, "Ultra-compact, low RF power, 10 Gb/s silicon Mach-Zehnder modulator.," *Opt. Express*, vol. 15, no. 25, pp. 17106–13, Dec. 2007.

[76]  S. Manipatruni, Q. Xu, B. Schmidt, J. Shakya, and M. Lipson, "High speed carrier injection 18 Gb/s silicon micro-ring electro-optic modulator," *LEOS 2007 - IEEE Lasers Electro-Optics Soc. Annu. Meet. Conf. Proc.*, pp. 537–538, Oct. 2007.

[77]  S. Manipatruni, Q. Xu, B. Schmidt, J. Shakya, and M. Lipson, "12.5 Gbit/s carrier-injection-based silicon microring silicon modulators," *LEOS 2007 - IEEE Lasers Electro-Optics Soc. Annu. Meet. Conf. Proc.*, vol. 15, no. 2, pp. 430–436, Oct. 2007.

[78]  M. R. Watts, D. C. Trotter, R. W. Young, and A. L. Lentine, "Ultralow power silicon microdisk modulators and switches," *5th IEEE Int. Conf. Gr. IV Photonics*, vol. 2, no. ll, pp. 4–6.

[79]  L. Chen, K. Preston, S. Manipatruni, and M. Lipson, "Integrated GHz silicon photonic interconnect with micrometer-scale modulators and detectors.," *Opt. Express*, vol. 17, no. 17, pp. 15248–56, Aug. 2009.

[80]  M. Ziebell, D. Marris-Morini, G. Rasigade, J.-M. Fédéli, P. Crozat, E. Cassan, D. Bouville, and L. Vivien, "40 Gbit/s low-loss silicon optical modulator based on a pipin diode.," *Opt. Express*, vol. 20, no. 10, pp. 10591–6, May 2012.

[81]  K. Ogawa, K. Goi, H. Kusaka, K. Oda, T.-Y. Liow, X. Tu, G.-Q. Lo, and D.-L. Kwong, "20-Gbps silicon photonic waveguide nested Mach-Zehnder QPSK modulator," *Natl. Fiber Opt. Eng. Conf.*, p. JTh2A.20, 2012.

[82]  D. Miller, "Optics for low-energy communication inside digital processors: quantum detectors, sources, and modulators as efficient impedance converters," *Opt. Lett.*, 1989.

[83]  K.-H. Koo, P. Kapur, and K. C. Saraswat, "Compact performance models and comparisons for gigascale on-chip global interconnect technologies," *IEEE Trans. Electron Devices*, vol. 56, no. 9, pp. 1787–1798, Sep. 2009.

[84]  E. Dulkeith, F. Xia, L. Schares, W. M. Green, L. Sekaric, and Y. a Vlasov, "Group index and group velocity dispersion in silicon-on-insulator photonic wires: errata.," *Opt. Express*, vol. 14, no. 13, p. 6372, Jun. 2006.

[85]  C. A. Barrios and M. Lipson, "Electrically driven silicon resonant light emitting device based on slot-waveguide.," *Opt. Express*, vol. 13, no. 25, pp. 10092–101, Dec. 2005.

[86]  P. Müllner and R. Hainberger, "Structural optimization of silicon-on-insulator slot waveguides," vol. 18, no. 24, pp. 2557–2559, 2006.

[87]  N. Feng, M. L. Brongersma, and L. D. Negro, "Metal – dielectric slot-waveguide structures for the propagation of surface plasmon polaritons at 1 . 55 m," *IEE J. Quantum Electron.*, vol. 43, no. 6, pp. 479–485, 2007.

[88]  D. K. Gramotnev and S. I. Bozhevolnyi, "Plasmonics beyond the diffraction limit," *Nat. Photonics*, vol. 4, no. 2, pp. 83–91, Jan. 2010.

[89]  W. L. Barnes, A. Dereux, and T. W. Ebbesen, "Surface plasmoncs subwavelength optics," *Nature*, vol. 424, no. August, pp. 824–830, 2003.

[90]  R. Zia, J. A. Schuller, A. Chandran, and M. L. Brongersma, "Plasmonics : the next chip-scale technology," *Mater. Today*, vol. 9, no. 7, pp. 20–27, 2006.

[91]  E. N. Photonics, "Surface Plasmon Resurrection," *Nat. Photonics*, vol. 6, 2012.

[92]  R. F. Oulton, V. J. Sorger, T. Zentgraf, R.-M. Ma, C. Gladden, L. Dai, G. Bartal, and X. Zhang, "Plasmon lasers at deep subwavelength scale.," *Nature*, vol. 461, no. 7264, pp. 629–32, Oct. 2009.

[93]  P. Berini and I. De Leon, "Surface plasmon–polariton amplifiers and lasers," *Nat. Photonics*, vol. 6, no. 1, pp. 16–24, Dec. 2011.

[94]  A. G. Brolo, "Plasmonics for future biosensors," *Nat. Photonics*, vol. 6, no. 11, pp. 709–713, Nov. 2012.

[95]  R. Yang and Z. Lu, "Subwavelength plasmonic waveguides and plasmonic materials," *Int. J. Opt.*, vol. 2012, pp. 1–12, 2012.

[96]  M. Kauranen and A. V. Zayats, "Nonlinear plasmonics," *Nat. Photonics*, vol. 6, no. 11, pp. 737–748, Nov. 2012.

[97]  V. Temnov, "Ultrafast acousto-magneto-plasmonics," *Nat. Photonics*, 2012.

[98]  a. N. Grigorenko, M. Polini, and K. S. Novoselov, "Graphene plasmonics," *Nat. Photonics*, vol. 6, no. 11, pp. 749–758, Nov. 2012.

[99]  and D. L. M. V. M. Agranovich, *Surface polaritons: electromagnetic waves at surfaces and interfaces*. Elsevier B.V., 1982.

[100]  R. Gordon, A. I. K. Choudhury, and T. Lu, "Gap plasmon mode of eccentric coaxial metal waveguide.," *Opt. Express*, vol. 17, no. 7, pp. 5311–20, Mar. 2009.

[101]  D. Sarid, "Long-range surface-plasma waves on very thin metal films," *Phys. Rev. Lett.*, vol. 47, no. 26, pp. 1927–1930, Dec. 1981.

[102]  S. Bozhevolnyi, *Plasmonic Nanoguides and Circuits*. 2008.

[103]  D. K. Gramotnev and D. F. P. Pile, "Single-mode subwavelength waveguide with channel plasmon-polaritons in triangular grooves on a metal surface," *Appl. Phys. Lett.*, vol. 85, no. 26, p. 6323, 2004.

[104]  S. Bozhevolnyi, *Plasmonic Nanoguides and Circuits*. 2008.

[105]  J. Dionne, L. Sweatlock, H. Atwater, and a. Polman, "Plasmon slot waveguides: Towards chip-scale propagation with subwavelength-scale localization," *Phys. Rev. B*, vol. 73, no. 3, p. 035407, Jan. 2006.

[106]  K. Tanaka and M. Tanaka, "Simulations of nanometric optical circuits based on surface plasmon polariton gap waveguide," *Appl. Phys. Lett.*, vol. 82, no. 8, p. 1158, 2003.

[107]  R. F. Oulton, V. J. Sorger, D. a. Genov, D. F. P. Pile, and X. Zhang, "A hybrid plasmonic waveguide for subwavelength confinement and long-range propagation," *Nat. Photonics*, vol. 2, no. 8, pp. 496–500, Jul. 2008.

[108]  S.-Y. S. and W.-G. L. Min-Suk Kwon, Jin-Soo Shin, "Characterization of realized metal insulator silicon insulator metal waveguides and nanochannel fabrication via insulator removal," *Opt. Express*, pp. 21875–21887, 2012.

[109]  G. Zhou, T. Wang, P. Cao, and H. Xie, "Design of plasmon waveguide with strong field confinement and low loss for nonlinearity enhancement," *Gr. IV Photonics ( …*, 2010.

[110]  M. Alam, S. Aitchison, and M. Mojahedi, "Hybrid plasmonic waveguide devices for silicon on insulator platform," *Integr. Photonics Res. Silicon Nanophotonics*, 2011.

[111]  D. D. and S. He, "Low loss hybrid plasmonic waveguide with double low index nano slots," *Opt. Express*, vol. 18, no. 17, pp. 17958–17966, 2010.

[112]  M. Kwon, "Metal-insulator-silicon-insulator-metal waveguides compatible with standard CMOS technology," *Opt. Express*, 2011.

[113]  S. Zhu, T. Liow, G. Lo, and D. Kwong, "Silicon-based horizontal nanoplasmonic slot waveguides for on-chip integration," *Opt. Express*, 2011.

[114]  E. Ozbay, "Plasmonics: merging photonics and electronics at nanoscale dimensions.," *Science*, vol. 311, no. 5758, pp. 189–93, Jan. 2006.

[115]  O. Integration, S. Zhu, G. Lo, and D. Kwong, "CMOS-compatible silicon nanoplasmonics for on chip integration," *Opt. Express*, pp. 486–493, 2012.

[116]  M. C. Gather, K. Meerholz, N. Danz, and K. Leosson, "Net optical gain in a plasmonic waveguide embedded in a fluorescent polymer," *Nat. Photonics*, vol. 4, no. 7, pp. 457–461, May 2010.

[117]  D. Dai, Y. Shi, S. He, L. Wosinski, and L. Thylen, "Gain enhancement in a hybrid plasmonic nano-waveguide with a low-index or high-index gain medium.," *Opt. Express*, vol. 19, no. 14, pp. 12925–36, Jul. 2011.

[118]  I. Avrutsky, R. Soref, and W. Buchwald, "Sub-wavelength plasmonic modes in a conductor-gap-dielectric system with a nanoscale gap.," *Opt. Express*, vol. 18, no. 1, pp. 348–63, Jan. 2010.

[119]  E. D. Palik, *Handbook of optical constants*. 1985.

[120]  H. S. Lee, C. Awada, S. Boutami, F. Charra, L. Douillard, and R. E. De Lamaestre, "Loss mechanisms of surface plasmon polaritons propagating on a smooth polycrystalline Cu surface," *Opt. Express*, vol. 20, no. 8, pp. 178–182, 2012.

[121]  D. Dai, S. He, and Y. Chowdhury, "A silicon-based hybrid plasmonic waveguide with a metal cap for a nano-scale light confinement.," *Opt. Express*, vol. 17, no. 19, pp. 16646–53, Sep. 2009.

[122] L. Chen, J. Shakya, and M. Lipson, "Subwavelength confinement in an integrated metal slot waveguide on silicon.," *Opt. Lett.*, vol. 31, no. 14, pp. 2133–5, Jul. 2006.

[123] A. Emboras, "CMOS integration of plasmon field effect devices," Institute National Polytechnique De Grenoble, 2012.

[124] and K. P. R. R. Rosenberg, D. C. Edelstein, C.-K. Hu, "Copper metallization for high performance silicon technology," *Annu. Rev. Mater. Sci.*, vol. 30, pp. 229–262, 2000.

[125] V. S. C. Len, R. E. Hurley, N. Mccusker, D. W. Mcneill, B. M. Armstrong, and H. S. Gamble, "An investigation into the performance of diffusion barrier materials against copper di € usion using metal-oxide- semiconductor ( MOS ) capacitor structures," *Solid. State. Electron.*, vol. 43, pp. 1045–1049, 1999.

[126] K. Holloway, P. M. Fryer, C. Cabral, J. M. E. Harper, P. J. Bailey, and K. H. Kelleher, "Tantalum as a diffusion barrier between copper and silicon: Failure mechanism and effect of nitrogen additions," *J. Appl. Phys.*, vol. 71, no. 11, p. 5433, 1992.

[127] J. Y. Park, J. Y. Kim, Y. Y. Do Kim, and H. Jeon, "Comparison of TiN and TiN / Ti / TiN Multilayer Films for Diffusion Barrier Applications," *J. Korean Phys. Soc.*, vol. 42, no. 6, pp. 817–820, 2003.

[128] T. C. Wang, Y. L. Cheng, Y. L. Wang, T. E. Hsieh, G. J. Hwang, and C. F. Chen, "Comparison of characteristics and integration of copper diffusion-barrier dielectrics," *Thin Solid Films*, vol. 498, no. 1–2, pp. 36–42, Mar. 2006.

[129] H. Shen and F. Pollak, "Generalized Franz-Keldysh theory of electromodulation.," *Phys. Rev. B. Condens. Matter*, vol. 42, no. 11, pp. 7097–7102, Oct. 1990.

[130] O. Madelung and Landot Bornstein, "Properties of group IV elements and III-V, II-V, and I-VII compounds," *Numer. Data Funct. Relationships Sci. Technol.*, vol. 17a, 1982.

[131] A. Frova and P. Handler, "Franz-Keldysh effect in the space-charge region of a germanium p-n junction," *Phys. Rev. Lett.*, vol. 788, no. 1958, 1961.

[132] a. Frova, P. Handler, F. Germano, and D. Aspnes, "Electro-absorption effects at the band edges of silicon and germanium," *Phys. Rev.*, vol. 145, no. 2, pp. 575–583, May 1966.

[133] B. Seraphin and R. Hess, "Franz-Keldysh effect above the fundamental edge in germanium," *Phys. Rev. Lett.*, vol. 14, no. 5, pp. 1–3, 1965.

[134] K. G. Ashar and R. L. Anderson, "Study of electroabsorption using differential photocurrent response," *Phys. Rev. Lett.*, vol. 154, no. 3, 1967.

[135] K.-S. Lee and J.-H. Jung, "Design of plasmonic slot waveguide with high localization and long propagation length," *J. Opt. Soc. Korea*, vol. 15, no. 3, pp. 305–309, Sep. 2011.

[136] G. Veronis and S. Fan, "Guided subwavelength plasmonic mode supported by a slot in a thin metal film.," *Opt. Lett.*, vol. 30, no. 24, pp. 3359–61, Dec. 2005.

[137]  X.-Y. Zhang, A. Hu, J. Z. Wen, T. Zhang, X.-J. Xue, Y. Zhou, and W. W. Duley, "Numerical analysis of deep sub-wavelength integrated plasmonic devices based on semiconductor-insulator-metal strip waveguides.," *Opt. Express*, vol. 18, no. 18, pp. 18945–59, Aug. 2010.

[138]  H.-S. Chu, E.-P. Li, P. Bai, and R. Hegde, "Optical performance of single-mode hybrid dielectric-loaded plasmonic waveguide-based components," *Appl. Phys. Lett.*, vol. 96, no. 22, p. 221103, 2010.

[139]  A. V Krasavin and A. V Zayats, "Silicon-based plasmonic waveguides.," *Opt. Express*, vol. 18, no. 11, pp. 11791–9, May 2010.

[140]  N. Berkovitch, M. Orenstein, and S. Lipson, "Why asymmetrical nanoscale plasmonic waveguides are guiding plasmons ?," *Lasers and Electro-Optics*, no. c, pp. 0–1, 2008.

[141]  L. Virot, L. Vivien, J. M. Hartmann, Y. Bogumilowicz, J. M. Fédéli, E. Cassan, C. Baudot, and F. Boeuf, "High Speed Waveguide Integrated Lateral P-I-N Ge on Si Photodiode with very Low Dark Current," *Opt. Express*, pp. 2–3, 2014.

[142]  A. B. Fallahkhair, S. Member, K. S. Li, T. E. Murphy, and S. Member, "Vector Finite Difference Modesolver for Anisotropic Dielectric Waveguides," *J. Light. Technol.*, vol. 26, no. 11, pp. 1423–1431, 2008.

[143]  A. P. Vasudev, J. Kang, J. Park, X. Liu, and M. L. Brongersma, "Electro-optical modulation of a silicon waveguide with an ' epsilon-near-zero ' material," *Opt. Express*, vol. 21, no. 22, pp. 123–129, 2013.

[144]  Y. Song, J. Wang, Q. Li, M. Yan, and M. Qiu, "Broadband coupler between silicon waveguide and hybrid plasmonic waveguide," vol. 18, no. 12, pp. 13173–13179, 2010.

[145]  P. Cheben, P. J. Bock, J. H. Schmid, J. Lapointe, S. Janz, D.-X. Xu, A. Densmore, A. Delâge, B. Lamontagne, and T. J. Hall, "Refractive index engineering with subwavelength gratings for efficient microphotonic couplers and planar waveguide multiplexers.," *Opt. Lett.*, vol. 35, no. 15, pp. 2526–8, Aug. 2010.